\documentclass[12pt,a4paper,twoside]{book}

\usepackage{graphicx}
\usepackage{geometry}
\usepackage[T1]{fontenc}
\usepackage[utf8x]{inputenc}
\usepackage[labelsep=period,font=footnotesize,labelfont=bf]{caption}
\usepackage{longtable}
\usepackage{fancyhdr}
\usepackage{url}
\usepackage[ngerman,english]{babel}
\renewcommand{\chaptermark}[1]{\markboth{Chapter \thechapter.}{}}	

\usepackage{sidecap}
\usepackage[clearempty]{titlesec}
\usepackage{natbib}
\usepackage{amsmath}

\usepackage{amssymb}
\usepackage{epstopdf}
\DeclareGraphicsExtensions{.pdf,.eps,.png,.jpg,.mps}

\usepackage{booktabs}
\usepackage{color}
\usepackage{rotating}
\usepackage{colortbl}
\usepackage{amssymb}
\usepackage[table]{xcolor}
\definecolor{gray}{rgb}{0.9,0.9,0.9}
\usepackage{subfig}
\usepackage{multirow}
\usepackage{expdlist}

\def\apjl{ApJ}%
\def\apjs{ApJS}%
\def\ao{Appl.~Opt.}%
\def\aap{A\&A}%
\DeclareMathAlphabet{\mathpzc}{OT1}{pzc}{m}{it}

\newcommand{\lOWLS}{\texttt{lOWLS~}}	
\newcommand{\hOWLS}{\texttt{hOWLS~}}	

\newcommand{\sqc}{cm$^{-2}$}                   
\newcommand{\cc}{cm$^{-3}$}                    
\newcommand{\kms}{km\,s$^{-1}$~}               

\newcommand{\Civ}{\mbox{C\,{\sc iv}~}}
\newcommand{\Cii}{\mbox{C\,{\sc ii}~}}
\newcommand{\Ciii}{\mbox{C\,{\sc iii}~}}

\newcommand{\Hi}{\mbox{H\,{\sc i}~}}
\newcommand{\Hii}{\mbox{H\,{\sc ii}~}}

\newcommand{\Lya}{\mbox{Ly-{$\alpha$}~}}
 \newcommand{\Lyb}{\mbox{Ly-{$\beta$}~}}

\newcommand{\Nv}{\mbox{N\,{\sc v}~}}
\newcommand{\Neviii}{\mbox{Ne\,{\sc viii}~}}
\newcommand{\Neix}{\mbox{Ne\,{\sc ix}~}}

\newcommand{\Ovi}{\mbox{O\,{\sc vi}~}}
\newcommand{\Ovii}{\mbox{O\,{\sc vii}~}}
\newcommand{\Oviii}{\mbox{O\,{\sc viii}~}}

\newcommand{\Siii}{\mbox{Si\,{\sc ii}~}}
\newcommand{\Siiii}{\mbox{Si\,{\sc iii}~}}
\newcommand{\Siiv}{\mbox{Si\,{\sc iv}~}}

\newcommand{\Mgii}{\mbox{Mg\,{\sc ii}~}}

\begin{document}
\title{
\foreignlanguage{ngerman}{
\centerline{\large{Institut f\"ur Physik und Astronomie}}
\centerline{\large{Astrophysik II}}
\centerline{\rule{\textwidth}{1pt}}\vspace{1.0cm} 
\centerline{\LARGE{\bf On the diversity of \Ovi absorbers}}
\centerline{\LARGE{\bf at high redshift}}\vspace{1.5cm} 
\centerline{\large{\bf Dissertation}}
\centerline{\large{\bf zur Erlangung des akademischen Grades}} 
\centerline{\large{\bf ``doctor rerum naturalium ''}}
\centerline{\large{(\bf Dr. rer. nat.)}} 
\centerline{\large{\bf in der Wissenschaftsdisziplin ``Astrophysik''}}\vspace{1.2cm}
\centerline{\large{\includegraphics[width=0.2\textwidth]{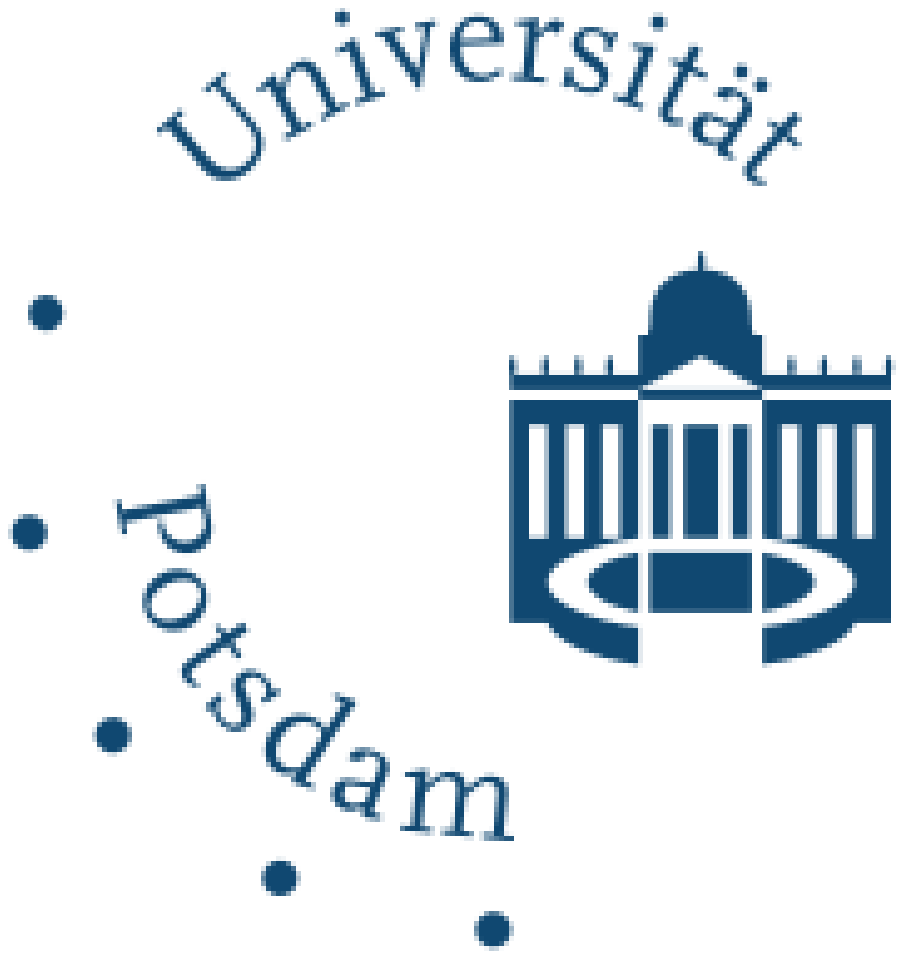}}}\vspace{1.2cm}
\centerline{\large{\bf eingereicht an der}} 
\centerline{\large{\bf Mathematisch-Naturwissenschaftlichen Fakult\"at}} 
\centerline{\large{\bf der Universit\"at Potsdam}}\vspace{1.0cm}
\centerline{\large{\bf von}}
\centerline{\Large{\bf Nadya Draganova}}\vspace{0.8cm}
\centerline{\large{\bf Potsdam, den 03.07.2013}}
}}
\date{}
\maketitle

\pagenumbering{roman}
{\Huge \bf Abstract}\vspace*{12pt}\\

\markboth{}{}

\hspace{0.6cm}The interstellar and intergalactic medium (ISM and IGM, respectively) can be observed in absorption in the spectra of bright background sources (e.g. stars and very distant quasars). Spectral observations of five-times ionized oxygen (\Ovi\!\!) play an important role in studies aimed to deepen our understanding of the ISM and the IGM at low and high redshifts. Assuming a collisional ionization mechanism, the ionization potential of \Ovi\!\! (138~eV) corresponds to temperatures $T\sim10^{5}-10^{6}$~K. Therefore, \Ovi is a potential tracer of intergalactic gas at $T>10^{5}$~K, i.e., the warm-hot intergalactic medium (WHIM). Moreover, the doublet \Ovi $\lambda\lambda$ 1031 \AA, 1037 \AA ~is strong and is expected to be detected and identified easily.

In this thesis, we systematically analyze the properties of intergalactic \Ovi absorbing gas structures at high redshift using optical spectra with intermediate ($\sim 6.6$ \kms\!\! FWHM) and high ($\sim 4.0$ \kms FWHM) resolution, obtained with the Ultraviolet and Visual Echelle Spectrograph (UVES) at the  Very Large Telescope (VLT). We complement our analysis with synthetic spectra obtained from extensive cosmological simulations that are part of the `OverWhelmingly Large Simulations' (OWLS) project \citep{Schaye10}. Our primary goal is to understand the origin and the physical properties of high-redshift \Ovi\!\! absorbing gas.

In the first part of the analysis (Chapter 3), we investigate in detail two \Ovi systems at redshift $z\approx2$ using two UVES data sets, both with intermediate and high spectral resolution. Comparing \Ovi absorbers observed at high and intermediate resolution, we find that the velocity structure of the absorbers is resolved in both data sets and we explore in detail the ionization conditions. The results indicate that the structure of the highly-ionized intergalactic gas at high redshift is complex and far more diverse then previously thought.

In the second part of the analysis (Chapter 4), we study a large sample of high-redshift \Ovi absorbers ($2.2\le z \le3.7$) along 15 quasar sight lines using UVES archival data with intermediate resolution. About 30 per cent of the intervening \Ovi systems turn out to be single-component absorbers, while the rest exhibit a more complex velocity structure. Absorption systems with small velocity offsets between \Ovi and neutral hydrogen (\Hi\!\!) -- i.e., aligned systems -- are simple, isolated gas domains, while those with a significant velocity offset seem to be embedded in structures with more complex kinematics and large internal velocity dispersions.

In the third part of the analysis (Chapter 5), we investigate a large sample of synthetic \Ovi spectra from a OWLS cosmological simulation at $z=2.5$, taking advantage of the direct knowledge of physical parameters of the absorbers. We find that aligned \Ovi\!\!/\Hi and \Ovi\!\!/\Civ pairs -- where \Civ is three-times ionized carbon -- trace gas at different temperatures, which hints to a multi-phase nature of the gas and different origins for most of the absorbers. Photoionization modeling shows that only about 30 per cent of the \Ovi\!\!/\Civ pairs arise in a photoionized, single gas-phase. 

Our main conclusions are the following:

1) Both the observations and simulations imply that \Ovi absorbers at high redshift arise in structures spanning a broad range of scales and different physical conditions. When the \Ovi components are characterized by small Doppler parameters, the ionizing mechanism is most likely photoionization; otherwise, collisional ionization is the dominant mechanism.

2) The baryon- and metal-content of the \Ovi absorbers at $z\approx2$ is less than one per cent of the total mass-density of baryons and metals at that redshift. Therefore, \Ovi absorbers do not trace the bulk of baryons and metals at that epoch.  

3) The \Ovi gas density, metallicity and non-thermal broadening mechanisms are significantly different at high redshift with respect to low redshift. In particular, non-thermal broadening mechanisms appear less important at high redshift as compared to low redshift, where the turbulence in the absorption gas might be significant. This, together with the result that \Ovi arises in different environments, embedded in small- and large-scale structures, indicates that \Ovi does not trace characteristic regions in the circumgalactic and intergalactic medium, but rather traces a gas phase with a characteristic transition temperature ($T\sim10^{5}$K). 

4) The \Ovi absorbers at high redshift arise in gas with metallicities significantly higher than the surrounding environment, which suggests an inhomogeneous metal enrichment of the IGM.

\newpage
{\Huge \bf Zusammenfassung}\vspace*{12pt}\\

\begin{otherlanguage}{ngerman}Das interstellare and intergalaktische Medium (ISM bzw. IGM) ruft eine Vielzahl von Absorptions"-linien in Stern- bzw. Quasarspektren hervor. Spektrale Beobachtungen des f\"unffach ionisierten Sauerstoffs (O\,{\sc vi}) sind von fundamentaler Bedeutung f\"ur unser Verst\"andnis des ISM bzw. IGM bei niedrigen sowie hohen Rotverschiebungen. Unter der Annahme, dass das Gas durch Teilchenst\"osse ionisiert wird, entspricht das Ionisationspotenzial von O\,{\sc vi} (138 eV) im Stoss-Ionisationsgleichgewicht Temperaturen von $T\sim10^5-10^6\,\textnormal{K}$. Demzufolge ist O\,{\sc vi} ein potenzieller Indikator f\"ur intergalaktisches Gas mit $T>10^5\,\textnormal{K}$, das sogenannte `warm-hei{\ss}e' intergalaktische Medium (WHIM). Zudem ist das Liniendoublet O\,{\sc vi} $\lambda\lambda$ $1031\,\textnormal{\AA}$, $1037\,\textnormal{\AA}$ relativ stark und in Spektren in der Regel leicht zu identifizieren.

In dieser Arbeit untersuchen wir systematisch die Eigenschaften von Gasstrukturen, die intergalaktisches O\,{\sc vi} aufweisen und bei hohen Rotverschiebungen auftreten. Hierf\"ur verwenden wir optische Spektren mit mittlerer ($\sim6.6\,\textnormal{km\,s}^{-1}$ FWHM) und hoher ($\sim4.0\,\textnormal{km\,s}^{-1}$ FWHM) spektraler Aufl\"osung, beobachtet mit dem `Ultraviolet and Visual Echelle' Spektrographen (UVES) des `Very Large Telescope' (VLT). Wir erg\"anzen unsere Untersuchungen mit synthetischen Spektren aus kosmologischen Simulationen. Unser vorrangiges Ziel ist das Verst\"andnis des Ursprungs und der physikalischen Eigenschaften des O\,{\sc vi} enthaltenden Gases bei hohen Rotverschiebungen.

Im ersten Abschnitt der Arbeit (Kapitel 3) untersuchen wir im Detail zwei O\,{\sc vi} Systeme bei einer Rotverschiebung von $z\approx2$ unter Verwendung zweier UVES Datens\"atze, jeweils mit mittlerer und hoher spektraler Aufl\"osung. Der Vergleich der beiden O\,{\sc vi} Absorptionssysteme bei hoher und mittlerer Aufl\"osung zeigt, dass die Geschwindigkeitsstruktur der Systeme in beiden Datens\"atzen aufgel\"ost ist. Des Weiteren untersuchen wir im Detail die Ionisationsbedingungen im Gas. Die Ergebnisse aus dieser Untersuchung deutet an, dass die Struktur des hochionisierten intergalaktischen Gases bei hoher Rotverschiebung komplex und wesentlich facettenreicher ist als bisher angenommen.

Im zweiten Abschnitt der Arbeit (Kapitel 4) analysieren wir eine gr\"o{\ss}ere Anzahl an O\,{\sc vi} Absorbern bei hohen Rotverschiebungen ($2.2\le z\le3.7$) entlang 15 Sichtlinien zu Quasaren unter Verwendung von UVES Archivdaten mit mittlerer Aufl\"osung. Etwa 30 Prozent der die Sichtlinien durchlaufenden O\,{\sc vi} Systeme stellen sich als Ein-Komponenten-Systeme heraus, wohingegen die verbleibenden Systeme eine komplexere Geschwindigkeitsstruktur aufweisen. Absorptionssysteme mit kleinen Geschwindigkeitsdifferenzen zwischen den Absorptionslinien von O\,{\sc vi} und neutralem Wasserstoff (H\,{\sc i}) repr\"asentieren einfache, isolierte Gasdom\"anen, wohingegen solche mit gr\"osseren Geschwindigkeitsdifferenzen in Gas-Strukturen mit einer komplexeren Kinematik und einer hohen internen Geschwindigkeitsdispersion eingebettet sind.

Im dritten Abschnitt der Arbeit (Kapitel 5) untersuchen schliesslich wir eine gro{\ss}e Auswahl synthetischer Spektren aus einer kosmologischen Simulation bei $z=2.5$. Die Simulation ist Teil des OWLS Projektes  (Schaye et al. 2010). Zur Analyse nutzen wir die unmittelbare Kenntnis \"uber die physikalischen Parameter der absorbierenden Gas-Strukturen in der Simulation aus. Die Mehrheit der kinematisch zusammenh\"angenden O\,{\sc vi}/H\,{\sc i} und O\,{\sc vi}/C\,{\sc iv} Paare (C\,{\sc iv} ist dreifach ionisierter Kohlenstoff) weist auf Gas mit verschiedenen Temperaturbereichen hin und somit eine mehrphasige Gasstruktur hin. Photoionisationsmodelle zeigen weiterhin, dass nur ca. 30 Prozent der O\,{\sc vi}/C\,{\sc iv} Paare in einer einzelnen, r\"aumlich koh\"arenten photoionisierten Gasphase entstehen.

Die Ergebnisse der Arbeit fassen wir wie folgt zusammen:

1) Die Beobachtungen und die Simulationen lassen darauf schlie{\ss}en, dass O\,{\sc vi} Absorptionssysteme bei hoher Rotverschiebung in Strukturen entstehen, die sich \"uber einen weiten Bereich an Gr\"o{\ss}enskalen und verschiedenartigen physikalischen Zust\"anden erstrecken. Wenn die O\,{\sc vi} Komponenten kleine Doppler-Parameter aufweisen, ist der Ionisationsmechanismus h\"ochstwahrscheinlich Photoionisation; anderenfalls ist Sto{\ss}ionisation der dominierende Mechanismus.

2) Der Baryonen- und Metallanteil der O\,{\sc vi} Absorber bei $z\approx2$ betr\"agt weniger als 1 Prozent der gesamten Massendichte der Baryonen bzw. Metalle bei dieser Rotverschiebung. Deshalb gehen O\,{\sc vi} Absorptionssysteme nicht mit dem Gro{\ss}teil der Baryonen bzw. Metalle in dieser Epoche einher.

3) Die O\,{\sc vi} Gasdichte, Metallizit\"at und nicht-thermischen Verbreiterungsmechanismen der O\,{\sc vi} Linien bei hoher Rotverschiebung unterscheiden sich erheblich von denen bei niedriger Rotverschiebung. Insbesondere scheinen nicht-thermische Verbreiterungsmechanismen bei hohen Rotverschiebungen weniger bedeutsam im Vergleich zu niedrigen Rotverschiebungen zu sein. Bei niedrigen Rotverschiebungen kann die Turbulenz im absorbierenden Gas signifikant sein. Zusammen mit dem Ergebnis, dass O\,{\sc vi} in verschiedenartigen Umgebungen entsteht, eingebettet in  klein- und gro{\ss}skaligen Strukturen, bedeutet dies, dass O\,{\sc vi} nicht typischen r\"aumlichen Regionen im zirkumgalaktischen und intergalaktischen Medium zuzuordnen ist, sondern vielmehr einer durch das jeweilige Umfeld definierten physikalischen Gasphase mit einer charakteristischen \"Ubergangstemperatur ($T\sim10^5\,\textnormal{K}$).

4) Die O\,{\sc vi} Absorptionssysteme bei hoher Rotverschiebung entstehen in Gas mit wesentlich h\"oheren Metallizit\"aten, als sie das Umfeld aufweist, was auf eine inhomogene Metallanreicherung des IGM hindeutet.

\end{otherlanguage}

\newpage 
\chapter*{Abbreviations}
\label{abbrev}
\begin{table}[!h]
\begin{tabular*}{1.\textwidth}{p{2cm}l}
CDDF   & Column Density Distribution Function \\
CIE    & Collisional Ionization Equilibrium \\
COS    & Cosmic Origins Spectrograph\\
DLA     & Damped \Lya (systems)  \\
ESO     & European Southern Observatory  \\
FUSE    & Far Ultraviolet Spectroscopic Explorer\\
FWHM    & Full Width at Half Maximum \\
GHRS    & Goddard High Resolution Spectrograph\\
HIRES   & High Resolution Echelle Spectrometer \\
HST     & Hubble Space Telescope \\
IGM	& InterGalactic Medium \\
ISM     & InterStellar Medium \\
LBGs	& Lyman Break Galaxies \\ 
$\Lambda$CDM    & Lambda Cold Dark Matter (cosmological model) \\
LLS	& Lyman Limit Systems \\
LTE     & Local Thermodynamic Equilibrium \\
OWLS    & OverWhelmingly Large Simulations \\
QSO     & Quasi Stellar Object \\
SBBN	& Standard Big-Bang nucleosynthesis (theory) \\
SFH	& Star Formation History \\
STIS    & Space Telescope Imaging Spectrograph \\
UVES    & Ultraviolet and Visual Echelle Spectrograph \\
VLT     & Very Large Telescope \vspace*{6pt}\\
WHIM	& Warm Hot Intergalactic Medium \\
WMAP    & Wilkinson Microwave Anisotropy Probe 
\end{tabular*}
\end{table}

\tableofcontents
\markboth{CONTENTS}{}
\listoffigures
\markboth{LIST OF FIGURES}{}
\addcontentsline{toc}{chapter}{List of Figures}
\listoftables
\markboth{LIST OF TABLES}{}
\addcontentsline{toc}{chapter}{List of Tables}


\chapter{Introduction}\label{introduction} 
\setcounter{page}{1}

\pagestyle{fancy}
\setlength{\headheight}{15pt}
\fancyhead{}
\fancyfoot{}
\fancyhead[LE,RO]{\bf{\thepage}}
\fancyhead[LO]{\nouppercase{\rightmark}}
\fancyhead[RE]{\nouppercase{\leftmark}}
\fancyheadoffset{0pt}
\pagenumbering{arabic}

\section{The Universe at high and low redshifts}\label{universe} 

\subsection*{Brief overview of the cosmic history}\label{history}

\hspace{0.6cm}According to the widely accepted Big Bang theory the Universe began its history from a point of infinitely small size and of infinitely high temperature, labeled {\it singularity}, which can not be described by the known physics.  It is believed that during the Big Bang, matter and antimatter were created in equal amounts. After its birth the Universe started to expand and went through different ``cosmological eras''\footnote{~This brief overview of the early evolution of the Universe down to the Radiation era is based on the information in \citet{Harrison00}.}. 

The first epoch, the {\it Planck era}, lasted only $10^{-43}$ seconds after the Big Bang. At that time the size of the Universe was $10^{-35}$ cm across while the density was enormous: $\sim 10^{94}$ g\cc. The particle energy was the Planck energy ($\sim 10^{19}$ GeV) that corresponds to a temperature of $\sim10^{32}$ K. The known fundamental forces -- gravity, electromagnetism, strong and weak nuclear forces -- were unified in single super-force. At the end of this era gravity became a distinct force while the other three forces were still unified in electronuclear force. This electronuclear force distinguished slightly (if at all) between matter and antimatter.

About $10^{-35}$ seconds after the Big Bang the temperature of the Universe fell down to $\sim10^{28}$ K, corresponding to a mean particle energy of $\sim10^{15}$ GeV. In the framework of the Grand Unified Theory, at those temperature and energy regimes the electronuclear force split into the strong nuclear and the electroweak forces. It was this phase transition that may have triggered the start of an exponential growth -- the Universe entered the so called {\it Inflationary era}. Its expansion was enormous, with an inflation factor (i.e. final scale factor, normalized to the initial one) from $5 \times 10^{21}$ to $3 \times 10^{43}$, according to various estimates. The inflation ended by some $10^{-33}$ seconds after the Big Bang. 

When the {\it Quark-lepton era} began, only three fundamental forces existed: gravity, strong nuclear force and electroweak force. The Universe was filled with a mixture of structureless particles and antiparticles in a state of thermal equilibrium. Particle-antiparticle pairs were continuously created and annihilated in this `quark plasma', consisting of quarks, leptons and gluons. It is believed that at some point in this era (if not earlier) the matter-antimatter symmetry broke due to unknown reaction. This baryogenesis led to a slight domination of matter over antimatter -- by one particle per billion. About $10^{-11}$ seconds after the Big Bang, when the Universe had cooled down to $\sim10^{15}$~K, corresponding to a mean particle energy of 100 GeV, the electroweak force split into the weak nuclear force and the electro-magnetic force. 
With the decrease of temperature, the annihilation processes started to prevail over those of creation of particles and antiparticles. About $10^{-7}$ seconds after the Big Bang, at temperature of $10^{13}$ K (mean particle energy $\sim$1 GeV), quarks vanished in the Universe; bound in pairs or triplets, they built up hadrons and antihadrons: pions (quark-antiquark pairs) and nucleons (quark triplets), mainly protons and neutrons. The hadronic matter was mixed also with a fraction of photons and leptons (light particles like electrons, positrons and neutrinos). As the mean particle energy dropped further and the temperature reached about $10^{12}$~K, protons, neutrons, their corresponding antiparticles and pions practically vanished in an enormous annihilation event. Because of the asymmetry in the particle-antiparticle ratio one in billion hadrons survived. 
The age of the Universe was $10^{-5}$ seconds and the density decreased to $10^{9}$ g\cc. The temperature was still high enough to produce lepton-antilepton pairs, which were continuously created and annihilated. But when it dropped down to $10^{10}$~K, one second after the Big Bang, light particles annihilated and only one per billion electrons survived. 

The {\it Radiation era} commenced. The density of electromagnetic radiation exceeded the one of matter by a factor of $10^{7}$. The temperature continued to decrease and at $T\sim 10^{9}$ K the mean energy of photons dropped below the binding energies of protons and neutrons in light atomic nuclei. That launched the so called {\it primordial nucleosynthesis}. Initially heavy hydrogen nuclei of deuterium (containing 1 proton and 1 neutron) were created. It was the ``fuel'' for a very important process that lasted around 200 seconds: deuterium reacted further with free protons and neutrons and formed helium nuclei (2 protons and 2 neutrons). About 25 per cent of the presently existing matter have been transformed to helium in this way. The rest of it consisted mostly of hydrogen nuclei (protons). Small amounts of deuterium, helium-3 (a nucleus of 2 protons and 1 neutron), and lithium-6 (a nucleus of 3 protons and 3 neutrons) were also created. The radiation era ended after 100 thousand years at temperature of 
order of $10^{3}$~K, and at comparable densities of radiation and matter. During that epoch, the Universe was in a state of thermal equilibrium and was opaque for radiation. The baryonic plasma was fully ionized.

\begin{figure}[h!]
\begin{center}
\resizebox{0.7\hsize}{!}{\includegraphics{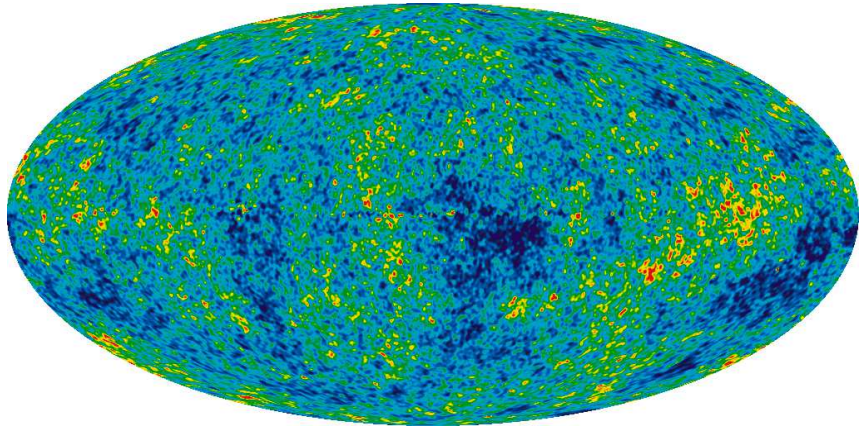}}
\caption[Map of the cosmic microwave background radiation.]{\footnotesize Detailed, all-sky picture of the infant Universe created from WMAP. The color scale illustrates initial temperature fluctuations of the CMB ($\sim13.77\times10^{9}$~yr old) that correspond to seeds which grew to become galaxies.This image shows a temperature range of $± 200$ microKelvin. Credit: NASA/WMAP Science Team}
\label{fig_cmb}
\end{center}
\end{figure}

The {\it Matter dominated era} began\footnote{~For the description of Matter dominated era we follow mainly \citep{Madau02}.} around 500 thousand years after the Big Bang ($z\sim1000$), when the temperature dropped to $\sim 3000$~K. 
The temperature was low enough to allow the binding of free electrons with hydrogen and helium nuclei (recombination) -- the baryonic gas became neutral. Matter and radiation decoupled and the Universe ``brightened'' becoming transparent to light. 
The photons at this recombination stage were free to propagate through the Universe. This process continues until nowadays -- as testified by the (ever-fading) cosmic blackbody radiation: cosmic microwave background radiation (CMB). The latter shifted gradually through the infrared to the radio spectral range. Its local fluctuations are measured by Wilkinson Microwave Anisotropy Probe (WMAP) (Fig.~\ref{fig_cmb}). The Universe entered the so called ``Dark ages'', period for which little or nothing is known.
The first stars and galaxies lit up the Universe and brought the end of the ``Dark ages''. In the period between redshifts 15 (age of $2.7 \times 10^{8}$ years) and 7 (age of $7.8 \times 10^{8}$ years\footnote{~These estimates were obtained  by use of the Cosmological calculator: \url{http://www.kempner.net/cosmic.php}.}) the first massive stars produced heavy elements which might have been ejected in the interstellar medium through supernovae explosions. The first stars and quasars probably generated enough ultraviolet radiation to reheat and re-ionize the surrounding Universe (the process known as reionization). The ionized gas containing primordial baryons was enriched by heavy elements, produced in stars. This gas environment is known as all pervading {\it intergalactic medium} (IGM). During the non-linear formation of structures, the IGM became clumpy and highly inhomogeneous. It provided material for star formation and, on the other hand, was an environment where heavy elements and energy were ejected.


\subsection*{The baryonic evolution of the Universe. Problems.}\label{baryons}

\hspace{0.6cm}According to the current Lambda cold dark matter ($\Lambda$CDM) cosmological model, the Universe consists of baryonic matter ($\approx$ 4 per cent) and ``cold'' dark matter ($\approx$ 25 per cent) that interacts only gravitationally but does not emit electromagnetic radiation. The remaining $\sim$ 70 per cent of the Universe is the fraction of dark energy conceived as anti-gravity that causes acceleration of the expansion of the Universe. The nature of dark matter and dark energy is still unclear and thus these components are matter of intensive research. It is believed that the distribution of baryonic matter follows the one of the dark matter in form of dark-matter halos. 

Although the fraction of baryonic matter is small, it is of key significance to understand the structure and the evolution of the Universe. The standard big-bang nucleosynthesis theory (SBBN) predicts the primordial abundances of helium-3, helium-4, deuterium and lithium-7 (a nucleus of 3 protons and 4 neutrons), depending on a single parameter: the dimensionless baryon density $\Omega_{\rm b}$, i.e. the baryon density normalized to the critical density of the Universe. There are various ways to estimate this parameter which put to test the value predicted by the SBBN and enable the study of the baryon evolution of the Universe. A large literature presenting different results on baryon density has been accumulated within the last decades.

One of the methods to estimate $\Omega_{\rm b}$ is through measurement of the deuterium-to-hydrogen ratio (D/H). In the early 1970s, \citet{Reeves73} pointed out that deuterium can be produced only during the Big Bang or other pre-galactic event and suggested that the Big bang is the main mechanism for its formation. 
Those authors concluded that the deuterium abundance sets an upper limit to the baryon density. A few years later \citet{Adams76} suggested for first time that the amount of deuterium in intergalactic clouds can be measured by use of Lyman absorption spectra (see Sect. \ref{absorbers}) of distant quasi stellar objects (QSOs; see Sect.~\ref{igm}). Afterwards, many efforts have been made to determine the baryon density from deuterium abundance \citep[e.g.][]{Carswell94, Songaila94, Tytler96}. About two decades later \citet{Burles98} measured the deuterium-to-hydrogen ratio by analysis of a Lyman limit absorption system (see Sect. \ref{absorbers}) at $z\sim3.6$ and asserted that their result is consistent with the primordial value from the SBBN. Using the obtained estimate of D/H in the framework of the SBBN theory, they derived the {\it total} baryon density in the Universe: $\Omega_{\rm b} h^{2}=0.0193\pm0.0014$ (where $h=H_{0}/100$~\kms$\rm Mpc^{-1}$). 

In the late 1990s, other attempts to determine the baryon density of the Universe have been made from measuring the \Lya alpha forest (see Sect.~\ref{types_lines}) flux decrement $D \equiv \langle 1- e^{-\tau}\rangle$, i.e. the mean absorbed fraction of the QSO continuum, by use of the relation between $\Omega_{\rm b}$ and the optical depth $\tau$ \citep[see Eq. 8 in][]{Weinberg97}. A flux decrement distribution was obtained in the work of \citet{Rauch97}, based on observations in the redshift range $2.52 \le z \le 4.55$ and on simulations. The derived value from the baryon content in the \Lya forest was $\Omega_{\rm b} h^{2}=0.021\pm0.017$. 

About 10 years ago \citet{Netterfield02} estimated the baryon density through analysis of the peaks of the CMB angular power spectrum. At $z=0$ the cosmic blackbody radiation has cooled down to temperature of $\approx 3$ K and thus has shifted to the microwave range. The amplitudes and the positions of CMB peaks are sensitive to photon and baryon content -- from their measured values one can derive the expected {\it total} baryon density and other fundamental cosmological parameters of the early Universe. \citet{Netterfield02} obtained $\Omega_{b} h^{2}=0.022\pm0.003$, in a good agreement with the SBBN prediction, and came to conclusion that their results confirm the standard cosmological model of structure formation. 
A very similar estimate of the total baryon density was done later by \citet{Huey04}: $\Omega_{b} h^{2}=0.025\pm0.002$. 
In addition, some very resent studies of the CMB angular power spectrum from WMAP yield values in the range $0.019\le\Omega_{b} h^{2}\le0.024$ ($0.039 \le \Omega_{b} \le 0.049$, for $h=0.7$), with 95 per cent confidence and in consistency with the SBBN \citep{Lahav10}. Comparing estimates of the total baryon density in the Universe \citep{Burles98, Netterfield02, Lahav10} and the measured baryon content at high redshifts \citep{Rauch97, Weinberg97}, it seems that the latter, as detected in the \Lya forest, accounts for nearly all of the total $\Omega_{b}$.

On the other hand, the baryon content at $z=0$ can be measured by summing up the amounts in different observable structures which are baryon tracers: stars, galaxies, galaxy groups and galaxy clusters. The mass of the latter can be directly estimated by use of the mass-to-light ratio. \citet{Fukugita04} argue that 6 per cent of the baryons at zero redshift are contained in stars and star remnants, 8 per cent are in galaxies and 4 per cent are concentrated in rich clusters of galaxies. \citet{Shull12} find that 5 per cent of the baryon content is in ionized circumgalactic medium (CGM), surrounding the galaxies, and 1.7 per cent is in cold neutral atomic \Hi gas, which can be detected through surveys in the 21 cm line.
 

Obviously, the richest ``containers'' of baryons in the Universe are to be sought among other objects. \citet{Shull12} estimate that 28 per cent of the baryons at $z=0$ are comprised in the photoionized \Lya forest ($T<10^{5}$). Besides that cool, photoionized gas, there exists a hot intergalactic gas with temperatures $T\sim10^{5}-10^{7}$~K and with low densities $n\sim10^{-6}-10^{-4}$~\cc, called {\it Warm-Hot Intergalactic Medium} (WHIM). The WHIM is believed to be a shock-heated and collisionally ionized gas, originating from a collapsing medium driven by the gravity of the large-scale filaments \citep{Cen99, Dave,Valageas02}. The contribution of WHIM at $z=0$ to the total baryon budget, estimated from \Ovi absorptions \citep[see][]{Danforth08, Tilton12} and broad \Lya (BLA) absorption lines \citep[see][]{Richter04, Richter06}, is $\approx$ 30 per cent \citep{Shull12}. On the other hand, summing up the fractions in the baryon budget at $z=0$ of galaxies, CGM, intracluster medium (ICM), cold neutral gas, 
photoionized \Lya IGM gas and the WHIM, \citet{Shull12} found that $\approx 30$ per cent of the baryons (in comparison to the {\it total} expected amount), are not observed. It seems that part of the baryons at low redshift is missing. This ``missing baryons problem'' was first presented by \citet{Fukugita98} who derived a value of $\Omega_{b}(z=0)= 0.021$, a factor of 2 lower than the total expected baryon content. 

Another still unresolved issue with matter content in the Universe is known as the ``missing metal problem'' at high redshift, formulated originally by \citet{Pettini99}. In astrophysics, all elements heavier than hydrogen and helium are traditionally labeled ``metals''. Observations of young stars in distant galaxies at various redshifts provide an opportunity to trace the evolution of star formation rate, i.e. the star formation history (SFH), up to $z\approx7$. Hence, assuming some initial stellar mass function, one can estimate the expected metal production rate and the density of cosmic metals $\Omega_{Z}^{\rm SFH}$ at given $z$ \citep{Ferrara05}. On the other hand, an observational estimate of metal content $\Omega_{Z}^{\rm obs}$ at high redshifts can be derived from studies of \Lya forest, damped \Lya absorbers (DLAs) (see \ref{types_lines}) and Lyman break galaxies (LBGs)\footnote{~Star-forming galaxies at high redshift ($z>2$), defined by means of colors (differing appearance in several imaging 
filters) near the Lyman limit (912 \AA) (see Sect.~\ref{types_lines}).} as demonstrated by \citet{Pettini99}. The comparison of $\Omega_{Z}^{\rm SFH}$ with $\Omega_{Z}^{\rm obs}$ shows that 80 per cent of the expected metal content is not detected at $z>2$, i.e. $\Omega_{Z}^{\rm obs}\le 0.2~ \Omega_{Z}^{\rm sfr}$ \citep{Ferrara05}. 


  \section{The importance of studying the IGM}\label{igm}

\hspace{0.6cm}The IGM is built up form filaments of ionized gas outside the galaxies. Recent studies demonstrate that it can be located inside as well outside the hosting dark matter halos \citep{Mo10}. Properties of intergalactic clouds can be studied best measuring their absorption of light from extragalactic background sources like QSOs. Some QSOs are very distant active galactic nuclei with sizes about the one of the Solar system and bolometric luminosities $\sim$100 times those of normal galaxies \citep{Hoyle00}. They 
exhibit a well defined, flat continuum spectrum within a very large spectral range. When an intergalactic gas cloud is located at the sightline toward the QSO, it is detected through absorption lines at certain wavelengths imposed on the QSO continuum. Since some QSOs are located at distances corresponding to redshifts beyond $z=6$, the analysis of their absorption spectra can reveal the properties of the IGM when the Universe was less that 10 per cent of its present age \citep{Mo10}.
\\

The study of IGM is important for several reasons:

\begin{enumerate}

 \item It can throw light on the problems of ``missing baryons'' and ``missing metals''. Predictions of some cosmological simulations show that~$\approx 30-40$ per cent of the ``missing baryons'' at $z=0$ are to be comprised in the WHIM \citep{Cen99, Dave}. Most of the ``missing metals'' at $z=2-3$ should be also found there \citep{Cen11}.

 \item The IGM provides the material for galaxy formation through large-scale gas condensations. After the formation of a galaxy, the IGM and the ISM have interacted actively. Gas from the IGM can be accreted into the galaxy and flow into the ISM. And vice versa, gas from the ISM can be ejected and flow back into the IGM. Moreover, the dark-matter halo can cause accretion from the IGM into the large-scale galactic environments \citep{Mo10}. Thus, the knowledge of the IGM is crucial for understanding galaxy formation and evolution.  

 \item Basic physical properties of the IGM like temperature, density, chemical composition, ionization state etc. are affected by radiative and gas-dynamical processes. Therefore the study of the IGM can provide information about the cosmological events after the recombination, during the Matter dominated Era.

 \item One should take into account the possible interaction of IGM gas particles with CMB photons and, hence, the possible distortion of the CMB spectrum. A good understanding of the IGM is necessary to extract correct information from the CMB. 

\item The IGM absorption spectra along the sightline of distant QSOs provide valuable information which can be used to test the evolution of fundamental physical constants like the fine-structure constant $\alpha$, comparing given redshifted atomic or molecular lines with the ones measured in earth bound laboratories \citep{Petitjean09, Srianand09, Uzan11}.

\end{enumerate}


  \section{Quasar absorption line systems as tracers of the IGM}\label{absorbers}

\subsection{Types of absorption line systems}\label{types_lines}

\hspace{0.6cm}Various types of absorbing systems are identifiable in QSO spectra that are characterized by a prominent \Lya emission line and a well-defined continuum from the background source. A typical QSO spectrum at $z=1.34$ is displayed in Fig.~\ref{fig_absorbers}. The QSO's redshifted emission lines \Lya and \Lyb are clearly visible at $\sim 2400$~{\AA} and $\sim 2850$~{\AA}; various absorption lines are observed blueward of \Lya emission line. 

The traditional classification of absorption systems distinguishes between {\it intrinsic} and {\it intervening} systems. The intervening systems are located randomly along the QSO sightline and are not related to it. On the other hand, broad and narrow absorption systems in the vicinity of the QSO ($z_{\rm abs} \approx z_{\rm qso}$), are believed to be intrinsic (physically connected) to it. Narrow intrinsic absorbers within separation velocity of $|\Delta v| \equiv (v_{\rm qso}-v_{\rm abs}) <  5000$ \kms are called in this work {\it associated} systems \citep{Weymann79,Foltz86,Anderson87}.

Since hydrogen is the most abundant element, the absorber type along the sightline can be characterized through the column density of the neutral hydrogen, $N(\Hi\!\!)$ [\sqc]:

 \subsubsection*{\Lya  Forest}

\begin{figure}[h!]
\begin{center}
\resizebox{0.75\hsize}{!}{\includegraphics{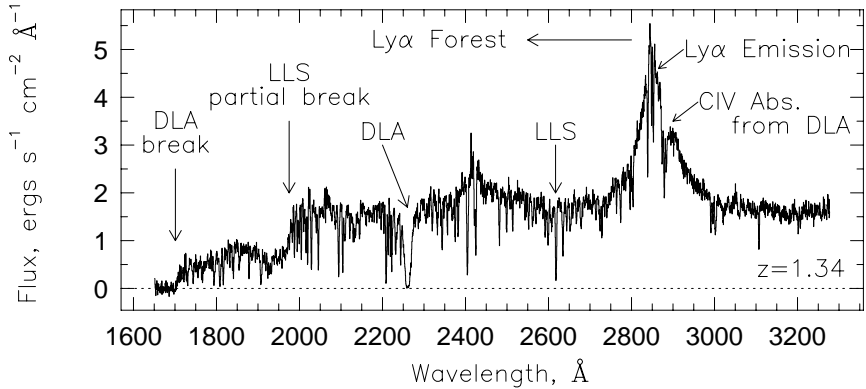}}
\caption[A typical QSO spectrum displaying various absorbing systems.]{\footnotesize
A typical QSO spectrum (PKS$0454+039$, at $z=1.34$) with absorption lines produced by galaxies and intergalactic material. The two strongest absorbers (associated with galaxies) are a damped {\Lya} absorber at $z=0.86$ and a Lyman limit system at $z=1.15$.  The former produces a Lyman limit break at $\sim1700$~{\AA} and the latter a partial Lyman limit break at $\sim1950$~{\AA} since the neutral hydrogen column density is not large enough for it to absorb all ionizing photons. Many absorption lines are produced by the DLA at $z=0.86$ ({\Civ}1548, for example, is redshifted onto the red wing of the quasar's {\Lya} emission line). [Adopted from \citet{Charlton00}]}
\label{fig_absorbers}
\end{center}
\end{figure}

\hspace{0.6cm}Clouds of neutral hydrogen which lie along the sightline absorb \Lya photons of wavelength 1215.67 \AA. Due to the various redshifts to these clouds, the corresponding absorption lines are redshifted by factor of $(z+1)$ and detected at various wavelengths blueward of the \Lya\!\! emission of the QSO. These lines are narrow and appear as a ``forest'' of lines (Fig.~\ref{fig_absorbers}). The \Hi column densities of \Lya forest absorbers span the range $10^{12}-10^{17}$ \sqc. The lower column density limit reflects the detection limits of the observations, while the upper limit is conditioned by the absorbers' optical depth -- systems with $N(\Hi\!\!) \approx 10^{17}$ are optically thick to Lyman continuum radiation and appear as Lyman limit systems (see below). \citet{Mo10} point out that, according cosmological simulations, \Lya forest systems with $N(\Hi\!\!)\ge10^{14.5}$ \sqc~ are associated with higher density filaments which connect collapsed objects, while \Lya systems with $N(\Hi\!\!)$ 
below this value inhabit low density regions. These authors suggest that \Lya forest absorbers with high column density are possibly enriched with metals from star formation processes in collapsed objects and their low column density peers seem to be more primordial in origin. 

The \Lya forest evolves from high to low redshifts. In the range $2 \le z \le 5$, the number of \Lya absorbers per unit redshift is large but rapidly decreasing, mostly due to the expansion of the Universe and partly due to the growth of the large-scale structures \citep{Charlton00}. At $z<2$, the intensity of extragalactic UV background radiation drops due to the decrease of star-formation rate and of the QSOs space density. (See Sect.~\ref{ionization_mechanisms} for more details on UV background radiation.). As a result, the fraction of neutral gas contained in \Lya absorbers increases. The decrease of UV radiation density counteracts the decrease of matter density and thus the absorbers' number in the \Lya forest decreases less rapidly than expected from the Universe expansion alone \citep{Mo10}.

 \subsubsection*{Lyman Limit Systems}

\hspace{0.6cm}{\it Lyman limit systems} (LLS) are rare narrow-line systems which are optically thick at wavelength 912 \AA , corresponding to the \Hi ionization potential of 13.6 eV. They can be detected as saturated \Lya lines which point to column densities distinctly higher than those of \Lya forest lines: $1.6 \times 10^{17} \le N(\Hi\!\!) \le 10^{19}$ \sqc. LLS cause a break at the rest wavelength of 912 \AA, which is well distinguishable (Fig.~\ref{fig_absorbers}). It is difficult, however, to determine accurately the column densities of LLS, since the associated \Lya lines are very saturated. A way to estimate $N(\Hi\!\!)$ is through the strength of the Lyman limit break, but such LLS are very rare, as pointed out by \citet{Mo10}. 
LLS are associated with strong metal absorption lines and are believed to arise in the gaseous halos of galaxies, which are respectively embedded in DM halos.

\subsubsection*{Damped \Lya Absorbers}

\hspace{0.6cm}{\it Damped \Lya systems} (DLAs) exhibit characteristic damping wings due to natural broadening (see \ref{line_broadening}), caused by uncertainty of the energy states involved in the transition (Fig.~\ref{fig_absorbers}). They are rare systems with column densities $N(\Hi\!\!)\ge 2 \times 10^{20}$ \sqc. This lower column density limit is historical in origin -- technically speaking, any absorber with $N(\Hi)\ge 2 \times 10^{19}$ \sqc~ will show damping wings. Systems with $10^{19 }\le N(\Hi\!\!) \le 10^{20}$ are called {\it sub-DLAs}. Hydrogen in DLAs at high redshifts is mostly neutral, while it is significantly ionized in sub-DLAs \citep{Mo10}. Various heavy elements (metals) are also associated with those systems. 

DLAs are referred to as the highest overdensity\footnote{~The ratio of density to the mean density.} absorbers -- it is believed that they form in DM halos. The total amount of gas in DLAs at $z\approx3$ is comparable to that of luminous matter in present day galaxies \citep{Mo10}. Along with the presence of metals, this fact suggests that high-redshift DLAs might have provided the material for galaxy formation. Thus, the detected species in DLAs could throw light on the chemical enrichment in protogalaxies and on the subsequent galaxy formation history. Chemical abundance in DLAs can be measured with a high accuracy, since their \Hi column densities can be precisely measured from analysis of the damping wings and normally no ionization corrections are necessary.

 \subsubsection*{Metal absorption line systems} 

\begin{figure}[h!]
\begin{center}
\resizebox{0.7\hsize}{!}{\includegraphics[angle=-90]{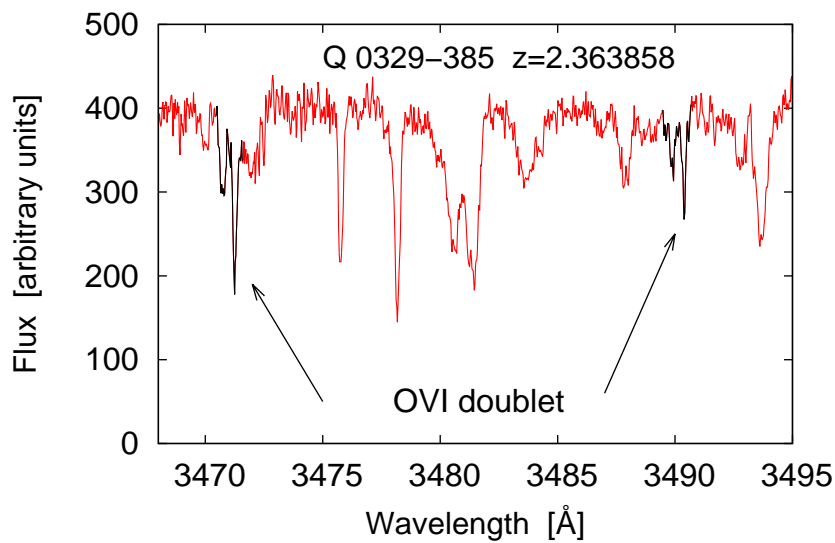}}
\caption[A typical \Ovi system at $z\approx2.4$: the doublet $\lambda\lambda~1031,1037$.]{\footnotesize A typical \Ovi system at $z\approx2.4$: the absorption lines of the doublet $\lambda\lambda~1031,1037$ along the sightline of a distant QSO.}
\label{fig_representation_ovi}
\end{center}
\end{figure}

\hspace{0.6cm}The QSO spectra often display absorption lines of metals. The most studied metal systems, which have significance for understanding of the IGM, are \Mgii$\lambda\lambda$\,2796,\,2800, \Civ$\lambda\lambda$\,1548,\,1550 and \Ovi$\lambda\lambda$\,1031,\,1037. As it is seen, the first two types have rest-frame wavelengths which are greater than the one of \Lya ($\lambda_{\footnotesize \Lya\!\!}=1216$\,\AA). Thus, their absorption lines lie redward of the QSO emission which enables easy identification of those systems. In contrast, \Ovi lines lie blueward of the QSO emission and are embedded in the \Lya forest. This leads to difficulties in identification and analysis, especially at high redshifts, where the \Lya forest is more densely populated. A typical \Ovi absorption system is shown in Fig.~\ref{fig_representation_ovi}.

The ionization potentials of \Mgii\!\! (15~eV), \Civ\!\! (65~eV) and \Ovi\!\! (138~eV) are substantially different. Therefore, the gaseous structures these species are associated with probably inhabit different environments. To produce \Mgii with particle collisions, a temperature of $\sim10^{4}$~K is necessary, while presence of \Civ and \Ovi in the spectrum requires higher temperatures: $T\sim10^{5}-10^{6}$~K. Thus lines from different metal ions can reflect the variety of physical conditions in the IGM.


    \subsection{Ionization mechanisms in the IGM}\label{ionization_mechanisms}

\hspace{0.6cm}Ionization of the IGM occurs mainly through two mechanisms: ionizing radiation (photoionization) and particle collisions (collisional ionization). We review them briefly below.

   \subsubsection*{Photoionization}

\hspace{0.6cm}Photoionization is a bound-free transition (i.e., a removal of an electron from an atom) due to photon absorption. The rate of this process, $\Gamma_{\rm ph}$, is an integral over all frequencies of the product between the photoionization cross section $\sigma_{\rm ph}(\nu)$ and the number density of ionizing photons at a given frequency. The latter is proportional to the energy flux of the radiation field, i.e., the metagalactic UV background radiation penetrating the IGM. The dominating source is the UV flux from QSOs and young star-forming galaxies, reprocessed and attenuated by the intergalactic gas. According to recent estimates, those objects provide sufficient UV flux to produce the observed ionization level at $z\le5$ \citep{Mo10,HM12}.

The typical lifetime of a hydrogen-like atom\footnote{~An atom with one valence electron.} 
is longer than the lifetime of an excited state at low densities. Therefore, the assumption that most photoionizations occur from the ground level is reasonable. The photoionization cross section can be estimated from the formula:

\begin{equation}
\label{photoion_cross_sect}
\sigma_{\nu, \rm ph} = \frac{C_{0}}{Z^{2}} \Big(\frac{\nu_{Z}}{\nu}\Big)^{3} g_{1}(\nu)~~~~~~~~~~({\rm for}~ \nu\ge\nu_{Z}),
\end{equation}
 
where $Z$ is the atomic number, $\nu_{Z}\equiv\nu_{t}(Z,1)$ is the threshold frequency\footnote{~Corresponding to the ionization potential.} at the 1$^{\rm st}$ energy level and $g_{1}$ is the bound-free Gaunt factor for the ground level, which accounts for quantum uncertainties and is close to unity at optical frequencies \citep{Mo10}. The constant $C_0$ does not depend on atomic number and frequency and takes different values on the different sides of the characteristic absorption limits \citep{Kramers23}. In the considered case, the constant $C_{0}=7.91\times10^{-18}$ \sqc~is determined at the Lyman edge of atomic hydrogen, i.e., it is the Kramers absorption cross section at $\lambda=912$~\AA. 

Absorption cross sections of multi-electron atoms are described by a more complicated formula. An useful approximation of the contribution of each threshold to the photoionization cross section is:

\begin{equation}
\label{photoion_cross_sect}
\sigma_{\nu, \rm ph} =a_{t} \Big[\beta \Big(\frac{\nu_{t}}{\nu}\Big)^{s}+(1-\beta)\Big(\frac{\nu_{t}}{\nu}\Big)^{s+1}\Big]~~~~~~~~~~({\rm for}~ \nu\ge\nu_{t}),
\end{equation}

where a list of numerical values of $\nu_{t}$, $a_{t}$, $s$ and $\beta$ for some atoms and ions can be found in \citet{Osterbrock89}. Then, the total cross section is the sum of individual thresholds. It achieves a maximal value at the threshold and declines with increase of energy. 

The photoionization rate of neutral hydrogen is calculated through the formula:

\begin{equation}
\label{photoion_rate}
\Gamma_{{\scriptsize \Hi\!\!},\rm ph} =4\pi \int\limits_{\nu_{\rm L}}^{\infty} \frac{\sigma_{\nu, \rm ph}J_{\nu}}{\rm h \nu} {\rm d}\nu \approx 2.5 \times 10^{-14} J_{-23}~~~~~[\rm s^{-1}],
\end{equation}

where $J_{\nu}$ is the mean intensity (see Sect.~\ref{radiative_transfer}) of the ambient (metagalactic) ionizing radiation field, $\nu_{\rm L}$ is the frequency at the Lyman limit and $\sigma_{\nu, \rm ph}$ is the photoionization cross section of hydrogen. $J_{-23}$ is the dimensionless mean intensity of the UV background intensity at the Lyman limit
in units $\rm 10^{-23}\,erg~cm^{-2}~s^{-1}~Hz^{-1}~sr^{-1}$ . $J_{-23}\sim1-2$ for $z=0$ and much higher ($J_{-23}\sim80$) for $z=3$.

\begin{figure}[h!]
\begin{center}
\resizebox{0.7\hsize}{!}{\includegraphics{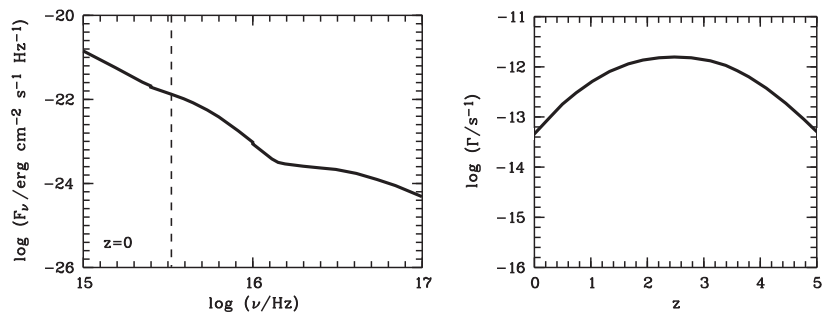}}
\caption[]{\footnotesize {\it Left panel:} Spectrum of the metagalactic UV background at $z=0$ \citep[from][]{HM96}. The Lyman limit is marked with A dashed line. {\it Right panel:} Photoionization rate as a function of redshift. [Adopted from \citet{Richter08}, see the references therein.]}
\label{fig_UV_background_rate}
\end{center}
\end{figure}

On the left panel of Fig.~\ref{fig_UV_background_rate} is shown the flux of the metagalactic UV background (see Sect.~\ref{radiative_transfer}) at $z=0$ against the frequency $\nu$. The redshift dependence of the hydrogen photoionization rate from the metagalactic UV background is shown on the right panel of the figure. 

\subsubsection{Collisional ionization} 

\hspace{0.6cm}Collisional ionization is another type of bound-free transition, a removal of an electron from an atom due to collision with other particles, predominantly electrons. It is a cooling process since part of the particle kinetic energy is used for ionization. The rate of collisional ionization, $\Gamma_{\rm col}$, is an integral over all velocities $v$ of the colliding electron of the product between the collisional ionization cross section $\sigma_{\rm col}(v)$, the number density of the colliding electrons (which does not depend on $v$) and the velocity distribution. If the velocity distribution is Maxwellian, then the rate and the cross section depend only on the electron energy.
The collisional ionization cross section is zero at the threshold energy, $I$, required to unbind the electron and increases with increase of the electron energy. For low energies $I < E \le3I$, it can be estimated by use of the approximate formula \citep{Draine11}:

\begin{equation}
\label{collion1_cross_sect}
\sigma_{\rm col}(E) \approx C\pi a^{2}_{0}\Big(1-\frac{I}{E}\Big),
\end{equation}

where $C$ is a constant of order unity (for hydrogen, $C=1.07$) and $a_{0}\equiv \hbar^{2}/4 \pi^{2}m_{e}q^{2}_{e}$ is the Bohr radius\footnote{~Where $\hbar_{p}$ is the reduced Planck constant, $m_{e}$ -- the electron mass and $q_{e}$ -- the electron charge.}. At higher, but still non-relativistic energies, the collisional ionization cross section decreases as $1/E$.


  

  \subsubsection{Recombination and ionization balance} 

\hspace{0.6cm}Radiative recombination is a process of free-electron capture by an ion after collision. The collided particles recombine into neutral or weakly-ionized atom, consuming part of their kinetic energy. Therefore radiative recombination is a cooling mechanism. The velocity distribution of electrons in local thermal equilibrium\footnote{~I.e. when temperature varies very slow in a small volume.} is Maxwellian and the recombination rate depends only on electron temperature and density. The radiative recombination coefficient $\alpha$ at given level is obtained as the average of recombination cross section over the velocity distribution \citep{Boardman_64}. For hydrogen-like atoms with atomic number $Z>1$, its physical behavior is described by \citep{Boer07}:

\begin{equation}
\label{recomb1_cross_sect}
\alpha_{\rm rad}(Z,T) \approx \frac{T_{e}^{-1/2}}{Z^{2}},
\end{equation}

In the process of recombination, the excess energy can be absorbed by another bound electron which moves to an excited level and frees its place for the captured one. This phenomenon is called {\it dielectronic recombination}. In that case, the recombination coefficient is given by the approximate formula \citep{Boer07}:

\begin{equation}
\label{recomb2_cross_sect}
\alpha_{\rm di}(T) = d_{1}T^{1/2}e^{d_{2}/T},
\end{equation}

where $d_{1}$ and $d_{2}$ are constants, available in a tabulated form.

The {\it total ionization rate}, $\Gamma_{\rm tot}$, is the sum of rates of photoionization, collisional ionization and charge exchange (in case the colliding atoms exchange charge):

\begin{equation}
\label{total_ion_rate}
\Gamma_{\rm tot}=\Gamma_{\rm ph}+\Gamma_{\rm col}+\Gamma_{\rm{charge~ exch.}} 
\end{equation} 

On the other hand, the {\it total recombination rate}, $[\alpha(T)n_{e}]_{\rm tot}$, is the sum of rates of radiative recombination, dielectronic recombination and of charge exchange:

\begin{equation}
\label{total_rec_rate}
[\alpha(T)n_{e}]_{\rm tot}=(\alpha_{\rm rad}+\alpha_{\rm di})n_{e}+\alpha_{\rm charge~exch.}n_{x}~,
\end{equation}

where $n_{e}$ is the volume density of electrons and $n_{x}$ -- the volume density of charge exchanging particles. The radiative (first) terms dominate in both considered total rates (Eqs. \ref{total_ion_rate} and \ref{total_rec_rate}). Then the {\it ionization balance} is governed by the ratio of total ionization and recombination rates \citep{Boer07}:  

\begin{equation}
\label{ion_balance}
\frac{n_{\rm ion}}{n_{\rm atom}} = \frac{\Gamma_{\rm tot}}{[\alpha(T)n_{e}]_{\rm tot}},
\end{equation}

The ionization balance is achieved under the state of local thermodynamic equilibrium (LTE).\footnote{~In LTE the local kinetic (Maxwellian) temperature is equal to the (Planckian) temperature of the radiation field.} 
In ionized regions of the ISM and the IGM the electron density depends mainly on the ionization state of hydrogen and $n_{e}\approx n_{p}$. At high electron densities, the ionization balance keeps metals at lower ionization stages.

\subsubsection{Collisional ionization equilibrium}

\hspace{0.6cm}Collisional ionization equilibrium (CIE) is the balance between the rates of collisional ionization from the ground level and of recombination from the higher ionization stages at given temperature. In a state of CIE, the fraction of free electrons $X=n_{e}/n$ depends only on the gas temperature\footnote{~Where $n$ is the volume density of all gas particles and $n_{e}$ is the volume density of the electrons.}. Charge-exchange reactions can be neglected in case of collision between a hydrogen atom and an electron and then the fraction of neutral hydrogen $f_{\scriptsize \Hi}=n_{\scriptsize \Hi}/n_{\scriptsize \rm \mbox H}$ is equal to:

\begin{equation}
\label{coll_ion_equilib}
f_{{\scriptsize \Hi\!\!},\rm col} = \frac{\alpha_{\rm H}(T)}{\beta_{\rm H}(T)},
\end{equation}

where $\alpha_{\rm H}(T)$ is the recombination coefficient and $\beta_{\rm H}(T)$ is the collisional ionization coefficient, which is the product of electron volume density, the collisional cross section and the velocity distribution of the colliding electrons. The latter in the considered case is Maxwellian and depends only on the electron temperature.

The neutral hydrogen fraction in the WHIM ($T=10^{5}-10^{7}$ K), assuming CIE, can be approximated by \citep{Richter08}:

\begin{equation}
\label{coll_ion_equilib_1}
\log f_{{\scriptsize \Hi\!\!},\rm col} \approx 13.9-5.4 \log T+0.33 (\log T)^{2}.
\end{equation}
 
That gives a vanishing neutral hydrogen fraction of $\sim 2.4 \times 10^{-7}$ in CIE at $T=10^{6}$~K.

\begin{figure}[h!]
\begin{center}
\resizebox{0.7\hsize}{!}{\includegraphics{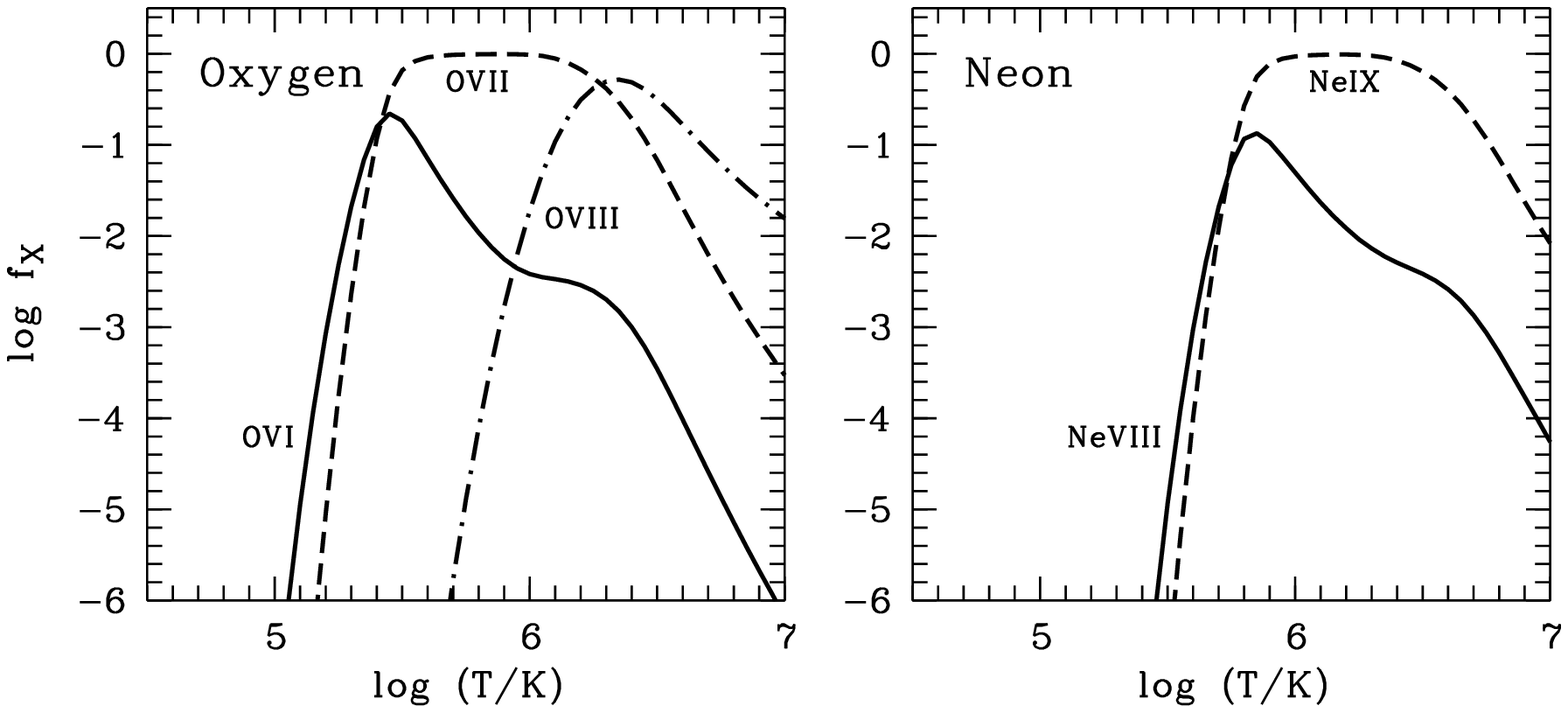}}
\caption[]{\footnotesize CIE ion fractions of selected high ions of oxygen (\Ovi, \Ovii, \Oviii; left panel) and neon (\Neviii, \Neix; right panel) in the WHIM temperature range log ($T/\rm K=4.5-7.0$). Adopted from \citet[][see the references therein]{Richter08}.}
\label{fig_ion_fr}
\end{center}
\end{figure}

The ionization states in CIE of typical tracers of WHIM like oxygen, neon, etc., depend only on the gas temperature. Then the fraction of given ion of such element (e.g., five-times ionized oxygen) is determined by the ionization potential of the corresponding ionization level. The temperature dependence of fractions of high ions of oxygen and neon in CIE \citep[based on calculations by][]{Sutherland93} is illustrated in Fig.~\ref{fig_ion_fr}.

\section{The case of \Ovi absorbers}

\subsection{Why it is particularly interesting to study intervening \Ovi\!\! absorbers?}\label{ovi}

\hspace{0.6cm}There are at least several reasons that make the study of five-times ionized oxygen (\Ovi\!\!) interesting and worth of effort:

\begin{enumerate}

\item It is believed that the ``missing baryons'' at low redshifts are hidden in the WHIM (cf. Sect. \ref{igm}), a diffuse medium with temperatures  $10^{5}<T<10^{7}$~K. Such warm-hot gas can be traced by highly ionized heavy elements -- the latter are usually not fully ionized under such conditions and still undergo electron transitions. In particular, \Ovi absorption is very important for studying the WHIM. The \Ovi ion fraction peaks at $T\approx 3 \times 10^{5}$~K in CIE and thus this species can probably trace the low temperature regions of the WHIM. 
  
\item The ``missing metals'' at high redshifts (cf. Sect.~\ref{igm}) are probably hidden in hot gaseous highly ionized halos around star-forming galaxies \citep{Ferrara05, Richter08}. The extragalactic UV background is intensive at high redshifts and most of the \Ovi systems are likely photoionized. However, collisional ionization due to galactic winds could also take place in them \citep{Fangano07}. This makes \Ovi systems good candidates for tracing the highly ionized metal enriched halos. 

\item Since \Ovi can be good tracers of metal-enriched ionized gas in the filamentary IGM and in the circumgalactic environment, the analysis of intervening \Ovi\!\! absorbers towards low- and high-redshift QSOs can be crucial for a better understanding of the physical nature, distribution and evolution of the IGM and its relation to galaxy evolution.

Let us mention as well two methodological benefits of study of \Ovi systems:

\item \Ovi is relatively abundant, while other highly ionized species which can be used as tracers of hot environments, like \Neviii\!\! or \Nv\!\!, have lower cosmic abundance and their detection is more difficult. 

\item The doublet \Ovi $\lambda\lambda 1031,\,1037$ is strong and can be identified relatively easily at low redshifts. Its identification is possible with a high accuracy even at high redshifts, although it is hampered by denser \Lya forest.

\end{enumerate}


 \subsection{Previous studies of O\,{\sc vi} absorbers: advance in our knowledge and unresolved problems}\label{ovi_advance}

\hspace{0.6cm}In general, the detected absorbers, including \Ovi\!\!, are classified either as {\it galactic}, i.e. associated with a galaxy, or as {\it intergalactic}, i.e. located in the IGM. It is commonly accepted that the \Hi absorbers with $N(\Hi\!\!)\le10^{16}$ \sqc~ are associated with intergalactic \Ovi absorbers, while the galactic \Ovi absorption is mainly seen in LLSs and DLSs. Although it should be noted that the discrimination galactic/intergalactic is not strict. For instance, some of the Galactic \Ovi high-velocity clouds (clouds moving with velocities $|v_{\rm LSR}| > 100$ \kms through the extended gaseous halo of the Milky Way\footnote{~$|v_{\rm LSR}|$ is the absolute local standard of rest velocity. It is a measure of the velocity of material with respect to the motion of the Sun.}) are probably intergalactic clouds in the Local Group rather than objects being associated with the Milky Way \citep{Richter08}. According to a review presented by \citet{Fox11}, 
at least 775 galactic and intergalactic \Ovi absorbers can be found in the literature, out of which 328 are low ($z<0.5$) and high ($z\ge2.0$) redshift intergalactic, both intervening and associated, absorbers (see Sect.~\ref{types_lines}). 

\begin{figure}[h!]
\begin{center}
\resizebox{0.7\hsize}{!}{\includegraphics{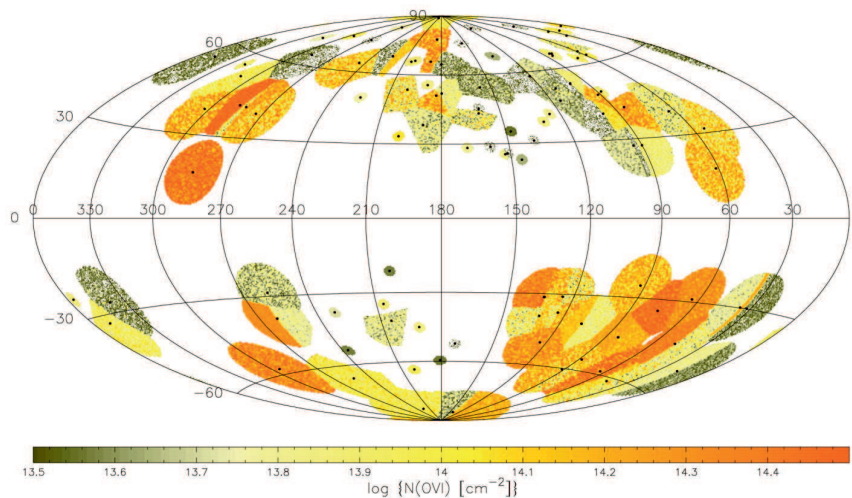}}
\caption[Map of the high-velocity \Ovi column densities at $z=0$]{\footnotesize All-sky Hammer-Aitoff projections of the high velocity \Ovi column densities at $z=0$ (the Galactic anticenter is at the center of the plot, and Galactic longitude increases from right to left). Detections are plotted as colored circles with a radius of $12^{\circ}$; in cases of overlapping (either along the same sightline or along adjoining sightlines), the shaded area size is adjusted accordingly. [Adopted from \citet{Sembach03}]}
\label{fig_OVI_lowz_absorbers}
\end{center}
\end{figure}

The \Ovi doublet $\lambda\lambda 1031,\,1037$ can be detected at low redshift with high-resolution UV spectrographs, such as Goddard High Resolution Spectrograph (GHRS), Space Telescope Imaging Spectrograph (STIS), Cosmic Origins Spectrograph (COS), installed on the {\it Hubble Space Telescope (HST)} or {\it Far Ultraviolet Spectroscopic Explorer (FUSE)}. The review of \citet[][see the references therein]{Fox11} lists 25 studies on low-redshift intergalactic intervening \Ovi absorbers, based on individual sightline detections through UV spectrographs, and a few more survey works by use of all available data. As already mentioned, such systems are often related to the WHIM and their analysis is used extensively to constrain the baryon content of the low-redshift WHIM, mostly under the assumption that they are collisionally ionized \citep[e.g.][see \citealt{Richter08} for a review]{Tripp, Savage02, Richter04, Sembach04}. However, the origin of the \Ovi absorbing gas phase is still not well known. An important 
clue to our understanding is the relation between intergalactic intervening \Ovi\!\! absorbers and the large-scale distribution of galaxies. \citet{Wakker09} pointed out that intergalactic \Ovi absorbers at low redshift preferably 
arise within 550 kpc of an $L>0.25L_{\star}$ galaxy, with expected metallicities higher than those far away from the galactic structure. \citet{Stocke06} found no evidence for \Ovi in intergalactic voids, i.e., at distances $>1.4$~Mpc from the nearest $L>L_{\star}$ galaxy. In view of the last result, \citet{Richter08} suggest that a local analogue of intergalactic intervening \Ovi\!\! might be the galactic \Ovi high-velocity clouds in the Local Group. \citet{Sembach03} estimated that 60 per cent of the sightlines contain high-velocity \Ovi clouds with $N(\Ovi\!\!)\ge 2.5 \times 10^{13}$, and 36 per cent have $N(\Ovi\!\!)\ge 10^{14}$~\sqc ~(see Fig.~\ref{fig_OVI_lowz_absorbers}) which corresponds to hydrogen ion densities $N(\Hii\!\!)\ge 1.2 \times 10^{18}$~\sqc ~and $N(\Hii\!\!)\ge 4.6 \times 10^{18}$~\sqc ~respectively, assuming a gas metallicity of $0.2~ Z_\odot$. Their results indicate that high-velocity \Ovi\!\! absorbers contain a significant fraction of baryons in the form of warm-hot circumgalactic 
gas, which might be the local counterparts of intergalactic intervening \Ovi absorbers at low redshift. \citet{Sembach03} suggest that collisions in hot gas are the dominating ionization mechanism being responsible for production of the high-velocity \Ovi\!\!. 

However, despite the apparent relation between \Ovi absorbers and the WHIM under the assumption of collisional ionization and the possible local analogue of \Ovi high-velocity collisionally ionized clouds, recent observational and theoretical studies indicate that part of the low-redshift intervening \Ovi\!\! absorbers in the IGM may trace low-density, photoionized gas rather than a shock-heated WHIM \citep {Tripp08}. Also, a simple estimate of the ionization state of the gas from the observed \Ovi\!\!/\Hi ratios in absorbers can lead to incorrect results because of the complex multi-phase character of the gas \citep {Tepper-Garcia11}.

At redshift $z\approx2$, the detection of \Ovi absorbers is possible in the optical regime and can be measured using high-resolution spectrographs installed on ground-based telescopes like Keck and the {\it Very Large Telescope} (VLT). There are three main surveys of high redshift ($z\ge2$) intergalactic \Ovi absorbers \citep{Simcoe02, Bergeron05, Fox08}. At least 14 systems have been studied by analysis of individual sightlines; several other works make use of the pixel optical depth method for search of \Ovi absorbers \citep[see the references in][]{Fox11}. Yet, the physics of \Ovi absorbers at high redshifts remains unclear until today. Photoionization seems to be a suitable ionization mechanism for most of them \citep{Carswell02, Bergeron02, Bergeron05}. On the other hand, some studies suggest that a significant fraction of the intergalactic \Ovi absorbers may be collisionally ionized \citep{Simcoe02, Simcoe06}. Numerical simulations indicate that shock-heating by collapsing large-scale 
structures is not efficient at  high redshifts to provide a widespread warm-hot intergalactic phase in the early Universe \citep[e.g.,][]{Theuns02,Oppenheimer08}. Instead, galactic winds probably represent the major source of collisionally ionized \Ovi absorbers at higher redshift, enriching the IGM with heavy elements at high gas temperatures \citep {Fangano07, Kawata07}. \citet{Bergeron05} found evidence of two distinct populations of intergalactic \Ovi absorbers: metal-poor systems that trace the large-scale IGM, and metal-rich, associated with star-forming and wind-blowing galaxies. However, these results need to be confirmed by use of a larger \Ovi sample and through cosmological simulations. We investigate this problem using both observational data and simulations in Sect.~\ref{discussion_paper1}, Sect.~\ref{metal-rich_and_metal-poor} and Sect.~\ref{abundance_owls}.

Recent studies on the evolution of intergalactic \Ovi absorbers from low to high redshifts indicate that the mean column density of these systems evolves only weakly over cosmic times, while the line widths are systematically broader at low $z$ compared to high redshift \citep{Fox11,Muzahid11}. The latter issue is addressed in Sect.~\ref{effect_redshift}, making use of observational and simulated data. The insensitivity of the mean $N(\Ovi\!\!)$ to the redshift can be explained by the dominance of photoionization as ionization mechanism only, if the average density of the \Ovi gas is approximately 20 times higher at $z=2.3$ than at $z=0.2$ and the high-redshift absorbers have smaller sizes, lower metallicities, or lower ionization fractions \citep{Fox11}. Another possible scenario that might explain this phenomenon is radiative cooling of initially hot shock-heated diffuse gas passing through the so-called ''coronal regime`` ($T\sim10^{5}-10^{6}$~K) and producing \Ovi\!\!. \citet{Heckman02} demonstrate that 
the column density of collisionally ionized and radiatively cooling coronal gas is independent of the total volume density. To explain the characteristic mean $N(\Ovi\!\!)$, the radiative cooling scenario requires a cooling flow speed of 20 \kms for single-phase gas, or multiphase gas with approximately 5 interfaces (between cooler and hotter gas) of $N\sim 10^{13}$ \sqc~ each \citep{Fox11}. A better understanding of the ionization mechanism of high-redshift \Ovi absorbers clearly is desirable. We investigate this issue in Sect.~\ref{ionization_modeling_paper1}, \ref{modeling_cloudy} and \ref{correlation_plots_owls}.

\section{Scientific objectives of the thesis}\label{scientific_objectives}

\subsubsection*{A) A case study of two intervening \Ovi\!\! absorbers from high-resolution observations.}  

\hspace{0.6cm}As mentioned in the previous section, the mechanisms of ionization for producing high-redshift \Ovi absorbers may be manifold. A plausible physical picture suggests that low density absorbers are photoionized by the UV background, while collisional ionization dominates in denser regions. The temperature of the \Ovi gas is the most important parameter related to the relevant ionization mechanism. In view of thermal line broadening, the upper limit $T_{\rm max}$ can be directly estimated from the line width as measured through the Doppler parameter $b$ (see Eq.~\ref{eq_temp_Doppler}). A comparison between $T_{\rm max}$ and the estimate $T_{\rm col}$ from collisional ionization models allows us to distinguish between photoionization and collisional ionization as dominant mechanisms. However, an accurate measurement of the Doppler parameter is necessary. Previous surveys of high-redshift \Ovi absorbers \citep{Bergeron02, Simcoe02, Carswell02, Bergeron05} have shown that many narrow absorbers with 
Doppler-parameters $b \leq 10$ km\,s$^{-1}$ do exist. They are to be associated with photoionized gas with $T<10^5$~K. However, many high-ion absorbers have a complex nature and are often composed of several velocity subcomponents. Therefore spectral resolutions  higher than $R\gtrsim45,000$ are required to detect narrow components. Hence, our {\it first goal} is to analyze data form {\it Ultraviolet and Visual Echelle Spectrograph} (UVES) with $R\approx75,000$ (Sect.~\ref{data_uves}) of a single QSO sightline to test whether high spectral resolution is crucial for full component decomposition of the overall structure of particularly complex \Ovi absorption systems and to obtain reliable results on their ionization conditions. Moreover, high-resolution data are important to achieve completeness of their Doppler-parameter distribution towards the lower end ($b<6$ km\,s$^{-1}$).

\subsubsection*{B) Detailed analysis of intergalactic \Ovi samples from UVES observations and OWLS simulations.}

\hspace{0.6cm}The {\it second main goal} of this thesis is to perform a detailed analysis of a large \Ovi sample along 15 high-redshift QSO sightlines using UVES spectra, and comparing it with a sample of \Ovi absorbers from high- and low-redshift cosmological OverWhelmingly Large Simulations (OWLS; Sect.~\ref{owls}). The objectives are to address the following issues:

\begin{itemize}
 \item Origin and nature of high-redshift \Ovi absorbers

The problem of the ionization mechanism of \Ovi absorbers still remains unsolved. Combining observables like column density and Doppler parameter from the large UVES sample with photoionization modeling with {\sc Cloudy} (see Sects.~\ref{cloudy}, \ref{modeling_cloudy} and \ref{aligned_absorbers_owls}) we aim to shed light on whether photoionization is the dominant ionizaion mechanism or not.
Further, in Sects.~\ref{metal-rich_and_metal-poor}, \ref{col_den_b_distr} and \ref{abundance_owls}, we want to investigate the possible existence of two \Ovi\!\! populations as proposed by \citet{Bergeron05}. Also, the OWLS synthetic spectra provide valuable information about physical parameters of the \Ovi absorbers like temperature, space density and metallicity. Therefore, a comparison between observational and simulated spectra is a powerful method to understand the nature and the origin of high-redshift \Ovi absorbers (see Sect.~\ref{simulations}). 

\item Metal and baryon fractions at high redshifts traced by intergalactic intervening \Ovi absorbers 

It is believed that the study of high-redshift \Ovi absorbers can lead to a solution of the ``missing metals problem'' \citep{Richter08}. A contribution of this work is to collect more information about \Ovi as tracers of matter (Sect.~\ref{baryons}). 
We aim to estimate the fractions of baryons and metals at high redshift that are traced by intergalactic \Ovi systems (Sect.~\ref{baryon_density} and \ref{baryons_owls}).  

\item  Possible differences in origin and nature between low- and high-redshift \Ovi\!\! absorbers

Cosmological evolution of \Ovi absorbers is a further issue that needs a better understanding. We address this issue by comparing observables (column density and Doppler parameter) and physical parameters like temperature, volume density and metallicity at $z=0$ and $z>2$ (Sect.~\ref{simulations}).

\end{itemize}



\chapter{Analysis techniques and spectral data}\label{framework}

  \section{Basics of absorption lines spectroscopy}\label{line_spectroscopy}

\hspace{0.6cm}We review briefly here some basic physical quantities and processes that are used in the analysis of absorbing systems. All information about the processes in the IGM is based on the detected radiation; in particular, on the absorption line spectra. In this chapter we present some elements of the theory of radiation, the important line broadening mechanisms and some methods and tools of absorption spectroscopy. 

\subsection{Elements of the theory of radiation}\label{radiative_transfer}

\hspace{0.6cm}Below we review some photometric quantities, the interaction between radiation and matter and the equation of radiative transfer\footnote{~The main source or this section is \url{http://zuserver2.star.ucl.ac.uk/~idh/PHAS2112/Lectures/Current} part1.pdf}.

\subsubsection{Basic notions}

\begin{itemize}
\item {\it Specific intensity} $I_{\nu}$ is defined as the rate of radiation energy flow $dE_{\nu}/dt$ per unit frequency interval $d\nu$, per unit area $dS$ and per unit solid angle $d\Omega(\theta,\phi)$:

\begin{equation}
\label{eq_Specific_intensity}
I_{\nu} (\theta,\phi) = \frac{dE_{\nu}}{dS ~dt~ d\nu ~d\Omega}=\frac{dE_{\nu}}{dA ~cos\theta ~dt~ d\nu ~d\Omega}   ~~~~~[\rm J ~m^{-2} ~s^{-1} ~Hz^{-1} ~sr^{-1}], 
\end{equation}

where $\theta$ is the (polar) angle between the sightline and the normal vector to $dS$, $\phi$ is the azimuthal angle and $dA$ is the unit area. Equation \ref{eq_Specific_intensity} defines the monochromatic specific intensity as a pencil of radiation. The total specific intensity is the integral over all frequencies:

\begin{equation}
\label{eq_total_Specific_intensity}
I =   \int\limits_{0}^{\infty}  I_{\nu}~d\nu             ~~~~~[\rm J ~m^{-2} ~s^{-1} ~sr^{-1}]. 
\end{equation}

        \item {\it Mean intensity} $J_{\nu}$ is the average of specific intensity over all solid angles:

\begin{equation}
\label{eq_mean_intensity}
J_{\nu} =   \frac{1}{4\pi}\oint  I_{\nu}~d\Omega =\frac{1}{4\pi} \int\limits_{0}^{2\pi}~d\phi~\int\limits_{0}^{\pi}~{I_{\nu}~\rm sin}\theta~ d\theta ~~~~~[\rm J ~m^{-2} ~s^{-1} ~Hz^{-1}~sr^{-1}]. 
\end{equation}

Using the designation $\mu={\rm cos}\theta$, one obtains $\oint d\Omega=-\int\limits_{0}^{2\pi}~d\phi \int\limits_{+1}^{-1} d\mu= \int\limits_{0}^{2\pi}~d\phi \int\limits_{-1}^{+1} d\mu$ and hence:

\begin{equation}
\label{eq_mean_mu}
J_{\nu} =   \frac{1}{4\pi}\int\limits_{0}^{2\pi}~d\phi\int\limits_{-1}^{+1}  I_{\nu}(\mu,\phi)~d\mu.  
\end{equation}

In plane-parallel media the radiation field is independent on $\phi$ (symmetry in regard to the $Z$-axis) and then the mean intensity can be written in the form:

\begin{equation}
\label{eq_mean_intensity_plane_parallel}
J_{\nu} =   \frac{1}{2}\int\limits_{-1}^{+1}  I_{\nu}(\mu)~d\mu.  
\end{equation} 

It follows from equation \ref{eq_mean_intensity_plane_parallel} that if the specific intensity is isotropic, i.e. independent of $\theta$, then $J_{\nu}=I_{\nu}$.

        \item  {\it Physical flux} $F_{\nu}$ is the net rate of radiation energy flow from all directions per unit area, per unit time and per unit frequency interval:     

\begin{eqnarray}
\label{eq_monochromatic_flux}
F_{\nu} & = &\oint I_{\nu}~{\rm cos}\theta~ d\Omega=\int\limits_{0}^{\pi}d\theta~\int\limits_{0}^{2\pi}~d\phi~I_{\nu} ~{\rm cos}\theta~{\rm sin}\phi    \nonumber\\
~ & = & \int_{0}^{2\pi}d\phi \int_{-1}^{+1}~I_{\nu}(\mu,\phi)~\mu~d\mu ~~~~~[\rm J ~m^{-2} ~Hz^{-1}].
\end{eqnarray}

In plane-parallel media, this formula is simplified:

\begin{equation}
\label{eq_monochromatic_flux_2}
F_{\nu}=2\pi\int_{-1}^{+1}~I_{\nu}(\mu)~\mu~d\mu.
\end{equation}

Specific intensity $I_{\nu}$ does not depend on the distance to the source $r$ while $F_{\nu}$ decreases as $r^{-2}$. 
The specific intensity can be measured only if the source is resolved; otherwise, only the physical flux can be measured.
\newpage
        \item Mean {\it radiation energy density $U_{\nu}$} is the energy density per unit frequency interval in a given volume $V$:

\begin{equation}
\label{eq_radiation_energy}
U_{\nu}~d\nu=\frac{1}{V}\oint_{V}\oint_{\Omega} dE_{\nu}=\frac{1}{c}\oint_{\Omega}I_{\nu~}d\nu~d\Omega,
\end{equation}

or:

\begin{equation}
\label{eq_radiation_energy_1}
U_{\nu}=\frac{1}{c}\oint_{\Omega}I_{\nu}~d\Omega=\frac{4\pi}{c}~J_{\nu}     ~~~~~~~~[\rm J~m^{-3}~Hz^{-1}] ~~~~~~ (\rm see ~Equation~ \ref{eq_mean_intensity}).
\end{equation}

The total radiation energy density is the integrated $U_{\nu}$ over all frequencies: \\$U=\int\limits_{0}^{\infty}~U_{\nu}~d\nu$.

        \item {\it Absorption}

When a radiation ray passes through gas clouds it loses part of its energy through scattering. Photons can be absorbed and re-emitted at different frequency -- this process of transformation of radiative energy into other forms is called ''true`` absorption. In another physical case, light intensity decreases due to scattering whereas photons are just redirected without destruction. The absorption coefficient gives the fraction of the total loss of energy in the pencil of radiation due to "true" absorption. It is related to the microphysics of particles -- how likely they absorb an incident photon. The change in intensity $dI_{\nu}$ due to true absorption along length unit $ds$ is given in a form:

\begin{equation}
\label{eq_energy_loss}
dI_{\nu}=-\sigma_{\nu}~n~I_{\nu}~ds,
\end{equation}

where $n$ is the volume density of absorbing particles and $\sigma_{\nu}$ is the absorption cross-section per particle, which is equivalent to the absorption coefficient in units of area. 

In a homogeneous medium at rest, the absorption coefficient is isotropic. However, if the medium is moving with respect to the observer, $\sigma_{\nu}$ depends on the angle between the photon direction and the radius vector at the point of observation, and on the frequency, due to the Doppler effect.

\item {\it Emission}

An increase in the pencil of radiation energy, due to de-excitation of atoms, is called "true" emission. The emission coefficient (or, monochromatic emissivity) $j_{\nu}$ is defined as the energy $dE_{\nu}$ generated per unit volume, per unit time, per unit frequency interval and per unit solid angle:

\begin{equation}
\label{eq_emission}
j_{\nu}=\frac{dE_{\nu}}{dV~dt~d\nu~d\Omega}     ~~~~~~~~~~[\rm J~m^{-3}~s^{-1}~Hz^{-1}~sr^{-1}].
\end{equation} 
\newpage
Hence the increase of the intensity along an elementary length $ds$ is:

\begin{equation}
\label{eq_emission}
dI_{\nu}=j_{\nu}(s)ds.
\end{equation}

Like in the case with absorption coefficient, emissivity is isotropic in a homogeneous medium at rest, but is angle dependent and anisotropic in moving medium, due to Doppler shift, aberration and advection. 

\end{itemize}

\subsubsection{Equation of Radiative Transfer}

\hspace{0.6cm}The change of specific intensity of radiation energy from point to point is expressed by the equation of radiative transfer. If a beam of radiation passes through intervening material along a path-length $ds$, its change  between the points $s$ and $s+ds$ is caused by emission and absorption effects in the medium:

\begin{equation}
\label{eq_rad_transfer}
dI_{\nu}=(j_{\nu}-k_{\nu}I_{\nu})ds
\end{equation}

or:

\begin{equation}
\label{eq_rad_transfer_1}
\frac{dI_{\nu}}{ds}=j_{\nu}-k_{\nu}I_{\nu}=-k_{\nu}(I_{\nu}-S_{\nu}),
\end{equation}

where $k_{\nu}\equiv \sigma_{\nu}n/\rho$ is the mass absorption coefficient (or, the opacity per unit mass), $\rho$ is the mass density and $S_{\nu}=j_{\nu}/k_{\nu}$ is the so called {\it source function}. In the case of LTE, the source function of the intervening material is equal to the intensity of radiation energy: $S_{\nu}=B_{\nu}(T)$. The latter is called {\it Planck function} and depends only on temperature. Hence, in LTE, Kirchhoff's law holds: $j_{\nu}=k_{\nu}B_{\nu}(T)$. 
    
The product of mass absorption coefficient $k_{\nu}(s)$ and length $ds$ is a measure of the optical thickness of the material, called {\it optical depth}: $d\tau_{\nu}=k_{\nu}(s)ds$. The cumulative effect is expressed through the integral along the line of sight: $\tau_{\nu}=\int\limits_{0}^{D}k_{\nu}(s)ds$, where $D$ is the distance of path length. Using this definition, the equation of radiative transfer can be re-written in the form:

\begin{equation}
\label{eq_rad_transfer_2}
\frac{dI_{\nu}}{d\tau_{\nu}}=S_{\nu}-I_{\nu}.
\end{equation}

The formal solution of the radiative transfer equation is:

\begin{equation}
\label{eq_rad_transfer_3}
I_{\nu}(s)=I(0)e^{-\tau(s,0)}+\int\limits_{0}^{s} S_{\nu}(s')e^{-\tau(s,s')}k_{\nu} ds',
\end{equation}

where $\tau_{\nu}(s,s')$ is the optical thickness (or depth) of absorbing material between points $s$ and $s'$. The equation shows that the radiation intensity at any point and in a given direction is sum of the emissions at all points $0\le s'\le s$, reduced by the factor $e^{-\tau(s,s')}$  which reflects the absorption by the intervening material \citep{Chandrasekhar60}.

\subsection{Line shape and broadening}\label{line_broadening}

\hspace{0.5cm}Let us consider\footnote{~The main source for this section is \url{http://www.ucolick.org/~krumholz/courses/spring10_ast230} Class 5} a population of particles of a given element $X$ in an energy state $X_{l}$ and of volume density $n_{l}$ which interacts with a population of photons with intensity $I_{\nu}$. If $E_{l}$ and $E_{u}$ are the energies of the lower and upper energy states of the element, respectively, photons with frequencies $h\nu=E_{u}-E_{l}$ can be absorbed. The relative probability that a photon with frequency $\nu$ will be absorbed can be expressed by definition with the line profile function, $\phi_{\nu}$, normalized so that:

\begin{equation}
\label{eq_line_profile}
\int \phi_{\nu} d\nu=1 .
\end{equation}

From the other hand, the cross section, $\sigma_{\nu}$, is given by:

\begin{equation}
\label{eq_line_profile_cross-section}
\int \sigma_{\nu}(lu)d\nu=\frac{g_{u}}{g_{l}}\frac{c^{2}}{8\pi\nu^{2}_{ul}}A_{ul} .      
\end{equation}

where $\nu_{ul}=(E_{u}-E_{l})/h$ is the frequency corresponding to the exact energy difference between the levels $u$ and $l$ and $g_{u}$ and $g_{l}$ are their statistical weights\footnote{~The number of different quantum states (sublevels) in a given energy level, i.e. the degree of degeneracy.}. The constant $A_{ul}$ is called {\it Einstein coefficient} and it is a measure of probability for spontaneous transition from upper to lower energy state. It is related to the intrinsic properties of the element energy levels. In astrophysics, the absorption-oscillator strength, $f_{lu}$, is often used instead of the Einstein coefficients. The relation between them is: 

\begin{equation}
\label{eq_line_profile_cross-section_1}
A_{ul}=\frac{8\pi^{2}e^{2}\nu^{2}}{m_{e}c^{3}}\frac{g_{l}}{g_{u}}f_{lu}  ~~~~~~[ s^{-1} ].
\end{equation}

Combining equations \ref{eq_line_profile}, \ref{eq_line_profile_cross-section} and \ref{eq_line_profile_cross-section_1}, one obtains:

\begin{equation}
\label{eq_line_profile_cross-section_2}
\sigma_{\nu}(lu)=\frac{g_{u}}{g_{l}}\frac{c^{2}}{8\pi\nu^{2}_{ul}}A_{ul}\phi_{\nu}=\frac{\pi e^{2}}{m_{e}c}f_{lu}\phi_{\nu} .      
\end{equation}

Clearly, the function $\phi_{\nu}$ contains all the information of how the absorption cross section depends on frequency. In other words, the line profile is simply a representation of this dependency.

\subsubsection{Natural broadening}

\hspace{0.6cm}The uncertainty principle says that the momentum (energy, velocity) and the position of a particle can not be precisely determined at the same time. Therefore an intrinsic quantum effect of line broadening takes place: the so called {\it natural broadening}. A good approximation of the line profile due to natural broadening is given through the Lorentz profile:
\newpage
\begin{equation}
\label{eq_natural_broadening}
\phi_{\nu}\approx\frac{4\gamma_{ul}}{16\pi^{2}(\nu-\nu_{ul})^{2}+\gamma_{ul}^{2}} ,      
\end{equation}

where $\gamma_{ul}=1/\tau_{u}+1/\tau_{l}=\sum\limits_{j<u}A_{uj}+\sum\limits_{j<l}A_{lj}$ has a dimensionality of frequency and $\tau_{u}$ and $\tau_{l}$ are the lifetimes of the upper and lower energy states, respectively. Thus the profiles of naturally broadened lines can be computed from the Einstein coefficients; their full width at half maxima are $\Delta\nu_{\rm FWHM}=\gamma_{ul}/2\pi$. Typical line widths for allowed optical and UV absorptions, produced by natural broadening, are $\sim0.01$ \kms\!\!.

\subsubsection{Doppler broadening}

\hspace{0.6cm}At finite temperature, the vast majority of a particle population has velocities (velocity dispersion) in some limited range. The Doppler effect allows absorption and emission processes in a range of frequencies around the frequency of given line. Therefore a second source of line broadening is the {\it Doppler broadening}. For a gas with Maxwellian velocity distribution, the fraction of particles $f_{v}$ in velocity interval $[v,\,v+dv]$ is: 


\begin{equation}
\label{eq_Maxwellian_distr}
f_{v} dv=\frac{1}{\sqrt{2\pi\sigma^{2}_{v}}}e^{-v^{2}/2\sigma^{2}_{v}} dv, 
\end{equation}

where 
$\sigma_{v}$ is (thermal) velocity dispersion, $\sigma_{v}=\sqrt{kT/m}$. In spectroscopy, the broadening (Doppler) parameter $b=\sqrt{2}\sigma_{v}=\sqrt{2kT/m}$ is widely used instead of the velocity dispersion. Obviously, if the line width is dominated by thermal motions, the gas temperature is directly derived from the Doppler parameter (in km~$\rm s^{-1}$) and the atomic mass number $A$ of the element:

\begin{equation}
\label{eq_temp_Doppler}
T(K) = \frac{mb^{2}}{2k} = A\Big(\frac{b}{0.129}\Big)^{2}~~~~[~{\rm K}~].
\end{equation}

The Maxwellian velocity distribution can be transformed in terms of frequency instead of velocity, using the Doppler width $\Delta\nu_{D}=(v/c)\nu_{ul}=(\sqrt{2}\sigma_{v}/c)\nu_{ul}=(\sqrt{2kT/m}/c)\nu_{ul}$ and thus a Gaussian line profile function with dispersion $\sigma_{\nu}=(\sigma_{v}/c)\nu_{ul}$ is obtained:


\begin{equation}
\label{eq_temp_Doppler_1}
\phi_{\nu} = \frac{1}{\sqrt{2\pi\sigma^{2}_{\nu}}}e^{-\nu^{2}/2\sigma^{2}_{\nu}}= \frac{1}{\sqrt{\pi}\Delta\nu_{D}}e^{-\nu^{2}/\Delta\nu_{D}^{2}}.
\end{equation}

Some physical conditions require to take into account, besides the thermal motion,  the bulk motions in the gas, e.g. {\it turbulent} flows. Then the {\it effective} Doppler width of a line is a sum of a thermal and a turbulent components:

\begin{equation}
\label{eq_temp_Doppler_2}
\Delta\nu_{D}^{\rm eff}=\frac{\nu_{0}}{c}\sqrt{\frac{2kT}{m}+v^{2}_{\rm turb}},
\end{equation}

where $\nu_{0}$ is the centroid frequency of the absorption line.

Under interstellar and intergalactic conditions the effect of Doppler broadening is much greater than that of natural broadening, because the gas speed is typically much higher than $\sim0.01$ \kms\!\!. 

\subsubsection{Voigt profile}\label{voigt_function}

\hspace{0.6cm}Both natural and Doppler broadening influence the profile of absorption or emission lines. Therefore the true line profile is a convolution of the Gaussian  (Eq.~\ref{eq_temp_Doppler_1}) and Lorentz (Eq.~\ref{eq_natural_broadening}) profiles. This convolution is the well-known {\it Voigt profile}:


\begin{equation}
\label{eq_Voigt}
\phi_{\nu} = \frac{1}{\sqrt{2\pi\sigma^{2}_{v}}}\int\limits_{-\infty}^{\infty}e^{-v^{2}/2\sigma^{2}_{v}}\frac{4\gamma_{ul}}{16\pi^{2}(\nu-(1-v/c)\nu_{ul})^{2}+\gamma_{ul}^{2}}dv.
\end{equation}

\begin{figure}[h!]
\begin{center}
\resizebox{0.7\hsize}{!}{\includegraphics[angle=-90]{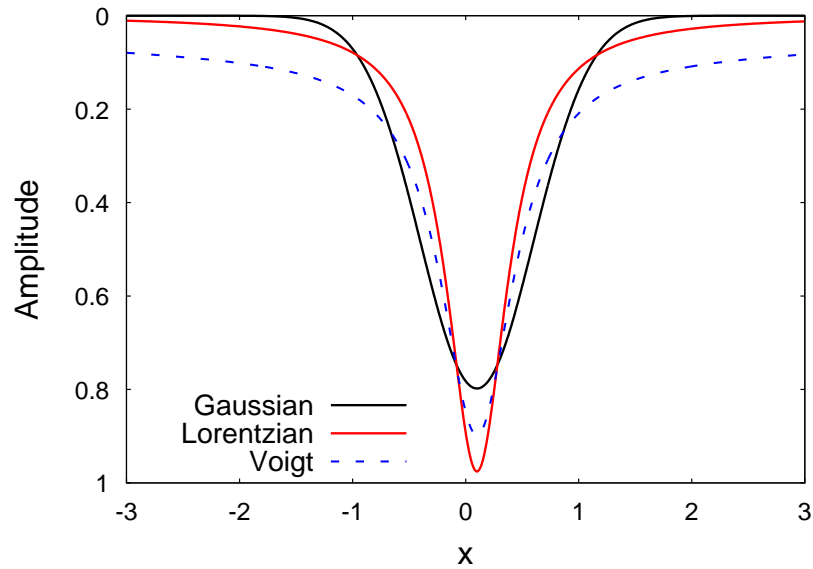}}
\caption[Doppler (Gaussian), Lorentz and Voigt line profiles.]{\footnotesize Doppler (Gaussian), Lorentz and Voigt line profiles.}
\label{fig_doppler_lorentz}
\end{center}
\end{figure}

The shape of the Voigt profile consists of a `core', dominated by a Doppler (Gaussian) profile ($\propto e^{-\nu^{2}/2\sigma^{2}_{\nu}}$) and broad `wings', described by a Lorentz profile ($\propto (\nu-\nu_{ul})^{-2}$). This behavior is illustrated in Fig.~\ref{fig_doppler_lorentz}.

\subsubsection{Other broadening mechanisms}
 
\begin{itemize}
\item Collisional (pressure) broadening

If the absorbing (or emitting) gas atoms or ions frequently collide with each other, their electron energy levels will be distorted. Subsequently, absorption or emission lines will undergo additional broadening, called {\it collisional or pressure broadening}. This effect depends on the frequency of collisions $\nu_{\rm col}=v_{\rm th}n\sigma_{\rm col}~~,$ where $v_{\rm th}=\sqrt{2kT/m}$ is the thermal velocity of the particles, $n$ is their volume density and $\sigma_{\rm col}$ is their cross section of collisions. The resulting line profile is a Lorentz one, like in the case of natural broadening. Thus both collisional and natural effects can be combined: $\Gamma=\gamma+2\nu_{\rm col}$. The effect of collisional broadening is even smaller than that of natural broadening and does not play a role in the low-density IGM.
\newpage
\item Stark and Zeeman effects

The Stark and Zeeman effects cause also distortion of the energy levels of gas atoms or ions, due to presence of an external static electric or magnetic fields, respectively. Those effects are not important in the IGM.

\end{itemize}

\subsection{Equivalent width and curve of growth}\label{curve_of_growth}

\hspace{0.6cm}Below we describe briefly some applications of the radiative transfer theory in absorption spectroscopy.\footnote{~The main source for this section is \url{http://www.ucolick.org/~krumholz/courses/spring10_ast230}, Class 8}

\subsubsection{Equivalent width}

\hspace{0.6cm}If a bright continuum point source with intensity $I_{\nu}(0)$ is observed (e.g., star or quasar) within a small solid angle $\Delta\Omega$, then the registered continuum flux $F_{\nu, \rm cont}$, free of any emission and absorption, is: $F_{\nu, \rm cont}=\int I_{\nu}(0)d\Omega=I_{\nu}(0)\Delta\Omega\equiv F_{\nu}(0).$ If the light from the point source passes through a gas cloud, uniform over $\Delta\Omega$, the radiative transfer equation of this system is: $I_{\nu}=I_{\nu}(0)e^{-\tau_{\nu}}+B_{\nu}(T_{\rm ext})(1-e^{-\tau_{\nu}})$,
where $T_{\rm ext}$ is the excitation temperature of the intervening material. Hence the actual registered flux is: 
$F_{\nu}=\int I_{\nu}d\Omega=F_{\nu}(0)e^{-\tau_{\nu}}+B_{\nu}(T_{\rm exc})\Delta\Omega(1-e^{-\tau_{\nu}})$,
assuming that the optical depth $\tau_{\nu}$ is constant. The ISM or IGM gas is usually cold, the rate of ionization is much lower than the rate of recombination ($n_{u}/n_{l}\ll1$) and thus $B(T_{\rm exc})\Delta\Omega\ll F_{\nu}(0)$. Therefore a good approximation of the actual flux is:

\begin{equation}
\label{eq_actual_flux}
F_{\nu} = F_{\nu}(0)e^{-\tau_{\nu}}.
\end{equation}

\begin{figure}[h!]
\begin{center}
\resizebox{0.7\hsize}{!}{\includegraphics{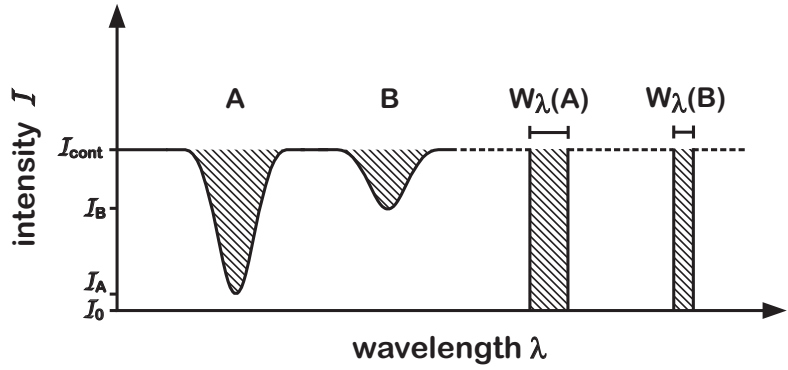}}
\caption[Equivalent widths of spectral absorption lines.]{\footnotesize Equivalent widths of two absorption lines. [The figure has been kindly provided by Philipp Richter.]}
\label{fig_eq_width}
\end{center}
\end{figure}

The optical depth $\tau_{\nu}$ is negligible except in a narrow frequency range and therefore $F_{\nu}(0)$ can be directly measured (outside this range). Knowing $F_{\nu}(0)$ on the two sides of an absorption line, one can estimate $F_{\nu}(0)$ through interpolation even in the range of strong absorption. Therefore, the equivalent width of a line can be defined as:

\begin{equation}
\label{eq_equivalent width}
W\equiv \int\limits_{0}^{\infty}\left(\frac{F_{\nu}(0)-F_{\nu}}{F_{\nu}(0)}\right)\frac{d\nu}{\nu_{0}}=\int\limits_{0}^{\infty}\left(1-\frac{F_{\nu}}{F_{\nu}(0)}\right)\frac{d\nu}{\nu_{0}}=\int\limits_{0}^{\infty}(1-e^{-\tau_{\nu}})\frac{d\nu}{\nu_{0}},
\end{equation}

where $\nu_{0}$ is the line center frequency. The equivalent width can be measured even if the absorption line is not resolved in frequency, since it is a measure of the integrated area of the absorption profile in respect to the unabsorbed local continuum, i.e., $F_{\nu}(0)$. The equivalent widths of two absorption lines of different optical depth is shown in Fig~\ref{fig_eq_width}.

There is a relation between $W$ and column density of absorbers through the optical depth, $\tau_{\nu}$. Using the definition of $\tau_{\nu}$ and \ref{eq_line_profile_cross-section_2}, one obtains:

\begin{equation}
\label{eq_tau_N}
\tau_{\nu}= \int\limits_{0}^{d}k_{\nu}ds=\int\limits_{0}^{d}\sigma_{\nu}n_{l}(s)ds=\frac{\pi e^{2}}{m_{e}c}f_{lu}\int\limits_{0}^{d}\phi_{\nu}(s)n_{l}(s)ds.
\end{equation}

If the profile function $\phi_{\nu}$ is independent of $s$, the equation above is modified:

\begin{equation}
\label{eq_tau_N_1}
\tau_{\nu}=\frac{\pi e^{2}}{m_{e}c}\phi_{\nu}f_{lu}\int\limits_{0}^{d}n_{l}(s)ds= \frac{\pi e^{2}}{m_{e}c}\phi_{\nu}N_{l}f_{lu},
\end{equation}

where $d$ is the length over which the absorption is present (i.e. the absorber size) and $N_{l}$ is the column density of the absorbing gas between the observer and the background source. As it is seen from the Eq.~\ref{eq_tau_N}, $N_{l}=\int\limits_{0}^{d}n_{l}ds$. Typically, the \Lya forest absorption lines have small column densities ($N_{\footnotesize \Hi\!\!}<10^{15}$\sqc) and undergo Doppler broadening, thus the Voigt profile can be approximated by a pure Gaussian (Eq.~\ref{eq_temp_Doppler_1}). The line profile $\phi_{\nu}$ in that case reaches a peak value of $1/\sqrt{2\pi\sigma_{v}^{2}}=1/\sqrt{\pi}b$ at $\nu_{0}=\nu_{ul}$, where $b=\sqrt{2}\sigma_{v}$ is the Doppler broadening parameter. Then the (maximal) optical depth  $\tau_{0}$ at the line center $\lambda_{ul}=c/\nu_{0}$ is

\begin{equation}
\label{eq_tau_N_2}
\tau_{0}=\sqrt{\pi}\frac{e^{2}}{m_{e}c}\frac{f_{lu}\lambda_{ul}N_{l}}{b},
\end{equation}

while the optical depth in the Doppler part (`core') of the line profile is described by

\begin{equation}
\label{eq_tau_Doppler}
\tau_{\nu}=\tau_{0}e^{-u^{2}/b^{2}},
\end{equation}

where $u=c(\nu-\nu_{0})/\nu_{0}$ is the required velocity shift for producing a frequency shift $\nu$.

\subsubsection{Curve of growth}

\hspace{0.6cm}The equivalent width of a line is an increasing function of $\tau_{0}$, depending on the Doppler parameter, the oscillator strength, the wavelength and the column density. The function $W(\tau_{0})$, or $W(N)$ is labeled {\it curve of growth}. Three limiting cases of the line profile lead to different behavior of the $W$:
\newpage
\begin{itemize}
\item {\it Optically thin lines} ($\tau_{0}\ll 1$)

In that case the factor $(1-e^{\tau_{\nu}})$ in Eq.~\ref{eq_equivalent width} can be approximated by use of Maclaurin series $e^{x}=\sum\limits_{0}^{\infty}x^{n}/n!$. Hence one obtains for the equivalent width to second order:

\begin{equation}
\label{eq_equivalent width_1}
W =\int\limits_{0}^{\infty}(1-e^{-\tau_{\nu}})\frac{d\nu}{\nu_{0}}\approx\int\limits_{0}^{\infty}\left(\tau_{\nu}-\frac{\tau_{\nu}^{2}}{2}\right)\frac{d\nu}{\nu_{0}},
\end{equation}

The absorption in optically thin lines can described almost always through a Doppler core and therefore $\tau_{\nu}$ can be approximated through a Doppler form (Eq.~\ref{eq_tau_Doppler}). Then the integral \ref{eq_equivalent width_1} is obtained straightforwardly:

\begin{equation}
\label{eq_equivalent width_2}
W =\sqrt{\pi}\frac{b}{c}\tau_{0}\left(1-\frac{\tau_{0}}{2\sqrt{2}}  \right).
\end{equation}

For small $\tau_{0}$ the second term in the brackets can be neglected and applying Eq.~\ref{eq_tau_N_2} the equivalent width in the limit of optically thin lines becomes:

\begin{equation}
\label{eq_equivalent width_3}
W =\sqrt{\pi}\frac{b}{c}\tau_{0}=\pi\frac{e^{2}}{m_{e}c^{2}}f_{lu}\lambda_{ul}N_{l}.
\end{equation}

The constant of proportionality depends only on atomic constants for the considered line. Thus, for a given equivalent width $W$, the corresponding column density $N_{l}$ is directly known. This behavior is illustrated in Fig.~\ref{fig_curve_of_growth}. The linear part of the curve of growth represents the case of optically thin absorption lines. Obviously, the equivalent width in this regime does not depend on the Doppler parameter.

\begin{figure}[h!]
\begin{center}
\resizebox{0.7\hsize}{!}{\includegraphics{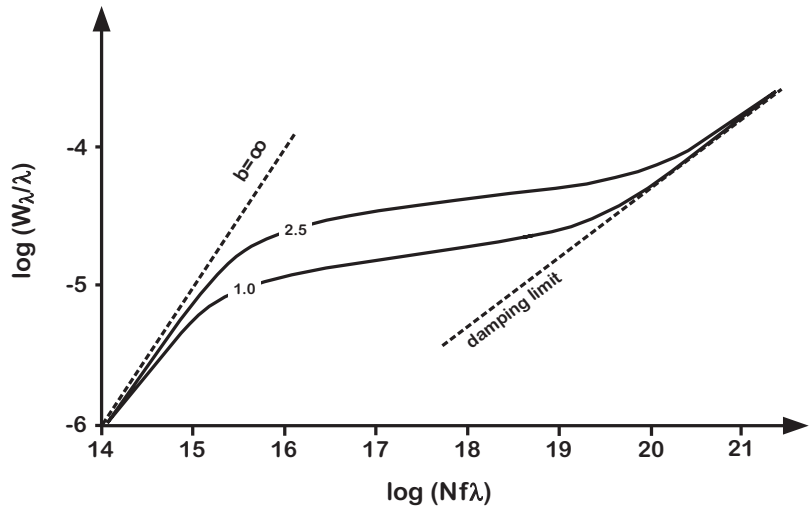}}
\caption[Curve of growth composed from 3 limiting cases of line profile.]{\footnotesize Curve of growth composed from 3 limiting cases of line profile. The solid curves are derived for $b =$ 1.0 and 2.5 \kms\!\!. The flat part of the curve corresponds to saturated profiles. For $N(\Hi\!\!) > 10^{20}$ \sqc\!\!, the profile develops damping wings, which dominate the equivalent width. [The figure has been kindly provided by Philipp Richter.]}
\label{fig_curve_of_growth}
\end{center}
\end{figure}

\item {\it Saturated lines} ($\tau_{0}> 1$)

In that case there are no photons with frequencies around the line center available to be absorbed. Therefore only a small fraction of gas particles with velocities essentially different than the most probable speed (away form the line center) contribute to the increase of the equivalent width -- adding more absorbers on the sightline would lead to a sub-linear increase of this quantity. The line-profile approximation through the Gaussian `core' (Eq.~\ref{eq_tau_Doppler}) is still possible, but the factor $(1-e^{\tau_{\nu}})$ in Eq.~\ref{eq_equivalent width} is replaced by an inverted `top-hat function': equal to 0 near $\nu_{0}$ and equal to 1, otherwise. Then the full width at half maximum (FWHM) is taken to be the hat width:

\begin{equation}
\label{eq_saturated_lines}
\frac{1}{2}=\exp(-\tau_{\rm FWHM})=\exp\big(-\tau_{0}e^{-[\Delta u_{\rm FWHM}/2]^{2}/b^{2}}\big).
\end{equation}

Solving this equation, one obtains for the equivalent width:

\begin{equation}
\label{eq_saturated_lines_1}
W=\frac{(\Delta u)_{\rm FWHM}}{c}=\frac{(\Delta \nu)_{\rm FWHM}}{\nu_{0}}=\frac{2b}{c}\sqrt{\rm ln(\tau_{0}/\rm ln 2)}.
\end{equation}

As expected, this formula shows that $W$ increases as the square root of the log $N_{l}$ (see Fig.~\ref{fig_curve_of_growth}).

\item {\it Optically thick (damped) lines} ($\tau_{0}\gg 1$)

In this case, the saturation around the line center extends beyond the Doppler `core'. Therefore, a good approximation of the line profile is to consider only its Lorentzian part:

\begin{equation}
\label{eq_damped_lines}
\tau_{\nu}=\frac{\pi e^{2}}{m_{e}c}N_{l}f_{lu}\frac{4\gamma_{lu}}{16\pi^{2}(\nu-\nu_{0})^{2}+\gamma_{lu}^{2}}.
\end{equation}

Inserting this relation into Eq.~\ref{eq_equivalent width} is obtained:

\begin{equation}
\label{eq_damped_lines_eqw}
W=\int\limits_{0}^{\infty}\left(1-{\rm exp} \left[-\frac{\pi e^{2}}{m_{e}c}N_{l}f_{lu}\frac{4\gamma_{lu}}{16\pi^{2}(\nu-\nu_{0})^{2}+\gamma_{lu}^{2}}  \right]\right)\frac{d\nu}{\nu_{0}},
\end{equation}

which solution is:

\begin{equation}
\label{eq_damped_lines_eqw_1}
W=\sqrt{\frac{e^{2}}{m_{e}c^{2}}N_{l}f_{lu}\lambda_{lu}\left(\frac{\gamma_{lu}\lambda_{lu}}{c} \right)}=\sqrt{\frac{b}{c}\frac{\tau_{0}}{\sqrt{\pi}}\frac{\gamma_{lu}\lambda_{lu}}{c}},
\end{equation}

Thus, in the limiting case of damped lines the equivalent width is proportional to the square root of column density: $W \propto \sqrt{N_{l}}$ (see Fig.~\ref{fig_curve_of_growth}).

\end{itemize}


\section{Methods and tools of analysis}\label{methods_and_tools}

\subsection{Absorption line measurement techniques}\label{methods}

\hspace{0.6cm}There are different ways to estimate the column density of absorbing gas, depending on the considered line characteristic: analyzing the optical depth, fitting Voigt profiles to absorption lines, or examine equivalent widths and constructing curve of growths. Below we present some of the most common techniques:

\begin{itemize}
      
\item {\it The curve-of-growth method}

The curve of growth can be used to measure column densities of different species. This method is efficient for low-resolution data wherein the line profile is not resolved. In principle, the recorded line shape is a convolution between the intrinsic shape and the instrumental broadening function. The instrumental broadening is caused by the imperfection of the optical systems of the telescope and the spectrograph. If it is larger than the intrinsic line width the information about the line width is lost. However, the equivalent width is independent on the instrumental broadening, since the latter effect only redistributes $W$ over frequency, without changing its value. Thus, by measuring $W$, it is possible to recover the column density $N$ of an absorber from the linear part and the square-root part of the curve of growth. The logarithmic part of the curve does not provide a good estimate of $N$. In general, if a line is not resolved and its shape can not be directly reproduced, the optical depth at the 
line center is unknown. Then it is unknown to which part of the curve of growth the considered species belong: the same value of $W$ can imply smaller or larger $N$, for larger or smaller values of the Doppler parameter, respectively.

However, the problem with unknown optical depth at the line center can be solved by use of doublet or multiplet transitions\footnote{~Several transitions, with the same atomic level and different $f\lambda$.}. A line doublet is observed when transitions are possible from an absorbing state $l$ to two different exited states $u_{1}$ and $u_{2}$ with a small energy separation due to the atomic fine structure. If the spectral resolution is good enough, both equivalent widths of the doublet lines can be measured. In the considered three limiting cases, their ratio is determined only by the known atomic constants\footnote{~Some exceptions can occur in case of saturated lines.}. Thus this ratio provides an information on what part of the curve of growth the doublet is.
If several transitions from the same atomic level and with different $f\lambda$ take place, an empirical curve of growth can be constructed and hence estimates of $N$ and $b$ can be obtained as well.

\item {\it The Voigt-profile-fitting method} 

In case the resolution of given absorption spectral lines is high enough, i.e., the lines are resolved, 
the Voigt-profile fitting technique can be applied. It provides the best-fit values of column density, Doppler parameter and redshift for each component of the absorption feature. To apply that technique, a polynomial fit of the QSO continuum is required (for the other methods as well), 
since the absorption lines are measured in relation to the continuum. Usually, a $\chi^{2}$-minimization is used to decompose the spectrum into several independent Voigt-profile components, as many as necessary in order to make the procedure free from effects of random fluctuations, i.e., to obtain the same value of the $\chi^{2}$-minimum many times with the same setup. The fitting procedure is quite general and can be used for any profile.

\item {\it The apparent-optical-depth method} 

This method was first introduced by \citet{Savage91}. It distinguishes between ``true'' and ``apparent'' optical depth. The ``true'' $\tau_{\nu}$ is the natural logarithm of the ratio of the continuum flux $F_{\nu}(0)$ and the actual absorbed flux $F_{\nu}$ (Eq.~\ref{eq_actual_flux}). However, the recording instrument has a finite resolution, defined by its spectral spread function, which leads to the already mentioned instrumental broadening. Therefore the actual absorption flux has to be convolved with the spectral spread function, in order to extract information about the observed absorption flux, $F_{\nu,\rm obs}$, which differs from the actual flux $F_{\nu}$. Then the ``apparent'' optical depth is the natural logarithm of the ratio of the continuum flux $F_{\nu}(0)$ and the absorption flux that includes the instrumental broadening $F_{\nu,\rm obs}$.

The apparent-optical-depth method is applicable if the absorption lines are weak and not (or, mildly) saturated. It can treat single unsaturated lines, as well doublets and multiplets.
The observational data are converted to apparent optical depth and further to apparent column density $N_{a}$ per unit velocity interval. A comparison of different $N_{a}$ for doublets and multiplets enables empirical estimates of the line saturation in the true line profile. 
The apparent optical depth method provides additional information on the velocity dependence of line saturation. 

\end{itemize}

\subsection{Absorption line fitting tools}\label{tools}

\hspace{0.6cm}In this section we describe briefly the two fitting tools that we have used for analysis of our observational data (see Sect.~\ref{data_uves}). We applied the tool {\sc Candalf} for handling the high- and intermediate- resolution samples of our case study of two \Ovi absorbers (see Chapter~\ref{paper1}). The other tool, {\sc Vpfit}, has been applied for fitting the intermediate-resolution observational spectra for our UVES survey (see Chapter~\ref{uves}).

\begin{itemize}

\item {\sc Candalf}

The {\sc Candalf} routine \footnote{~Written by Robert Baade, Hamburger Sternwarte} is aimed at fitting absorption spectra with a Gaussian profile by use of a standard Levenberg-Marquard minimization algorithm. The program simultaneously fits the continuum and the absorption lines, producing as output ion column densities $N$ and Doppler parameters $b$ for each absorption component. The continuum is modeled as a Legendre polynomials of order up to 4. The one-$\sigma$ fitting uncertainties of the obtained estimates of $N$ and $b$ are calculated using the diagonals of the Hessian matrix. 
\newpage
\item {\sc Vpfit}

The {\sc Vpfit} program \citep{Carswell03} is created to fit observed normalized absorption spectra with multi-component Voigt profiles\footnote{~More technical information is available at: \url{http://www.ast.cam.ac.uk/~rfc/vpfit.html} and \url{http://www.ast.cam.ac.uk/~rfc/vp_errest.html}}. The Voigt profiles are convolved with the instrumental profile and then iteratively fitted to the absorption spectra, until the $\chi^{2}$ value is minimized. 




\end{itemize}

\subsection{Ionization modeling with  Cloudy}\label{cloudy}

\hspace{0.6cm}The software package {\sc Cloudy} is a spectral-synthesis code designed for simulation of the physical conditions in astrophysical plasma and of its emission spectra \citep{Ferland}, with emphasis on achieving reliable estimates of abundances and luminosities of galactic and extragalactic objects. The emitting gas is not in thermodynamic equilibrium and thus an analytical approach for its physical description cannot be used. Therefore numerical simulations of micro-level processes responsible for the observed spectra are performed. As pointed out by \citet{Ferland}, the structure and the method of such simulation codes (including {\sc Cloudy}) are similar: an optically thick slab of gas is divided into zones with approximately constant physical conditions. A given ionization level is maintained by balancing ionization and recombination processes. The former include photoionization, Auger ionization and collisional ionization and the latter are radiative recombination, low- and high-temperature 
dielectronic recombination, three-body recombination and charge transfer. The important assumptions are that the velocity distribution of the free electrons is Maxwellian and that the kinetic temperature of electrons is determined by balance between heating and cooling processes. The heating sources can be various: mechanical or photoelectric processes, cosmic rays etc. Main cooling mechanism are the inelastic collisions between electrons and other particles. 

{\sc Cloudy} solves simultaneously the radiative transfer equations in the continuum and in the lines. To predict line intensities and column densities, it is required to specify the incident continuum, the gas density and the chemical composition. The code is widely used for analysis of emission and absorption line spectra.


\section{Spectral data used}\label{data}

\subsection{Observational data (VLT/UVES)}\label{data_uves}

\hspace{0.6cm}The observational data set used in this work is composed of {\it intermediate}- and {\it high}-resolution spectra taken with the UVES spectrograph at the VLT. The VLT, installed at Cerro Paranal (Chile), is designed for visible and infrared observations. It consists of four 8.2-meters Unit Telescopes; when they operate in a combined mode, the VLT provides a light collecting power of 16-meter telescope. 
The UVES is a high-resolution optical spectrograph, installed on the UT2 of the VLT. The telescope light beam is separated in two arms: UV-to-Blue and Visual-to-Red with maximum resolutions of 80,000 and 110,000, respectively \citep[for more details see ][]{Dekker00}. 


The {\it intermediate}-resolution data set consists of 15 QSO sightlines of spectra with resolution $R\approx45 000$, corresponding to velocity resolution $\sim 6.6$ \kms FWHM. The data were obtained and reduced as part of the former ESO Large Programme "The Cosmic Evolution of the IGM" which was aimed at study of the IGM at $z=1.5-5$ and along a large number of lines of sight, in order to improve the statistics on intervening absorption line systems (see Sect.~\ref{types_lines}) and to derive their physical properties \citep{Bergeron02}. The wavelength coverage of the intermediate resolution data is $3050-10,400$~\AA. The signal-to-noise (S/N) ratio varies between $15$ and $90$ per spectral resolution element.

The {\it high}-resolution data set consists of spectra of the quasar PKS\,1448$-$232 ($z_{\rm em}=2.208$; $V=16.9$), observed at the VLT with the UVES spectrograph in 2007, in an independent run (program ID 079.A$-$0303(A)). The spectral resolution is $R\approx75,000$ which corresponds to $\sim 4$ km\,s$^{-1}$ FWHM velocity resolution. The full wavelength coverage is $3000-6687$~\AA. The raw data were reduced using the UVES pipeline implemented in the ESO-MIDAS software package. The pipeline reduction includes flat-fielding, bias- and sky-subtraction and a relative wavelength calibration. Then the individual spectra have been corrected to vacuum wavelengths and co-added. The S/N ratio of the high-resolution data is $20-70$ per resolution element.

\subsection{Numerical simulations (OWLS)}\label{owls}

\begin{figure}[h!]
\begin{center}
\resizebox{0.9\hsize}{!}{\includegraphics{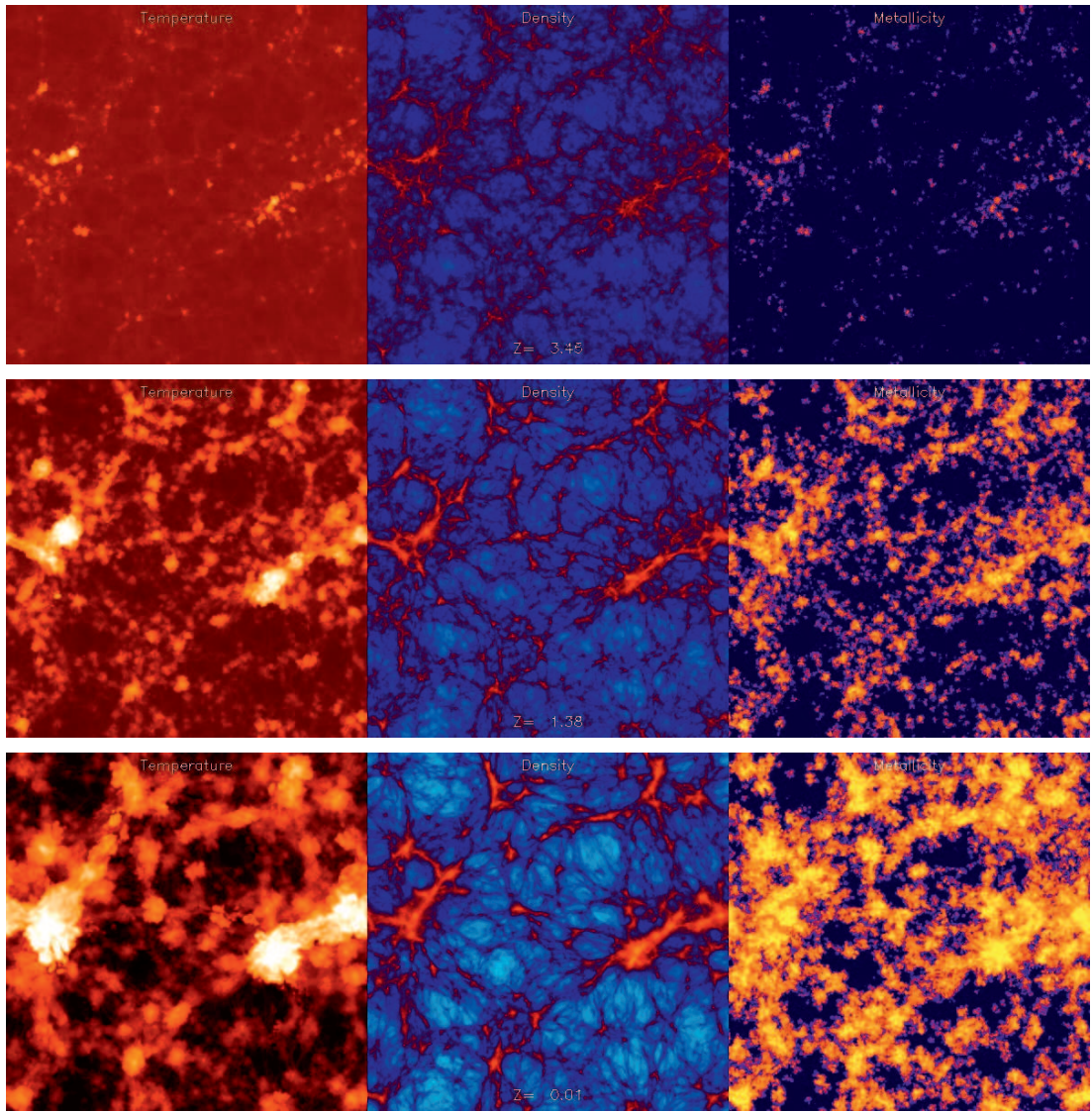}}
\caption[The evolution of the `cosmic web' according to the OWLS.]{The evolution of the `cosmic web' according to the OWLS: in terms of temperature, volume density and metallicity (left to right) and at three different redshifts (top to bottom). Credit: Craig Booth}
\label{fig_owls}
\end{center}
\end{figure}

\hspace{0.6cm}The {\it synthetic} spectra, analyzed in this work, are generated from a run of the OWLS. The OWLS are a large set of N-body, Smoothed Particle Hydrodynamical (SPH) simulations of structure formation in the Universe \citep{Schaye10}. The {\sc Gadget III} code\footnote{~More information on the {\sc Gadget} code can be found at: \url{http://www.mpa-garching.mpg.de/galform/gadget/index.shtml.}} is used for calculating the gravitational and hydrodynamical forces on the system of particles. The values of cosmological parameters adopted in the runs are typical for a flat $\Lambda$CDM cosmological model and in agreement with the results from 3-year WMAP data: $\Omega_{\rm m}$=0.238, $\Omega_{\rm b}$=0.0418, $\Omega_{\rm \Lambda}$=0.762, $\sigma_{\rm 8}$=0.74, $n{\rm s}$=0.95, $h$=0.73 \citep{Spergel07}. The OWLS use starformation recipe by \citet{Vecchia08}. 
Galactic winds from core-collapse SNe are also implemented as described by \citet{Vecchia08}. The descriptions of radiative cooling and heating used in OWLS are described by \citet{Wiersma09a}. Briefly, the cooling rates are computed in the presence of CMB and a \citet{HM01} model of the UV background radiation. An illustration of the temperature, density and metallicity evolution from $z=3.45$ to $z=0.01$ as predicted by OWLS is shown in Fig.~\ref{fig_owls}. More details about OWLS can be found in \citet{Schaye10} and the references therein. 

the OWLS we refer the reader to \citet{Schaye10}. 

Detailed information about the computed synthetic spectra can be found in \citet{Tepper-Garcia11}. The spectra were computed by use of the package {\sc Specwizard}\footnote{~Written by Joop Schaye, Craig Booth and Tom Theuns.} and following the method described by \citet{Theuns98}. {\sc Specwizard} calculates the optical depths from the velocities along a given sightline and then they are transformed to fluxes at the wavelength of each transition. In that way the observational spectra are modeled and comparable results are achieved. The synthetic spectra, used in this work, are obtained to mimic HST/STIS observations by convolving the spectra with an instrumental Gaussian Line-Spread Function with a FWHM $=7$~\kms and by resampling them onto a 3.5 \kms pixel. Additionally, a Gaussian noise is added by normalizing the flux through different values of the S/N ratio. The reference model in this work (${\rm REF\_L050N512}$) adopts the fiducial value S/N$= 50$ per pixel. More details about different reference 
models are given in \citet{Tepper-Garcia11}; the model, used here, is \# 4 in their Table 1 at $z=2.5$, rather that $z=0.25$. In particular, we analyse the physical properties of \Ovi absorbers found in the synthetic spectra at redshift $z=2.5$.



\chapter{A case study of two O\,{\sc vi} absorbers \\at $z\approx2$ towards PKS\,1448$-$232}\label{paper1}

In this Chapter\footnote{~The results presented in this chapter were published in A\&A 538, A85}, we study in detail two O\,{\sc vi} absorption-line systems at $z\approx2$ in the direction of the quasar PKS\,1448$-$232, using two different UVES data sets with different spectral resolutions. Because of its brightness ($V=16.9$), PKS\,1448$-$232 ($z_{\rm em}=2.208$) can be observed with UVES at high S/N with relatively moderate observing times. Two prominent O\,{\sc vi} systems with redshift $z\approx2$ have been identified previously (Philipp Richter, priv.\,comm.), each exhibiting a well-defined velocity-component structure with no major line blending problems. Therefore these two systems are particularly well-suited for a case study of high-redshift \Ovi absorption.

\section{Observations and absorption-line analysis}\label{observations_analysis}

\subsection{VLT/UVES observations}\label{observations}

\hspace{0.6cm}Our data set for this study consists of intermediate- ($R\approx 45\,000$) and high-resolution ($R\approx 75\,000$) spectra of PKS\,1448$-$232 (see Sect.~\ref{data_uves}). 
In Table~\ref{uves_data} we provide a summary of the observations of PKS\,1448$-$232.

\begin{table}[h!]
\begin{center}                  
\caption[Log of the UVES observations of PKS\,1448$-$232]{Log of the UVES observations of PKS\,1448$-$232: signal-to-noise ratio (S/N), resolution $R$, exposure time, $t_{\rm exp}$, and the wavelength range are listed for both observation runs.}
\begin{small}
\begin{tabular}{lrr}
\hline
\hline
& high res. & intermediate res. \\  
\hline    
$R$                            & 75,000      & 42,000 \\
S/N                            & 20$-$70     & 15$-$90 \\
$t_{\rm exp}$ [min]            & 780         & 720    \\
$v_{\rm res}$ [km\,s$^{-1}$]   & 4           & 7      \\
$\lambda$ range [\AA]          & 3000$-$6687 & 3050$-$10400 \\
\hline
\end{tabular} 
\end{small}
\label{uves_data}
\end{center}
\end{table}


\begin{table*}[]
\caption[Fit parameters for the absorbing system at $z=2.1098$]{Fit parameters for the absorbing system at $z=2.1098$}
\begin{tiny}
\begin{tabular}{cccccccccc}
\hline
\hline
\\
& \multicolumn{3}{c}{\small $z$}& \multicolumn{2}{c}{\small \Ovi} &
\multicolumn{2}{c}{\small C\,{\sc iv}} &  \multicolumn{2}{c}{\small H\,{\sc i}}\\ 
\hline
 & O\,{\sc vi} & C\,{\sc iv} & H\,{\sc i} & log[$N$(cm$^{-2}$)] & 
$b$\,[km\,s$^{-1}$] & log[$N$(cm$^{-2}$)] & $b$\,[km\,s$^{-1}$] &       
log[$N$(cm$^{-2}$)] & $b$\,[km\,s$^{-1}$] \\ \hline
\multicolumn{10}{c}{\it high resolution data} \\ \hline
1 & 2.10982 & 2.10982 & 2.10981 & 14.27($\pm$0.01) & 10.7($\pm$0.2) & 13.12($\pm$0.01) & 7.5($\pm$0.1) & 13.38($\pm$0.04) & 19.6($\pm$0.7) \\
2 & 2.11011 & 2.11008 & 2.11018 & 13.50($\pm$0.02) &  8.4($\pm$0.4) & 12.23($\pm$0.03) & 6.1($\pm$0.6) & 13.37($\pm$0.04) & 28.6($\pm$1.7) \\ \hline
\multicolumn{10}{c}{\it intermediate resolution data} \\ \hline
1 & 2.10984 & 2.10983 & 2.10981 & 14.32($\pm$0.02) &10.1($\pm$0.2)  & 13.12($\pm$0.01) & 7.1($\pm$0.1) & 13.39($\pm$0.01) & 20.6($\pm$0.3) \\
2 & 2.11014 & 2.11008 & 2.11019 & 13.49($\pm$0.20) & 5.4($\pm$0.5)  & 12.23($\pm$0.03) & 5.3($\pm$0.6) & 13.35($\pm$0.01) & 26.4($\pm$0.8) \\ \hline
 \end{tabular}
\end{tiny}
\label{fit_data_2.1098}
\end{table*}

\begin{figure*}[!th]
\begin{center}
\resizebox{0.55\hsize}{!}{\includegraphics{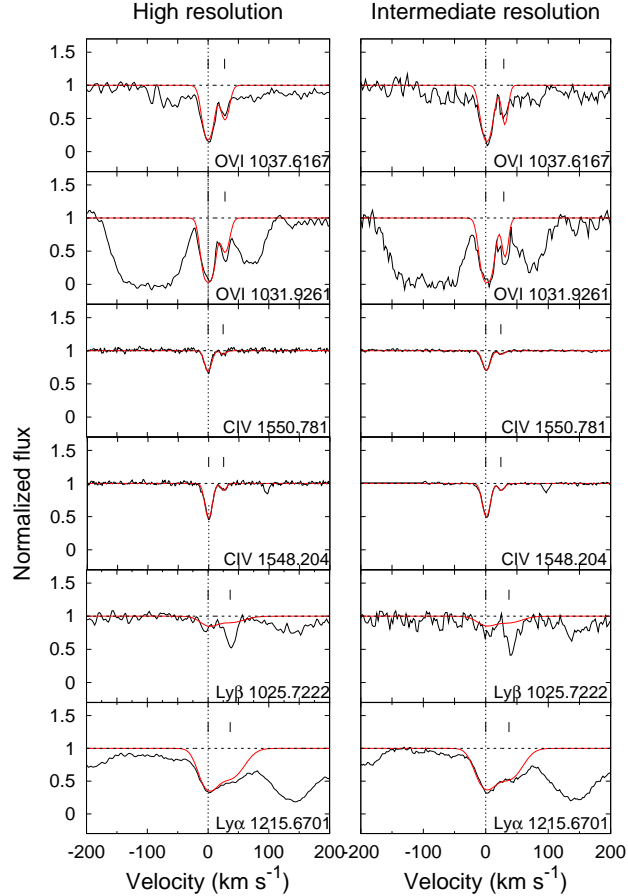}}
\caption[Absorption profiles for an \Ovi absorber at $z=2.1098$]{Absorption profiles for the \Ovi absorber at $z=2.1098$ in the high-resolution data (left panel) and the intermediate-resolution data (right panel).}
\label{abs_profile_2.1098}
\end{center}
\end{figure*}

The detected absorption features that are associated with the two absorbers at $z_{\rm abs}=2.1098$ and
$z_{\rm abs}=2.1660$ were fitted independently in both spectra (at intermediate and high resolution) 
with Gaussian profiles using the {\sc Candalf} fitting routine outlined in Sect.~\ref{tools}.
\begin{figure*}[!th]
\begin{center}
\resizebox{0.77\hsize}{!}{\includegraphics{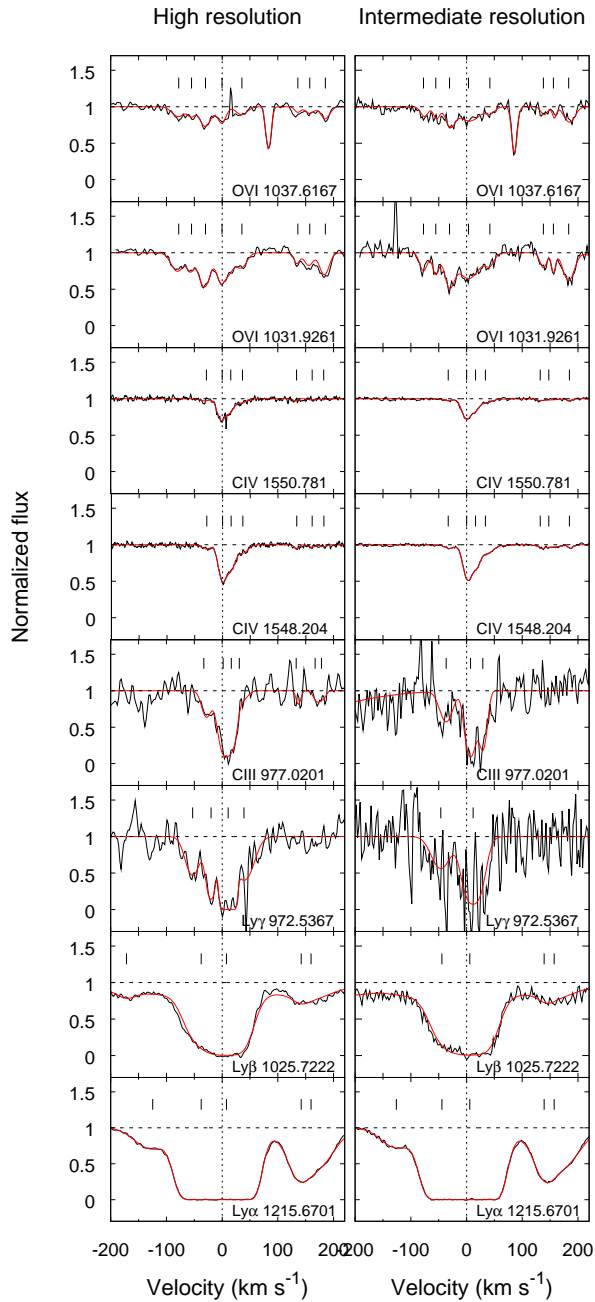}}
\caption[Absorption profiles for an \Ovi absorber at $z=2.1660$]{Absorption profiles for the \Ovi absorber at $z=2.1660$ in the high-resolution data (left panel) and the intermediate-resolution data (right panel). The strong absorption observed in \Ovi $\lambda 1037.62$ plot is a \Siiii line at $z=1.7236$.}
\label{abs_profile_2.1660}
\end{center}
\end{figure*}


\subsection{The O\,{\sc vi} system at $z=2.1098$}\label{sys_2.1098}

\hspace{0.6cm}Fig.~\ref{abs_profile_2.1098} shows the velocity profiles of \Ovi($\lambda\lambda 1031.9,1037.6$),
\Civ($\lambda\lambda 1548.2,1550.8$), and \Hi\!\!, \Lya\!\!, and
\Lyb($\lambda\lambda 1215.7,1025.7$) for the $z=2.1098$
absorber in the high-resolution data (left panel) and the intermediate-resolution
data (right panel). A visual inspection of both panels displays
no significant differences between the two data sets. The
S/N is on average higher for the high resolution data, 
except for the \Ovi region
where this ratio is slightly higher at intermediate resolution. 
Therefore, the differences in the values for $N$, $b$, and $z$ derived for the 
individual absorption components in the intermediate and high-resolution 
spectra are a result of the different S/N values for the two data sets.

Two absorption components are detected considering each of these ions. The \Ovi\!\! absorption is relatively strong compared to \Civ\!\!. The \Hi\!\! absorption is weak in comparison with other \Ovi\!\! absorbers at similar redshift \citep[e.g.,][]{Bergeron02} with a central absorption depth in the \Hi line of less than 70 per cent. We note that the second, weaker, component of \Hi \Lya and \Lyb absorption associated with the high-ion absorption is blended, so that the true component structure of \Hi as well its relative column density and Doppler parameter values remain somewhat uncertain. The blending aspect is not taken into account in the formal error estimate for $N$ and $b$ given in
Table~\ref{fit_data_2.1098}, which is based on the profile fitting. While the \Ovi\!\!, \Civ and \Hi absorption is well aligned for the stronger component, there appears to be a small ($< 10$ km\,s$^{-1}$) 
velocity shift between \Hi and the high ions in the weaker one component (see Table~\ref{fit_data_2.1098}). If real, this shift may indicate that \Hi and the metal ions may not trace the same gas phase in the weaker absorption 
component. Owing to the blending of the \Hi absorption, however, the reality of this shift remains unclear.

For the stronger component, we obtain for column densities log $N$(\Ovi\!\!)$\approx 14.3$,
log $N$(\Civ\!\!)$\approx 13.1$, and log $N$(H\,{\sc i}$)\approx 13.4$ as listed in Table~\ref{fit_data_2.1098}. 
The resulting ion-to-hydrogen ratios of $N$(\Ovi\!\!)/$N$(\Hi\!\!)$\sim 8$ and 
$N$(\Civ\!\!)/$N$(\Hi\!\!)$\sim 0.5$ already indicate 
that the metallicity of this absorber must be fairly high \citep{Bergeron05}. We note that -- because of the 
blending problem in the \Lya and \Lyb lines, --  the \Hi column density may be regarded as an upper limit and
thus the ratios given above might be even higher.


\begin{table}[t!]
\begin{center}
\caption[\Lya fit parameters for the absorbing system at $z=2.1660$]{\Lya fit parameters for the absorbing system at $z=2.1660$}
\label{lya_fit_2.1660}
\begin{small}
\begin{tabular}{ccccc}
\hline
\hline
 & $z$ & log\,[$N$(H\,{\sc i}) (cm$^{-2}$)] & $b$ [km\,s$^{-1}$]  \\
\hline
\multicolumn{4}{c}{\it high resolution data} \\ \hline
1 &  2.16466 & 13.27 ($\pm$0.01) & 40.0 ($\pm$1.2) \\
2 &  2.16552 & 14.49 ($\pm$0.07) & 29.9 ($\pm$1.2) \\
3 &  2.16605 & 15.18 ($\pm$0.03) & 34.9 ($\pm$0.5) \\
4 &  2.16746 & 13.16 ($\pm$0.03) & 19.2 ($\pm$0.6) \\
5 &  2.16765 & 13.77 ($\pm$0.01) & 45.6 ($\pm$0.3) \\
\hline
\multicolumn{4}{c}{\it intermediate resolution data}   \\ 
\hline
1 & 2.16467  & 13.21 ($\pm$0.02) & 37.4 ($\pm$1.3) \\
2 & 2.16559  & 14.57 ($\pm$0.12) & 31.9 ($\pm$1.7) \\
3 & 2.16607  & 15.13 ($\pm$0.05) & 35.5 ($\pm$0.8) \\
4 & 2.16749  & 13.19 ($\pm$0.02) & 18.8 ($\pm$0.5) \\
5 & 2.16768  & 13.76 ($\pm$0.01) & 46.8 ($\pm$0.3) \\
\hline
\end{tabular}
\end{small}
\end{center}
\end{table}

\begin{table*}[t!]
\caption[{Fit parameters for metal lines in the absorber at $z=2.1660$}]{Fit parameters for metal lines in the absorber at $z=2.1660$}
\label{fit_metal_2.1660}
\begin{tiny}
\begin{tabular}{ccccllllll}
\hline
\hline
\\
& \multicolumn{3}{c}{\small $z$} & \multicolumn{2}{c}{\small O\,{\sc vi}} &
\multicolumn{2}{c}{\small C\,{\sc iv}}&
\multicolumn{2}{c}{\small C\,{\sc iii}} \\
& O\,{\sc vi} & C\,{\sc iv} & C\,{\sc iii} & log[$N$(cm$^{-2}$)]
& $b$\,[km\,s$^{-1}$] & log[$N$(cm$^{-2}$)] & $b$\,[km\,s$^{-1}$]
& log[$N$(cm$^{-2}$)] & $b$\,[km\,s$^{-1}$] \\
\hline
\multicolumn{10}{c}{\it high resolution data} \\ \hline
1 & 2.16518 & ---     & ---     & 13.35($\pm$0.03)     & 15.4($\pm$1.3)
& ---         & ---            & ---                   & ---           \\
2 & 2.16542 & ---     & ---     & 13.02($\pm$0.08)     &  8.0($\pm$1.2)
& ---         & ---            & ---                   & ---           \\
3 & 2.16569 & 2.16569 & 2.16566 & 13.63($\pm$0.02)     & 13.5($\pm$0.7)
& 12.18($\pm$0.05) & 11.1($\pm$1.7)      & 12.68($\pm$0.07)           & 11.1$^{\rm a}$ \\
4 & 2.16600 & 2.16600 & 2.16602 & 13.41($\pm$0.03)     &  9.9($\pm$0.6) & 13.17($\pm$0.05)
&  9.3($\pm$0.5)      & 13.49($\pm$0.05)       & 15.1($\pm$1.5)     \\
5 & ---     & 2.16616 & 2.16618 & ---             & ---
& 12.96($\pm$0.10) & 10.0($\pm$1.6)      & 12.78($\pm$0.23)           &  8.0($\pm$3.7)     \\
6 & 2.16638 & 2.16638 & 2.16633 & 13.21($\pm$0.04)     & 13.7($\pm$0.8)
& 12.53($\pm$0.09) & 12.6($\pm$2.2)      & 12.22($\pm$0.47)       & 12.6$^{\rm a}$\\
7 & 2.16744 & 2.16741 & 2.16740 & 12.75($\pm$0.15) &  6.8($\pm$0.8)
& 11.90($\pm$0.05) &  4.8$^{\rm b}$ & 12.17($\pm$0.14) &  4.8$^{\rm a}$\\
8 & 2.16766 & 2.16770 & 2.16780 & 13.01($\pm$0.16) & 10.6($\pm$0.4)
& 11.84($\pm$0.08) &  9.2$^{\rm c}$ & 12.32($\pm$0.15) &  9.2$^{\rm a}$\\
9 & 2.16796 & 2.16792 & 2.16789 & 13.30($\pm$0.08) & 10.9($\pm$1.3)
& 11.50($\pm$0.12) &  4.8$^{\rm b}$ & 11.85($\pm$0.36) &  4.8$^{\rm a}$\\
\hline
\multicolumn{10}{c}{\it intermediate resolution data} \\ \hline
1 & 2.16521 & ---     & ---     & 13.12($\pm$0.03)     &  8.3($\pm$1.0)
& ---         &  ---           & ---                   & ---           \\
2 & 2.16544 & ---     & ---     & 13.13($\pm$0.04)     &  6.4($\pm$0.9)
& ---         &  ---           & ---                   &  ---          \\
3 & 2.16600 & 2.16567 & 2.16557 & 13.53($\pm$0.05)     & 10.6($\pm$0.9)
& 12.16($\pm$0.05) & 11.1($\pm$1.5)      & 12.87($\pm$0.11)           & 14.4($\pm$3.7)     \\
4 & 2.16606 & 2.16602 & 2.16603 & 13.73($\pm$0.04)     & 24.8($\pm$3.1)
& 13.23($\pm$0.02) & 10.4($\pm$0.4)      & 13.39($\pm$0.08)           & 10.4$^{\rm a}$\\
5 & ---     & 2.16619 & 2.16626 & ---             & ---
& 12.73($\pm$0.09) &  7.5($\pm$0.9)      & 13.12($\pm$0.15)           &  7.5$^{\rm a}$\\
6 & 2.16646 & 2.16637 & 2.16635 & 12.97($\pm$0.11)     & 10.4($\pm$2.0)
& 12.67($\pm$0.08) & 15.3($\pm$2.6)      & ---                   &  ---   \\
7 & 2.16748 & 2.16742 & ---     & 12.98($\pm$0.03)     &  7.3($\pm$0.9)
& 11.67($\pm$0.17) &  4.8($\pm$2.3)      & ---                   &  ---   \\
8 & 2.16767 & 2.16758 & ---     & 12.93($\pm$0.05)     &  2.8($\pm$0.9)
& 12.24($\pm$0.07) & 22.2($\pm$3.1)      & ---                   &  ---   \\
9 & 2.16796 & 2.16797 & ---     & 13.50($\pm$0.01)     & 12.7($\pm$0.6)
& 11.85($\pm$0.06) &  4.8($\pm$1.2)      & ---                   &  ---   \\
\hline
\end{tabular}
\end{tiny}\\
\tiny$^{\rm a}$ Fixed to $b_{\rm C\,{III}} = b_{\rm C\,{IV}}$ \\
\tiny$^{\rm b}$ Fixed to $b$-value derived from the intermediate resolution data \\
\tiny$^{\rm c}$ Lower limit, fixed to the minimal value \\
\end{table*}


\subsection{The O\,{\sc vi} system at $z=2.1660$}\label{sys_2.1660}

\hspace{0.6cm}The \Ovi system at $z=2.1660$ exhibits a significantly more complex absorption pattern than the absorber at $z=2.1098$, as can
be seen in the velocity profiles presented in Fig~\ref{abs_profile_2.1660}. We detect \Ovi absorption in eight individual absorption
components, spanning a velocity range as large as $\sim 300$ km\,s$^{-1}$. From our visual inspection, it is evident that
the absorption pattern of \Ovi differs than those of the other detected intermediate and high ions (\Ciii\!\!, \Civ\!\!)
and \Hi\!\!, although some of the components appear to be aligned in velocity space. As for the system at $z=2.1098$, there are no
significant differences between the absorption characteristics of the high-resolution data and the intermediate-resolution data.
However, the S/N ratio is somewhat lower in the latter for lines that are located in the blue part of the spectrum,
hence the resulting fitting values for $N$, $b$, and $z$ for the individual absorption components differ slightly
(Tables \ref{lya_fit_2.1660} and \ref{fit_metal_2.1660}). 

We modeled the \Hi absorption by simultaneously fitting \Lya and \Lyb in four absorption 
components (components $2-5$; see Table~\ref{lya_fit_2.1660}), obtaining column densities 
in the range $13.2 <$ \,log $N$(\Hi\!\!)$< 15.2$. One additional component (component 1) is present in the \Lya
absorption, but is blended in \Lyb (see Fig~\ref{abs_profile_2.1660}), so that $N$(\Hi\!\!)
was derived solely from \Lya. We note that for the \Hi fit we did not attempt to link the
\Hi component structure to the structure seen in the the metal ions, as this requires knowledge about the physical
conditions in the absorber. We discuss this aspect in detail in Sect.~\ref{model_2.1660}, where we try to reconstruct the \Hi
absorption pattern based on a photoionization model. We fitted the \Hi absorption with the minimum 
number of absorption components required to match the observations (Fig.~\ref{fit_metal_2.1660}, lowest panel) and to obtain an
estimate on the total \Hi column in the absorber.

By summing over the column densities in the individual absorption components, we derive total column 
densities of log $N$(\Ovi\!\!)$\approx 14.2$, log $N$(\Ciii\!\!)$\approx 13.7$, log $N$(\Civ\!\!)$\approx 13.5$,
and log $N$(\Hi\!\!)$\approx 15.3$. The resulting ion-to-hydrogen ratios of 
$N$(\Ovi\!\!)$/N$(\Hi)$\sim 0.1$ and $N$(\Civ\!\!)$/N$(\Hi\!\!)$\sim 0.02$ 
(representing the average over all components) are substantially smaller than in the $z=2.1098$ system,
which is indicative of a lower (mean) absorber metallicity.

The complexity of the absorption patterns for the various species in this system and the large velocity spread 
suggests that this absorber arises in an extended multi-phase gas structure.


\begin{table*}[]
\caption[Modelled column densities for the absorber at $z = 2.1098$]{Modelled column densities for the absorber at $z = 2.1098$}
\begin{small}
\begin{tabular}{crcccccccc}
\hline
\hline
\\
&  $v$\,[km\,s$^{-1}$]& \multicolumn{3}{c}{log [$N$\,(cm$^{-2}$)]} &  log [$n_{\rm H}\,$(cm$^{-3}$)] & log $Z$
& log [$T$(K)] & $L$ [kpc] & $f_{\rm HI}$ \\
& & C\,{\sc iv} & O\,{\sc vi} & H\,{\sc i}  \\
\hline
1 &     0 & 13.12 & 14.27 & 13.38           & $-$4.20
& $-$0.24 & 4.54 & 19.9 & $-$5.21 \\
2 & $+$25 & 12.23 & 13.50 & 13.37           & $-$4.25
& $-$1.02 & 4.64 & 30.5 & $-$5.35 \\
\hline
2 & $+$25 & 12.23 & 13.50 & 12.57$^{\rm a}$ & $-$4.28
& $-$0.24 & 4.57 &  4.7 & $-$5.32 \\
\hline
\label{tabl_model_2.1098}
\end{tabular}
\end{small}
\\
\tiny$^{\rm a}$ Our best H\,{\sc i} guess in the model for the second component with fixed metallicity\\
\end{table*}

\section{Ionization modeling and physical conditions in the gas}\label{ionization_modeling_paper1}

\hspace{0.6cm}To infer information about the physical properties of the two \Ovi absorbers towards PKS\,1448$-$232, we modeled 
in detail the ionization conditions in these systems. Since the two absorbers at $z=2.1098$ and $z=2.1660$ have redshifts
close to the quasar redshift ($z_{\rm QSO}=2.208$), it is necessary to check whether the two systems lie in the 
proximity of the background quasar and are influenced by its ionizing radiation.

With the above given redshifts, the two absorbers have velocity separations from the QSO of $\delta v_{2.1098}\approx 9000$ km\,s$^{-1}$ 
and $\delta v_{2.1660}\approx 4000$ km\,s$^{-1}$, thus the absorber at $z=2.1660$ can be regarded (depending on the definition)
as an associated system (see Sect.~\ref{types_lines}). With a (monochromatic) luminosity at the Lyman limit of
$L_{912}=3.39 \times 10^{31}$ erg\,s$^{-1}$\,Hz$^{-1}$, the size of the sphere-of-influence of the ionizing radiation from
PKS\,1448$-$232 is known to be $6.7$ Mpc, corresponding to a velocity separation of $\sim 1400$ km\,s$^{-1}$ \citep{Fox08}.  
Therefore, it is safe to assume that the ionizing radiation coming from PKS\,1448$-$232 itself has no measurable influence on the 
ionization conditions in the two \Ovi systems.

The small values of the Doppler parameter $b$ measured for \Ovi\!\!, \Civ\!\!, and \Hi indicate that collisional
ionization is not the source of the \Ovi existence in the gas. Using Eq.~\ref{eq_temp_Doppler}, we find that
the measured Doppler parameters of the O\,{\sc vi} components in the two absorbers (all with $b<16$ km\,s$^{-1}$ and many with $b<10$ km\,s$^{-1}$; see Tables \ref{fit_data_2.1098} and \ref{fit_metal_2.1660}) indicate temperatures
$T<10^5$ K. This value is below the peak temperature of \Ovi in CIE 
\citep[$T\sim 3\times 10^5$ K;][]{Sutherland93}; it is also lower than the temperature range expected for
\Ovi arising in turbulent mixing layers in the interface regions between cold and hot gas 
\citep[$T=10^5-10^6$ K;][]{Kwak10}. Consequently, photoionization by the hard UV background remains as the only 
plausible origin of \Ovi in the two high-ion absorbers towards PKS\,1448$-$232.

In consideration of this physical picture, we modeled the ion column densities in the two \Ovi systems using the photoionization code {\sc Cloudy} 
\citep[version C08;][]{Ferland} (for details see Sect.~\ref{cloudy}). We assumed a solar abundance pattern of O and C and an optically thin plane-parallel geometry 
in photoionization equilibrium exposed to a \citet{HM01} UV background spectrum at $z = 2.16$, which had been
normalized to $\log~J_{912} = -21.15$ \citep{Scott2000} at the Lyman limit.
 
We assumed that each of the observed velocity components is produced by a "cloud", which is modeled as an individual entity.
As input parameters, we considered the measured column densities of \Ciii (for only the $z=2.1660$ absorber), \Civ\!\!,  
\Ovi\!\!, the metallicity $Z$ (in solar units), and the hydrogen volume density $n_{\rm H}$. The metallicity 
of each cloud and the hydrogen density were varied across a range appropriate to intergalactic clouds  (i.e., $-3\leq \,$log$~Z\,\leq 0$
and $-5\leq \,$log$ ~n_{\rm H} \leq 0$).  

We then applied the following iterative modeling procedure. In a first step, we derived models using {\sc Cloudy} for a set of values of $Z$, $n_{\rm H}$, and 
$N$(\Hi\!\!), where $N$(\Hi\!\!) has been constrained by the observations. In a second step, the corresponding
values of $N$(\Ciii\!\!), $N$(\Civ\!\!), and $N$(\Ovi\!\!) were calculated. The output was compared with the observed 
column densities and, in the case of a mismatch, the input parameters $Z$ and $n_{H}$ were adjusted before the next iteration step. 
This process was repeated until the differences between the output column densities and the observed values became negligible
and we obtained a unique solution. In addition to the ion column densities, our {\sc Cloudy} model provides 
information about the neutral hydrogen fraction, $f_{\rm HI}$, the gas temperature, $T$, and the absorption path-length,
$L=N$(H\,{\sc i}$)/(f_{\rm HI}\,n_{\rm H})$. 

\subsection{The system at $z=2.1098$}\label{model_2.1098}

\hspace{0.6cm}As mentioned earlier, absorption by \Ovi and \Civ is well-aligned in both components in this system, while the true
component structure of \Hi is uncertain because of blending effects in the \Lya and \Lyb lines.
Owing to the alignment of \Ovi and \Civ\!\!, we assumed a single-phase model, in which each of the two components
(clouds) at $v=0$ and $+25$ km\,s$^{-1}$ in the $z=2.1098$ rest frame hosts \Ovi\!\!, \Civ\!\!, and \Hi of
column densities similar to those derived from the profile fitting. Consequently, we assumed log $N$(\Hi\!\!)$=13.37$ and $13.38$
as input for the {\sc Cloudy} modeling and followed the procedure outlined above. The results of the {\sc Cloudy} modeling
of the $z=2.1098$ absorber are summarized in Table~\ref{tabl_model_2.1098}. Our model closely
reproduces the observed \Ovi and \Civ column densities in both components, if the clouds have a density of log $n_{\rm H}
\approx -4.2$, a temperature of log $T\approx 4.6$, and a neutral hydrogen fraction of log $f_{\rm HI}\approx -5.3$. However,
to match the observations, the second component (at $+25$ km\,s$^{-1}$) in our initial model (Table~\ref{tabl_model_2.1098}, 
first two rows) needs to have a metallicity of log $Z=-1.02$, which is $\sim 0.8$ dex lower than that of the other component 
(log $Z=-0.24$). The absorption path-lengths are $\sim 20$ kpc for the component at $0$ km\,s$^{-1}$ and $\sim 30$ kpc for the 
component at $+25$ km\,s$^{-1}$. 

Taking into account the blending problem for the \Hi \Lya and \Lyb absorption, which affects in particular the estimate of $N$(\Hi\!\!)
in the cloud at $+25$ km\,s$^{-1}$ (Fig.~\ref{abs_profile_2.1098}), we set up a second {\sc Cloudy} model in which we tied the metallicity of the $+25$ km\,s$^{-1}$
component to the metallicity of the other component (log $Z=-0.24$), but left the $N$(\Hi\!\!) of this component as a free parameter. 
From this, we derived a value of log $N$(\Hi\!\!)$=12.57$ for the cloud at $+25$ km\,s$^{-1}$ and the absorption path-length decreased to
$L=4.7$ kpc. In terms of the blending, we regard this model as more plausible than the model with two different metallicities
and to have a larger absorption path-length.

In summary, our {\sc Cloudy} modeling suggests that the $z=2.1098$ absorber towards PKS\,1448$-$232 represents a relatively simple, metal-rich \Ovi
absorber in which the highly ionized \Ovi and \Civ states coexist in a single gas-phase.


\begin{table*}[t!]
 \caption[Modelled column densities for the \Ciii/\Civ phase at $z=2.1660$]{Modelled column densities for the \Ciii/\Civ\!\! absorbing phase in the $z=2.1660$ absorber}
\begin{scriptsize}
\begin{tabular}{clccccccccccccccc}
\hline
\hline
\\
& $v$\,[km\,s$^{-1}$]& \multicolumn{4}{c}{log [$N$\,(cm$^{-2}$)]}& log [$n_{\rm H}\,$(cm$^{-3}$)] & log $Z$ & log [$T$(K)] & $L$ [kpc] & $f_{\rm H}$ \\
&  & C\,{\sc iii} & C\,{\sc iv} & O\,{\sc vi} &
H\,{\sc i}$^{\rm a}$  \\
\hline
\\
3 & $-$28  & 12.68 & 12.18 & 10.97          & 14.51 & $-$2.74 & $-$1.7 & 4.42 & 0.3   & $-$3.68  \\
4 & $+$0   & 13.49 & 13.17 & 12.25          & 14.18 & $-$2.97 & $-$1.7 & 4.46 & 4.1   & $-$3.95  \\
5 & $+$16  & 12.78 & 12.96 &  ---           & 14.51 & $-$3.56 & $-$1.7 & 4.58 & 16.3  & $-$4.63  \\
6 & $+$37  & 12.22 & 12.53 & 12.86          & 14.08 & $-$3.71 & $-$1.7 & 4.61 & 12.3  & $-$4.79  \\
7 & $+$134 & 12.17 & 11.90 & 10.92          & 13.26 & $-$2.93 & $-$1.0 & 4.38 & 0.04  & $-$3.84  \\
8 & $+$162 & 12.32 & 11.84 & 10.49          & 13.57 & $-$2.67 & $-$1.0 & 4.34 & 0.02  & $-$3.55  \\
9 & $+$182 & 11.85 & 11.50 & 10.38          & 12.99 & $-$2.84 & $-$1.0 & 4.37 & 0.01  & $-$3.73  \\
\hline
\label{tabl_model_2.1660}
\end{tabular}
\end{scriptsize}
\\
\tiny$^{\rm a}$ Our best H\,{\sc i} guess in the models.\\
\end{table*}

\subsection{The system at $z=2.1660$}\label{model_2.1660}

\hspace{0.6cm}We started to model this system with {\sc Cloudy}, again under the assumption of a single gas-phase hosting the observed intermediately and highly ionized 
\Ciii\!\!, \Civ\!\!, and \Ovi states in the various subcomponents. However, during the modeling process 
it quickly turned out that it is impossible to match the observed column densities of \Ciii and \Ovi by assuming a single gas-phase
in the components, when these two ions are aligned in velocity space. Our modeling indicates that 
the \Ciii absorption must instead arise in an environment that has a relatively high gas density and is 
spatially distinct from the \Ovi phase. In a second step, we tried to tie the highly ionized \Civ and
\Ovi states in a single gas phase (like in the case of $z=2.1098$ system) in the relevant absorption components, ignoring the \Ciii phase. 
However, this approach did not deliver satisfying results, as we obtained for some components, for which \Civ\!\!/\Ovi 
was constrained by observations, very low gas densities and very large absorption path-lengths on Mpc scales, which
are highly unrealistic. Given that the overall component structures of \Ovi and \Civ 
differ substantially from each other in this system (Fig.~\ref{abs_profile_2.1660}), this result is not really surprising.

The only modeling approach for which we obtain realistic results for gas densities, temperatures, and absorption path-lengths in this
system and its subcomponents is a two-phase model, in which \Ciii coexists with \Civ and part of the \Hi in one (spatially 
relatively confined) phase, and \Ovi and the remainder of the \Hi in a second (spatially relatively extended) phase. The coexistence
of \Ciii and \Civ in one phase is also suggested by the \Ciii and \Civ absorption, which is well- 
aligned in velocity space (see Fig.~\ref{abs_profile_2.1660}). The results of this  two-phase model are presented in 
Tables \ref{tabl_model_2.1660} and \ref{tabl_model_o6_2.1660}. A critical issue
for the modeling of this complex multi-phase absorber with its many absorption components is the assumption of a neutral
gas column density in each subcomponent (and phase). Since in the \Hi \Lya and \Lyb absorption, most subcomponents 
are smeared together to one large absorption trough, the observational data provide little information about the distribution
of the \Hi column densities among the individual components. Nevertheless, the data provide a solid estimate of the {\it total} \Hi column
density in the absorber (log $N\approx 15.3$; see Sect.~\ref{sys_2.1660}), which must match the sum of $N$(\Hi\!\!) over all subcomponents considered
in our model. Consequently, we included in our iteration procedure the constraints on $N$(\Hi)$_{\rm tot}$ and the {\it shape}
of the (total) \Hi absorption profile. The latter aspect also concerns the choice of the gas temperature
in the model, as $T$ regulates the thermal Doppler-broadening and thus the width of the modeled \Hi lines. We modeled the \Hi
width following the approach of \citet{Ding}.

With these various constraints, we first modeled the \Ciii\!\!/\Civ phase in the absorber. However, owing to the extremely complex parameter
space, we did not find a unique solution for $(T,n_{\rm H},Z)$ among the individual components, but had to make additional constraints. Since
the individual components observed in \Ciii\!\!/\Civ are very close together in velocity space, we assumed they all have
the same metallicity and, based on the $Z$ range allowed in the model, we set log $Z=-1.5$ for all subcomponents.
This model was able to match the observed column densities of these two ions in the individual subcomponents, but did not match 
closely the overall shape of the overall \Hi absorption, implying that the metallicity in this absorber is non-uniform among the
individual absorption components. Therefore, we refined our model by using two different 
metallicities, log $Z=-1.7$ for the saturated \Hi components and log $Z=-1.0$ for the weaker \Hi components 
(see Tables \ref{tabl_model_2.1660} and \ref{tabl_model_o6_2.1660} for details). 
Although imperfect, this model delivers a satisfying match between the modeled spectrum and the UVES data.

Adopting this model, we found that the \Ciii\!\!/\Civ absorbing components have temperatures between
log $T=4.3$ and $4.6$, densities between log $n_{\rm H}=-3.7$ and $-2.7$, and neutral gas fractions between log $f_{\rm HI}=-4.8$
and $-3.6$ (see Table~\ref{tabl_model_2.1660}). The absorption path lengths were found to vary between $0.3$ and $16.3$ kpc 
for the components with log $Z=-1.7$, and between $0.01$ and $0.04$ kpc for the components with log $Z=-1.0$. These numbers 
suggest that the \Ciii\!\!/\Civ absorbing phase resides in relatively small and confined 
gas clumps. This scenario is consistent with the small turbulent $b$-values of $<6$ km\,s$^{-1}$ for the subcomponents 
that we derive in our model. We note that in Table~\ref{tabl_model_2.1660}, we also list the predicted column densities for \Ovi\!\!, 
which are typically $1-2$ orders of magnitude below the observed ones in this absorber. This, again, indicates
that \Ciii\!\!/\Civ and \Ovi must reside in different gas phases with different physical conditions 
to explain the observed column densities. 

Finally, we modeled the \Ovi absorbing phase in the $z=2.1660$ absorber, based on the observed \Ovi 
column densities. Since we had information for no ions other than \Hi and \Ovi that could provide information
about the physical conditions in this phase, we fixed the metallicity of the gas to log $Z=-1.7$ and 
log $Z=-1.0$ (equal to the  phase) and constrained the temperature range [$T_{\rm min},T_{\rm max}$] 
in the {\sc Cloudy} models based on the observed line widths of \Ovi (giving $T_{\rm max}$) and the 
modeling results of the \Ciii\!\!/\Civ phase (giving $T_{\rm min}$ for all components except the 
first two). The results of this model are shown in Table~\ref{tabl_model_o6_2.1660}. We derived gas densities in the range log $n_{\rm H}=-4.6$ to 
$-3.2$ and neutral gas fractions in the range log $f_{\rm HI}=-5.8$ to $-4.6$. The absorption path length was found to vary 
between $19.8$ and $83.3$ kpc for the components with log $Z=-1.7$, and between $1.3$ and $38.3$ kpc for those with log $Z=-1.0$.
The mismatch between $N$(\Ovi\!\!) of the model and the data for components one and nine (see Table~\ref{tabl_model_o6_2.1660}) implies that the
metallicity distribution among the individual absorption components is even more complex than the one 
assumed in our model. Despite this (minor) concern, our {\sc Cloudy} modeling for \Ovi provides clear
evidence that the \Ovi absorbing phase has substantially lower gas densities than the \Ciii\!\!/\Civ absorbing
phase and is spatially more extended.

In summary, our {\sc Cloudy} modeling of the $z=2.1660$ absorber suggests that this system has a complex
multi-phase gas structure, in which a number of cooler, \Ciii\!\!/\Civ\!\! absorbing cloudlets are embedded
in a spatially more extended, \Ovi absorbing gas phase spanning a total velocity range of $\sim 300$ km\,s$^{-1}$.
Although the metallicity is not well-constrained in our model, it appears that log $Z\leq -1$ in the absorber, which
is $\sim 0.8$ dex below the value obtained for the system at $z=2.1098$.

\begin{table*}[t!]
\caption[Modelled column densities for the \Ovi phase at $z=2.1660$]{Modelled column densities for the \Ovi absorbing phase in the $z=2.1660$ absorber}
\begin{scriptsize}
\begin{tabular}{clcccccrc} 
\hline
\hline 
& $v$\,[km\,s$^{-1}$]& \multicolumn{2}{c}{log [$N$\,(cm$^{-2}$)]}& log $Z$ & log [$T$(K)] & $L$ [kpc] & $f_{\rm H}$  \\
&  & O\,{\sc vi} &
H\,{\sc i}$^{\rm a}$ &  log [$n_{\rm H}\,$(cm${-3}$)] \\
\hline
1 & $-$78    & 12.96$^{\rm b}$ & 13.17 & $< -$3.76 & $-$1.7 & $<$5.36  & $<$19.8  &  $>-$5.86  \\
2 & $-$55    & 13.02       & 14.10 & $< -$3.77 & $-$1.7 & $<$4.79  & $<$25.8  &  $>-$5.03  \\  
3 & $-$30    & 13.63       & 14.51 & $-$3.91 ... $-$3.31 & $-$1.7 & 4.42 ... 5.24  &58.4  ... 67.7  & $-$5.50 ... $-4.84$  \\   
4 & $+$0     & 13.41       & 14.18 & $-$3.99 ... $-$3.88 & $-$1.7 & 4.46 ... 4.98  &42.6  ... 83.3  & $-$5.36 ... $-$4.95  \\ 
6 & $+$36    & 13.21       & 14.08 & $-$3.92 ... $-$3.23 & $-$1.7 & 4.61 ... 5.26  &32.1  ... 21.5  & $-$5.52 ... $-$5.00  \\
7 & $+$136   & 12.75       & 13.26 & $-$3.72 ... $-$3.72 & $-$1.0 & 4.38 ... 4.64  & 1.3  ... 2.0   & $-$4.82 ... $-$4.61  \\ 
8 & $+$157   & 13.01       & 13.57 & $-$3.70 ... $-$3.52 & $-$1.0 & 4.34 ... 5.03  & 2.2  ... 6.1   & $-$5.19 ... $-$4.56  \\ 
9 & $+$185   & 13.25$^{\rm c}$ & 12.99 & $-$4.60 ... $-$4.25 & $-$1.0 & 4.37 ... 5.06  & 32.3  ... 38.3  & $-$5.76 ... $-$5.48  \\ 
\hline  
\label{tabl_model_o6_2.1660} 
\end{tabular} 
\end{scriptsize}\\
\tiny$^{\rm a}$ Our best H\,{\sc i} guess in the models\\ 
\tiny$^{\rm b}$ Observed {log} $N$(O\,{\sc vi}) $= 13.35$\\ 
\tiny$^{\rm c}$ Observed {log} $N$(O\,{\sc vi}) $= 13.30$\\  
\end{table*}  


\section{Discussion}\label{discussion_paper1}

\hspace{0.6cm}Our detailed analysis of the two \Ovi absorbers at $z=2.1098$ and $z=2.1660$ towards the quasar 
PKS\,1448$-$232 has clearly illustrated the large diversity and complexity of high-ion absorbers at high redshift. 

A number of studies based on both optical observations \citep[e.g.,][]{Bergeron02, Carswell02, 
Simcoe02, Simcoe04, Simcoe06, Bergeron05, Aguirre08} and numerical simulations \citep[e.g.,][]{Fangano07, 
Kawata07} have investigated the properties of high-redshift \Ovi systems and their relation to galaxies.

As a result of their survey of \Ovi absorbers at redshifts $z=2.0-2.6$, \citet{Bergeron05}
suggested that \Ovi systems may be classified into two different types: metal-rich absorbers (``type 1'')
that have large $N$(\Ovi\!\!)/$N$(\Hi\!\!) ratios and that appear to be linked to both galaxies and 
galactic winds, and metal-poor absorbers (``type 0'') with small $N$(\Ovi\!\!)/$N$(\Hi\!\!) ratios, which are
embedded in the IGM. The two absorbers observed towards PKS\,1448$-$232 that we have discussed in this Chapter
do not match the classification scheme of \citet{Bergeron05}. The absorber at $z=2.1098$ has a very large $N$(\Ovi\!\!)/$N$(\Hi\!\!)
ratio of $\sim 8$ (i.e., it is of type 1): it is a simple, single-phase, metal-rich system with a metallicity slightly below the solar value.
Nevertheless, this system is completely isolated with no strong \Hi \Lya absorption within $1000$ km\,s$^{-1}$. 
In contrast, the absorber at $z=2.1660$ has a $N$(\Ovi\!\!)/$N$(\Hi\!\!) ratio
of only $\sim 0.1$ and a metallicity of $0.1$ solar or lower \citep[i.e., it is of type 0 according to][]{Bergeron05}.
This absorber is a complex multi-phase system with a non-uniform metallicity, suggesting that it originates
in a circumgalactic environment.  While this mismatch with the \citet{Bergeron05} classification
scheme certainly has no statistical relevance for the general interpretation of \Ovi absorbers at high redshift,
the results suggest that for a thorough understanding of highly-ionized gas at high redshift
the absorption characteristics of \Ovi systems may be too diverse to permit
a simple classification scheme based solely on the observed (and partly averaged) column density ratios of \Ovi\!\!,
\Hi\!\!, and other ions. 

One critical drawback of many previous \Ovi surveys at high redshift is that they 
often considered only simplified models for the ionization conditions in their sample of highly ionized absorbers, so that 
the multi-phase character of the gas and the possible ionization conditions far from a photoionization equilibrium 
are insufficiently taken into account. As pointed out by \citet{Fox11}, single-phase, 
single-component ionization models, if applied, will provide physically irrelevant results for most
of the \Ovi systems at high redshift. This implies that previous estimates of the
baryon- and metal-content of \Ovi absorbers at low and high redshift are possibly affected by large systematic
uncertainties. We investigate this problem in Sect.~\ref{aligned_absorbers_owls}

One firm conclusion from many previous observational and theoretical studies of highly ionized absorbers is that a considerable 
fraction of the \Ovi systems at low and high $z$ must arise in the metal-enriched circumgalactic
environment of (star-forming) galaxies \citep[e.g.,][]{Wakker09, Prochaska11, Fox11a, Tepper-Garcia11, Fangano07}. Thus, the complex
absorption pattern observed in the $z=2.1660$ system towards PKS\,1448$-$232 and many other \Ovi absorbers
at high $z$ may reflect the complex gas distribution of enriched gaseous material that was ejected from galaxies into the IGM during
their wind-blowing phase \citep[e.g.,][]{Kawata07}. In this context, \citet{Schaye07} suggested that the intergalactic 
metals were transported from galaxies by means of galactic winds and reside in the form of dense, low- and high-metallicity patches within large
hydrogen clouds. These authors point out that much of the scatter in the metallicities derived for high-redshift absorbers could be explained by
the spatially varying number of the metal-rich patches and the different absorption path lengths through the surrounding 
metal-poor intergalactic filament instead of an overall (large-scale) metallicity scatter in the IGM. In this scenario, 
the substantial differences in the metallicities of the two \Ovi systems towards PKS\,1448$-$232, and 
even the intrinsic metallicity variations within the $z=2.1660$ system, could be explained by the different geometries of the absorbing structures,
suggesting that much of the \Hi that is associated with the metal absorption in velocity space, arises in a spatially distinct region. In Sect.~\ref{aligned_absorbers_owls} 
we further investigate that problem, using a set of cosmological OWLS.  A similar 
conclusion was drawn by \citet{Tepper-Garcia11}, who studied the nature of \Ovi absorbers at low redshift using the same set of 
cosmological simulations. We note that absorbers with larger \Hi column densities,
such as LLS and DLAs, sometimes exhibit abundance variations among
the different velocity subcomponents \citep[e.g.,][]{Richter05, Prochter10}. This indicates that the metals in the 
gas surrounding high redshift galaxies have not been well-mixed.

The observed velocity differences between \Ovi and other ions and the multi-phase nature of the gas provide
further evidence of an inhomogeneous metallicity and density distribution in intervening highly ionized absorbers.
It is an interesting that the velocity misalignment appears to concern only
the \Ovi absorbing phase in highly ionized absorbers at high redshift, while other highly ionized states such as \Nv and \Civ generally
appear to be well-aligned with \Hi\!\!, even in systems that exhibit a complex velocity-component structure \citep{FR09}.
This puzzling aspect underlines that additional detailed studies of individual \Ovi absorption
systems could be very important to our understanding of intergalactic and circumgalactic gas at
high redshift, as this ion traces a metal-enriched gas phase that cannot be observed by other means.

\section{Conclusions}\label{summary_paper1}

\hspace{0.6cm} In this chapter, we investigated two \Ovi absorbers at $z=2.1098$ and $z=2.1660$ towards the QSO PKS\,1448$-$232. For this, we have used high- ($R\approx 75,000$) and intermediate-resolution ($R\approx 45,000$) optical spectra obtained with the VLT/UVES instrument together with {\sc Cloudy} photoionization models. From our study, we draw the following conclusions: 

\begin{itemize}

\item

The first \Ovi system at $z=2.1098$ is characterized by strong \Ovi absorption and weak \Hi absorption in a relatively simple, two-component absorption pattern. The absorptions by \Ovi\!\!, \Civ\!\!, and \Hi are well aligned in the velocity space, indicating that they trace the same gas phase. From a detailed photoionization modeling of this system, we derive a metallicity of $\sim 0.6$ solar, a characteristic density of log $n_{\rm H} \approx -4.2$, a temperature of log $T\approx 4.6$, and a total absorption path length of $\sim 30$ kpc. The absorber is isolated with no strong H\,{\sc i} Ly\,$\alpha$ absorption detected within $1000$ km\,s$^{-1}$.

\item

The second \Ovi absorber at $z=2.1660$ represents a complex, multi-component absorption system with eight relatively weak and narrow
\Ovi absorption components spanning almost $300$ km\,s$^{-1}$ in radial velocity. The \Ovi components are accompanied by a strong \Hi absorption, and \Ciii and \Civ absorption features. The \Ovi component structure differs from that of \Hi and \Civ\!\!, indicating the absorber contains a multi-phase IGM. Our photoionization modeling with {\sc Cloudy} suggests that there are (at least) two distinct gas phases in this system. The first consists of \Ciii\!\!, \Civ\!\!, and most of the \Hi\!\!, which appear to coexist in several relatively compact cloudlets at gas densities from 
log $n_{\rm H}\approx -3.7$ to $-2.7$, temperatures of log $T\approx 4.3-4.6$, and absorption path-lengths of $<16$ kpc. We have found that \Ovi appears to reside in a highly ionized, more spatially extended gas phase at densities in the range log $n_{\rm H}\approx -4.6$ to $-3.2$, temperatures between log $T\approx 4.3$ and $5.3$, and absorption path-lengths up to $83$ kpc. While the precise metallicity of the absorber is not well-constrained, our modeling is most consistent with a non-uniform metal abundance in the individual absorption components with (at least) two different metallicities of log $Z=-1.7$ and log $Z=-1.0$.

\item

Our study illustrates the large diversity and complexity of \Ovi systems at high redshift. We speculate that some of the observed differences between the studied two highly ionized absorbers towards PKS\,1448$-$232 could result from an inhomogeneous metallicity and density distribution in the photoionized IGM. Our study indicates that multi-phase, multi-component highly ionized absorbers similar to the one at $z=2.1660$ can be described by a detailed ionization modeling of the various subcomponents to obtain information about the physical conditions
and metal-abundances in the gas. 
\end{itemize}

We conclude that much effort will be required to achieve a more complete view of the nature of \Ovi absorbers at high redshift. Therefore, the next logical step in this project is to analyze a larger sample of \Ovi absorbers using high-quality UVES archival data. This analysis will be presented in the following chapter.


\chapter{An UVES survey of \Ovi absorbers at high redshift}\label{uves}  

\hspace{0.6cm}The case study of two \Ovi absorbers towards PKS\,$1448-232$ presented in the previous chapter indicates the complexity and diversity of highly-ionized intergalactic gas structures at high redshift. However, a much larger sample of \Ovi absorbers is required to study the nature of these objects on a statistically secure basis and to pinpoint their role in the cosmological gas circulation processes in the early Universe. 

In this chapter, we therefore study a larger sample of \Ovi absorbers from high-quality UVES archival data (see Sect.~\ref{data_uves}). These data are publicly available in the UVES data archive and have a spectral resolution of $R\approx 45\,000$ (see Sect.~\ref{data_uves}). We further use the \Ovi absorber list of Dr. Tae-Sun Kim (priv. comm.), who has fitted intervening absorption-lines systems from these data using the {\sc Vpfit} fitting program (see Sect.~\ref{tools}).

  \section{Overview of the sample}\label{overview}

\hspace{0.6cm}Following \citet{Songaila05} and \citet{Scannapieco06}, we define a {\it system} as an \Ovi absorber with at least two components as the separation of each of them from its closest neighbor is $\le100$ \kms\!\!. 

We found 93 candidates for \Ovi systems in total in 15 inspected sightlines in the composed UVES sample. Fifty three systems exhibit a strong \Lya contamination and were therefore excluded from further consideration. Thus, the final sample consists of 40 systems and is divided into three categories marked by integer between {\it 2} to {\it 0}, according to the quality of the detected \Ovi and \Civ features.

\begin{itemize}
\item[-] {\it 2} (very good): The \Ovi lines are clearly seen (although, in some cases, partly blended) and \Civ lines are present\footnote{~The single exception is the system towards PKS 0329-255 at $z$=2.66.}.
\item[-] {\it 1} (good): Some of the systems are partly blended, but the features of both lines from the \Ovi doublet are seen and accompanying \Civ lines are present\footnote{~The single exception is the system towards Q 0109-3518 at $z$=2.40, where no \Civ is detected.}.
\item[-] {\it 0} - (reliable): Weak and/or blended lines are present with still distinguishable \Ovi features. 
\end{itemize} 

\begin{figure}[h!]
\begin{center}
\resizebox{0.7 \hsize}{!}{\includegraphics[angle=-90]{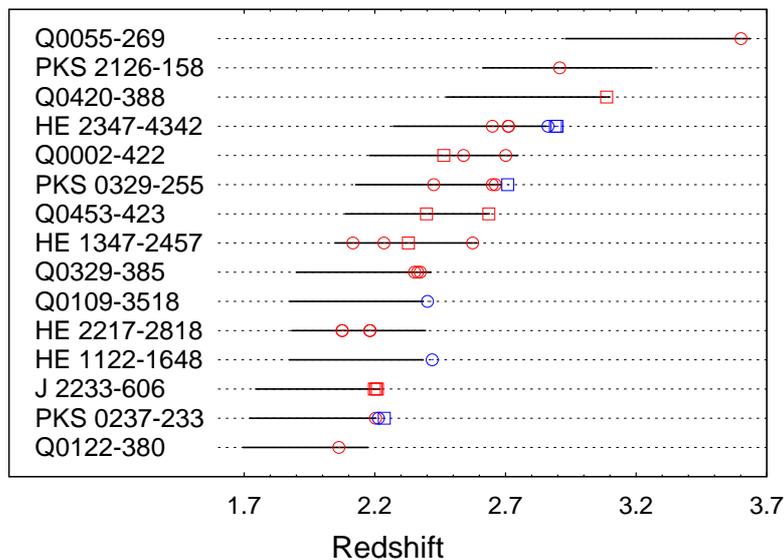}}
\caption[Location of the \Ovi absorbers from the UVES sample in redshift space.]{Schematic diagram showing the positions of the intervening (red) and associated (blue) absorbers in redshift space along the lines of sight of the sampled QSOs. The multiplicity of the system is denoted by different symbols: up to 3 components (circles) and more than 3 components (squares). 
The solid horizontal lines indicate redshift coverage $[z_{\rm min},~z_{\rm max}]$ along each line of sight (dotted).}
\label{z_coverage}
\end{center}
\end{figure}

The redshift coverage of the QSOs $[z_{\rm min},~z_{\rm max}]$ and the redshifts of the individual \Ovi absorbers are displayed in Fig.~\ref{z_coverage}. The lower redshift limit $z_{\rm min}$ is defined as the redshift of the \Lyb emission, while $z_{\rm max}$ is the redshift at which the velocity difference from the QSO emission redshift reaches $ \ge 5000$ \kms\!\!, i.e. corresponds to distances from the QSO where its influence is negligible (see Sect.~\ref{types_lines}). Imposing a separation velocity criterion of 5000 \kms\!\!, 32 systems from the sample are found to be intervening ($|\Delta v| > 5000$ \kms\!\!) while 8 are associated ($|\Delta v| < 5000$ \kms\!\!). Detailed information on the redshift limits of the selected QSOs as well on the redshift and the category of the \Ovi absorbers is given in Appendix~\ref{appendix_systems}, Table ~\ref{observed_quasars}. Examples of \Ovi absorption systems, according their category, are given in Appendix~\ref{examples_uves}. 

The total probed redshift path in our sample is $\Delta z = 8.34$, in the range $2.06 \le z \le 2.91$. The corresponding redshift-invariant absorption path length (also called `absorption distance') can be calculated by solving the following integral:
\begin{equation}
\label{absorption_path_length}
X = \int\limits_{z_{\rm min}}^{z_{\rm max}} \frac{(1+z)^{2}}{\sqrt{{\Omega_{m}}(1+z)^{3}+\Omega_{\Lambda}}} dz.
\end{equation}

The absorption path length for each QSO is specified in Appendix~\ref{appendix_systems}, Table ~\ref{observed_quasars} (column 4). 
By summing up the path lengths at each sightline, we obtain a total absorption path length for the whole \Ovi sample $\sum\limits_{i} X_{i}=\Delta X = 27.3$.  

To minimize the uncertainties in determination of column densities and Doppler parameters we imposed strict selection criteria when composing the final \Ovi sample, removing those systems that are heavily blended by other intergalactic lines. This means, however, that the rate of incidence $d \mathcal N/dz$ (the number of absorbers per redshift path length interval) is underestimated when using these strict selection criteria for the analysis of the \Ovi line density. Therefore, we  added other 11 intervening \Ovi systems from the original (unrestricted) sample, only for the determination of $d \mathcal N/dz$ and $d \mathcal N/dX$. In these 11 systems, both \Ovi lines are clearly detected, but partly blended, so that these systems can be used only to constrain the \Ovi incidence rate, but not the physical conditions in the \Ovi systems.

The calculated rate of incidence for the extended sample of 43 intervening \Ovi systems is $d \mathcal N/dz = 5.2 \pm0.79$. This estimate was obtained by probing the full range of column densities ($11.5 \le \log N \le 15.5$), while the proximity zones of $\le$ 5000 \kms for each line of sight were excluded. The corresponding rate of incidence per absorption path length is $d \mathcal N/dX = 1.6 \pm0.24$. For comparison, \citet{Muzahid11} found $d \mathcal N/dz = 10.1 \pm1.15$ and $d \mathcal N/dX = 3.1 \pm0.35$, respectively; i.e., roughly twice higher values, using similar data and a similar redshift range. However, their sample includes some systems where one of the \Ovi lines is blended by saturated \Lya or both \Ovi and dubious absorption features that might not be at all associated with intervening \Ovi. Therefore, our estimate of the rate of incidence  might be considered as a lower limit, while the one by \citet{Muzahid11} -- as an upper limit. 

\begin{table*}[th!]
\begin{center}
\caption[Number of components in the intervening and associated \Ovi systems.]{Number of components in the intervening and associated \Ovi systems with or without \Ovi-\Hi alignment.}
\begin{small}
\begin{tabular}{lcccccccc}
 \hline
 \hline
\multicolumn{6}{c}{Systems} \\
\hline
Type & Total & \multicolumn{2}{c}{ Intervening systems} & \multicolumn{2}{c}{ Associated systems}   \\
~~&~& \Ovi\!\!/\Hi al.& No al.& \Ovi\!\!/\Hi al.& No al.\\
 \hline
 1 comp.   & 10 & 6 & 3& 1& 0  \\
 2-3 comp.& 18 & 7& 7 & 3& 1  \\
 multicomp. & 12 & 4& 5 & 0& 3  \\
  \hline  
\end{tabular}
\end{small}
\label{number_of_components}
\end{center}
\end{table*}

\begin{figure}[h!]
\begin{center}
\resizebox{0.8 \hsize}{!}{\includegraphics[angle=-90]{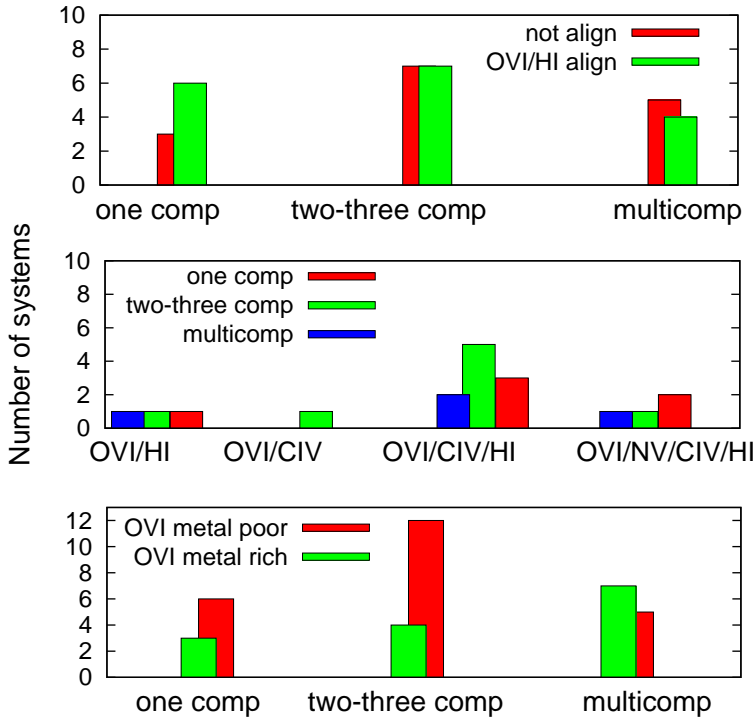}}
\caption[Statistics of the \Ovi absorbers from the UVES sample]{Statistics of the \Ovi absorbers from the UVES sample: {\it (top panel)} Fraction of absorbers with at least one \Ovi component aligned with \Hi (within 10 \kms\!\!) in the single-, two-three and multicomponent sample; {\it (middle panel)} Number of \Ovi systems aligned with \Civ\!\!, \Nv and/or \Hi; {\it (bottom panel)} Statistics of the considered \Ovi systems according to the number of components.}
\label{number_systems}
\end{center}
\end{figure}

An important characteristic parameter of intervening absorption-line systems is the number of absorption components (i.e., velocity components) per system. Detailed information on the number of components in each studied \Ovi system is presented in Table~\ref{observed_quasars}, column 7. The results are summarized in Table~\ref{number_of_components}. Two subsamples can be distinguished: (1) absorbers with at least one \Ovi component aligned in velocity space with \Hi within 10 \kms\!\!, and (2) absorbers without any such alignment. Fig.~\ref{number_systems} (top panel) displays the distribution of \Ovi intervening systems, according to the number of their components and to their alignment with \Hi\!\!. Evidently, one third of the components in the single-component systems do not exhibit alignment with \Hi while this fraction grows to 50 per cent in two-three component systems. The tendency persists as well in the multicomponent case where systems without alignment slightly dominate. The existence of 
alignment between \Ovi and \Hi is important for the performed ionization modeling, as demonstrated later in this Chapter (Sect.~\ref{modeling_cloudy}). Lack of alignment indicates that the \Ovi and \Hi absorption arise in physically distinct regions and gas phases. 

A component alignment in velocity space between \Ovi and other metal lines such as \Civ is another characteristic of the absorbers that deserves special attention, because the combined use of ions with different ionization potentials provide important constraints for the photoionization modeling. Thirty-three \Ovi systems (27 intervening and 6 associated) out of 40 in our sample exhibit \Civ absorption and 11 of them (9 intervening and 2 associated) show \Nv absorption. Fifteen \Civ\!\!/\Ovi pairs in the intervening systems are aligned (i.e. their component velocity offsets are within 10 \kms\!\!). The statistics on \Ovi absorbers aligned with \Civ\!\!, \Nv and/or \Hi lines according to the number of their components is shown in Fig.~\ref{number_systems} (middle panel). It appears that alignment between the components of the three ions \Civ\!\!, \Nv\!\!, and \Ovi occurs most frequently in systems with one to three absorption components (one fourth of this subsample).

  \section{Key observables}\label{observables}

\subsection{Metal-rich and metal-poor \Ovi absorbers}\label{metal-rich_and_metal-poor}

\hspace{0.6cm}As already mentioned in Sect.~\ref{discussion_paper1}, \citet{Bergeron05} suggested a classification of \Ovi absorbers in two populations: metal-rich absorbers (``type 1''; $N(\Ovi\!\!)/N(\Hi\!\!)\\>0.25$) and metal-poor (``type 0''; $N(\Ovi\!\!)/N(\Hi\!\!)<0.25$). We follow their approach to investigate the physical reasoning behind this distinction and independently study the distributions of Doppler parameter and column density for both absorber classes. The total \Ovi and \Hi column densities of the sampled systems are defined as the sums of the column densities of their components. The extremal values of the ratio are estimated as follows:

\[\sigma^{\rm max}_{\rm ratio} = \frac{N(O\,\text{\sc  vi})_{\rm tot} + \sigma^{+}(N(O\,\text{\sc  vi}))_{\rm tot}}{N(H\,\text{\sc i})_{\rm tot} - \sigma^{-}(N(H\,\text{\sc i}))_{\rm tot}}-\frac{N(O\,\text{\sc vi})_{\rm tot}}{N(H\,\text{\sc i})_{\rm tot}}, \]

\[\sigma^{\rm min}_{\rm ratio} = \frac{N(O\,\text{\sc vi})_{\rm tot}}{N(H\,\text{\sc i})_{\rm tot}} - \frac{(N(O\,\text{\sc vi})_{\rm tot} - \sigma^{-}(N(O\,\text{\sc vi}))_{\rm tot}}{(N(H\,\text{\sc i})_{\rm tot} + \sigma^{+}(N(H\,\text{\sc i}))_{\rm tot}}. \]\\

The results and the estimated $1\sigma$ deviations $\sigma^{\pm}$ are presented in Appendix \ref{appendix_systems}, Table~\ref{column_density_ratio} and \ref{appendix_errors}, respectively. 

 
In Fig.~\ref{number_systems} (bottom panel) is shown the distribution of metal-rich and metal-poor systems, according to the number of their components. Evidently, most of the metal-poor absorbers ($\approx$ 79 per cent) inhabit in systems with one to three components, whereas the metal-rich systems (about one third of the whole sample) tend to be more dispersed in the component space. 

\subsection{Distributions of Doppler parameter and column density}\label{col_den_b_distr}

\begin{figure}[!h]
\begin{center}
\resizebox{0.75\hsize}{!}{\includegraphics[angle=-90]{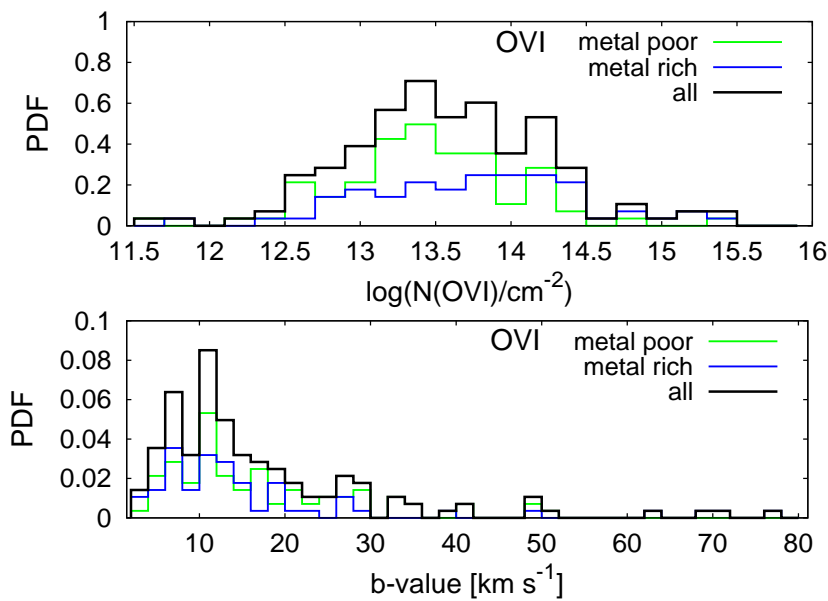}}
\caption[Distributions of N(\Ovi\!\!) and $b$ from the UVES sample.]{Column density (top) and Doppler parameter (bottom) distributions for all (black), metal-rich (blue) and metal-poor (green) \Ovi components.}
\label{N_b_rich_poor}
\end{center}
\end{figure}

\hspace{0.6cm}
In this section we present briefly the obtained Doppler parameter and column density distributions of the observed \Ovi absorbers. (More detailed analysis and comparison with the results from OWLS are provided in Sect.~\ref{doppler_parameter_owls}.) The distributions 
of the whole UVES sample are shown in Fig. \ref{N_b_rich_poor}. The median value for the Doppler parameter is 13.0 \kms\!\!, which corresponds to $T_{\rm UVES}\sim 1.6 \times 10^{5}$~K, assuming purely thermal broadening. This temperature is consistent with the result of \citet{Muzahid11}: $T \sim 1.8 \times 10^{5}$~ K, obtained from a similar UVES data set.  

The fractional column density and Doppler-parameter distributions for the populations of metal-rich and metal-poor absorbers are compared in Fig.~\ref{N_b_rich_poor}. No significant difference between the fractional $b$ distributions is seen. In contrast, the column density distributions appear to be different. The distribution peaks at log N(\Ovi\!\!)$\approx 14.0$ for metal-rich systems, whereas the column densities of their metal-poor peers span a narrower range, centered at log N(\Ovi\!\!)$\approx 13.4$. The latter value is close to the median of the total sample: log N(\Ovi\!\!)$\approx13.6$.


    \subsubsection{The column density distribution function}\label{cddf}

\hspace{0.6cm}The column density distribution function (CDDF), $f(\text{O\,{\sc vi}})$, is defined as the number of absorption systems ${\mathcal N} (N_\text{O\,{\sc vi}})$ per column density bin $\Delta N_\text{O\,{\sc vi}}$ and per unit absorption path length $\Delta X$ \citep{Petitjean93}: 

\begin{equation}
\label{cddf}
f(N_\text{O\,{\sc vi}}) = \frac{\mathcal N} {\Delta N_\text{O\,{\sc vi}} \times \sum\limits_{i=1}^n  X_{i}},
\end{equation}

\begin{figure}[h!]
\begin{center}
\resizebox{0.7\hsize}{!}{\includegraphics[angle=-90]{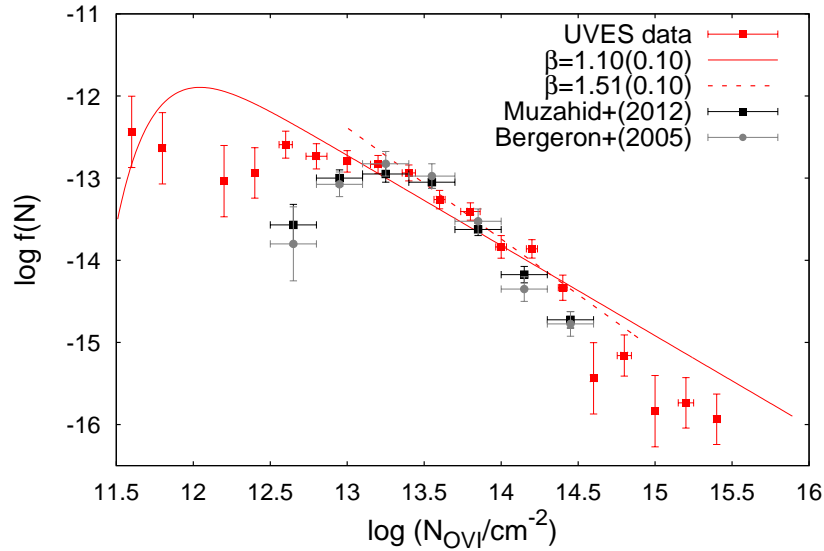}} 
\caption[CDDF from the UVES sample, fitted with power law functions.]{\small CDDF from the UVES sample, fitted with power law functions: corrected for completeness (solid  line) and with lower threshold $N(\rm \Ovi) \approx 13.3$ (dashed). The data points obtained by \citet{Muzahid11} and by \citet{Bergeron05} are plotted for comparison.}
\label{fig_cddf}
\end{center}
\end{figure}

From the total absorption path length, obtained in Sect.~\ref{overview}, we derive the CDDF from the UVES data. The result is plotted in Fig. \ref{fig_cddf}. The uncertainties on the $Y$ axis are estimated by the Poisson noise of the CDDF values while those on the $X$ axis, $\sigma(N_{\footnotesize \rm \Ovi\!\!})$, are calculated as follows: 

\begin{itemize}

\item[i)] Each column density value in a given bin is taken with a weight $\omega_{i}$ which reflects the fitting error $\sigma_{i}$:

\[\omega_{i} = 1/\sigma_{i}^2 \]

\item[ii)] The weighted mean for each bin is calculated from the column densities $N_{i}$ that fall within it:

\[ N_\text{bin} = \frac{\displaystyle{\sum_{i}\omega_{i}N_{i}}}{\displaystyle{\sum_{i}\omega_{i}}},\]

\item[iii)] Finally, the dispersion $D(N_\text{bin})$, is obtained from:

\[ D(N_\text{bin}) = \sigma(N_\text{bin}) = \sqrt{<N_{i}-N_\text{bin}>^{2}} ,\]

where 

\[{<N_{i}-N_\text{bin}>}^{2} = \frac{\displaystyle{\sum_{i}\omega_{i}(N_{i}-N_\text{bin})^{2}}}{\displaystyle{\sum_{i}\omega_{i}}}.\]

\end{itemize}

The data were fitted with a single power-law and with a power-law, multiplied by a completeness function (Fig. \ref{fig_cddf}, dashed and solid lines). In the second case we followed the procedure of \citet{Bekhti12}:

\begin{itemize}
 \item Calculation of the signal-to-noise (S/N) ratio. For normalized spectra and per resolution element, it can be computed through:

\[\text{S/N} =\sqrt \frac{\Delta \lambda_{r}} {\Delta \lambda_{p}} \frac{1}{\sigma_{p}^{\rm RMS}}~,\]

where $\Delta \lambda_{r}$ and $\Delta \lambda_{p}$ are the wavelength separations per resolution and per pixel element and ${\sigma_{p}^{\rm RMS}}$. We calculated the S/N ratio for 36 out of 40 O\,{\sc vi} systems with UVES spectra. It was not possible to estimate $\sigma_{p}^{\rm RMS}$ and the S/N ratio for 4 systems. In three cases, two nearby systems turned out to have identical S/N ratio.

The S/N ratio was calculated as well for \Civ and \Hi systems. The S/N distribution for spectral bands around the \Ovi\!\!, \Civ and \Hi lines is shown in Fig.~\ref{fig_SN_hist}. 

\begin{figure}[h!]
\begin{center}
\resizebox{0.7\hsize}{!}{\includegraphics[angle=-90]{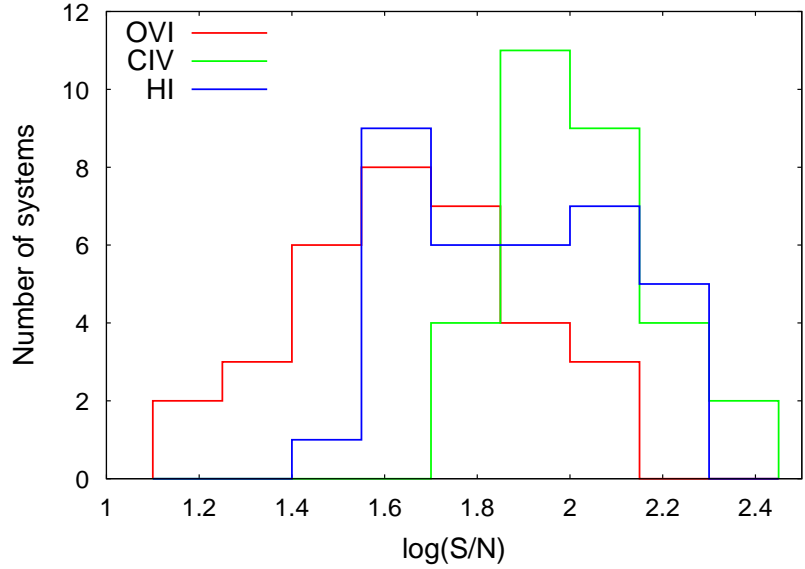}}
\caption[Distribution of S/N for \Ovi, \Civ and \Hi systems from the UVES sample.]{\small Distribution of S/N for the studied \Ovi absorbers (red), compared to those for \Civ and \Hi systems (green and blue, respectively).}
\label{fig_SN_hist}
\end{center}
\end{figure}

\item Calculation of the column density detection limit for all 36 \Ovi systems: the 3$\sigma$ limiting equivalent width 
of an unresolved line at wavelength $\lambda_{0}$ (see \citet{Tumlinson02}, their Eq. 1) is given by:

\[ W_{\lambda, \text{limit}} =\frac{\sigma \lambda_{0}} {(\lambda/\Delta \lambda)\text{(S/N)}},\]  

where $\lambda/\Delta \lambda$ is the spectral resolution. The relation between the equivalent width and the column density for small 
optical depths ($\tau\ll1$) is linear (see Eq.~\ref{eq_equivalent width_3}).



Replacing $W_{\lambda}$ from Eq.~\ref{eq_equivalent width_3} with $W_{\lambda,\text{limit}}$, one can estimate the column density detection limit $N_{\text{limit}}$:

\begin{eqnarray}
 N_{i,\text{limit}} & = & \frac{{\rm m_{e}} c^{2}} {\pi \rm e^{2}} \frac{\sigma}{(\lambda/\Delta \lambda)\text{(S/N)} f_{ij} \lambda_{0}}\nonumber\\
~ & =& 1.13\times 10^{20} \frac{\sigma}{\frac{\lambda}{\Delta \lambda}\big(\frac{\rm S}{\rm N}\big)f_{ij}\frac{\lambda_{0}}{\AA}} ~~~~~~~~~~~~~~~~~\rm[cm^{-2}] \nonumber.
\end{eqnarray}

Taking into account the UVES spectral resolution $(\lambda/\Delta \lambda) = 45\,000$ and adopting a $3\sigma$ threshold level (i.e., $\sigma = 3$), we 
calculated $N_{i,\text{limit}}$ for \Ovi\!\!, \Civ and \Hi absorption in the sampled 36 systems. The cumulative normalized 
distribution of \Ovi\!\!, \Civ and \Hi absorbers per column density bin, when certain $N_{i,\text{limit}}$ thresholds are imposed, is shown in Fig. \ref{fig_complet}. 

\begin{figure}[h!]
\begin{center}
\resizebox{0.7\hsize}{!}{\includegraphics[angle=-90]{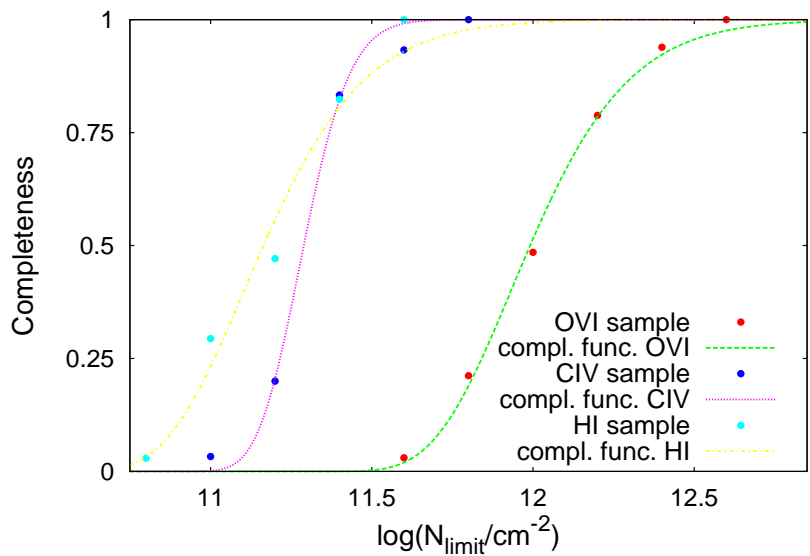}}
\caption[Column-density completeness of the UVES sample.]{\small Column-density completeness of the UVES sample.}
\label{fig_complet}
\end{center}
\end{figure}

\item Fitting of CDDF by use of an error function: following \citet{Bekhti12}, we introduce the parametrized error function: 

\[C(\text{log}N_{i,\text{limit}};a,b) = \frac{1}{2} \text{erf} [a(\text{log}(N_{i,\text{limit}}-b))]+ \frac{1}{2}.\]

This function can be understood as a completeness function, as it returns the fraction of sightlines which potentially can reveal an 
absorber with column density $N_{i,\text{limit}}$. The parameter $a$ describes the shape of the function and the parameter $b$ accounts 
for the offset.

Fig. \ref{fig_cddf} demonstrates that the CDDF drops at lower column densities due to the increasing incompleteness of the sample. 
To describe this behavior the completeness function $C$ is incorporated into the CDDF-fitting function: 

\[f(\text{log}N_{i}) = \beta \text{log}N_{i} + n + \text{log}C(\text{log}N_{i};a,b) ,\]

where $\beta$ is the slope of the power-law part and $n$ is a normalization constant. The shape parameter $a$ is fixed to a constant value as determined from the curves in Fig. \ref{fig_complet} and varied only the offset $b$ of the completeness function. 

\end{itemize}

The obtained slope of the fitted completeness function for the full range of column densities is $\beta \approx 1.10 \pm0.10$. It changes sensibly up to $\beta \approx 1.51 \pm0.10$ when the data are fitted by a single power-law function (without completeness corrections) and an apparent threshold of log$N(\rm \Ovi\!\!) \approx 13.3$ is imposed as a lower limit. Obviously the slope is very sensitive to the choice of a threshold. Next, we compare in Fig.~\ref{fig_cddf} our results of CDDF with this derived by \citet{Bergeron05} (gray) and \citet{Muzahid11} (black) from the same sample of UVES data. In both works a steeper slope is obtained; e.g. $\beta = 1.71 \pm0.48$ \citep{Bergeron05}. However, within the uncertainties, all derived distributions are in a good agreement in the column density range $13.3 \le \log N(\rm \Ovi\!\!) \le 14.3$. Neither \citet{Bergeron05} nor \citet{Muzahid11} find any components with column densities higher than $\log N(\rm \Ovi\!\!) \approx 14.5$. Components with such values in 
our study originate from four 
multicomponent systems that include other species, such as \Civ and \Nv\!\!. They are found along two sightlines: HE 2347-432 and J 2233-606. The latter QSO is not included in the sample of \citet{Muzahid11}. Three of these systems turn out to be saturated (HE 2347-432 at $z=2.897458$, HE 2347-432 at $z=2.891062$, and J 2233-606 at $z=2.198231$) which leads to large uncertainties for the determination of $N$. The fourth system J 2233-606 at $z=2.204508$ is not saturated, but blended. On the other hand, the two bins of lowest column density in our sample include only one component each (see Fig.~\ref{fig_cddf}). These components are part of the systems HE 2347-432 at $z=2.897458$ and HE 2347-432 at $z=2.891062$, respectively. \citet{Muzahid11} do not include them in the sample used for derivation of the CDDF.


\subsection{Contribution to cosmic baryon density}\label{baryon_density}

\hspace{0.6cm}The contribution of the \Ovi absorbers to the cosmic baryon density, $\Omega_{\footnotesize \rm \Ovi}$, can be estimated through the formula: 

\begin{equation}
\label{omega_OVI}
\Omega_{{\footnotesize \rm \Ovi}} = \frac{1}{\rho_{c}}m_{{\footnotesize \rm \Ovi}}\frac{\Sigma~w_{i}~ N_{{\footnotesize \rm \Ovi\!\!},i}}{(c/H_{0})\Sigma~ \Delta X},
\end{equation}

where $\rho_{c}$ is the critical density, $m_{\footnotesize \rm \Ovi\!\!}$ is the mass of the oxygen atom, $\Delta X$ is the absorption path length (see Eq.~\ref{absorption_path_length}), $w_{i}$ is the completeness correction as calculated in the previous subsection (see Fig.~\ref{fig_complet}), and $N_{{\footnotesize \rm \Ovi\!\!},i}$ are column densities of the individual absorbers \citep{Burles96}. Here, the value of the Hubble constant is taken to be: $H_{\rm 0}=73$ \kms\!\!Mpc.

To achieve a statistically significant result, we calculate the cosmic density of \Ovi\!\! absorbers from the extended sample of 43 intervening systems in the redshift range $2.04 \leq z \leq 3.60$. The obtained estimate is $\Omega_{{\footnotesize \rm \Ovi\!\!}} = 3.3 \times 10^{-7}$, for the full column density range $11.5 \le \log N \le 15.5$. For comparison, the result of \citet{Bergeron05} is a factor of 2 less: $\Omega_{{\footnotesize \rm \Ovi\!\!}} = 1.5 \times 10^{-7}$, while \citet{Muzahid11} obtain an even lower value of $\Omega_{{\footnotesize \rm \Ovi\!\!}} = 1.0 \times 10^{-7}$, with a  lower column density cutoff of 13.7 in log scale.

The cosmological density of baryons, associated with the \Ovi absorbing phase, can be estimated by use of a similar formula:

\begin{equation}
\label{omega_OVI}
\Omega^{{\footnotesize \rm \Ovi}}_{{\tiny \rm IGM}} = \frac{1}{\rho_{c}}\mu m_{{\tiny \rm H}}\frac{\Sigma~w_{i}~ N_{{\footnotesize \rm \Ovi\!\!},i}}{(c/H_{0})(f_{{\footnotesize \rm \Ovi}}Z(\rm O/\rm H)_{\odot})\Sigma~ \Delta X},
\end{equation} 

where the mean atomic weight is taken to be $\mu=1.3$, $m_{\rm H}$ is the mass of the hydrogen atom, $f_{\footnotesize \rm \Ovi}$ is the ionization fraction of \Ovi, and $Z$ is the gas metallicity \citep{Tripp}. Following \citet{Muzahid11}, we adopt $f_{\footnotesize \rm \Ovi\!\!}= 0.2$ and $Z = 0.1 Z_{\odot}$ and obtain $\Omega^{{\footnotesize \rm \Ovi\!\!}}_{{\rm IGM}}/\Omega_{\rm b}= 0.00856$ with $\Omega_{\rm b}=0.0418$ \citep{Spergel07}. The fraction of baryons associated with \Ovi absorbers turns out to be less than 1 per cent of the total baryon density. This value -- as we shall see later -- is in excellent agreement with the predictions from the cosmological simulations (Sect.~\ref{baryons_owls}). For comparison, \citet{Muzahid11} find a slightly higher value. According to their estimate, the contribution of the \Ovi absorbers to the total baryon density in the Universe is 2.8 per cent.


    \subsection{Velocity widths and velocity offsets}\label{velocity}

\hspace{0.6cm}We investigated the velocity widths (or spreads) of the \Ovi absorbers in our UVES sample and their velocity offsets to the corresponding \Hi and \Civ lines. The weighted mean velocity for a given spectral profile is defined as:

\begin{equation}
\label{eq_mean_velocity}
\langle  v_{\rm w}\rangle = \frac{\displaystyle {\sum_{v_{\rm min}\le v_{i}\le{v_{\rm max}}}v_{i}(F_{c}-F_{i}})}{\displaystyle {\sum_{v_{\rm min}\le v_{i}\le{v_{\rm max}}}(F_{c}-F_{i})}}~,
\end{equation} 

where $F_{c}\equiv 1$ is the continuum flux, $F_{i}$ is the normalized flux per pixel, $v_{i}$ is the  velocity per pixel and $v_{\rm min}$ and $v_{\rm max}$ are the velocities corresponding to the lower and upper limit (or left and right absorption sides) of the fitting Voigt profile ($F(v_{\rm min})=F(v_{\rm max})=F_c$). The absorption-velocity spread $\delta v$ is calculated simply as $[v_{\rm max}-v_{\rm min}]$ which is a good first-order approximation of this quantity when the line wings are not very wide \citep{Ledoux06}. 

The velocity spreads and the weighted mean velocities for each \Ovi\!\!, \Civ and \Hi\!\! absorption feature were calculated in 34 out of 40 considered \Ovi systems. Six \Ovi systems were excluded from the sample due to a very high uncertainty of $\langle v_{\rm w}\rangle$. Table~\ref{fig_velocity_spread} (Appendix \ref{velocity_spread}) lists lower, $v_{\rm min}$, and upper, $v_{\rm max}$, velocity limits (or left and right sides of the absorption feature) for the species \Ovi\!\!, \Civ\!\!, \Nv and \Hi\!\!, where the velocity zero point $v=0$ is set at the center of the strongest \Ovi line. The \Ovi and \Hi velocity spreads for each absorber are also given. In some cases, two different sets $(v_{\rm min},~v_{\rm max},~\delta v)$ are calculated: one for the associated \Ovi absorber and a second one that characterizes the system itself. 

\begin{figure}[h!]
\begin{center}
\resizebox{0.75\hsize}{!}{\includegraphics[angle=-90]{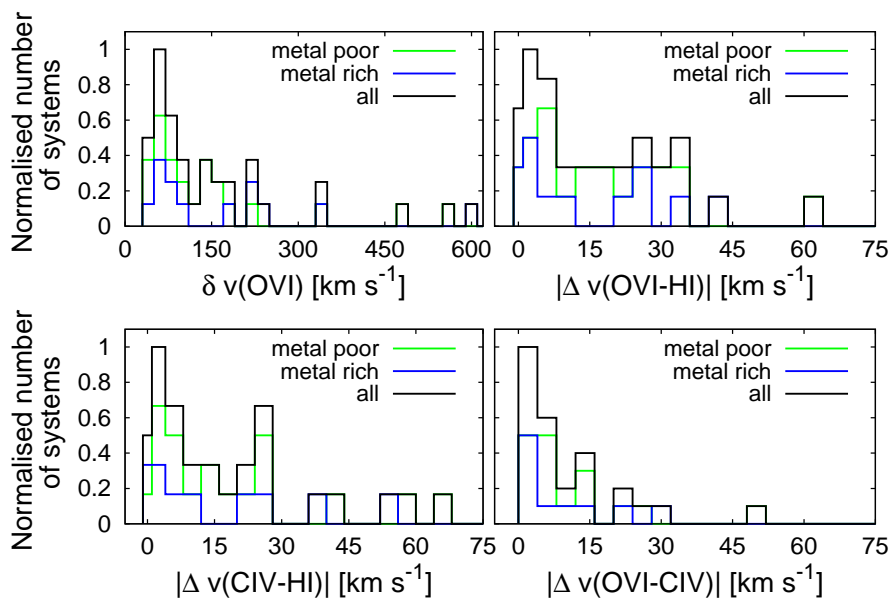}}
\caption[Distributions of \Ovi velocity spread and of the velocity offsets \Ovi-\Hi, \Civ-\Hi and \Ovi-\Civ.]{\small Distributions of \Ovi velocity spread (top left panel) and of the velocity offsets between the absorption features \Ovi-\Hi (top right panel), \Civ-\Hi (bottom left panel) and \Ovi-\Civ (bottom right panel).}
\label{fig_vel_off_ovi_civ_hi}
\end{center}
\end{figure}

The distribution of the \Ovi velocity spread for metal-rich and metal-poor absorbers is plotted in Fig.~\ref{fig_vel_off_ovi_civ_hi} (top left panel). Evidently, there is a clear maximum at about 40 \kms\!\!. The large spread in the distribution indicates the complex kinematics of spatially separated metal-enriched gas phases in the absorbing gas structures. 

Then the velocity offset $\Delta v$ between \Ovi\!\!, \Civ and \Hi absorbers was obtained through:

\[ \lvert \Delta v(\text{X}_{1}-\text{X}_{2})\rvert = \lvert \langle {v_{\rm w}(\text{X}_{1})}\rangle - \langle {v_{\rm w}(\text{X}_{2})}\rangle \rvert~,\]

where $\text{X}_{1}$ and $\text{X}_{2}$ are the studied species (ions). The normalized distributions of the velocity offsets between 
\Ovi\!\!, \Civ and \Hi weighted mean velocities are shown in Fig.~\ref{fig_vel_off_ovi_civ_hi}. In all considered cases, the systems were classified as metal-rich or metal-poor, according to the criterion of \citet{Bergeron05} (see Sect.~\ref{discussion_paper1}). The distributions of the velocity offsets $\Delta v(\text{O\,{\sc vi} - H\,{\sc i}})$ (top right panel) and $\Delta v(\text{C\,{\sc iv} - H\,{\sc i}})$ (bottom left panel) are evidently bimodal, while the distribution of the velocity offset $\Delta v(\text{O\,{\sc vi} - C\,{\sc iv}})$ (bottom right panel) is not. Looking for some specific characteristics that might be related to this bimodality, we delineated two subsamples with velocity offsets in ranges $\sim10-20 ~\rm km~s^{-1}$ around the peaks: 
$\lvert\Delta v_{\rm min}(\text{O\,{\sc vi} - H\,{\sc i}})\rvert = [0 - 13] ~\rm km~s^{-1}$, $\lvert\Delta v_{\rm max}(\text{O\,{\sc vi} - H\,{\sc i}})\rvert= [22 - 38] ~\rm km~s^{-1}$, and $\lvert\Delta v_{\rm min}(\text{C\,{\sc iv}} - \text{H\,{\sc i}})\rvert= [0 - 12] ~\rm km~s^{-1}$, $\lvert\Delta v_{\rm max}(\text{C\,{\sc iv}} - \text{H\,{\sc i}})\rvert= [20 - 30] ~\rm km~s^{-1}$, respectively. 

\begin{figure}[h!]
\begin{center}
\resizebox{0.8\hsize}{!}{\includegraphics[angle=-90]{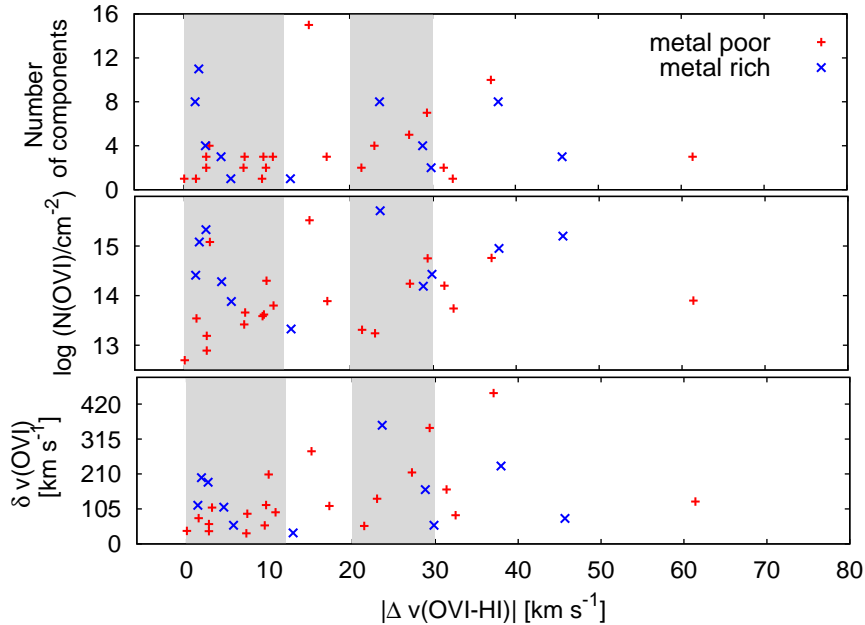}}
\caption[Diagram \Ovi-\Hi velocity offset vs. number of components, column density and absorption width in a given \Ovi system]{\small Diagram \Ovi-\Hi velocity offset vs. number of components (top panel), column density (middle panel) and absorption width (bottom panel) in a given \Ovi system. Zones of peaks in the velocity offset distributions (shadowed areas) are shown.}
\label{fig_vel_spread_N_ovi}
\end{center}
\end{figure}

A characteristic of the absorbers which seems to be related to the bimodality of the $\lvert\Delta v(\text{O\,{\sc vi}} - \text{H\,{\sc i}})\rvert$ distribution is the number of components in a given system. As seen in Fig.~\ref{fig_vel_spread_N_ovi}, top panel, the majority of objects within range of the first peak (13 out of 17) exhibit a simple structure with 1 up to 3 absorption components. The population around the second peak range is apparently different: 7 out of 12 systems are multicomponent and have a more complex structure. 

This trend implies that absorbers with small velocity offsets between \Ovi and \Hi (i.e., absorbers within range of the first peak) represent kinematically simple, isolated gas structures. \Ovi absorbers that have a significant velocity offset compared to \Hi instead appear to be embedded in a kinematically more complex host structure with a larger internal velocity dispersion, as expected, for instance, for gaseous material expelled from galactic winds and outflows.  


\begin{figure}[h!]
\begin{center}
\resizebox{0.7\hsize}{!}{\includegraphics[angle=-90]{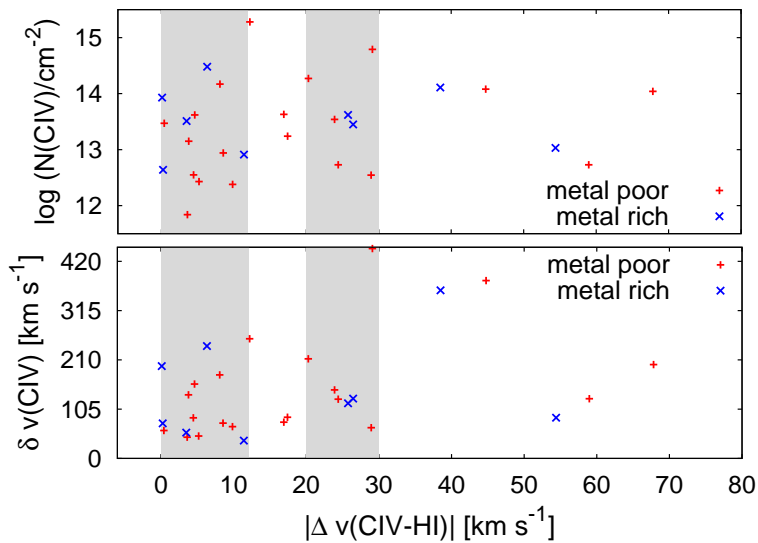}}
\caption[Diagrams \Civ-\Hi velocity offset vs. \Civ absorption width and column density.]{\small Velocity offset between the weighted mean velocities of \Civ and \Hi absorption features versus \Civ absorption width (bottom panel) and \Civ column density (top panel). Zones of peaks in the velocity offset distributions (shadowed areas) are shown.}
\label{fig_vel_spread_N_civ}
\end{center}
\end{figure}

Other properties of \Ovi\!\!, \Civ and \Hi absorption features, that can be studied in relation to the evident bimodality of the velocity offset distributions, are the velocity spreads and the total column density. Although the interpretation of the observed trends for these quantities is not straight-forward, we nevertheless discuss their properties to gain an insight into the kinematics of the absorbing gas structures. Our tests are performed and illustrated in Figs.~\ref{fig_vel_spread_N_ovi}, middle and bottom panels, and \ref{fig_vel_spread_N_civ}, respectively. In both cases, the zone of the low velocity offset seem to be populated mainly by metal-poor systems.

Weak correlations between the velocity width, the column density and the velocity offset are seen in Fig.~\ref{fig_vel_spread_N_ovi}: systems with 
$\lvert\Delta v(\text{O\,{\sc vi}} - \text{H\,{\sc i}})\rvert \le 13 ~\rm km~s^{-1}$ are mostly systems with low velocity spread 
$\delta v(\text{O\,{\sc vi}}) \le 105  ~\rm km~s^{-1}$. On the other hand, O\,{\sc vi} systems with velocity offsets in the range of the second peak 
($\lvert\Delta v(\text{O\,{\sc vi}} - \text{H\,{\sc i}})\rvert \ge 22 ~\rm km~s^{-1}$) are mostly systems of higher column density $\log N$(O\,{\sc vi})$\ge 14$. A similar correlation is found as well considering the H\,{\sc i} absorption in those systems. Among the systems from the subsample with low velocity offsets, 82 per cent exhibit column densities $\log N(\text{H\,{\sc i}}) \le 14.9$ while systems with high velocity offsets are typically objects with high column density: 75 per cent of them have $N(\text{H\,{\sc i}}) \ge 15.0$. 

The same tendency concerning the \Hi column density is evident when $\lvert\Delta v(\text{C\,{\sc iv}} - \text{H\,{\sc i}})\rvert$ 
is considered: 73 per cent of the absorbers in the first subsample have $\log N(\text{H\,{\sc i}}) \le 14.9$ vs. $\log N(\text{H\,{\sc i}}) \ge 14.9$ for 80 per cent of the absorbers in the one with high velocity offsets. 

\begin{figure}[h!]
\begin{center}
\resizebox{0.7\hsize}{!}{\includegraphics{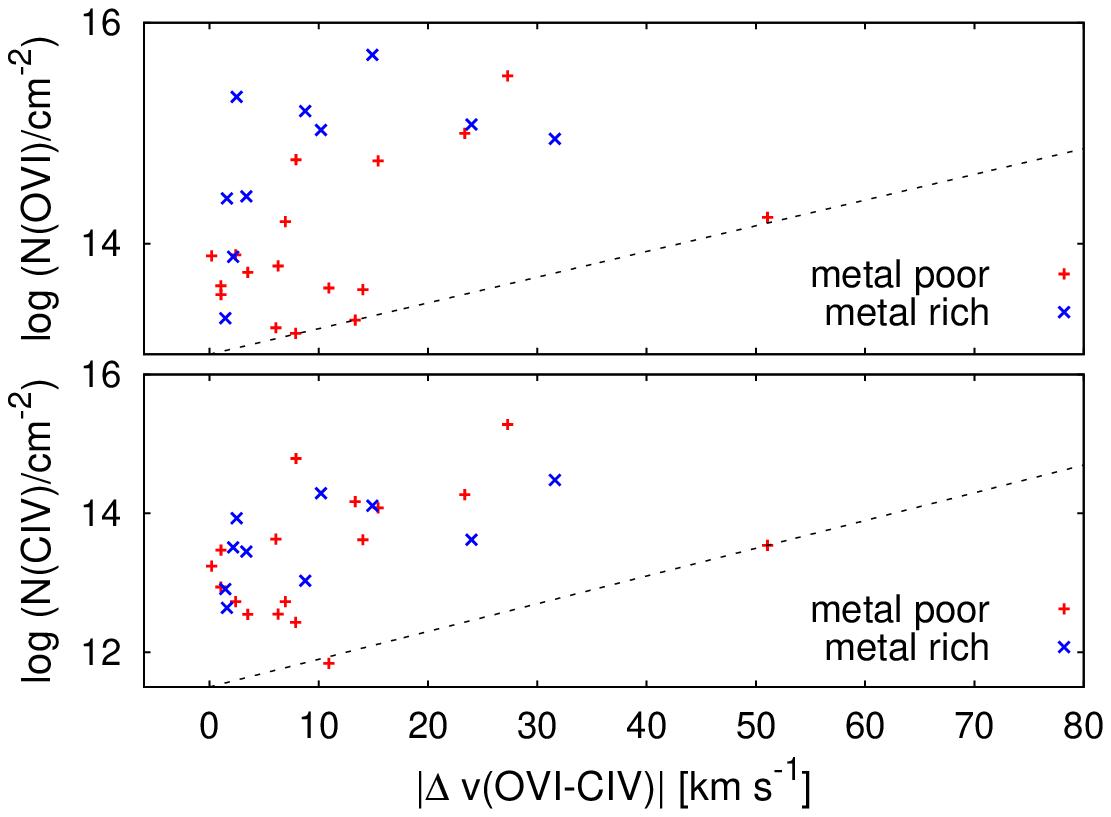}}
\caption[Diagrams \Ovi-\Civ velocity offset vs. \Civ and \Ovi column density.]{\small Velocity offset between the weighted mean velocities of \Ovi and \Civ absorption features versus column density of \Civ (bottom panel) an \Ovi (top panel). The possible presence of a lower envelope is indicated by black line.}
\label{fig_N_civ_ovi}
\end{center}
\end{figure}

\begin{figure}[h!]
\begin{center}
\resizebox{0.7\hsize}{!}{\includegraphics{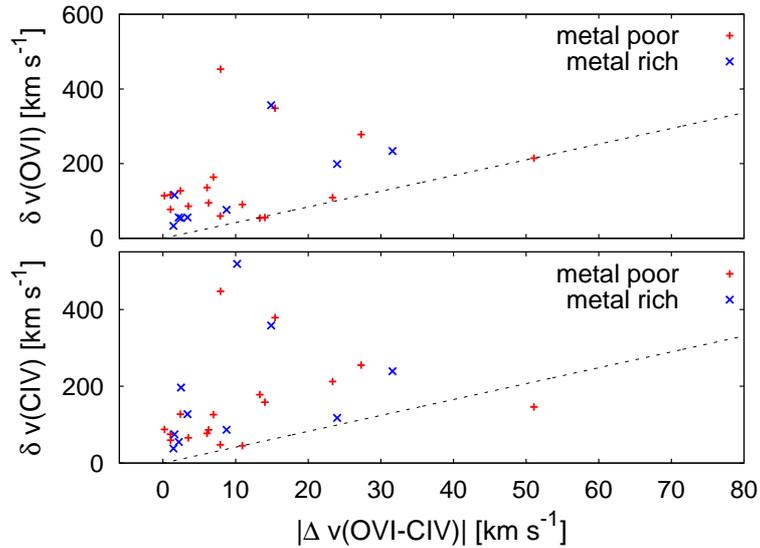}}
\caption[Diagrams \Ovi-\Civ velocity offset vs. \Civ and \Ovi velocity spread.]{\small Velocity offset between the weighted mean velocities of \Ovi and \Civ absorption features versus velocity spread of \Civ (bottom panel) and  \Ovi (top panel). The possible presence of a lower envelope is indicated by the black line.}
\label{fig_vel_spread_civ_ovi}
\end{center}
\end{figure}

Finally we study the possible relation of velocity offsets between \Ovi and \Civ\!\! absorption features and the corresponding column densities $N$(X) and velocity widths $\delta v(\text{X})$ (Figs.~\ref{fig_N_civ_ovi} and \ref{fig_vel_spread_civ_ovi}). It seems that metal-poor systems have preferably lower column densities: $\log N$(O\,{\sc vi})$\le$14 (Fig.~\ref{fig_N_civ_ovi}, upper panel). Most of the systems are characterized by low velocity offset:
$\lvert\Delta v(\text{O\,{\sc vi}} - \text{C\,{\sc iv}})\rvert \le 20 ~\rm km~s^{-1}$. Systems with higher \Ovi - \Civ velocity offset are those with higher column density ($\log N$(O\,{\sc vi})$\ge14$) and velocity spread
$\delta v(\text{O\,{\sc vi}}) \ge 100  ~\rm km~s^{-1}$ (Figs.~\ref{fig_N_civ_ovi} and~\ref{fig_vel_spread_civ_ovi}, upper panels). 
A similar trend is observed when the velocity spread of \Civ is considered (Fig.~\ref{fig_vel_spread_civ_ovi}, lower panel). These findings are consistent with the study of \citet{Muzahid11}.


    \section{Ionization modeling with {\sc Cloudy}}\label{modeling_cloudy}

\hspace{0.6cm}As shown in the previous section, only a fraction of the \Ovi absorbers in our sample is kinematically aligned with \Hi and \Civ\!\!. Only for such aligned systems it is reasonable to assume that the above ions trace the same physical regions within the host structure and that an ionization model provides meaningful results. 

To investigate the physical conditions in kinematically aligned \Ovi\!\!/\Hi absorbers, we modeled the ion column densities of selected absorbers from the UVES sample using the photoionization code {\sc Cloudy} \citep[version C08;][]{Ferland}. For this, we follow a strategy that is similar to that adopted for the case study of 2 individual absorbers presented in Sect.~\ref{ionization_modeling_paper1}. A solar abundance pattern of O and C was assumed. We adopt an optically thin plane-parallel geometry of systems in photoionization equilibrium exposed to a \citet{HM01} UV background spectrum at $z=2.16$, normalized to $\log~J_{912} = -21.15$ \citep{Scott2000} at the Lyman limit.

The number of \Ovi\!\!/\Hi pairs aligned in velocity space ($\Delta v_{\rm comp.}\le$ 10 \kms\!\!) is given in Table~\ref{number_of_components}. We modeled 6 single-component intervening systems, 7 intervening systems with $2-3$ absorption components, 4 intervening milticomponent systems and 1 intervening system with 3 components and a velocity shift of $\approx$11 \kms\!\!. Five associated \Ovi systems with up to 3 components were also modeled as a test sample. The models in the latter case give limits of the output physical parameters, since only an UV background spectrum \citep{HM01} was included as input spectrum without taking into account the effect of the local quasar spectrum. Details on each model are specified in Appendix \ref{ionization_modeling_uves}.

We consider the following input model parameters: 

\begin{itemize}

 \item Measured column densities of \Ciii and/or \Civ\!\!, when present;
 \item Column density of \Ovi\!\!;
 \item Metallicity $Z$ (in solar units);
 \item Hydrogen volume density $n_{\rm H}$.

\end{itemize}

The parameter ranges for the metallicity and hydrogen density were $0.001\leq ~Z/Z_{\odot} \leq 1$ and $-5\leq \,$log$ ~n_{\rm H} \leq 0$, respectively. In some cases, we derive super-solar metallicities together with very low corresponding volume densities. We included this results in the tables as possible solutions of {\sc Cloudy}, although the physical meaning remains unclear. The column density of the  \Hi component that is closest to the modeled \Ovi has been taken directly from observations as a stopping criterion for modeling. In two individual cases, PKS 0237-233 at $z=2.202783$ and Q 0453-423 at $z=2.636236$, \Siiii and \Siiv were modeled as well.  

The first step in the modeling procedure was to derive models using {\sc Cloudy} for a discrete set of values of $Z$ and $n_{\rm H}$. In case that \Ciii and/or \Civ were present, the output column densities $N$(\Ciii\!\!), $N$(\Civ\!\!) and $N$(\Ovi\!\!) were compared to the observed ones. The input parameters $Z$ and $n_{H}$ then were adjusted before the next iteration step, until the differences between the output column densities and the observed values became negligible. Thus an unique solution was obtained.
 
In cases, where no ions other than \Hi and \Ovi have been detected, we constrained the gas temperature from the observed \Ovi line widths and used $T$ as a fixed input parameter (see Eq.~\ref{eq_temp_Doppler}). This leads to more than one valid solution for $Z$ and $n_{H}$.

In the cases, where the components of the absorption were close to each other in velocity space, we modeled the sum of the 2 column densities: $N_{\rm sum} = N_{1}+N_{2},~~ \sigma_{\rm sum}=\sqrt{\sigma_{1}^2+\sigma_{2}^2}$. Whenever necessary, we calculated the temperature from the weighted mean $b$-value of the two components: $<b>=\frac{N_{1}b_{1} + N_{2}b_{2}}{N_{1} + N_{2}}$, ~~ $\sigma_{<b>}=\sqrt{\frac{(N_{1}(b_{1}-<b>)^{2}+N_{2}(b_{2}-<b>)^{2}}{N_{1}+N_{2}}}$. 

In addition to the ion column densities, our {\sc Cloudy} models provide information on the gas temperature, $T$, the neutral hydrogen fraction, $f_{\rm HI}$, and from the latter the absorption pathlength (the thickness of an absorber), $L=N$(H\,{\sc i}$)/(f_{\rm HI}\,n_{\rm H})$. 

\begin{figure}[h!]
\begin{center}
\resizebox{0.9\hsize}{!}{\includegraphics[angle=-90]{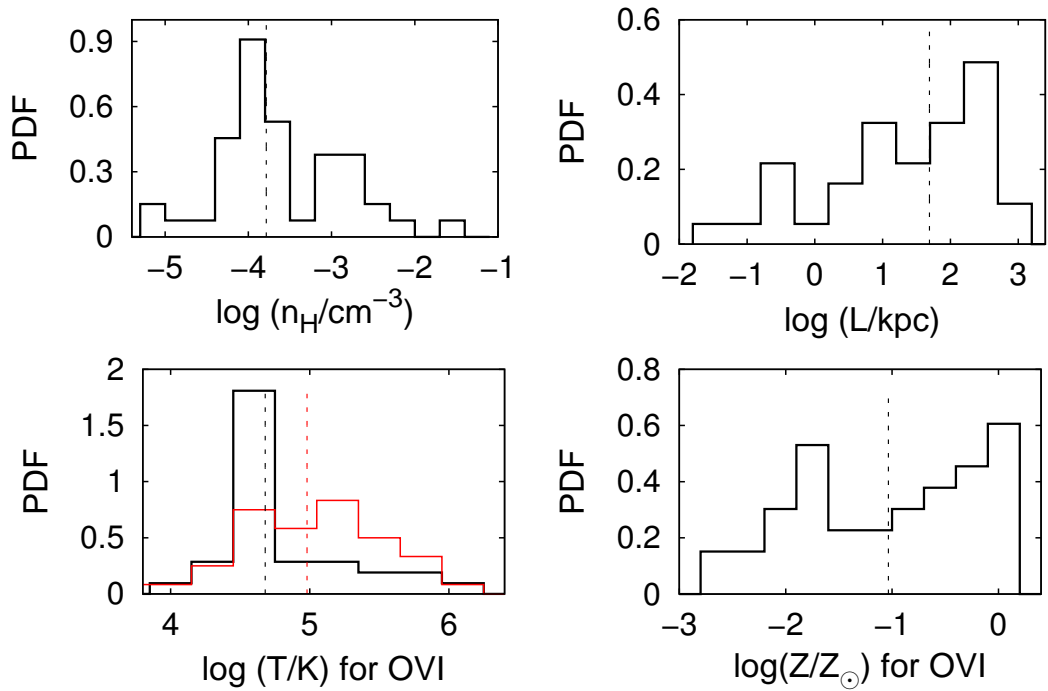}}
\caption[Distributions of physical parameters of \Ovi absorbers obtained from {\sc Cloudy} modeling.]{Distributions of density, thickness, temperature and metallicity of \Ovi absorbers obtained from {\sc Cloudy} modeling. In the lower left panel, the temperature distribution derived from {\sc Cloudy} models (black) is compared with that of upper temperature limit $T_{\rm max}$ (red) derived from the Doppler parameter. The dashed vertical lines in each panel mark the medium values in the samples.}
\label{fig_uves_hist_cloudy}
\end{center}
\end{figure}

The distributions of volume density $n_{\rm H}$,  temperature $T$, metallicity $Z$ and absorption path length $L$ through a cloud, as derived through {\sc Cloudy}, are plotted in Fig.~\ref{fig_uves_hist_cloudy}. The density distribution (top left panel) is rather broad with a weak evidence for bimodality, with peaks at log $n_{\rm H} \approx -4.0$ and log $n_{\rm H} \approx -2.8$. This apparent bimodality might be caused by incompleteness effects that we are not able to treat statistically at this point. However, as one can see in Sect.~\ref{density_owls}, a small second peak at log $n_{\rm H} \approx -3.0$ is indeed expected from the OWLS which are more complete in a statistical sense. The median value of $\log\tilde n_{\rm H} = -3.8$  is typical for intergalactic absorbers.  

Two temperature distributions of the \Ovi sample are plotted in Fig.~\ref{fig_uves_hist_cloudy}, bottom left panel. The distribution derived from {\sc Cloudy} modeling is juxtaposed with the one of the upper temperature limit $T_{\rm max}$ derived from the \Ovi Doppler parameter values assuming purely thermal line broadening (Eq.~\ref{eq_temp_Doppler}). The temperatures calculated from {\sc Cloudy} are systematically lower than $T_{\rm max}$ -- note the difference between the median values $\log \tilde T \approx 4.7$ and $\log \tilde T_{\rm max} \approx 5$. This result will be discussed in more details in Sect.~\ref{aligned_absorbers_owls}, where it is compared to the results from the OWLS.

Note that the {\sc Cloudy} modeling procedure mimics the observational approach: the metallicity is calculated from the total $N(\rm \Ovi\!\!)$/$N(\rm \Ovi\!\!)$ ratio. Later, in Sect.~\ref{metallicity_owls}, we will introduce a second measure of the metallicity based on the OWLS and will discuss in more details the results from applying both metallicity definitions. The metallicity distribution of \Ovi systems, as derived from {\sc Cloudy}, is characterized by median value (in log scale) of $-1.04$ dex (Fig.~\ref{fig_uves_hist_cloudy}, bottom right panel). It appears bimodal, with peaks at $\log Z/Z_{\odot} \approx -1.8$  and $\log Z/Z_{\odot} \approx 0.0$. A similar bimodality in metallicity is found by \citet{Simcoe06}. However, their study includes {\it all}  heavy elements, modeled with {\sc Cloudy}. The authors suggest that the systems populating the zone around the higher-metallicity peak are probably produced by recent galaxy formation and feedback, while those around the lower-metallicity peak might 
trace 
older galactic debris. Interestingly enough, the bimodal tendency in the metallicity distribution of \Ovi absorbers is not confirmed by the OWLS (see Sect.~\ref{metallicity_owls}). 

Finally, we study the distribution of the absorption path length $L$ in the \Hi structure that contains \Ovi (Fig.~\ref{fig_uves_hist_cloudy}, top right panel). It spans about five orders of magnitude: $0.01\le L \lesssim 10^3$ kpc, with a median value $\tilde L\simeq 50$~kpc. Generally, \Ovi absorbers obviously arise in large-scale cosmological structures ($\log L\gtrsim2$) as well as in local, small-scale structures. There is, however, an clear tendency that \Ovi absorbers with absorption path lengths $>10$ kpc are far more numerous than absorbers with smaller absorption path lengths. 

\begin{figure}[h!]
\begin{center}
\resizebox{0.9\hsize}{!}{\includegraphics[angle=-90]{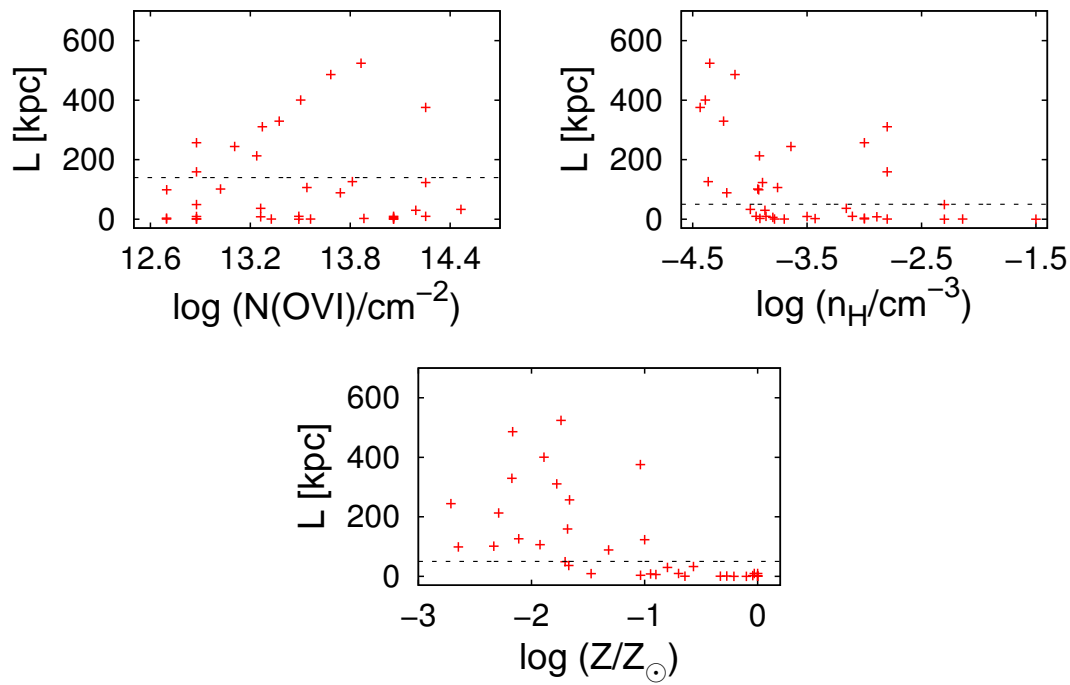}}
\caption[Correlation plots of physical parameters of \Ovi absorbers obtained from {\sc Cloudy} modeling.]{Correlation plots of some physical parameters obtained from the {\sc Cloudy} models. The dashed line on the top left panel denotes a critical value for the absorption path length of $L=140$ kpc, while in the other cases it is plotted at $L=50$ kpc (see text).}
\label{fig_uves_plots_cloudy}
\end{center}
\end{figure}

We examine possible correlations between the parameters derived from {\sc Cloudy} modeling in (Fig.~\ref{fig_uves_plots_cloudy}). The \Ovi column density is put into relation with the absorber thickness $L$ on the top left panel. Clouds with smaller absorption path lengths (critical value: $L\approx140$ kpc) seem to be homogeneously spread over the full column density range, while systems of high $L$ values exhibit a positive correlation between the thickness of the \Hi cloud and the \Ovi column density. The chosen critical value $L\approx140$ demarcates two groups of systems: a population of local (small) structures and a population of the large scale structures. The most noticeable result from Fig.~\ref{fig_uves_plots_cloudy} is the absence of points in the upper right corners. If there were systems with high column densities/volume densities/metallicities and high absorption path lengths, they should be detectable. Systems with metallicities higher than 10 per cent of the solar solar value or 
with volume densities higher than $2 \times 10^{-3}$\cc~ display a maximum absorption path length of 50 kpc. A similar result is found by \citet{Simcoe06}, who suggest that the lack of data points in this parameter range might be explained by the fact that optically thick metal-enriched systems with higher gas densities are rare and have a small absorption-cross section.
 
\section{Conclusions}\label{conclusions_uves_survey}

\hspace{0.6cm}In this chapter, we analyzed the statistical and physical properties of intervening \Ovi absorbers at high redshift based on a sample of 40 ($+$ 11) \Ovi systems observed with VLT/UVES. Our main conclusions could be summarized as follows:

\begin{itemize}

\item 
The calculated rate of incidence for the extended sample of 43 intervening \Ovi\!\! systems is $d \mathcal N/dz = 5.2 \pm0.8$. The corresponding rate of incidence per absorption path length is $d \mathcal N/dX = 1.6 \pm0.2$. Because of the imposed strong selection criteria, these values are slightly lower than previous estimates of the incidence rate of \Ovi absorbers at high redshift.

\item
About 30 percent of the basic (not extended) sample of 32 intervening \Ovi absorbers are single-component absorbers, while the remaining systems exhibit a more complex component structure. One third of the components in the single-component systems do not exhibit alignment with \Hi\!\!, while this fraction increases with the number of \Ovi subcomponents.

\item
The median value of the Doppler parameter distribution from the UVES sample is 13.0 \kms\!\!, which corresponds to temperature $T_{\rm UVES}\sim 1.6 \times 10^{5}$~K, assuming purely thermal broadening. The median column density value is log$N(\Ovi\!\!)=13.6$.

\item
The obtained power-law slope of CDDF at high redshift (using a completeness function) is $\beta \approx 1.10 \pm0.10$, when all points are included. The slope changes sensibly to $\beta \approx 1.51 \pm0.10$, imposing an apparent threshold of log$N(\rm \Ovi\!\!) \approx 13.3$ as a lower limit. Obviously the slope is very sensitive to the choice of a threshold.

\item
We calculate the cosmic density of \Ovi\!\! absorbers from the extended sample of 43 intervening systems in the redshift range $2.06 \leq z \leq 2.91$. We derive a value of $\Omega_{{\footnotesize \rm \Ovi\!\!}} = 3.3 \times 10^{-7}$ for the full column-density range $11.5 \le \log N \le 15.5$. The cosmological density of baryons, associated with the \Ovi absorbing phase is $\Omega^{{\footnotesize \rm \Ovi\!\!}}_{{\rm IGM}}/\Omega_{\rm b}= 0.00856$ with $\Omega_{\rm b}=0.0418$. The fraction of baryons associated with \Ovi absorbers turns out to be less than 1 per cent of the total baryon density.

\item
We investigate the velocity widths $\delta v$ of the \Ovi absorbers in our UVES sample and their velocity offsets $\Delta v$ to the corresponding \Hi and \Civ lines. We find that absorbers with small velocity offsets between \Ovi and \Hi ($\Delta v \le 13$~\kms\!\!) represent kinematically simple, isolated gas structures. In contrast, \Ovi absorbers that have a significant velocity offset compared to \Hi appear to be embedded in a kinematically more complex host structure with a larger internal velocity dispersion, as expected, for instance, for gaseous material expelled from galactic winds and outflows.  

\item
To investigate the physical conditions in kinematically aligned \Ovi\!\!/\Hi absorbers, we model the ion column densities of selected absorbers from the UVES sample using the photoionization code {\sc Cloudy}. The density distribution is rather broad with a weak evidence for a bimodality with peaks at log $n_{\rm H} \approx -4.0$ and log $n_{\rm H} \approx -2.8$. The temperatures derived under the assumption of photoionization are typically less than $10^5$ K. They are systematically lower than what is estimated from the \Ovi Doppler parameters, assuming pure thermal broadening. The absorption path length $L$ in the \Hi structures containing \Ovi spans about five orders of magnitude ($0.01\le L \lesssim 10^3$ kpc), with a median value $\tilde L\simeq 50$~kpc. In general, \Ovi absorbers evidently arise in large-scale cosmological structures, as well as in local, small-scale structures.

\end{itemize}

In summary, the study of the UVES sample of \Ovi absorbers presented in this chapter demonstrates that their physical properties at high redshift are highly diverse, which indicates multiple origins of the \Ovi absorbing gas: in large-scale intergalactic gas structures as well as in small-scale interface regions between hot and cools gas clouds in multi-phase structures. The implications of these findings for the understanding of the origin of highly-ionized gas at high redshift will be discussed in Chapter~\ref{conclusions}. In the following Chapter \ref{simulations}, we describe the next steps in our systematic investigation of \Ovi systems: a study of a very large sample of \Ovi synthetic absorption spectra, generated from the cosmological OWLS that include star-forming galaxies and their intergalactic environment.


\chapter{A systematic study of O\,{\sc vi} absorbers in the cosmological OWLS}\label{simulations}

\hspace{0.6cm}This chapter is dedicated to another important aspect of our systematic investigation of intervening \Ovi\!\! absorbers at high redshift: namely, the use of cosmological simulations that include a proper treatment of the dynamics and physics of intergalactic and circumgalactic gas. We here study a large sample of \Ovi absorbers at high redshift in synthetic absorption spectra generated from the OverWhelmingly Large Simulations (OWLS) \citep{Schaye10}. Details on the set-up of the synthetic absorption spectra are presented in Sect.~\ref{owls}. To understand the nature of high-redshift \Ovi absorbers in the context of the evolution of the distribution of hot, metal-enriched gas from high to low redshifts, we also compare the statistical properties of high-$z$ \Ovi absorbers with those at $z\approx0$. To avoid repetitions, the low- and high-redshift runs of the OWLS are abbreviated hereafter as \lOWLS ($z=0.25$)and \hOWLS ($z=2.50$), respectively.
 
There are many advantages in using synthetic spectra from cosmological simulations to study absorption systems in the IGM. In particular, physical parameters like temperature, gas density and metal abundance can be obtained directly from the simulations. Here, we will focus on the analysis of the physical properties of a sample of \Ovi absorbers obtained from a fiducial run of the OWLS set dubbed {\it REF\_L100N512}. The statistical results from the simulations will be compared to the properties of the \Ovi absorbers, as derived from the observations presented in the previous chapter.

  \section{Key observables}\label{key_observables_owls}

    \subsection{Column-density distribution}\label{column_density_owls}

\begin{figure}[h!]
\begin{center}
\resizebox{0.7\hsize}{!}{\includegraphics[angle=-90]{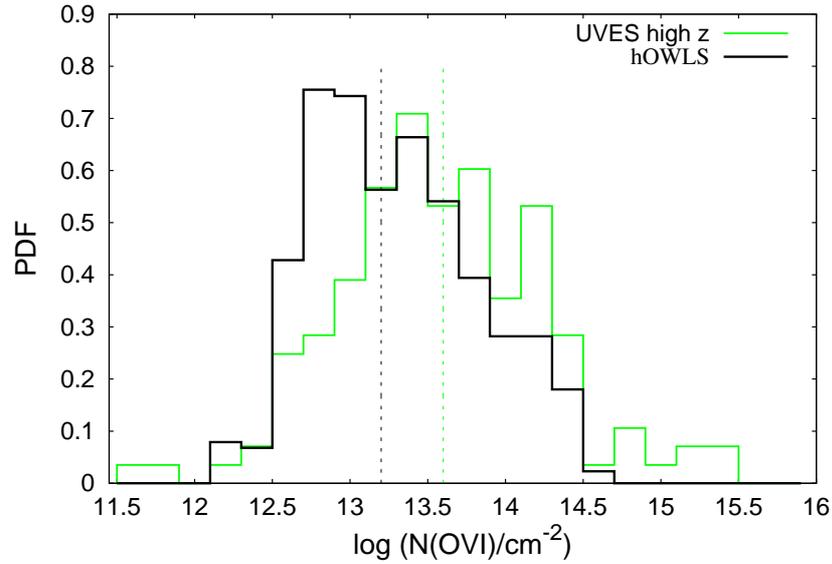}}
\caption[Distributions of $N(\Ovi\!\!)$ for observed and simulated spectra at high redshift.]{\small Column density distributions in the observational (blue) and simulated (black) high redshift sample. The positions of the medians are denoted with dashed lines.}
\label{fig_N_hist_obs_sim}
\end{center}
\end{figure}

\hspace{0.6cm}The column density distributions of observed (UVES) and simulated (OWLS) \Ovi\!\! components are 
shown in Fig.~\ref{fig_N_hist_obs_sim}. The corresponding median values (in logarithmic scale) are 13.6 and 
13.2, respectively. The higher median value in the observed sample can be explained by the lack of higher-column-density components 
in the OWLS (cf. Fig. \ref{fig_cddf}). Another reason for this difference can be the higher S/N and the resulting lower \Ovi detection limit in the spectra from the
simulations ($\log N$(\Ovi\!\!) = 12.8). 

\begin{figure}[h!]
\begin{center}
\resizebox{0.7\hsize}{!}{\includegraphics[angle=-90]{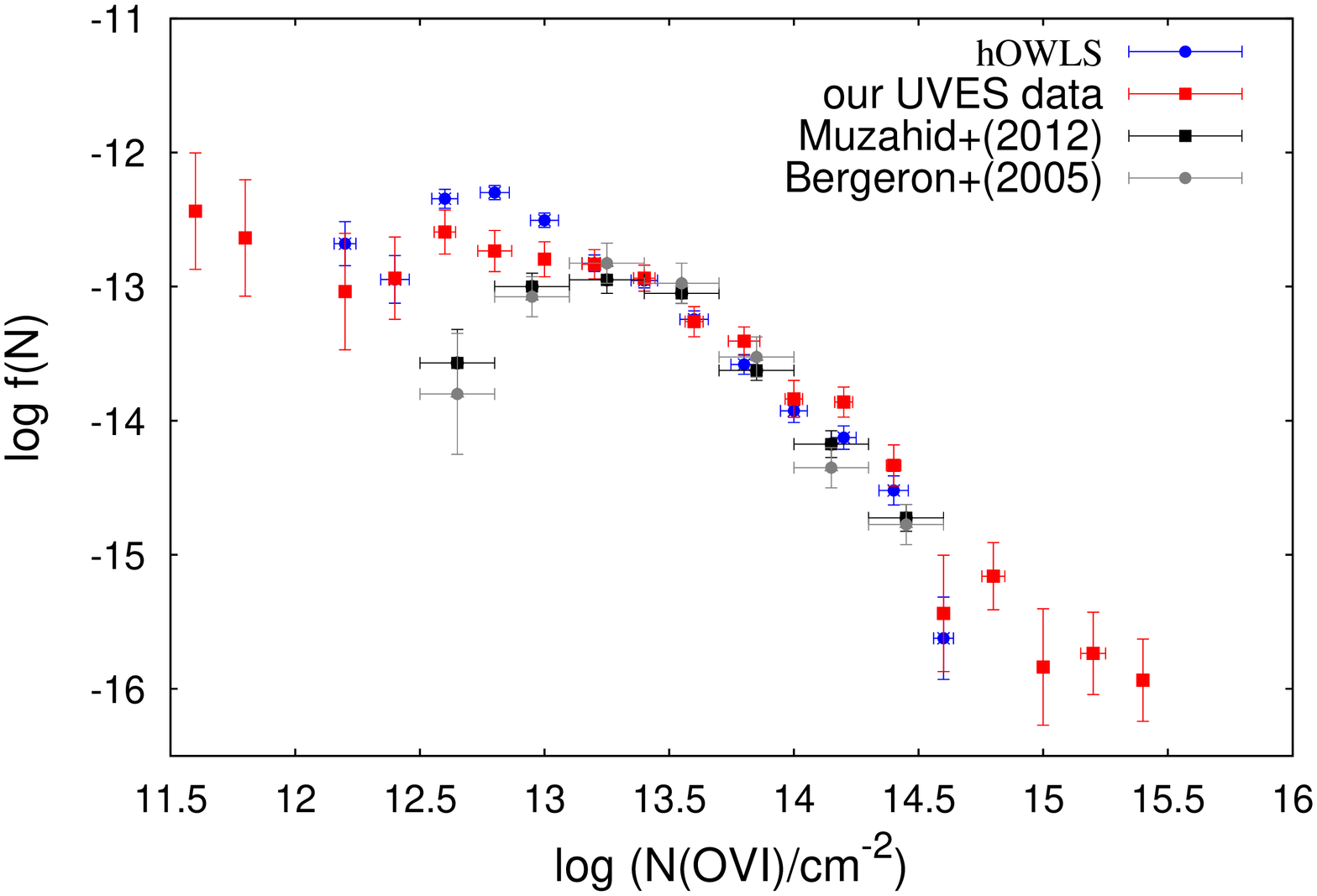}} 
\caption[Comparison of CDDF derived from observations and simulations.]{\small CDDFs derived from observations and simulations: from our UVES sample (red), from our sample of synthetic spectra (blue), from the UVES samples of \citet[][grey]{Bergeron05} and \citet[][black]{Muzahid11}.}
\label{fig_cddf_owls}
\end{center}
\end{figure}

In Fig.~\ref{fig_cddf_owls} we compare the CDDF derived from our UVES sample, as described in Sect.~\ref{cddf}, with the one obtained from the simulations. The results of 
\citet{Bergeron05} and \citet{Muzahid11}, who used a UVES data sample comparable to ours, are included for reference. The absorption path of the UVES data is $\Delta X = 27.3$ (15 sight-lines), 
while that of the OWLS systems is $\Delta X = 83.8$ (100 sightlines).  

Our UVES sample appears to be complete down to $\log~N$(\rm \Ovi\!\!) $\approx 13.3$. Taking this threshold as a lower limit in the fitting procedure, we obtain a 
power-law slope $\beta_{\rm OVI} = 1.51 \pm 0.10$ (see Sect.~\ref{cddf}). The detection limit for the synthetic spectra is $\log\,N$(\Ovi\!\!) $= 12.7$. Adopting this threshold, we obtain $\beta_{\rm OVI} = 1.37 \pm 0.07$. However, adopting a threshold of $ \log~N$(\Ovi\!\!) $\approx 13.3$  yields $\beta_{\rm OVI} = 1.60 \pm 0.11$ for the OWLS data, in agreement  with the observations. As can be seen in the figure, all distributions agree within the uncertainties in the range  $13.3 \le \log N$(\Ovi\!\!)$\le 14.6$. The power-law slopes of the CDDF derived by \citet{Bergeron05} and \citet{Muzahid11} are steeper, $\beta_{\rm OVI} = 1.71 \pm^{0.48}_{0.47}$ and $\beta_{\rm OVI} = 2.4 \pm 0.2$, respectively. The CDDF derived from our synthetic spectra is very similar to the one in \citet{Muzahid11}, if the bin of highest column density ($ \log N$(\Ovi\!\!) = 14.4) is excluded. 

We perform power-law fits to the CDDF using our observed and simulated samples and those from \citet{Bergeron05} and \citet{Muzahid11} in the range $13.3 \le \log (N_{\rm OVI})\le 14.6$ to assess the sensitivity of the slope to the adopted threshold. For this range, the slope of the data used by \citet{Muzahid11} yields $\beta =1.56 \pm 0.17$, close to the value from the OWLS: $\beta_{\rm OVI} = 1.60 \pm 0.11$. On the other hand, the slope, obtained from our UVES data is shallower: $\beta_{\rm OVI} = 1.35 \pm 0.16$, but still consistent with the other values. This is because of the relatively high number of \Ovi components with column densities $\log (N_{\rm OVI})\ge 14$ in our UVES data set compared to the \Ovi sample of \citep{Muzahid11} (see Fig.~\ref{fig_cddf_owls}).   

Also at higher column densities $\log (N_{\rm OVI})> 14.6$, the \hOWLS and UVES samples used by \citet{Bergeron05} and \citet{Muzahid11} exhibit a lack of components with respect to our observed UVES sample (Fig.~\ref{fig_cddf_owls}, red squares).
A similar trend apparently also exists in the \lOWLS sample, as discussed in Sect.~\ref{doppler_parameter_owls}.  


    \subsection{Distribution of Doppler parameters}\label{doppler_parameter_owls}

\hspace{0.6cm}Because of the expected turbulent motions in the gas, the value of the Doppler parameter derived from a Voigt-profile fit yields only an upper limit to the kinetic temperature of the gas (see also Sect.~\ref{line_broadening}). Therefore, instead of setting constraints to the physical state of \Ovi systems from individual values of $b$ from the \Ovi absorption components, we consider their distribution. 

\begin{figure}[h!]
\begin{center}
\resizebox{0.7\hsize}{!}{\includegraphics[angle=-90]{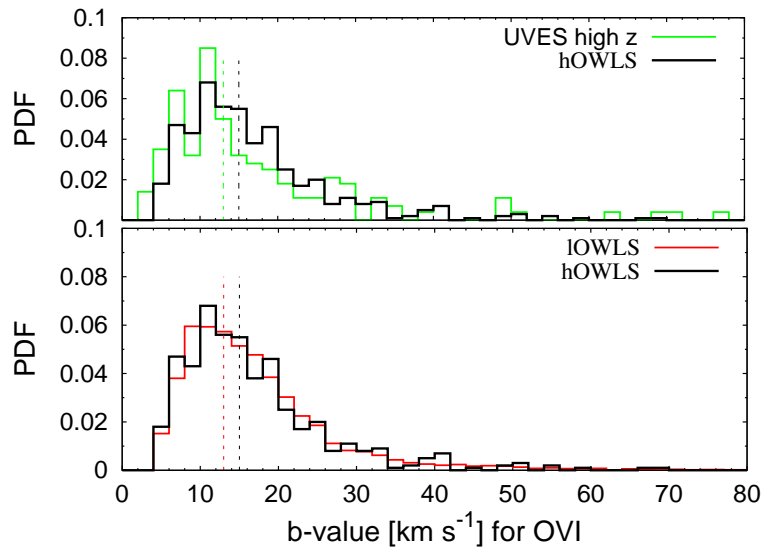}}
\caption[Doppler parameter distributions for observed and simulated spectra.]{\small Comparison between the distributions of the Doppler parameter in the observed and simulated samples at high $z$ (top panel), and the Doppler parameter in \lOWLS and \hOWLS samples (bottom panel). The position of the median is indicated by dashed lines.}
\label{fig_b_hist_obs_sim}
\end{center}
\end{figure}

The top panel of Fig.~\ref{fig_b_hist_obs_sim} displays the Doppler parameter distributions for the sample of observed and simulated \Ovi 
components. The median values are 
$\tilde{b}_{\rm UVES}=13~\rm km~s^{-1}$ and $\tilde{b}_{\rm OWLS}=15~\rm km~s^{-1}$ which correspond, assuming purely thermal broadening, 
to $T_{\rm UVES}\sim 1.6 \times 10^{5}$~K and $T_{\rm OWLS}\sim 2.2 \times 10^{5}$~K, respectively. The latter temperature is consistent with the 
estimate of \citet{Simcoe02}: $T \sim 2.1 \times 10^{5}$ K, obtained from high-$z$ Keck I/HIRES observations. On the other hand, as mentioned in Sect.~\ref{col_den_b_distr},
$T_{\rm UVES}$ is close to the result of \citet{Muzahid11}: $T \sim 1.8 \times 10^{5}$~ K. 

In the case of collisional ionization equilibrium (CIE), the temperature in \Ovi\!\! systems would peak around $T \approx 3 \times 10^{5}$ K \citep{Gnat07,Sutherland93}, which corresponds to $b \approx 17$ \kms\!\!. We note that 66 per cent of the components in OWLS sample and 68 per cent in the UVES sample have values of the Doppler parameter $<17$ \kms\!\!. This result suggests that photoionization is the main ionizing mechanism in these absorbers. At high metallicities, non-equilibrium collisional ionization might be an alternative mechanism.  

\citet{Gnat07} point out that in CIE the \Ovi fraction must be less than 0.01 at temperatures $T \ge 6 \times 10^{5}$ K ($b$(\Ovi\!\!)$ \ge 25.5$ \kms\!\!). Therefore, one should not expect higher $b$(\Ovi\!\!) values in case of purely thermal broadening unless the metallicity and/or $N(\rm H)$ are very high. We find 12 per cent of the components in the OWLS sample and 18 per cent of the components in the UVES sample with higher values of $b$ than 25.5 \kms\!\!. 

\subsection{Redshift-dependence of the line width}\label{effect_redshift}

\hspace{0.6cm}The Doppler parameter distributions from the \lOWLS and \hOWLS samples are compared in the bottom panel of Fig.~\ref{fig_b_hist_obs_sim}. Evidently, both distributions are similar, with median values $\tilde b$ (\Ovi\!\!)$= 13$ \kms for low $z$ and  $\tilde b$ (\Ovi\!\!)$= 15$ \kms for high redshift samples. 

\begin{figure}[h!]
\begin{center}
\resizebox{0.7\hsize}{!}{\includegraphics{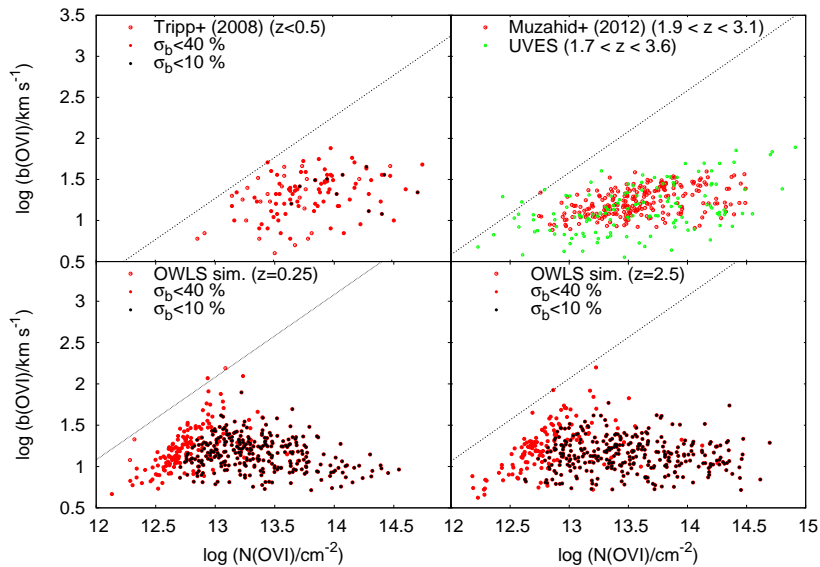}}
\caption[Diagram $N$ vs. $b$ from observed and simulated spectra.]{$N$ vs. $b$ from observed (bottom panel) and simulated (top panel) spectra at low $z$ (left panel) and high $z$ (right panel).}
\label{fig_N_b_obs_sim}
\end{center}
\end{figure}

A similar comparison using observations has been done by \citet{Fox11} and \citet{Muzahid11}. \citet{Fox11} compares the low-$z$ sample of \citet{Tripp08} with the high-$z$ sample of \citet{Bergeron05}. \citet{Muzahid11} further compare the same sample of \citet{Tripp08} with a two-times larger UVES high-$z$ sample. The median Doppler parameters of the \Ovi spectra found by these authors and other relevent parameters of their data are listed in Table~\ref{observational_technical_parameters}. \citet{Tripp08} obtain a median $\tilde b_{\rm STIS} \sim 26$ \kms for the low $z$ sample. \citet{Fox11} and \citet{Muzahid11} obtain  $\tilde b_{\rm UVES}\sim 14$ \kms at high $z$ and conclude that the \Ovi components are, on average, almost twice as broad at low-$z$ than at high-$z$. 
This finding does not agree with the result from our OWLS synthetic spectra. To investigate this, we have composed diagrams $N$(\Ovi\!\!) vs. $b$ for the OWLS, UVES and STIS data (Fig.~\ref{fig_N_b_obs_sim}). 

\begin{table*}[th!]
\caption[Observational and technical parameters of reference works.]{Observational and technical parameters of reference works.}
\begin{small}
\begin{tabular}{lrcrrrccc}
 \hline
 \hline
Study&\multicolumn{3}{c}{Technical parameters} &\multicolumn{2}{c}{Observational parameters} \\
\hline
\scriptsize ~ &\scriptsize Spectrograph &\scriptsize Resolution &\scriptsize mean $S/N$  &\scriptsize Redshift &\scriptsize Median $b$-value   \\
~&~&\scriptsize \,[\kms\!\!] ~& ~&~&\scriptsize \,[\kms\!\!]&\\
 \hline
\scriptsize \citet{Tripp08}   &\scriptsize STIS/HST &\scriptsize 7.0 &\scriptsize 13 at 1300 \AA    &\scriptsize $z<$ 0.5            &\scriptsize 26.0$\pm$14.0   \\
\scriptsize \citet{Bergeron05}&\scriptsize UVES/VLT &\scriptsize 6.6 &\scriptsize 30-40 at 3300 \AA &\scriptsize 2.0 $\le z \le$ 2.6 &\scriptsize 14.0$\pm$7.0     \\
\scriptsize \citet{Muzahid11} &\scriptsize UVES/VLT &\scriptsize 6.6 &\scriptsize 30-40 at 3300 \AA &\scriptsize 2.0 $\le z \le$ 2.6 &\scriptsize 13.8$\pm$14.0    \\
 \hline
\end{tabular}
\end{small}
\label{observational_technical_parameters}
\end{table*}

Brief eye inspection already indicates the differences between the low-$z$ samples. There is a lack of components with high column densities and large $b$-values in \lOWLS in 
comparison to the sample of \citet{Tripp08} (left panels). Since the OWLS synthetic spectra mimic the spectral properties of STIS data, this difference cannot be attributed to the resolution or other instrumental characteristics. The reason for that mismatch between simulations and observations is still unclear. As pointed out by \citet{Tepper-Garcia11}, one can rule out the explanation that the observed \Ovi systems arise in gas at higher temperatures (producing higher $b$-values), as this would result in a too low \Ovi ion fraction for either CIE or non-equilibrium ionizaion conditions. 

Similarly, a lack of systems with large $b$-values at low $z$ is found in the simulations of \citet{Oppenheimer09}. Because the resolution achieved by current simulations does not allow to properly model turbulence at the smallest scales, these authors added a sub-resolution turbulent component to the thermal $b$-value. This brings the predicted distribution of equivalent widths into a better agreement with observations and suggests that turbulence plays a significant role in the line broadening at low $z$.

The physical picture at high $z$ is different. The similarity between the Doppler parameter distributions from synthetic and observed spectra suggests that non-thermal mechanisms at small scales do not significantly affect the \Ovi line widths. This will later allow us to combine the results from the simulations and observations to study the thermal evolution of \Ovi systems (see Sect.~\ref{aligned_absorbers_owls}). 


  \section{Physical conditions: high $z$ vs. low $z$}\label{physical_conditions_owls}

    \subsection{Temperature}\label{temperature_owls} 


\hspace{0.6cm}The maximum kinetic temperature of clouds in photoionization equilibrium is $T\sim 40~000$~K. Below this temperature, collisional ionization is not significant. The latter process dominates the ionization state of gas at $T \gtrsim 10^{5}$~K, regardless of the mean density. As mentioned in Sect.~\ref{doppler_parameter_owls}, the ionization fraction of \Ovi peaks at $T \approx 3 \times 10^{5}$~K in CIE \citep{Sutherland93,Gnat07}. 

\begin{figure}[h!]
\begin{center}
\resizebox{0.7\hsize}{!}{\includegraphics[angle=-90]{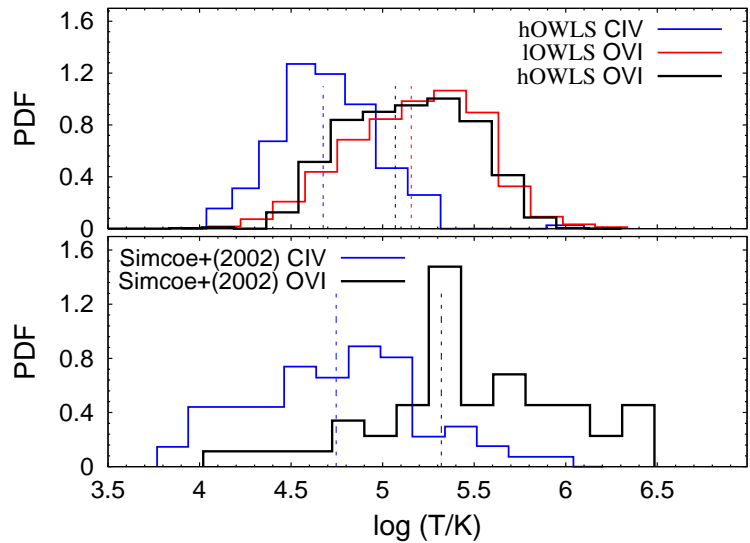}}
\caption[Temperature distributions from simulations and observations]{\small Temperature distributions of \Ovi absorbers from simulations (top panel) and observations (bottom panel). The distribution for \Civ absorbers (blue) from the \hOWLS sample (top panel) and from the sample of \citet[][bottom panel]{Simcoe02} is plotted for comparison.}
\label{fig_hist_T}
\end{center}
\end{figure}

Simulations by \citet{Tepper-Garcia11} reveal a bimodal temperature distribution of the \Ovi bearing gas at low $z$, conditioned by photoionization at $T \approx 3 \times 10^{4}$~K and by collisionally ionization at $T \approx 3 \times 10^{5}$~K. However, the lower temperature regime is not significantly detected in \Ovi absorption. A possible explanation is a low oxygen abundance in the cool gas and, hence, a low \Ovi column density that prevents from tracing this gas phase in absorption. 

Following \citet{Tepper-Garcia11}, we consider optical-depth weighted physical quantities ({\it e.g.}, temperature, density, metallicity) of the gas traced by the \Ovi absorption lines in the synthetic spectra\footnote {~The according computations in the OWLS data were kindly provided by Thorsten Tepper-Garc{\'{\i}}a, Universit$\rm\ddot{a}$t Potsdam.}.
The optical-depth weighting relates an absorption-line profile to the absorbing gas in the simulation and allows us to compare the fitted parameters (Doppler parameter, column density) with the physical quantities given in the simulations. 

The optical-depth weighted temperature distributions of \Ovi absorbers from the \hOWLS \\(black) and \lOWLS(red) samples are shown in Fig.~\ref{fig_hist_T}. Additionally, both distributions are compared with the \Ovi temperature distribution by \citet{Simcoe02} (black) in the redshift range 2.2 $< z <$ 2.8 . In the work of \citet{Simcoe02} the temperature is calculated from formula (\ref{eq_temp_Doppler}), neglecting non-thermal broadening. Therefore it 
should be considered as an upper limit $T_{\rm max}$ of the gas temperature.

As seen in Fig.~\ref{fig_hist_T}, the \Ovi temperature distributions derived from synthetic spectra at low and high $z$ are very similar, with almost identical median values: 
$1.4 \times 10^{5}$~K and $1.2 \times 10^{5}$~K, respectively. Sixty-five per cent of the \lOWLS \Ovi seen in absorption traces mainly the hot gas phase at temperatures 
$T \ge 10^{5}$~K which is the {\it low} temperature regime of the WHIM \citep{Tepper-Garcia11}. The same trend is evident considering the \hOWLS sample where 62 per cent 
of the components are hotter than $10^{5}$~K. Apparently, there is (on average) no temperature evolution of the \Ovi absorbers from high to low redshift. The \Ovi distribution from 
the \hOWLS sample is in agreement with the result of \citet{Simcoe02} who obtain from high $z$ observations a median value of $\widetilde T_{\rm max}$ = 2.1 $\times 10^{5}$~K. These authors also find that 62 per cent of all systems fall in the range $5.0 \le \log T_{\rm max} \le 6.0$. A plausible mechanism that is able to produce \Ovi absorbers with such temperatures is collisional ionization in hot winds expelled from galactic environments.

We further compare the temperature distributions of \Ovi (black) and \Civ (blue) absorbers (\hOWLS sample, top panel; \citet{Simcoe02}, bottom panel). The difference is notable: 
about $0.6$~dex in log $T$ between the median \hOWLS values. The same trend is found in the distributions by \citet{Simcoe02}. A comparison between the \Civ temperature distributions from \hOWLS and \citet{Simcoe02} shows that they have almost identical median values: $4.6 \times 10^{4}$~K and $4.7 \times 10^{4}$~K, respectively. This implies that, considering both observed and synthetic spectra, the \Ovi lines are broader and arise in gas with physical conditions that are different from that of the \Civ absorbing phase. The gas traced by the \Ovi absorption is hotter, while \Civ is indicator of a cooler gas phase.

\subsection{Density}\label{density_owls} 

\begin{figure}[h!]
\begin{center}
\resizebox{0.7\hsize}{!}{\includegraphics[angle=-90]{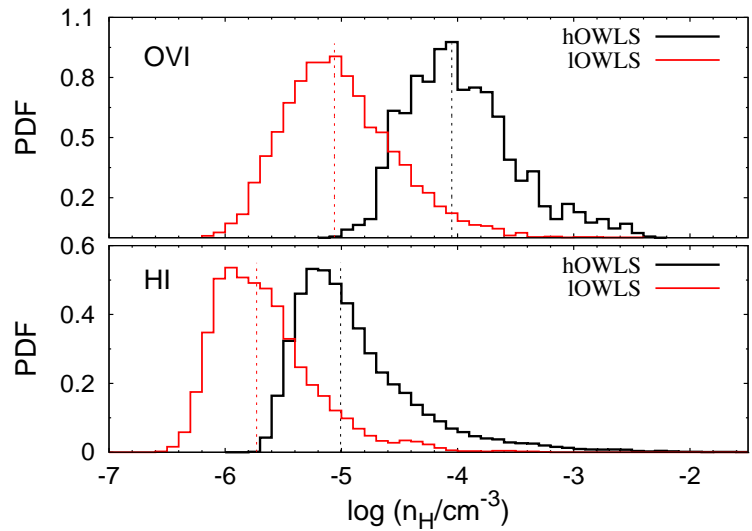}}
\caption[Hydrogen density distributions of simulated \Ovi and \Hi absorbers]{\small Hydrogen density distributions of all simulated \Ovi (top) and \Hi absorbers (bottom), at high (black) and low (red) redshift.}
\label{fig_hist_n}
\end{center}
\end{figure}

\hspace{0.6cm}
As seen in Fig.~\ref{fig_hist_n} (top), the median values of 
the hydrogen densities in all \Ovi\!\! absorbers from the \lOWLS and \hOWLS samples differ by one order of magnitude: $8.7\times 10^{-6}$~cm$^{-3}$ vs. 
$8.7 \times 10^{-5}$~cm$^{-3}$. This result suggests that \Ovi absorbers at low redshifts are produced in a more diffuse environment than those at high $z$.

We also compare the hydrogen gas density distributions derived for all \Hi absorbers at low and high $z$ (bottom) with those of \Ovi absorbers. The pattern is the 
same -- \Hi absorbers at low redshifts populate more diffuse regions in comparison to high $z$ systems. Most probably, this is an ionization effect: the photoionization rate $\Gamma_{-12}$ is an order of magnitude higher at $z=2$ than at $z=0$.
Yet, in both cases, \Ovi absorption traces the densest regions of the \Hi absorbers. 

A small second peak in the \hOWLS \Ovi density distribution is seen at log $n_{\rm H} = -3.0$, which does not appear in the \lOWLS \Ovi distribution. However, such a second peak is also seen in the \Ovi density distribution from the UVES observations (see Sect.~\ref{modeling_cloudy}).

\subsection{Overall metallicity}\label{metallicity_owls} 


\hspace{0.6cm}The OWLS track the abundance of 11 individual elements (H, He, C, N, O, Ne, Mg, Si, Fe, Ca and S). They simultaneously keep record of the metallicity of each SPH particle \citep{Wiersma09b}. The metallicity, as a parameter in the simulations, can be defined in two ways. The fraction of metals to the total gas mass in each SPH particle  $Z_{\rm part} \equiv m_{Z}/m$ is called `particle metallicity'. The alternative definition of metallicity is $Z_{\rm sm} \equiv \rho_{Z}/\rho$, i.e. the ratio of the SPH smoothed metal mass density to the SPH smoothed gas mass density \citep[see e.g.][]{Okamoto05, Tornatore07}. Hereafter, we will refer to this metallicity as `smoothed metallicity', following \citet{Wiersma09b}.

\subsubsection{Smoothed metallicity}\label{local_metallicity_owls}

\hspace{0.6cm}In this section, we analyze the distribution of the optical-depth weighted smoothed metallicity $Z_{\rm sm}$(\Ovi\!\!) of the gas as traced by \Ovi absorbers at high $z$ and compare it with that of the \lOWLS sample. We stick to this definition of metallicity as it is more appropriate for the analysis of SPH simulations \citep{Wiersma09b}. Moreover, as pointed out by \citet{Tepper-Garcia11}, this choice takes into account the metallicity of each SPH particle that contributes to the \Ovi absorption. In other words, this quantity is a measure of the metallicity in small regions around the spatial resolution limit of the simulations while the observable $Z$ (\Ovi\!\!), derived from the ratio of the total $N$(\Ovi\!\!) to $N$(\Hi\!\!), corresponds to metallicity of more extended host structures and can be substantially less than $Z_{\rm sm}$(\Ovi\!\!). Hereafter, we call $Z$ (\Ovi\!\!) the `mean metallicity', like in \citet{Wiersma09b} and \citet{Tepper-Garcia11}.

\begin{figure}[h!]
\begin{center}
\resizebox{0.7\hsize}{!}{\includegraphics[angle=-90]{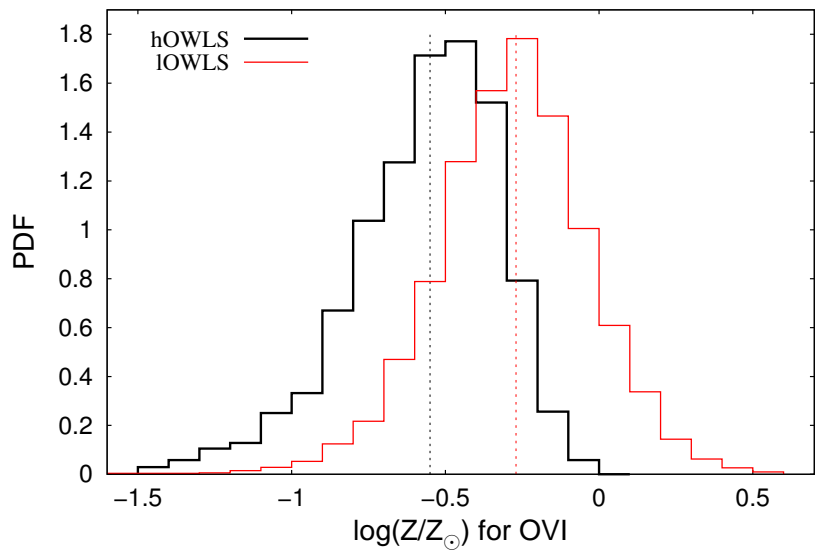}}
\caption[Metallicity distributions of \Ovi absorbers from OWLS simulations.]{\small Metallicity distributions of simulated O\,{\sc vi} absorbers from the \hOWLS (black) and \lOWLS (red) data samples.}
\label{fig_hist_metal}
\end{center}
\end{figure}


The metallicity distributions of \Ovi absorbers from the \hOWLS and \lOWLS samples are plotted in Fig.~\ref{fig_hist_metal}. Apparently, \Ovi absorbers at high $z$ trace metal-enriched material with $\log (Z_{\rm sm}$(\Ovi\!\!)/$Z_{\odot}) > -1.5$ and a median value $\widetilde Z_{\rm sm}$(\Ovi\!\!)$\approx 0.3~ Z_{\odot}$. The median for the \lOWLS sample is closer to the solar value $\widetilde Z_{\rm sm}$(\Ovi\!\!) $\approx 0.6~ Z_{\odot}$. The increase of $Z$ from high to low redshifts is consistent with most models of cosmic chemical evolution that predict a rise of the global mean interstellar metallicity in galaxies with time, reaching quasi-solar values at $z=0$ \citep[see e.g.][]{Pei95, Malaney96, Pei99, Somerville01}.

\subsubsection{Mean metallicity}\label{average_metallicity_owls}
 
\hspace{0.6cm}Structures with measurable \Ovi absorption are traced by well-aligned \Ovi\!\!/\Hi\!\! absorber pairs within some small velocity range $\Delta v$. Following \citet{Tepper-Garcia11}, we adopt $\Delta v \pm10~$\kms and select by this criterion a subsample of aligned absorbers to investigate their metallicity distribution. Proximity in velocity space between \Hi and \Ovi lines most probably indicates that these species arise in the same gas phase or, at least, in the same overall gas structure. The mean metallicity $Z$(\Ovi\!\!) of the absorber is calculated according to the formula:

\begin{equation}
\label{mean_metallicity}
Z(\small{\rm OVI)} = \Big (\frac{N_{\rm \small{\rm OVI}}}{N_{\rm \small{\rm HI}}}\Big)\Big (\frac{f_{\rm \small{\rm HI}}}{f_{\rm \small{\rm OVI}}}\Big) \Big (\frac{m_{\rm O}}{m_{\rm H}}\Big) X_{\rm H}  
\end{equation}

where $m_{\rm H}$ and $m_{\rm 0}$ are the masses of hydrogen and oxygen atoms, $f_{\rm \small{\rm HI}}$ and $f_{\rm \small{\rm OVI}}$ are the optical-depth weighted ionization fractions, $N_{\rm \small{\rm HI}}$ and $N_{\rm \small{\rm OVI}}$ are the column densities of \Hi and \Ovi, and $X_{\rm H}$ is the hydrogen fraction.

\begin{figure}[h!]
\begin{center}
\resizebox{0.7\hsize}{!}{\includegraphics[angle=-90]{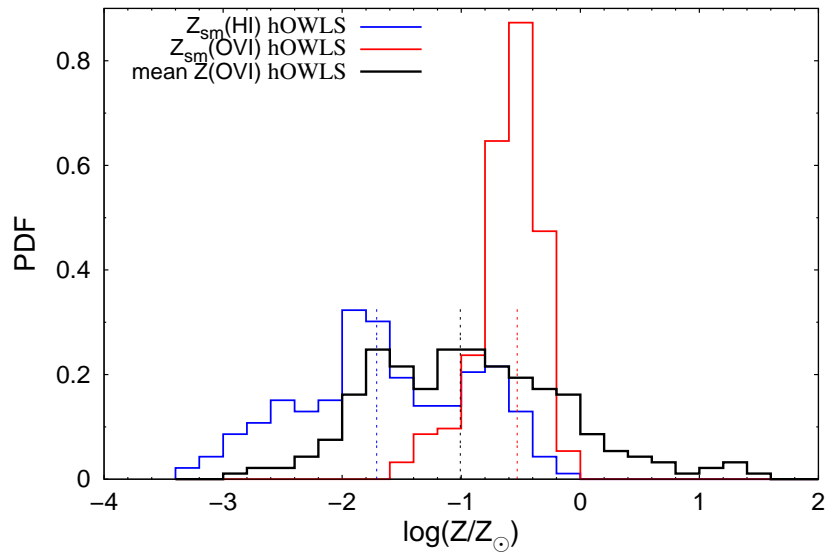}}
\caption[Metallicity distributions of simulated well-aligned \Hi-\Ovi absorbers.]{\small Metallicity distributions of simulated well-aligned \Hi(blue) -\Ovi(red) absorber pairs at high $z$. The metallicity distribution based on an estimate of $Z$ from the \Ovi\!\!/\Hi column density ratio is plotted for 
comparison (black).}
\label{fig_hist_Z_sol}
\end{center}
\end{figure}


The distributions of the resulting metallicities for all well-aligned pairs $Z_{\rm sm}$(\Hi\!\!) and $Z_{\rm sm}$(\Ovi\!\!) are shown in Fig.~\ref{fig_hist_Z_sol}. It is evident that the distribution of $Z$(\Ovi\!\!) has a larger dispersion than that of $Z_{\rm sm}$(\Ovi\!\!). The median values are 10 per cent of the solar metallicity and  30 per cent of the solar metallicity, respectively. The peak of the $Z$(\Ovi\!\!) distribution is at higher metallicities in comparison to the distribution of $Z_{\rm sm}$(\Hi\!\!), but similar in width. These results are similar to those of \citet{Tepper-Garcia11} from low-redshift OWLS: the mean metallicity of the absorbing structures can be very different from the smoothed metallicities due to the possible concentration of the metals in small, tiny subregions as represented by the quantity $Z_{\rm sm}$(\Ovi\!\!). As suggested by these authors, a possible interpretation might be a highly inhomogeneous distribution of $Z$, where the \Ovi absorbers arise in high-
metallicity regions embedded in a lower-mean-metallicity gas phase. 

\subsection{Oxygen abundance}\label{abundance_owls}

\begin{figure}[h!]	
\begin{center}
\resizebox{0.7\hsize}{!}{\includegraphics[angle=-90]{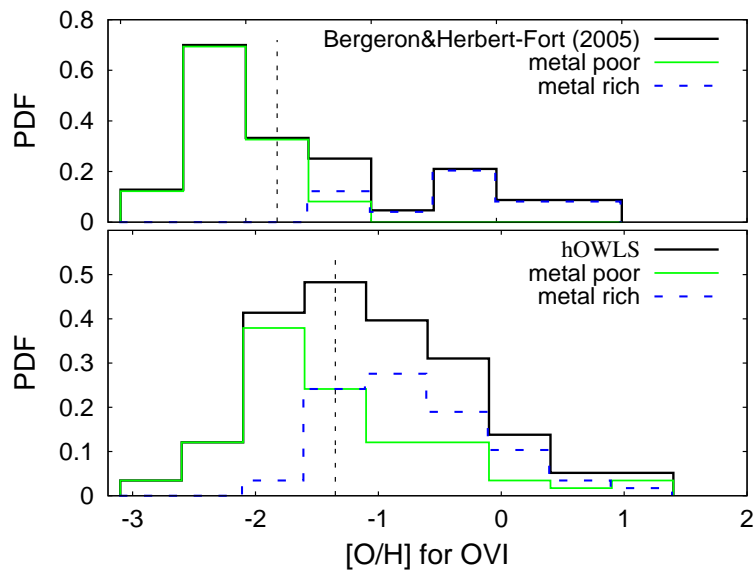}}
\caption[Abundance distribution of simulated \Ovi absorbers at high redshift.]{\small Distribution of oxygen abundance in the sample of simulated well-aligned \Hi\!\!/\Ovi\!\! absorber pairs at high $z$ (bottom panel). The observational result of \citet{Bergeron05} for redshift range $2.0 \le z \le 2.6$ is plotted for comparison (top panel). The distribution is divided in the two populations -- 'metal-rich' and 'metal-poor' absorbers. The position of the median is denoted with dashed lines.}
\label{fig_hist_abundance_rich_poor}
\end{center}
\end{figure}

\hspace{0.6cm}In addition to the overall metallicity we compute the oxygen abundance, [O/H], of the well-aligned \Hi and \Ovi\!\! synthetic absorption components, derived from their column-density ratios and ion fractions. In order to compare our result with the oxygen abundance from observations, obtained by \citet{Bergeron05} we assume the same solar relative abundanses used by this authors \citep{Anders89}. 
\citet{Bergeron05} analyze a sample of \Ovi absorbers in the redshift range $2.0 \le z \le 2.6$, using spectral data of 10 quasars from UVES observations. Their estimate of the oxygen abundance is calculated under the assumption of photoionization as the only ionization mechanism, while the simulations we use include both photoionization and collisional ionization. The two distributions are shown in Fig.~\ref{fig_hist_abundance_rich_poor}. The median value for OWLS synthetic spectra is  $-1.35$ dex, which is higher than the median value of \citet{Bergeron05} ($-1.75$ dex). The ranges of both abundance distributions are around $-3.0\le[\rm O/H]\le 1.0$.

As a next step, we continue our investigation of metal-rich and metal-poor absorbers, proposed by \citet{Bergeron05} (see Sect.~\ref{metal-rich_and_metal-poor}). 
The result of this authors is plotted in Fig.~\ref{fig_hist_abundance_rich_poor}, top panel. The [O/H] distributions of the two populations overlap slightly. From this, the authors concluded that \Ovi absorbers indeed represent two distinct populations. To reevaluate this result, we use their observational criteria and divide our absorbers in two sub-samples: metal-rich and metal-poor (see Sect.~\ref{metal-rich_and_metal-poor}), based on the \Ovi and \Hi column-density ratios. Our result is plotted in Fig.~\ref{fig_hist_abundance_rich_poor}, bottom panel. In contrast to the result of \citet{Bergeron05}, we find a significant overlap between the abundance distributions of metal-rich and metal-poor populations. We also derive their $b$-value distributions (not shown) and find again an overlap, including the high velocity tails. Our results from the simulations suggest that there is no compelling evidence for the existence of such two distinct populations of \Ovi absorbers. 


  \section{Discussion on physical parameters and observables of the OVI absorbers}\label{correlation_plots_owls}

\hspace{0.6cm}In the first part of this section we analyze possible correlations between global physical parameters of the absorbers (weighted by optical depth; e.g., gas 
temperature, gas density and smoothed metallicity) and key observables such as column density and Doppler parameter. Such correlations can provide clues to our understanding 
of the nature and the origin of the \Ovi absorbers at high redshift.

In the second part of this section a subsample of well-aligned (in velocity space) ~\Ovi\!\!/\Hi and \Ovi\!\!/\Civ pairs is considered. This part of the study will be directly connected to the results from our UVES observations, as these provide key observables like column density and Doppler parameter, whereas the physical properties of the absorbers can be constrained only based on certain assumptions (see also Chapter~\ref{paper1}). One of these assumptions is, that the different ions coexist in the same gas-phase. 
A common approach for observers therefore is, to look for an alignment in 
velocity space, which usually is interpreted as evidence that the observed species reside in the same physical region. However, this assumption introduces systematic uncertainties in modeling the physical state of the gas: an alignment in velocity space does not {\it necessarily} imply a true spatial alignment and a common gas-phase, but
could be just the result of overlapping radial velocities and similar bulk motions. Therefore, we take the advantage of using cosmological simulations to extract information about true physical conditions in the absorbing gas. Then, we are able to systematically investigate, whether the usual assumption of the coexistence of \Ovi, \Hi, and \Civ in a single gas phase for well-aligned absorbers in observational data is really justified. 

    \subsection{All absorbers}\label{all_absorbers_owls} 
\subsubsection*{Doppler parameter vs. column density}

\hspace{0.6cm}One possible correlation to look for is the one between the Doppler parameter and the column density in the absorber. Such a correlation might be present because of a number of different effects: undetected velocity structure in absorbers of higher column density, collisional ionization that links column densities in highly ionized systems and temperature, or selection effects (since broad, weak lines are hardly detectable in the noise). 

Different research groups had searched for such a correlation at low redshifts. \citet{Heckman02} found a correlation between $b$ and $N$ in their sample of observed \Ovi systems in a variety 
of environments: Milky Way disk and halo, High-Velocity Clouds, the Magellanic Clouds, and the IGM. These authors suggested that above $\log b \approx 1.6$ the column density increases 
linearly with line width, while the relation steepens for lower linewidths. However, \citet{Danforth06} did not find such correlation from the analysis of their low $z$ data. 
\citet{Tripp08} found a weak correlation in their observed \Ovi sample, but noted that the significance is not high enough to support the model of \citet{Heckman02}. 
\citet{Tepper-Garcia11} studied this issue further, considering a low-redshift \Ovi sample from OWLS synthetic spectra. They do not find a significant correlation between the \Ovi linewidths and column densities (see their Fig. 2, bottom right panel).

\begin{figure}[h!]
\begin{center}
\resizebox{0.7\hsize}{!}{\includegraphics[angle=-90]{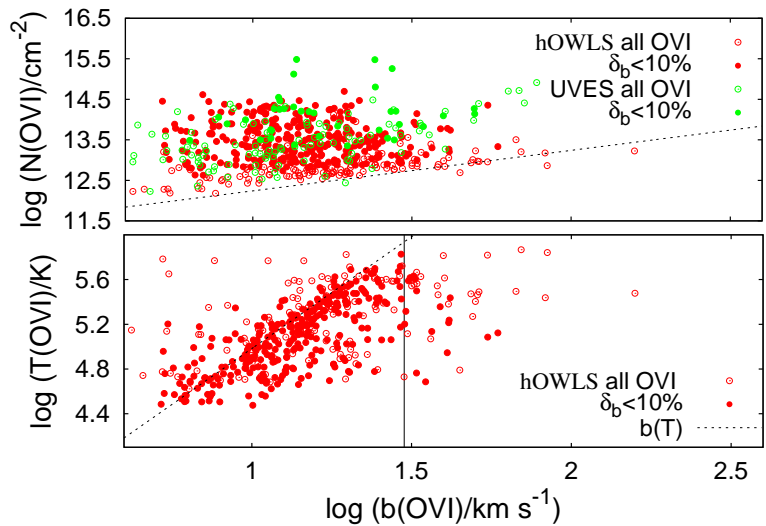}}
\caption[Correlation plots of the Doppler parameter from observational and synthetic spectra]{\small Correlation plots of the Doppler parameter from observational and synthetic spectra. A diagram $b$ vs. $N$ for the full OWLS sample of \Ovi absorbers (top panel) with its detection limit (dashed line) and a diagram $b$ vs. $T$ for 
the full samples of \Ovi absorbers (bottom panel) with UVES (green) and OWLS (red) data are shown. The subsamples with $\delta b\le10$ per cent are shown with filled symbols. The upper temperature limit (assuming purely thermal broadening) is drawn with a dashed line. The solid vertical line points to the the $b$-value above which the non-thermal broadening becomes the dominating factor.}
\label{fig_T_N_b_o6}
\end{center}
\end{figure}


A possible correlation between the Doppler parameter and column density at {\it high} redshifts has not been investigated intensively so far. \citet{Carswell02} do not find a 
correlation in their UVES \Ovi data for two sightlines (see their Fig. 15). \citet{Muzahid11} find a statistically insignificant correlation, considering an extended UVES 
\Ovi sample (18 sightlines). 

Motivated by these studies, we have searched for a correlation between the Doppler parameter and column density in our simulated \Ovi absorber sample. The results are shown in Fig.~\ref{fig_T_N_b_o6} (top panel). {\it No evidence for a correlation between the two quantities is found.}

\subsubsection*{Doppler parameter vs. temperature}

\hspace{0.6cm}The temperature is another important physical quantity that governs the ionizaion state of the gas. A standard approach to estimate gas temperature, used by us in Sect.~\ref{modeling_cloudy} and Appendix \ref{ionization_modeling_uves} for the analysis of the high-redshift UVES data, is, to consider the linewidths of observed absorber pairs \Ovi\!\!/\Hi that are well aligned in velocity space. It should be pointed out that velocity alignment does not {\it necessarily} indicates that the considered absorbers arise in the same gas phase. We revisit the applicability of this method in Sect.~\ref{aligned_absorbers_owls}. Due to its ambiguity, different authors obtain different temperature estimates in low-redshift absorbers \citep{Thom08b,Tripp08,Danforth08}. Therefore, it is interesting to look for (and to study) a possible correlation between \Ovi linewidth and gas temperature in our data. The correlation plot is shown in Fig.~\ref{fig_T_N_b_o6}, bottom panel. Points that lie above the upper 
temperature limit (under assumption of purely thermal broadening) are caused by resolution effects and the fitting procedure \citep{Tepper-Garcia11}. There is a clear tendency of increase of gas temperature with the Doppler parameter. For $b > 30$ \kms (solid line in Fig.~\ref{fig_T_N_b_o6}) non-thermal broadening becomes the dominating factor, leading to a larger $b$ for a given temperature than expected from purely thermal broadening. A qualitative conclusion is evident: generally, linewidths are representative for the real temperature, at least in a statistical sense. For temperature estimates of {\it individual} absorbers, however, one needs additional information, e.g. from supplementary data of other related ions.
 
\subsubsection*{Temperature vs. density}

\begin{figure}[h!]
\begin{center}
\resizebox{0.7\hsize}{!}{\includegraphics[angle=-90]{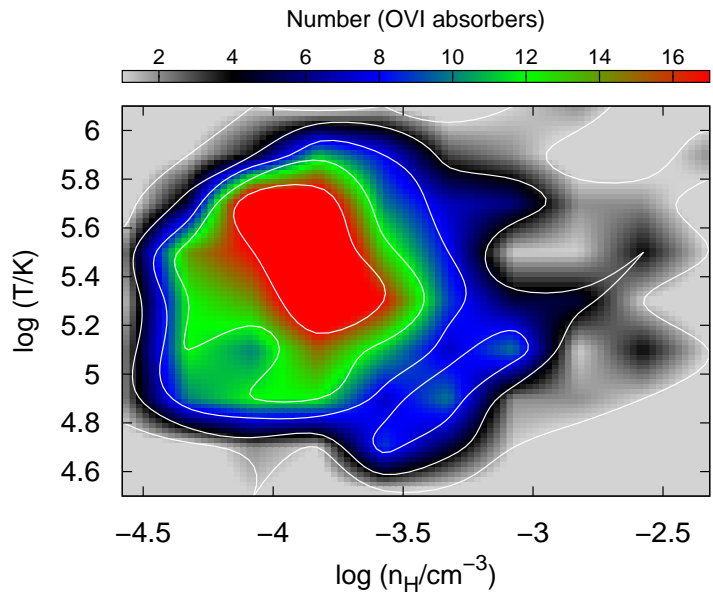}}
\caption[Distribution of the \hOWLS sample on the plane $T$ vs. $n_{\rm H}$]{\small Distribution of all absorbers from the OWLS sample on the plane $T$ vs. $n_{\rm H}$ (see text). The colour scale indicates the distribution amplitude. The red/green/blue/black/gray areas and the white contours enclose 25/50/75/90/99 per cent of the total number of absorbers, respectively.}
\label{fig_3dmap_n_T}
\end{center}
\end{figure}

\hspace{0.6cm}The distribution of \Ovi absorbers at $z= 2.5$ on the gas density-temperature plane is shown in Fig.~\ref{fig_3dmap_n_T}. The absorbers populate a region with corresponding densities of $10^{-5} \le  n_{\rm H} \le 4 \times 10^{-3}$\cc (overdensities\footnote{The relation between the gas density $n_{\rm H}$ and the
baryonic overdensity $\Delta$ is given by \\ $n_{\rm H}=\frac{\langle \rho_{\rm b}\rangle}{m_{\rm H}}~X_{\rm H}~(1+z)^{3}~\Delta$, where 
$\Delta \equiv \frac{\rho_{\rm b}}{\langle \rho_{\rm b}\rangle}$} $1.23 < \Delta < 4.9 \times 10^{2}$) and 
temperatures $4 \times 10^{4} \le \rm T \le 1 \times 10^{6}$~K. Such overdensities reach four to five times higher values than the typical value for the WHIM ($ \Delta \approx 0.1 - 100$). More than 60 per cent of all \Ovi absorbers are found in the 
low temperature regime of the WHIM, i.e. $10^{5} < \rm T < 10^{6}$ K. The high temperatures indicate shock-heated material. Twenty-five per cent of the sample (red area) 
exhibit a tendency of decreasing of temperature with increasing gas density, although the correction is weak. 

\begin{figure}[h!]
\begin{center}
\resizebox{0.7\hsize}{!}{\includegraphics[angle=-90]{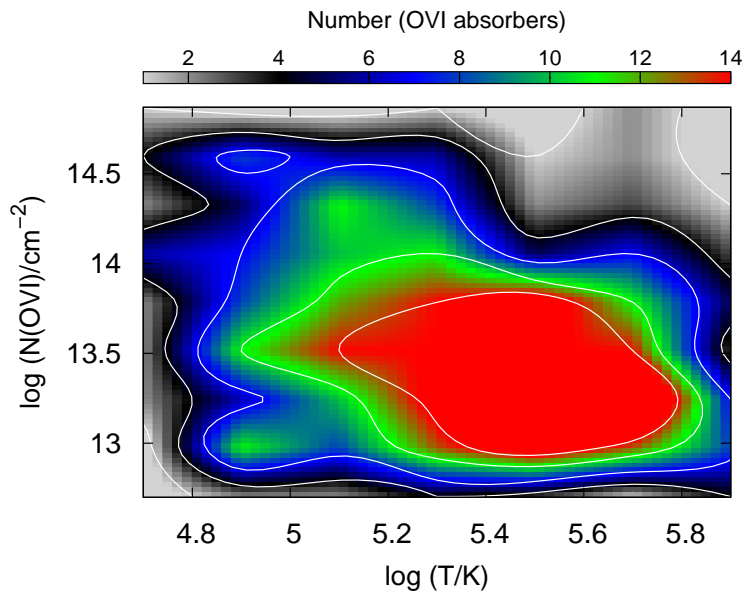}}
\caption[Distribution of the \hOWLS sample on the plane $N_{\rm OVI}$ vs. $T$]{\small Distribution of all absorbers from the \hOWLS sample on the plane $N_{\rm OVI}$ vs. $T$. The colour scales and contours are similar to the one used in the previous figure.}
\label{fig_3dmap_T_N}
\end{center}
\end{figure}

We study also the possible anti-correlation between temperatures and {\it column} densities of \Ovi absorbers, since it links an observed quantity with a physical parameter that is known directly from the simulations. 
Fig.~\ref{fig_3dmap_T_N} shows the distribution of all \Ovi absorbers from the \hOWLS sample on the temperature-column density plane.

The bulk of \Ovi absorbers is found in the column-density range $ 10^{13} \le  N$(\Ovi\!\!) $\le 10^{14}$\sqc. The temperatures $10^{5} \le \rm T \le  10^{6}$~K hint at shock-heated material. A high fraction (>50 per cent) of all \Ovi absorbers exhibit temperatures  $T > 10^{5}$~K and 
column densities $N$(\Ovi\!\!) <  $10^{14}$\sqc. Systems with lower column densities tend to have higher temperatures, but no definitive conclusion about individual absorbers can be reached.

\subsubsection*{Column density vs. metallicity}

\begin{figure}[h!]
\begin{center}
\resizebox{0.7\hsize}{!}{\includegraphics[angle=-90]{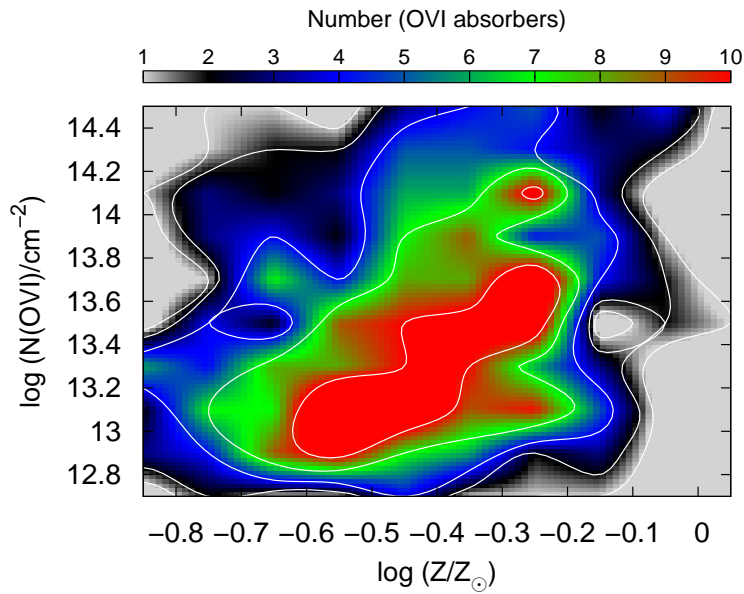}}
\caption[Distribution of the whole OWLS sample on the plane $N_{\rm OVI}$ vs. $Z$]{\small Distribution of all absorbers from the OWLS sample on the plane $N_{\rm OVI}$ vs. $Z$. The colour scales and contours are similar to the one used in the previous figure.}
\label{fig_Z_N}
\end{center}
\end{figure}

\hspace{0.6cm}
We inspect a possible correlation between the metallicity $Z_{\rm sm}$(\Ovi\!\!) and the column density of \Ovi, considering the full OWLS sample of \Ovi absorbers. Their distribution on the column density-metallicity plane is shown in Fig.~\ref{fig_Z_N}. 

Such a correlation is indeed present for a subsample of absorbers that includes 25 per cent of the systems (red area). With a rise of column density in the range 
$12.8 \le \rm log(N$(\Ovi\!\!)/$\rm cm^{-2}) \le 14.2$ the metallicity increases as well from $\sim-0.7$ dex to about $-0.2$ dex. It is clear from the figure that \Ovi synthetic 
absorbers trace enriched material with metallicities $\rm log(Z_{\rm sm}$(\Ovi\!\!)/$Z_{\odot}) \ge -1.5$, as already discussed in Sect.~\ref{local_metallicity_owls}. 

\subsubsection*{Temperature vs. metallicity}

\begin{figure}[h!]
\begin{center}
\resizebox{0.7\hsize}{!}{\includegraphics[angle=-90]{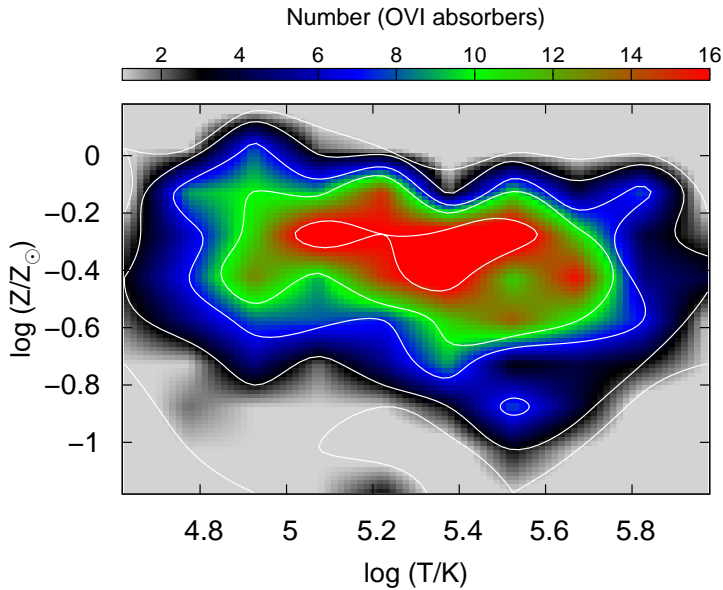}}
\caption[Distribution of the \hOWLS sample on the plane $Z$ vs. $T$]{\small Distribution of all absorbers from the OWLS sample on the plane $Z$ vs. $T$. The colour scales and contours are similar to the one used in the previous figure.}
\label{fig_T_Z}
\end{center}
\end{figure}

\hspace{0.6cm}Finally, we plot the distribution of \Ovi absorbers at $z= 2.5$ on the temperature-metallicity plane (Fig.~\ref{fig_T_Z}). The majority of the absorbers lie within the temperature range 
$4 \times 10^{4} \le \rm T \le 10^{6}$~K and in the metallicity range $-1.0 \le \rm \log (Z_{\rm sm}$(\Ovi\!\!) $\le 0.0$. About half of the sample (red and green areas) is constrained within a rather narrow metallicity range $-0.6 \le \rm \log (Z_{\rm sm}$(\Ovi\!\!) $\le -0.1$ and exhibit a tendency toward slight decrease of $\rm Z_{\rm sm}$(\Ovi\!\!) with increasing temperature. 

\begin{figure}[h!]
\begin{center}
\resizebox{0.7\hsize}{!}{\includegraphics[angle=-90]{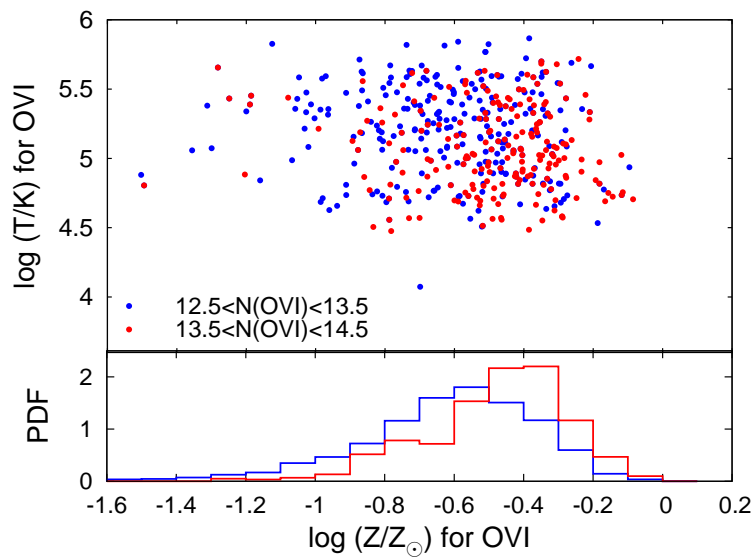}}
\caption[Metallicity and temperature of \Ovi absorbers from the \hOWLS sample.]{\small Metallicity vs. temperature for all \Ovi absorbers from the \hOWLS sample. {\it Top panel}: two subsamples with $12.5 \le \rm log(N(\Ovi\!\!/{\rm cm}^{-2}) \le 13.5$ (blue) and $13.5 \le \rm log(N(\Ovi\!\!)/{\rm cm}^{-2}) \le 14.5$ on the $Z$ vs. $T$ diagram. {\it Bottom panel}: the metallicity PDFs of the subsamples.}
\label{fig_T_Z_N}
\end{center}
\end{figure}



In an attempt to distinguish a possible effect of column density on metallicity and/or temperature of the absorbers, we divided the \hOWLS sample in two subsamples, according the column densities, and plotted them on the $T$ vs. $Z$ plane (Fig. \ref{fig_T_Z_N}, top panel). Evidently, systems with higher column densities tend to have higher metallicities as illustrated as well by the PDF distributions of metallicity (bottom panel). The Kolmogorov-Smirnov test yields a 0.42 probability that the two distributions have the same origin.    


\subsection{Aligned absorbers}\label{aligned_absorbers_owls} 

\subsubsection{Thermal and non-thermal components of the Doppler parameter in the aligned \Ovi absorbers}

\hspace{0.6cm}To explore the contribution of the non-thermal processes in the line broadening at high and low redshift (see Sec.~\ref{doppler_parameter_owls}), we consider a 
subsample of \Ovi absorbers that exhibit velocity alignment between \Ovi and \Hi ($\le 10$ km~$\rm s^{-1}$).

The Doppler parameter $b$, as introduced in Sect.~\ref{line_broadening}, provides only an upper limit of the temperature on the absorbing gas. If the line broadening 
is influenced also by non-thermal mechanisms such as turbulence or peculiar velocities, the true temperature $T$ in an absorber is lower. The thermal and the non-thermal 
component of the Doppler parameter as defined by Eqs \ref{eq_temp_Doppler} and \ref{eq_temp_Doppler_2} are:

\begin{equation}
\label{eq_b_comp}
b^{2} = b^{2}_{\rm th} + b^{2}_{\rm nt} = (0.129)^{2}~\frac{T}{A} + b^{2}_{\rm nt}. 
\end{equation}

For ions located in the same physical region, the nonthermal component is assumed to be identical, whereas the thermal component 
scales inversely with the mass of the ion.

\begin{figure}[h!]
\begin{center}
\resizebox{0.7\hsize}{!}{\includegraphics[angle=-90]{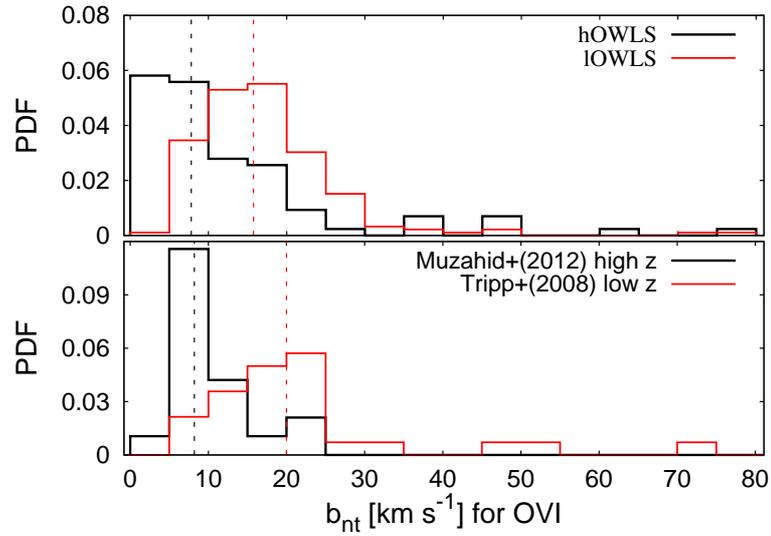}}
\caption[Distributions of non-thermal Doppler parameter for well-aligned components.]{Distributions of $b_{\rm nt}$ for the well-aligned system components ($\Delta v = 10 ~\rm km~s^{-1}$) in 5 $\rm km~s^{-1}$ bins detected in simulated (top) and observed (bottom) spectra at low and high $z$.}
\label{fig_hist_b_nth_obs_sim}
\end{center}
\end{figure}
 

Since the \lOWLS and \hOWLS data provide information about the temperature for each ion, one can obtain $b_{\rm nt}$ (Eq.~\ref{eq_b_comp}) 
for $b_{\rm th}$ of the aligned components (Eq.~\ref{eq_temp_Doppler}). In Fig.~\ref{fig_hist_b_nth_obs_sim} the derived distributions of $b_{\rm nt}$ for low and high $z$ are compared with results from observational studies of \citet{Tripp08} and \citet{Muzahid11}. In both observational studies $b_{\rm nt}$ was calculated 
from the total Doppler parameter using the following approach. First,  \Ovi and \Hi components that are well aligned in velocity space were selected. The underlying assumption is, that these elements arise in the same gas phase. Second, two versions of Eq.~\ref{eq_b_comp} have been employed for \Hi and \Ovi to 
solve for $T$ and $b_{nt}$.  

As seen in Fig.~\ref{fig_hist_b_nth_obs_sim}, the median $b_{\rm nt}$ for the low $z$ observations \citep{Tripp08} is $\sim$20 $\rm km~s^{-1}$, which is similar to the one obtained from the \lOWLS synthetic spectra (15.75 $\rm km~s^{-1}$). For high $z$, the medians from \hOWLS and from observations of \citet{Muzahid11} are also almost identical: 7.8 $\rm km~s^{-1}$ and 8.2 $\rm km~s^{-1}$, respectively. Despite the similarity between the $b_{nt}$ distributions in the synthetic and the observed spectra at high and low $z$,  
we note that interpretation of this result is not straight-forward. The non-thermal Doppler parameter components are estimated by use of two very different approaches, as outlined above. Nevertheless, the result suggests that the $b_{\rm nt}$ distributions at high and at low $z$ are essentially different and it seems that this supports our previous scenario, namely that the nonthermal processes play a more important role at low redshifts (see Sect.~\ref{effect_redshift}). Therefore, the previous assumption,
that the mean \Ovi line widths at high $z$ provide a rough estimate of the temperature of the absorbing gas, is justified.

\begin{figure}[h!]
\begin{center}
\resizebox{0.7\hsize}{!}{\includegraphics[angle=-90]{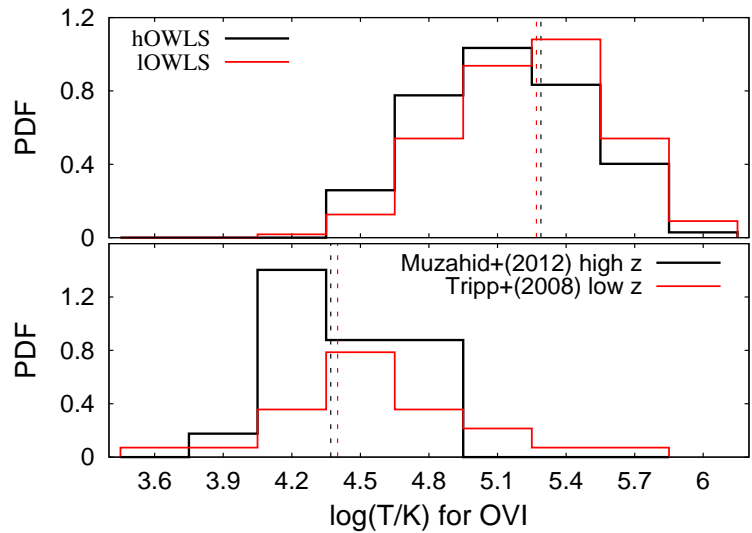}}
\caption[Temperature distribution of well-aligned \Hi\!\!-\Ovi absorbers.]{\small Temperature distribution of the \Hi\!\!/\Ovi absorber pairs ($\Delta v = 10 ~\rm km~s^{-1}$) at low (red) and high (black) $z$ from observed (bottom panel) and simulated (top panel) spectra.}
\label{fig_hist_T_obs_sim_10}
\end{center}
\end{figure}


The corresponding temperature distributions for the well aligned \Ovi absorbers at low and high $z$ are plotted in Fig.~\ref{fig_hist_T_obs_sim_10}. The temperature from the 
observed spectra is calculated by consideration of $b(\Hi\!\!)-b(\Ovi\!\!)$ pairs, as described briefly in the beginning of the section. The conclusion drawn by \citet{Muzahid11} is, that the temperature distribution for the low $z$ observational sample does not differ significantly from the one derived in their high $z$ observational study -- the probability that the difference is statistical artifact is estimated to be $\approx$ 38 per cent. These authors point out that 42 per cent of their sample exhibits temperatures 4.6 $\le$ log $T$ $\le$ 5.0, which are higher than those expected from photoionization equilibrium. The suggested explanation is, that higher temperatures can be reached in a rapidly cooling over-ionized gas phase that was shock-heated through mechanical processes such as galactic winds. 

However, the median of the temperature distribution in both observational samples is $\sim 3 \times 10^{4}$ K, while the median 
value of the temperature distribution in both simulational samples is $\sim 1.9 \times 10^{5}$ K, i.e., about one order of magnitude higher than found from observations.  

\subsubsection{Temperature distribution of the aligned \Ovi absorbers}

\begin{figure}[h!]
\begin{center}
\resizebox{0.7\hsize}{!}{\includegraphics[angle=-90]{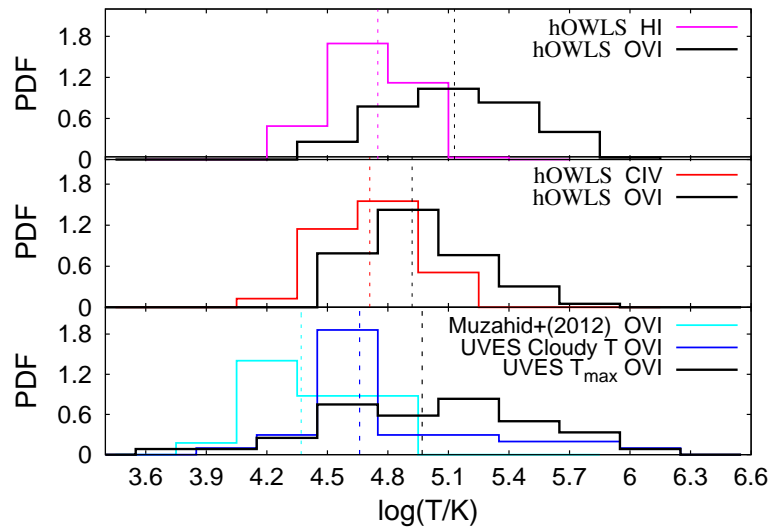}}
\caption[Temperature distributions of the well-aligned \Ovi components]{Temperature distributions of the well-aligned \Ovi components ($\Delta v = 10 ~\rm km~s^{-1}$) at high $z$ as derived from different data. {\it Bottom panel}: our UVES sample, assuming purely thermal broadening (black) and from the {\sc Cloudy} modeling (blue), and the UVES sample of \citet[][violet]{Muzahid11}; {\it Middle panel}: OWLS synthetic \Ovi components (black) and OWLS 
synthetic \Civ components aligned with them (red); {\it Top panel}: OWLS synthetic \Ovi components (black) and OWLS synthetic \Hi components aligned with them (green). The dashed lines indicate the corresponding median values.}
\label{fig_hist_T_obs_sim_10_o6_c4}
\end{center}
\end{figure}

\hspace{0.6cm}The difference between the temperature distributions found from observational and simulated spectra of well-aligned \Ovi absorbers at high $z$ is puzzling. 
Looking for an explanation, we recall our previous conclusion that the non-thermal processes do not significantly affect the line broadening at high $z$ 
(see Sect.~\ref{doppler_parameter_owls}). Note that we calculated the upper limit of temperature, $T_{\rm max}$, from the $b$ values in the UVES subsample 
of well aligned \Ovi absorbers (see Sect.~\ref{modeling_cloudy}), assuming purely thermal broadening (Eq.~\ref{eq_temp_Doppler}).  In Fig.~\ref{fig_hist_T_obs_sim_10_o6_c4} (bottom panel) the distribution of $T_{\rm max}$ is plotted together with the distribution of temperatures obtained from the {\sc Cloudy} modeling for the same subsample of \Ovi absorbers. We added, for comparison, the \Ovi temperature distribution derived by \citet{Muzahid11} (violet). 
The temperature distributions of \Ovi absorbers derived from the OWLS synthetic spectra, together with those of \Civ and \Hi systems that are well aligned with 
them, are shown in the middle and top panels of this figure, respectively.

As can be seen in this figure, the samples of \Civ and \Hi synthetic components aligned with \Ovi exhibit similar temperature distributions with median 
values $\log\,T = 4.71$ and $\log\,T = 4.75$, respectively. In contrast, the \Ovi distributions differ from them, depending on whether \Ovi is aligned 
with \Civ or with \Hi\!\!. The distribution of oxygen absorbers, aligned with \Hi (Fig.~\ref{fig_hist_T_obs_sim_10_o6_c4}, top panel) is shifted to higher temperatures with 
a median value $\log\,T = 5.13$. This result hints at different physical conditions and different gas phases, regardless of the velocity alignment. Better 
agreement, marked by a significantly closer median value $\log\,T = 4.92$ is found with the distribution of the aligned \Civ components (middle panel). In 
this case, it seems much more likely that the components of both species trace the same physical conditions. 

A similar pattern in the temperature distribution of \Ovi absorbers is seen, when the UVES subsample of well aligned \Hi\!\!/\Ovi pairs (Fig.~\ref{fig_hist_T_obs_sim_10_o6_c4}, bottom panel) is considered. 
In case the temperature is calculated from the observed \Ovi $b$-values, assuming purely thermal line broadening, the median value of the \Ovi distribution is $\log\,T \approx 5$ (black line). However, when one assumes that \Ovi originates in the same gas phase as \Hi (and \Civ\!\!, in some cases) and models the physical conditions with {\sc Cloudy}, 
the predicted median value $\log\,T \approx 4.66$ for the same UVES \Ovi subsample (blue line) is close to those of the \Civ and \Hi distributions 
from the synthetic spectra. The distribution itself becomes more similar -- although with a more extended high-temperature tail, -- to the one derived by \citet[][violet line]{Muzahid11}, whose temperature calculations were based also on the assumption of a single gas phase for the observed \Ovi and \Hi systems. Apparently, the temperature distribution of the \Ovi components depends strongly on the chosen model, namely, whether the \Ovi lines arise in gas phase that is similar or different from that of \Civ and \Hi\!\!. Therefore, 
it is important to explore further, how reliable the assumption is that well-aligned components \Ovi\!\!, \Civ and \Hi actually arise in a single gas phase.  

\begin{figure}[h!]
\begin{center}
\resizebox{0.9\hsize}{!}{\includegraphics[angle=-90]{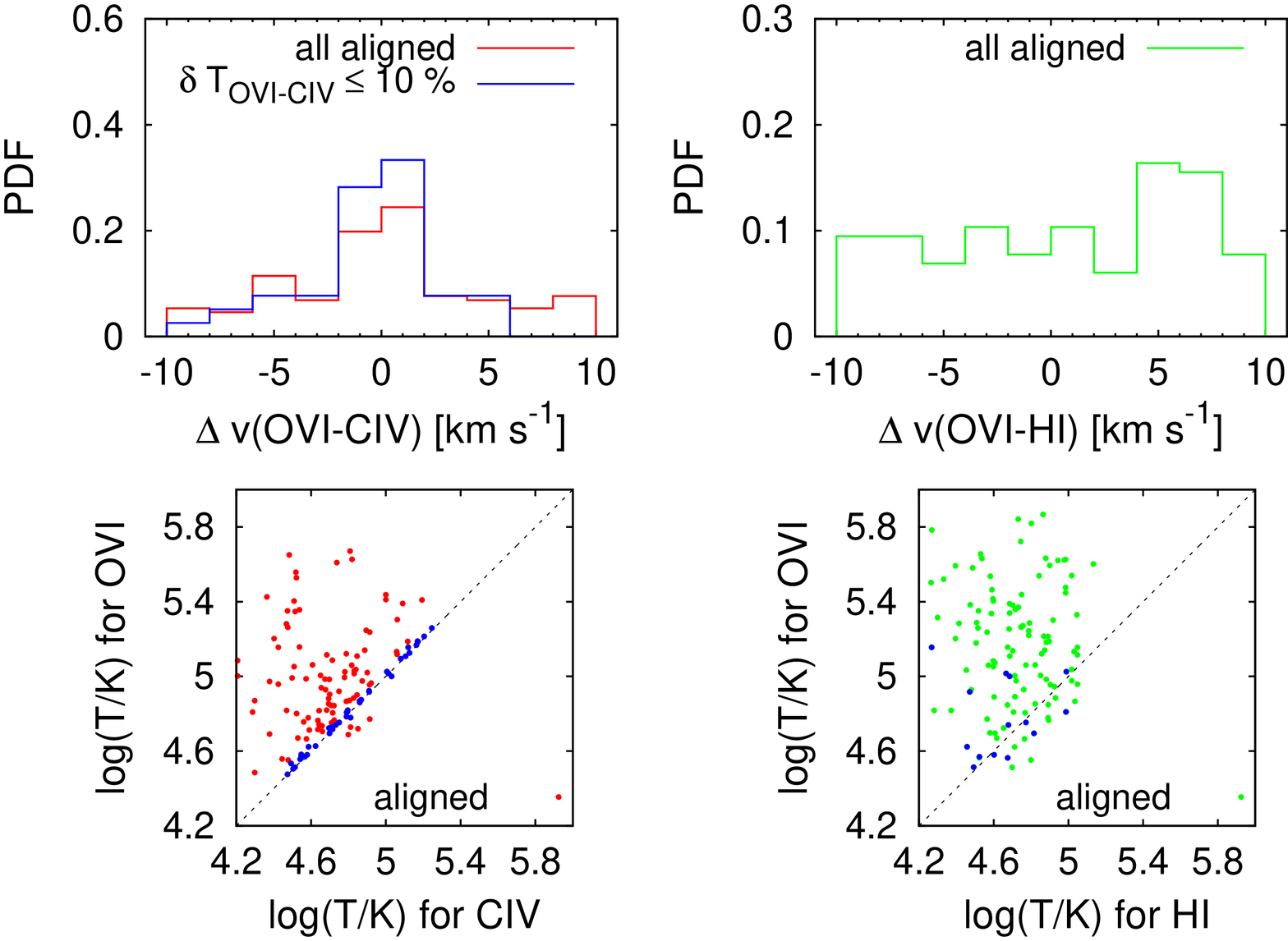}}
\caption[Temperatures and velocity offsets of \Ovi components in \hOWLS spectra.]{\small Temperatures 
and velocity offsets of \Ovi components in high-$z$ OWLS spectra that are well aligned with \Civ and \Hi. {\it Bottom panel}: temperature/temperature plots of the well-aligned ($\Delta v = 10 ~\rm km~s^{-1}$) pairs \Ovi\!\!/\Civ (left) and \Ovi\!\!/\Hi (right). Blue dots denote the subsample of \Ovi\!\!/\Civ pairs that are aligned in velocity space and that have nearly the same temperature ($\Delta T/T\le10$~\%). The dashed line indicates the identity of the two quantities. {\it Top panel}: the velocity offset distributions are shown. The subsample mentioned above is denoted by the blue line.}
\label{fig_T_vel_o6_c4_aligned_sim}
\end{center}
\end{figure}

Clearly, if two or more ions are in a state of local ionization equilibrium and arise in the same gas phase, their corresponding absorbers trace the same physical region with one and the same gas temperature. Therefore, a comparison between the temperatures derived for \Ovi\!\!, \Civ and \Hi in well-aligned absorption components in the OWLS synthetic spectra would provide valuable information on this scenario. Such temperature/temperature diagrams for well-aligned \Ovi\!\!/\Civ 
and \Ovi\!\!/\Hi components are plotted in Fig.~\ref{fig_T_vel_o6_c4_aligned_sim}, bottom panel. In the top panels the corresponding distributions of the velocity offsets are shown. 

\subsubsection*{\Ovi\!\!/\Hi aligned absorbers}

\hspace{0.6cm}In general, the temperatures of the \Ovi components are systematically higher (up to one order of magnitude) than those of the \Hi components aligned with them 
(Fig.~\ref{fig_T_vel_o6_c4_aligned_sim}, bottom right panel). As said earlier, the velocity alignment might indicate that both ions trace the same overall structure, but possibly arise in physically distinct regions within the host structure with different physical conditions. It is possible, however, that next to the dominating \Hi absorption component that obviously traces cooler gas there exists an underlying, broad and weak \Hi component that is physically associated with the warmer \Ovi absorbing region. For instance, this hot phase 
could be an interface layer on the surface of a cooler cloud \citep{Boehringer87}. In other words, \Ovi and \Hi arise in multi-phase gas, where \Ovi traces the hotter phase while the strong \Hi traces the cooler one. Considering more complex, multi-component, multi-phase systems, it is to be expected that some broad \Hi components associated with 
hot \Ovi absorbers can be found in spectra with very high S/N. Overall, this result confirms the suggestion of \citet{Fox11}, namely that the alignment between \Ovi and \Hi components should not be taken as a proof of single-phase photoionization.   

Note that the velocity offset distribution of \Ovi\!\!/\Hi aligned absorbers (Fig.~\ref{fig_T_vel_o6_c4_aligned_sim}, top right panel) has a weak peak at $\approx$ 6 \kms\!\!. 

\begin{figure}[h!]
\begin{center}
\resizebox{0.7\hsize}{!}{\includegraphics[angle=-90]{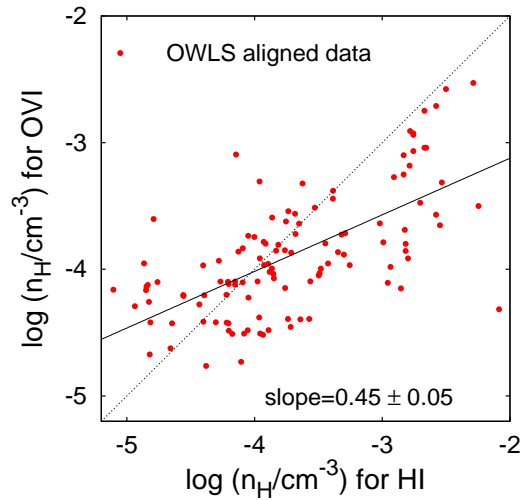}}
\caption[Volume density of hydrogen in \Ovi absorbers and in aligned \Hi components.]{\small Volume density of hydrogen in \Ovi absorbers and in \Hi components aligned with them from the \hOWLS data. The dashed line indicates the identity of the two quantities.}
\label{fig_n_aligned}
\end{center}
\end{figure}


If the aligned components would reside in single phase, they should exhibit roughly the same volume density of hydrogen, $n_{\rm H}$. However, as seen in Fig.~\ref{fig_n_aligned}, this is apparently not the case,`  
suggesting that the aligned \Ovi and \Hi components rather arise in different gas phases. The hydrogen density in the \Ovi absorbers tends to increase slowly (slope $\approx 0.45$) with $n_{\rm H}$ in the aligned \Hi components. This can be interpreted as an indicator for the existence of multi-phase gas. 

\begin{figure}[h!]
\begin{center}
\resizebox{0.8\hsize}{!}{\includegraphics[angle=-90]{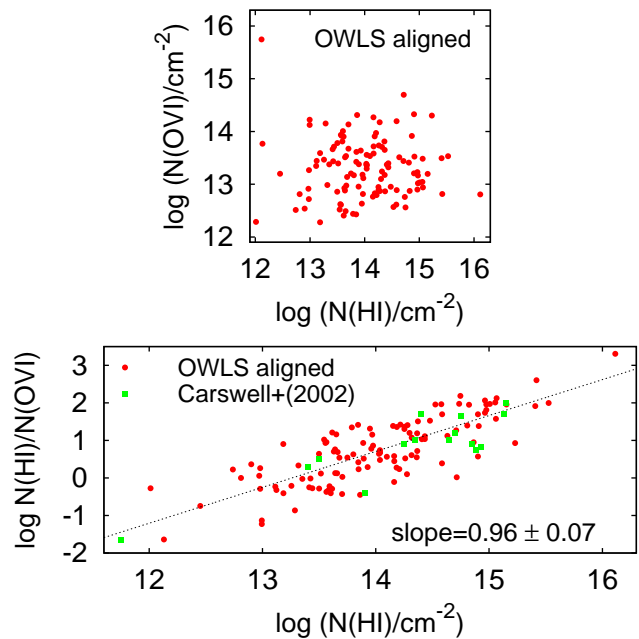}}
\caption[Column densities in \Ovi absorbers and in aligned \Hi components.]{\small Column densities in \Ovi 
absorbers and in \Hi components aligned with them from \hOWLS and from observations. {\it Top panel}: $N$(\Ovi\!\!) vs. $N$(\Hi\!\!) from \hOWLS. {\it Bottom panel}: $N$(\Hi\!\!) vs. $N$(\Hi\!\!)/N(\Ovi\!\!). Observational measurements from the \Ovi\!\!/\Hi sample of \citet{Carswell02} are plotted with green symbols.}
\label{fig_N_aligned}
\end{center}
\end{figure}


A supplementary comparison of column densities in aligned \Ovi and \Hi components is shown in Fig.~\ref{fig_N_aligned}. The upper panel shows that there is no correlation between the column densities of \Ovi and \Hi\!\!, i.e., $N$(\Ovi\!\!) does not change with increasing $N$(\Hi\!\!). A different visualization of this behavior is shown in the lower panel on this figure, where the slope of the distribution $N$(\Hi\!\!)/$N$(\Ovi\!\!) vs. $N$(\Hi\!\!) is close to unity.  

\subsubsection*{\Ovi\!\!/\Civ aligned absorbers}

\hspace{0.6cm}Two populations of aligned \Ovi and \Civ components can be distinguished in Fig.~\ref{fig_T_vel_o6_c4_aligned_sim}. The one with nearly identical temperatures 
(difference of $\le$ 10 per cent) represents about 30 per cent of the whole sample of aligned absorbers (bottom left panel, blue points). The other population contains the rest of 
the components with a significantly larger scatter in temperatures, suggesting a multi-phase structure of the gas (bottom left panel, red points). The velocity offset distributions 
(top left panel) for the two populations are similar and do not provide any additional information on the gas phase.

\begin{figure}[h!]
\begin{center}
\resizebox{0.7\hsize}{!}{\includegraphics[angle=-90]{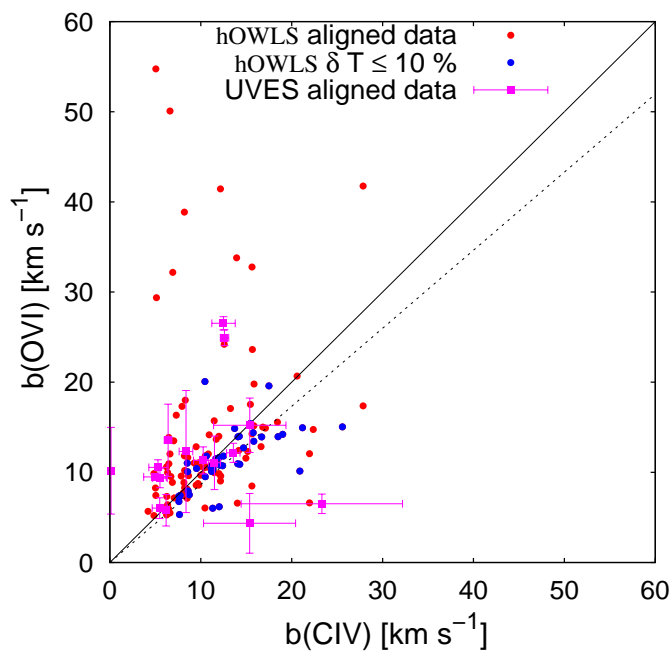}}
\caption[Doppler parameters of \Ovi absorbers and in aligned \Hi components.]{\small Doppler parameters of \Ovi absorbers and in the \Hi components aligned with them. The identity of the two quantities is plotted with the solid line, while the dashed line indicates the identity of $b$ under assumption of purely thermal line broadening. The population of aligned absorbers with nearly identical temperatures from the \Ovi and \Civ components is denoted with blue symbols. The observational sample of aligned \Ovi and \Civ components from UVES is shown with violet symbols.}
\label{fig_b_aligned}
\end{center}
\end{figure}

\begin{figure}[h!]
\begin{center}
\resizebox{0.7\hsize}{!}{\includegraphics[angle=-90]{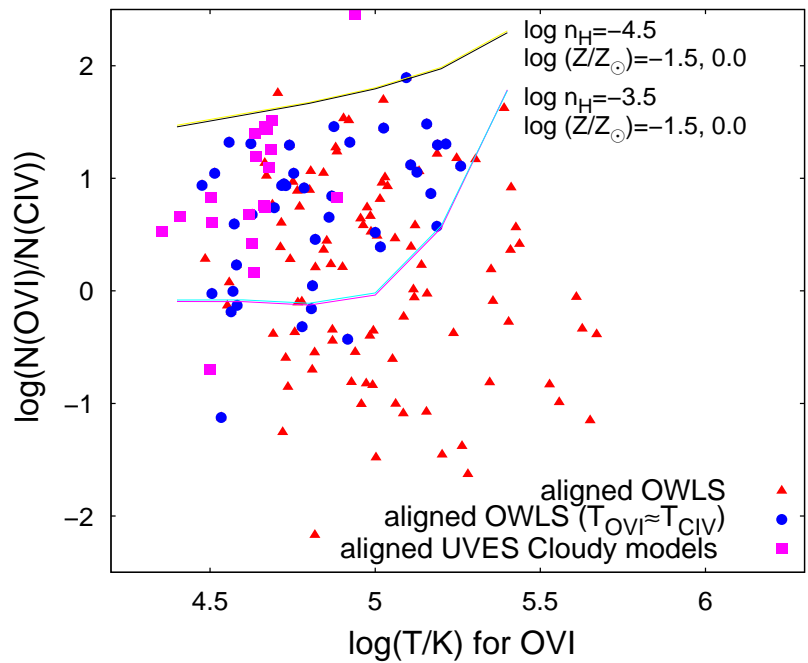}}
\caption[Temperature dependence of $N(\Ovi\!\!)/N(\Civ\!\!)$ of aligned components.]{\small Column-density ratio of aligned \Ovi and \Civ components from \hOWLS spectra (red triangles) as a function of temperature. The subsample with nearly identical Doppler parameters (temperature differences of $\le$ 10 per cent) is shown with blue dots. The observational sample of aligned \Ovi and \Civ components from UVES (violet) also are shown;
their temperature is obtained from the modeling with {\sc Cloudy}. {\sc Cloudy} models for two fixed values of hydrogen density and for two choices of the metallicity (in solar units) are plotted with colored lines.}
\label{fig_T_o6_c4_ratio_aligned_sim}
\end{center}
\end{figure}



In an attempt to identify {\it observational} criteria to distinguish between the two populations of aligned \Ovi\!\!/\Civ absorbers, we compose a comparative diagram of 
the Doppler parameters of these ions (Fig.~\ref{fig_b_aligned}). The identity of the two quantities is plotted with the solid line, while the dashed line indicates the identity of $b$ under assumption of purely thermal line broadening. About 83 per cent of the absorbers with similar temperatures seem to be mostly thermally broadened with comparable $b$-values for \Ovi and \Civ. The Pearson coefficient of correlation\footnote{~More about the Pearson coefficient can be found in \citet{Dorogovtsev10}.} between the components of that population is 0.62, which points to a strong correlation. On the other hand, the value for the second population (with different temperatures) is only 0.19, indicating a weak correlation between the components. The likelihood of the coefficients occurring for fully uncorrelated data is $2 \times 10^{-5}$ and $0.06$, respectively. The UVES sub-sample of \Ovi absorbers aligned in velocity with \Civ is added in Fig.~\ref{fig_b_aligned} for comparison. 

A general conclusion that can be drawn from the Doppler parameters derived for the synthetic \Ovi\!\!/\Civ pairs aligned in velocity space is, that, if the \Ovi absorbers have smaller or equal $b$-values compared to \Civ absorbers, 42 per cent have the same temperature and arise in a single gas phase. The remaining 58 per cent of the \Ovi\!\!/\Civ pairs have different temperatures and arise in multi-component structures.

To gain further insight into the physical state of \Ovi absorbers aligned with \Civ\!\!, we search for a correlation between the \Ovi\!\!/\Civ column density ratio and \Ovi 
temperature. As can be seen in Fig.~\ref{fig_T_o6_c4_ratio_aligned_sim}, such a correlation indeed appear to exist for the absorber population with similar temperatures
(temperature differences of $\le$ 10 per cent), as well for the UVES subsample of \Ovi absorbers aligned with \Civ and having the same temperatures as modeled 
with {\sc Cloudy} (see Sect.~\ref{modeling_cloudy}). We plot the predicted $N$(\Ovi\!\!)/$N$(\Civ\!\!) ratio as a function of temperature from {\sc Cloudy} for two fixed 
metallicities (solar and 0.03 $Z_\odot$) and for two choices of the hydrogen density ($\log n_{\rm H}=-3.5$ and $\log n_{\rm H}=-4.5$). Remarkably, $\approx$ 90 per cent of the aligned 
population with similar temperatures and most of the absorbers from the UVES subsample fall into a narrow density 
range $-4.5\le \log n_{\rm H}\le -3.5$, regardless of the adopted metallicity. We have assumed solar relative abundances of O and C and thus the predicted column density ratio 
from {\sc Cloudy} does not depend on the absolute metallicity. The ratio $N$(\Ovi\!\!)/$N$(\Civ\!\!) for both samples exceeds unity and varies within $\sim 2$ orders of magnitude. The temperature range in these models (see Fig.~\ref{fig_T_o6_c4_ratio_aligned_sim}) hints at photoionization as the relevant ionizating mechanism in these components.

\section{Baryon and metal content}\label{baryons_owls}

\hspace{0.6cm}As final step of our analysis of high redshift \Ovi absorbers in the OWLS, we determine the baryon and metal content in these systems. The relative abundances of baryons and metals in different gas phases and their total densities as estimated from the \hOWLS are specified in Table \ref{baryon_metal_content}. We find that $\approx$ 38 per cent of the metals in the gas are in the temperature regime $\log T > 5 \times 10^{4}$~K, whereas the \Ovi absorbers contribute with  $< 1$ per cent to the total metal budget. Hence, the conclusion is that \Ovi absorbers arise in metal-enriched gas, but they do not represent tracers of the main metal reservoirs at high redshift. The baryonic density in \Ovi systems at high redshift $\Omega_{{\tiny \rm IGM}}^{{\tiny \rm OVI}}$ is less than 1 per cent in the OWLS, in excellent agreement with our estimate from the UVES observations (Chapter~\ref{uves}). 

{\it Therefore, our results indicate that high-redshift \Ovi absorbers neither trace the bulk of the metals at that epoch, nor do they host a significant fraction of the baryonic mass in the Universe.}

\begin{table*}[th!]
\begin{center}
\caption[Relative mass densities of baryons and metals in the \hOWLS.]{Relative mass densities of baryons and metals in \hOWLS for different temperature regimes and overdensities.}
\begin{small}
\begin{tabular}{llll}
 \hline
 \hline
~~&Baryons &Metals \\   \\
 \hline
 log (T/K) $>$ 4.7, $\log\Delta <$2.0 & 18.30 \%   & 35.53 \%   \\
 log (T/K) $<$ 4.7, $\log\Delta <$2.0 & 73.65 \% & 0.1 \%    \\
 \hline
 ~ & $\Omega_{\rm b}(\rm OVI)/\langle \Omega_{\rm b}\rangle$=9.8 $\times 10^{-3}$ & $\Omega_{Z}(\rm tot)$= 7.22 $\times 10^{-4}$ \\  
\hline
\end{tabular}
\end{small}
\label{baryon_metal_content}
\end{center} 
\end{table*}

\section{Conclusions}\label{conclusions_owls}

\hspace{0.6cm}In this chapter, we studied the properties of \Ovi absorbers at high redshift using artificial absorption spectra obtained from numerical simulations (OWLS). We compared in details the results from OWLS with those from UVES observations presented in the previous chapter. Below we summarize separately the conclusions about the total \Ovi absorber  sample and about the subsample of \Ovi systems that are aligned in velocity space with the associated \Hi absorption.

\subsection*{Total \Ovi sample}

\begin{itemize}

\item 

The \Ovi CDDF obtained from the OWLS \Ovi absorber sample is in good agreement with that derived from the UVES observations. This implies that the simulations successfully reproduce the statistical properties of metal-enriched, highly ionized gas structures observed in the early Universe. The Doppler parameter distributions obtained from the synthetic and observed \Ovi spectra suggest that turbulence does not affect significantly the \Ovi line broadening at high redshift.

\item 

Most of the \Ovi absorbers from the \hOWLS sample populate the region on the density-temperature plane $10^{-5} \le  n_{\rm H} \le 4 \times 10^{-3}$~\cc~(overdensities $1.23 < \Delta < 4.9 \times 10^{2}$ at $z=2.5$) and $4 \times 10^{4} \le \rm T \le 1 \times 10^{6}$~K. These (over)densities are roughly one order of magnitude higher than those expected for the diffuse warm-hot WHIM produced by cosmological accretion shocks. They exceed also by $\sim 1$ magnitude the values predicted by the simulations for \Ovi absorbers at $z=0$. More than 60 per cent of all \Ovi absorbers at high redshift are found in the temperature regime $10^{5} < \rm T < 10^{6}$ K. These relatively high temperatures indicate shock-heated material, as expected for highly-ionized gas expelled by galactic winds and outflows. 

\item 

\Ovi absorbers at high redshift trace metal-enriched material with $\log (Z_{\rm \small{\Ovi}}/Z_{\odot}) > -1.5$ and a median value $\approx 0.3~ Z_{\odot}$. The median for the low-redshift sample (\lOWLS\!\!\!) is closer to the solar value: $\approx 0.6~ Z_{\odot}$. The increase of metallicity from high to low redshifts is consistent with the standard models of cosmic chemical evolution, that predict a rise of the global mean interstellar metallicity in galaxies with time, reaching quasi-solar values at $z=0$.

\item 

The OWLS imply that $\sim$ 38 per cent of the metals in the gas at high redshift are in the temperature regime $\log T > 5 \times 10^{4}$~K, while \Ovi absorbers contribute by only $<1$ per cent to the metal budget. We conclude that \Ovi absorbers arise in metal-enriched gas structures that have a large absorption cross section, but they do not trace the main metal reservoirs. The baryonic density fraction of \Ovi systems at high redshifts $\Omega_{{\tiny \rm IGM}}^{{\tiny \rm OVI}}$ is less than 1 per cent, so that high-redshift \Ovi absorbers do not trace the bulk of the baryons either. 

\end{itemize}

\subsection*{Aligned \Ovi absorbers}

\begin{itemize}

\item 

The comparison between non-thermal components of the Doppler parameters in subsamples of well-aligned ($\Delta v \leq 10$ \kms\!\!) 
\Ovi\!\!/\Hi absorber pairs in the \hOWLS and \lOWLS spectra shows that both distributions are essentially different. At high redshift, non-thermal broadening contributes poorly to the total width of the \Ovi lines and thus the Doppler parameter distribution of \Ovi absorbers can be used (in a statistical sense) to roughly constrain the gas temperature in the absorbers. In contrast, the non-thermal contribution to $b$(\Ovi\!\!) at low redshift is significant, indicating a higher velocity dispersion in \Ovi absorbing gas structures in the local Universe.

\item 

The temperatures of \Ovi components are systematically higher, within an order of magnitude, than those of \Hi components aligned with them in velocity space. This clearly hints at a multiphase gas in the host structure. We speculate that weak, broad \Hi components, possibly related to the \Ovi absorbing phase, could be present, but observations with very high S/N ratio would be required to detect them. \Ovi lines evidently arise in a hot gas phase that is closely aligned in velocity with a cooler phase producing the bulk of the detected \Hi absorption. 

\item 

Two populations of aligned \Ovi\!\!/\Civ pairs are identified. One of them contains $\approx 30$ per cent of the whole subsample of aligned \Ovi\!\!/\Civ absorbers and displays nearly identical temperatures in the two ions. This population could be well explained by photoionization as the dominating ionization mechanism, with the two ions arising in a single gas phase. The other population, containing $\approx70$ per cent of the subsample of aligned \Ovi\!\!/\Civ\!\!, displays significantly  different temperatures in \Ovi and \Civ\!\!, which suggests a multiphase structure of the gas. 
\end{itemize}


\chapter{Conclusions}\label{conclusions}

\hspace{0.6cm}Our study of high-redshift \Ovi absorbers presented in the context of this thesis unveils the diverse nature of highly-ionized absorption-line systems in the early Universe. The use of intermediate- and high-resolution optical spectra and cosmological simulations turns out to be a powerful combination that enables us to gain insight into the statistical properties {\it as well as} into the physical conditions of these systems. The results from our study that has been described in detail in the previous chapters now can be combined to characterize the complex nature of \Ovi absorbers at high redshift. The following conclusions can be drawn from our investigations:

\begin{itemize}

\item[1)] {\it High-redshift \Ovi absorbers do not have a common origin.}

There are many different indicators for the diverse origin of \Ovi absorbers at high redshift. Firstly, the spectral analysis of observed and simulated \Ovi absorbers demonstrates that the kinematics in the absorbers is non-uniform. There are simple \Ovi absorbers with only one or a few velocity components as well as highly complex \Ovi systems that span a large velocity range with a large number of subcomponents. Secondly, the missing velocity alignment between \Ovi\!\!, \Civ\!\!, and \Hi\!\!, seen in a large fraction of the \Ovi absorbers, and the temperature differences of {\it aligned} \Ovi\!\!/\Hi and \Ovi\!\!/\Civ pairs, found in the synthetic spectra, both indicate multi-phase gas that contains hotter and cooler gas regions that are spatially separated from each other, and that are traced by different ions. Other absorbers, however, do not show such a velocity signature, indicating a different spatial alignment of the different gas phases. Thirdly, the small Doppler parameters of some of the \Ovi 
absorbers indicate that they have temperatures less that $10^5$ K and are photoionized, while collisional ionization appears to dominate in other \Ovi systems. Therefore, the physical conditions, that govern the ionization state of these systems, appear to vary substantially among the \Ovi absorber population, indicating that they arise in a wide variety of environments. Fourthly, the results that we obtain for the gas densities and absorption path lengths of the \Ovi absorbers demonstrate that measurable \Ovi absorption occurs both in large-scale structures of several hundred kpc and in small-scale structures with sizes on the order of a few parsec. 

The physical pictures emerging from this facts is that the observed \Ovi absorption traces a transition gas phase of temperature $T\sim 10^5$ K which is characteristic for many small-and large-scale gaseous environments in the Universe. As discussed by \citet{Fox11}, the most plausible scenario is that many \Ovi absorbers arise in conductive, turbulent, or shocked boundary layers between warm ($\sim10^4$ K) and hot ($\sim10^6$ K) gas, supporting the unified cooling-flow model proposed by \citep{Heckman02}. Thus the \Ovi absorption in such transition regions would be a natural result of the multi-phase nature of the circumgalactic and intergalactic gas. Including the photoionized systems, \Ovi absorbers possibly trace galactic winds and outflows, transition layers between infalling intergalactic gas and hot coronal halo gas, but also photoionized metal enriched gas patches that had been injected previously into the intergalactic filaments, and gas  with transition-temperature has been stripped by galaxy 
interactions.  


The summarizing conclusion is that \Ovi reflects, like no other ion, the extreme physical and spatial complexity of gas in the circulgalactic and intergalactic environment of galaxies at low and high redshift. The interpretation of \Ovi absorption in high-redshift QSO spectra thus requires a careful case-by-case analysis including other intermediate and high ions and a detailed comparison of the velocity-component structure in the different ions, using both, cosmological simulations and observations.

\item[2)] {\it High-redshift \Ovi absorbers neither trace the bulk of the baryons, nor do they trace the bulk of the metals at that epoch.}

The analyzed UVES and OWLS samples both clearly indicate that the highly-ionized gas as traced by intervening \Ovi absorbers at high redshift cannot host a significant fraction of the baryons and metals at $z=2-3$. Instead, our results imply that the baryon- and metal-content of these absorbers is expected to be less than 1 percent of the total mass-density of baryons and metals at that epoch. Therefore, our results contradict to some previous \Ovi studies wherein the \Ovi ion is identified as an important tracer of the cosmic baryons and metals in the early Universe \citep[e.g.,][]{Bergeron05}. On the other hand, our results are in excellent agreement with more recent results of other groups, e.g., \citet{Muzahid11}, that find similar (low) \Ovi mass/metal densities from their analysis of high-redshift high-ion absorbers.

\item[3)] {\it The physical conditions in \Ovi absorbers change over the cosmic time scale.}

The comparison between \Ovi absorbers at low and high redshift from our study and others \citep{Tepper-Garcia11, Muzahid11, Tripp08} unveils interesting aspects on the time-evolution of the \Ovi absorbing gas in the Universe. Clearly, the physical properties of \Ovi absorbers at high redshift are different from the low redshift \Ovi absorber population. The OWLS imply that the gas density in the \Ovi absorbers at high redshift is (on average) by one order of magnitude higher than at low redshift, while the smoothed metallicity is twice less.
Non-thermal line-broadening mechanisms appears to be irrelevant for high redshift \Ovi systems, while they are important for \Ovi absorbers at low redshift, possibly indicating  higher turbulence in the absorbing region, caused, for instance, by large-scale accretion shocks. Again, this indicates that \Ovi does not trace characteristic {\it regions} in the circumgalactic 
and intergalactic medium in the Universe, but rather a characteristic {\it gas phase} which results from the ambient physical conditions that govern the ionization structure in the absorbers. Thus, the observed frequency of \Ovi absorption in QSO spectra at different epochs is conditioned by the absorption-cross section of the widespread gas with  transition temperature in the course of the on-going structure evolution of the Universe from high to low redshift.

\item[4)] {\it The kinematic and physical properties of high-redshift \Ovi absorbers suggest an inhomogeneous metal enrichment of the IGM.}

As mentioned above, the kinematic displacement between \Ovi\!\!, \Civ\!\!, \Hi and other ions observed in the optical and the synthetic spectra, considered together with the temperature constraints for the different ions obtained from simulations, indicate that the hosts of \Ovi absorbers at high redshift represent multi-phase gas absorbers with substantial substructure. Moreover, the OWLS further indicate that the metals are not uniformly distributed along the absorber host, but are concentrated in confined regions (i.e., \Ovi absorbing regions) that exhibit a substantially higher local metallicity than in the surrounding medium. Such a scenario is supported by the observational study of \citet{Schaye07}, who argued that the intergalactic metals were transported into the IGM through galactic winds in the form of highly-enriched gas patches (metal ``bullets'') that do not (fully) mix with the ambient hydrogen gas. A similar conclusion was drawn by \citet{Tepper-Garcia11}, who studied \Ovi absorbers at {\it 
low} redshift using OWLS. Therefore, we point out that derived metallicities in high redshift intergalactic and circumgalactic medium from observational \Ovi\!\!/\Hi ratios, without knowledge of the 3D structure of the absorbing gas phases, are afflicted with large systematic uncertainties.


\end{itemize}

Our summarizing conclusion is that the \Ovi ion is indeed an important species that provides substantial information on the physical conditions in multi-phase gas around galaxies, but our understanding of the complex interplay between the different gas phases will remain incomplete if derived from \Ovi data {\it alone}. Therefore, future studies of {\it other} tracers of hot gas, such as the broad-line absorbers \citep[e.g.,][]{Richter06} and Ne\,{\sc viii} \citep[e.g.,][]{Savage05}, together with high-resolution cosmological simulations and simulations of galaxy formation, will substantially improve our understanding of the distribution and physical conditions of highly-ionized gas in the circumgalactic and intergalactic medium and its role in the ongoing structure formation in the Universe.

\begin{appendix}\label{appendix1}

\newpage

\renewcommand{\chaptermark}[1]{\markboth{Appendix \thechapter.}{}}		  

\addcontentsline{toc}{chapter}{Appendix}
\chapter{Details of OVI systems from the UVES sample} \label{appendix_systems}


\section{Basic information about the sample}\label{basic_info}

%

\begin{longtable}{lcccccr}
\caption[The selected sample of \Ovi absorbers from the UVES survey]{The selected sample of \Ovi absorbers from the UVES survey. Columns 1-3 give the designations of the associated QSO, its emmision redshift and proximity redshifts, correspondingly. Column 4 gives the absorption lenght of each QSO.
Columns 5 and 6 contain information about the redshift of the \Ovi absorbers in each QSO sightline and specify their number of components.}
\label{observed_quasars}      \\
\hline \\ [-3.ex]
\hline 
\multicolumn{4}{c}{QSO}  &\multicolumn{3}{c}{\Ovi absorption systems} \\
\endfirsthead

\multicolumn{5}{c}%
{{\tablename\ \thetable{c} -- continued from previous page}} \\
\hline 
\endhead

\hline \multicolumn{7}{c}{{Continued on next page}} \\ \hline
\endfoot

\hline \\ [-3.ex] \hline
\endlastfoot

Name & $z_\text{em}$& $z_\text{prox}$ & $X$ & $z_\text{abs}$&category &comp. \\ 
\hline     \\
HE 1122-1648  &  2.404  &  2.387  &  1.60312 & 2.4193$^{a}$ &  2 &3  \\ 
HE 1347-2457  &  2.609  &  2.592  &  1.76352 & 2.1162       &  0 &2 \\
        &         &         &        &  2.2347              &  2 &3\\
        &         &         &        & 2.3287               &  1 &4 \\ 
	&	  &         &	     & 2.5745               &  2 &1 \\       
HE 2217-2818  &  2.413  &  2.396  & 1.60999&  2.0747        &  0 &1 \\
	&         &         &       &  2.0755               &  2 &1\\
        &         &         &       &   2.1806              &  2 &2 \\ 
	&         &         &       &  2.1818               &  2 &3 \\ 
HE 2347-4342  &  2.874  &  2.857  &1.97706 &  2.6498        &  1 &3 \\ 
        &         &         &       &  2.7105               &  0 &1\\
        &         &         &       &    2.7119             &  0 &2 \\ 
	&         &         &       &    2.8625$^{a}$       &  2 &3\\
        &         &         &       &    2.8911$^{a}$       &  2 &8 \\
        &         &         &       &    2.8975$^{a}$       &  2 &15 \\ 
J 2233-606   &  2.250  &  2.233  & 1.48524  &   2.1982      &  1 &4 \\
        &         &         &       &    2.2045             &  1 &6\\
        &         &         &       &    2.2099             &  2 &4  \\ 
PKS 0237-233 &  2.223  &  2.206  & 1.46482 &   2.2028       &  1 &2 \\
        &         &         &       &     2.2135$^{a}$      &  1 &3\\
        &         &         &       &    2.2364$^{a}$       &  1 &5 \\
        &         &         &       &    2.2378$^{a}$       &  2 &3 \\
PKS 0329-255 &  2.704  &  2.687  & 1.83925 &   2.4252       &  1 &3 \\
        &         &         &       &     2.6494            &  2 &2 \\
        &         &         &       &    2.6610             &  2 &2 \\
        &         &         &       &    2.7089             &  1 &12  \\ 
PKS 2126-158 &  3.279  &  3.262  & 2.3164  &2.9074          &  0 &1 \\ 
Q 0002-422   &  2.767  &  2.750  & 1.88999  &2.4640          &  1 &10 \\
        &         &         &       &    2.5395             &  0 &1\\
        &         &         &       &    2.7011             &  0 &1  \\ 
Q 0055-269   &  3.655  &  3.638  &2.64486 &3.6015            &  0 &2 \\ 
Q 0109-3518   &  2.405  &  2.388  &  1.60378 &2.4012$^{a}$   &  1 &1 \\ 
Q 0122-380   &  2.193  &  2.176  & 1.4422&2.0626             &  2 &1 \\ 
Q 0329-385   &  2.434  &  2.417  & 1.62627&2.3521            &  1 &1 \\
        &         &         &      &     2.3639             &  2 &3 \\
        &         &         &    &2.3737                    &  2 &2 \\ 
Q 0420-388   &  3.116  &  3.099  &  2.17793 &3.0872          &  1 &7  \\ 
Q 0453-423   &  2.658  &  2.641  & 1.80252&2.3978            &  0 &8 \\
        &         &         &     &2.6362                   &  1 &4  \\
        &         &         &     &2.6405                   &  2 &2  
\end{longtable}
\small$^{\rm a}$ Associated \Ovi systems. The rest are intervening \Ovi systems.\\

\newpage
\section{OVI/HI column density ratio}

\begin{longtable}{lrr}
\caption[\Ovi\!\!/\Hi column density ratio of the absorbers from the UVES sample.]{\Ovi\!\!/\Hi column density ratio of the absorbers from the UVES sample.} \\
\hline \\ [-3.ex]
\hline 
\multicolumn{3}{c}{Absorption systems}  \\ 
\endfirsthead

\multicolumn{3}{c}%
{{\tablename\ \thetable{c} -- continued from previous page}} \\ 
\hline 
\endhead

\hline \multicolumn{3}{c}{{Continued on next page}} \\ \hline  
\endfoot

\hline \\ [-3.ex] \hline
\endlastfoot

QSO name& $z_\text{abs}$(comp.)& N(\Ovi\!\!)/N(\Hi\!\!) \\ 
\hline    \\
HE 1122-1648  & 2.4193(3) & 2.48 ($+0.125895\atop -0.119031$)\\
HE 1347-2457  & 2.1162(2) & 1.36 ($+2.115240\atop -0.629865$)\\             
              & 2.2347(3) & 0.30 ($+0.038379\atop -0.028066$)\\
              & 2.3287(2) & 0.01 ($+0.006326\atop -0.002888$)\\
              & 2.5745(1) & 0.04 ($+0.002690\atop -0.002534$)\\ 
HE 2217-2818  & 2.0747(1) & 0.17 ($+0.026903\atop -0.027346$)\\
              & 2.0755(1) & 1.32 ($0.232336\atop -0.239987$)\\
              & 2.1818(5) & 0.01 ($+0.003321\atop -0.002649$)\\
HE 2347-4342  & 2.6498(3) & 0.05 ($+0.004540\atop -0.004138$)\\
              & 2.7105(1)/2.7119(2) & 0.09 ($+0.024777\atop -0.020612$)\\           
              & 2.8625(3) & 0.20 ($+0.032029\atop -0.023465$)          \\
              & 2.8911(8) & 16.29 ($+2.861322\atop -4.189452$)         \\
              & 2.8975(15)&  0.23 ($+0.039362\atop -0.030874$)         \\ 
J 2233-606    & 2.1982(4) & 42.52 ($+9.125267\atop -7.319435$)         \\
              & 2.2045(7) & 9.45 ($+13.220907\atop -3.972540$)         \\
              & 2.2099(4) & 69.29 ($+65.221252\atop -30.303299$)\\ 
PKS 0237-233  & 2.2028(2) & 0.004 ($+0.021299\atop -0.002665$)\\
              & 2.2135(3) & 0.25 ($+0.166593\atop -0.052706$) \\
              & 2.2364(8) & 1.16 ($+0.473380\atop -0.417801$) \\ 
PKS 0329-255  & 2.4252(3) & 0.02 ($+0.004236\atop -0.003700$) \\
              & 2.6494(2) & 0.04 ($+0.026871\atop -0.015146$) \\
              & 2.6610(2) & 0.13 ($+0.015739\atop -0.013707$) \\
              & 2.7089(11)& 1.48 ($+2.164616\atop -0.946345$) \\ 
PKS 2126-158  & 2.9074(1) & 0.002 ($+0.000177\atop -0.000160$)\\ 
Q 0002-422     & 2.4640(10)& 0.007 ($+0.000952\atop -0.000719$)\\ 
              & 2.5395(1) & 0.14 ($+0.083378\atop  -0.072789$)\\  
              & 2.7011(1) & 0.071 ($+0.008505\atop -0.007606$)\\ 
Q 0055-269     & 3.6015(2) & 0.02 ($+0.001966\atop  -0.001821$)\\  
Q 0109-3518    & 2.4012(1) & 0.11 ($+0.013681\atop  -0.012222$)\\  
Q 0122-380     & 2.0626(1) & 3.96 ($+1.747881\atop  -1.255565$)\\  
Q 0329-385     & 2.3521(1) & 5.94 ($+1.304771\atop  -1.143078$)\\  
              & 2.3639(3) & 0.11 ($+0.095031\atop  -0.053387$)\\  
              & 2.3737(2) & 0.086 ($+0.015742\atop -0.013308$)\\ 
Q 0420-388     & 3.0872(7)& 0.00002 ($+0.000001\atop -0.000001$)\\ 
Q 0453-423     & 2.3978(8) & 0.75 ($+0.183549\atop -0.201516$)\\
              & 2.6362(4) & 0.096 ($+0.018013\atop -0.015103$)\\ 
              & 2.6403(2) & 0.04 ($+0.024809\atop -0.039332$)            
\label{column_density_ratio}
\end{longtable}
\section{Error estimates} \label{appendix_errors}

\[N_{tot}(O\,\text{\sc vi}) = \displaystyle{\sum_{c} N_{c}(O\,\text{\sc vi})}\] 

\[\sigma^{+}_{tot}(N(O\,\text{\sc vi})) =\sqrt {\displaystyle{\sum_{c} \sigma^{+}_{c}(N(O\,\text{\sc vi}))^{2}}}\]

\[\sigma^{-}_{tot}(N(O\,\text{\sc vi})) = \sqrt {\displaystyle{\sum_{c} \sigma^{-}_{c}(N(O\,\text{\sc vi}))^{2}}}\]

$N_{c}(O\,\text{\sc vi})$ and $\sigma_{c}(N(O\,\text{\sc vi}))$ are the $\log$ of the column densities
and their errors for each component for a given system, derived form the fit. Similar is for \Hi. 

\scriptsize\[ \sigma^{+}_{c}(N(O\,\text{\sc vi})) = (10^{\sigma_{c}(N(O\,\text{\sc vi}))} -1)\times 10^{N_{c}(O\,\text{\sc vi})}\]
\[\sigma^{-}_{c}(N(O\,\text{\sc vi})) = (1- 10^{-\sigma_{c}(N(O\,\text{\sc vi}))})\times  10^{N_{c}(O\,\text{\sc vi})}\]


\section{Fitting results}\label{fitting_results}



\begin{center}
\begin{longtable}{lccccrr}
\caption[Voigt Profile parameters of the absorption lines (HE 1122-1648).]{Voigt Profile parameters of the absorption lines (HE 1122-1648).} \\
\hline \\ [-3.ex]
\hline 
\multicolumn{7}{c}{$z = 2.419294$}  \\ 
\endfirsthead

\multicolumn{7}{c}%
{{\tablename\ \thetable{} -- continued from previous page}} \\
\hline 
\endhead

\hline \multicolumn{7}{c}{{Continued on next page}} \\ \hline
\endfoot

\hline \\ [-3.ex] \hline
\endlastfoot

\small $\rm Ion$ & \small $z$ &\small $\rm Error$ & log[N (\sqc)] &\small $\rm Error$ & \small $b$ [\kms] &\small $\rm Error$\\    
\hline \\   \vspace{0.15cm}
\small O VI &\small 2.419290 &\small 0.000001 &\small 13.992 &\small 0.010 &\small  9.49 &\small 0.18\\ \vspace{0.15cm}
\small O VI &\small 2.419631 &\small 0.000030 &\small 13.499 &\small 0.030 &\small 27.84 &\small 2.22\\ \vspace{0.15cm}
\small O VI &\small 2.419846 &\small 0.000006 &\small 12.837 &\small 0.087 &\small  7.58 &\small 1.30\\ \vspace{0.15cm}
\small C IV &\small 2.419229 &\small 0.000007 &\small 11.537 &\small 0.054 &\small  4.96 &\small 1.24\\ \vspace{0.15cm}
\small N V  &\small 2.419232 &\small 0.000010 &\small 12.127 &\small 0.027 &\small 17.10 &\small 1.32\\ \vspace{0.15cm}
\small H I  &\small 2.419248 &\small 0.000004 &\small 13.650 &\small 0.009 &\small 28.60 &\small 0.19\\ \vspace{0.15cm}
\small H I  &\small 2.419724 &\small 0.000020 &\small 13.013 &\small 0.040 &\small 31.50 &\small 1.66\\ 
\end{longtable} 
\end{center}    


\begin{center}
\begin{longtable}{lccccrr}
\caption[Voigt Profile parameters of the absorption lines (HE 1347-2457).]{Voigt Profile parameters of the absorption lines (HE 1347-2457).} \\
\hline \\ [-3.ex]
\hline  
\multicolumn{7}{c}{$z = 2.574499$}  \\ 
\endfirsthead

\multicolumn{7}{c}%
{{\tablename\ \thetable{} -- continued from previous page}} \\
\hline 
\endhead 

\hline
\multicolumn{7}{c}{{Continued on next page}} \\ \hline
\endfoot

\hline \\ [-3.ex] \hline
\endlastfoot

\small $\rm Ion$ & \small $z$ &\small $\rm Error$ & log [N (\sqc)] &\small $\rm Error$ & \small $b$  [\kms] &\small $\rm Error$ \\   
 \hline \\ \vspace{0.15cm}
\small O VI  &\small 2.574503  &\small 0.000006&\small 13.540&\small 0.011 &\small 26.53&\small 0.74 \\ \vspace{0.15cm}
\small C IV &\small 2.574516  &\small 0.000028&\small 13.122&\small 0.172 &\small 12.50&\small  1.29 \\ \vspace{0.15cm}
\small C IV &\small 2.574586  &\small 0.000006&\small 12.758&\small 0.162 &\small  3.26&\small  1.51 \\ \vspace{0.15cm}
\small C IV &\small 2.574808  &\small 0.000279&\small 12.249&\small 0.695 &\small 18.84&\small 21.96 \\ \vspace{0.15cm}
\small C IV &\small 2.574408  &\small 0.000005&\small 13.033&\small 0.089 &\small  3.91&\small  0.72 \\ \vspace{0.15cm}
\small C III &\small 2.574512 &\small 0.000004 &\small 14.504&\small 0.163&\small  7.72 &\small 0.48 \\
\end{longtable}                                                                                          
\end{center}                                                                                             
                                                                                                          
                                                                                                          
\begin{center}                                                                                              
\begin{longtable}{lccccrr}                                                                                 
\caption[Voigt Profile parameters of the absorption lines (HE 2217-2818)]{Voigt Profile parameters of the absorption lines (HE 2217-2818)} \\
\hline \\ [-3.ex] \hline 
\endfirsthead

\multicolumn{7}{c}%
{{\tablename\ \thetable{} -- continued from previous page}} \\
\hline 
\endhead

\hline \multicolumn{7}{c}{{Continued on next page}} \\ \hline
\endfoot

\hline \\ [-3.ex] \hline
\endlastfoot

\hline
\multicolumn{7}{c}{$z = 2.074720, ~~~~2.075462$} \\
\hline        \\    \vspace{0.15cm}
\small O VI &\small 2.075464&\small 0.000004 &\small 14.253 &\small 0.029&\small 12.15&\small 0.55 \\   \vspace{0.15cm}
\small O VI &\small 2.074719&\small 0.000010 &\small 14.194 &\small 0.042&\small 15.02&\small 1.40 \\   \vspace{0.15cm}
\small C IV &\small 2.075352&\small 0.000049 &\small 12.134 &\small 0.323&\small 20.40&\small 6.62 \\    \vspace{0.15cm}
\small C IV &\small 2.074687&\small 0.000011 &\small 12.674 &\small 0.054&\small 10.18&\small 0.99 \\    \vspace{0.15cm}
\small C IV &\small 2.074835&\small 0.000010 &\small 12.408 &\small 0.099&\small  7.47&\small 1.11 \\    \vspace{0.15cm}
\small C IV &\small 2.075398&\small 0.000003 &\small 12.737 &\small 0.082&\small 10.09&\small 0.96 \\
\end{longtable}
\end{center} 


\begin{center}
\begin{longtable}{lccccrr}
\caption[Voigt Profile parameters of the absorption lines (HE 2347-4342).]{Voigt Profile parameters of the absorption lines (HE 2347-4342).} \\
\hline \\ [-3.ex]
\hline 
\multicolumn{7}{c}{$z = 2.649846$}  \\ 
\endfirsthead

\multicolumn{7}{c}%
{{\tablename\ \thetable{} -- continued from previous page}} \\
\hline 
\endhead

\hline \multicolumn{7}{c}{{Continued on next page}} \\ \hline
\endfoot

\hline \\ [-3.ex]  \hline
\endlastfoot

\small $\rm Ion$ & \small $z$ &\small $\rm Error$ & log [N (\sqc)] &\small $\rm Error$ & \small $b$  [\kms] &\small $\rm Error$  \\  
\hline      \\ \vspace{0.15cm}
\small  O VI &\small 2.649554  &\small 0.000003 &\small 13.538 &\small 0.032  &\small 11.40  &\small 0.57\\    \vspace{0.15cm}
\small  O VI &\small 2.649849  &\small 0.000002 &\small 13.486 &\small 0.011  &\small  6.34  &\small 0.27\\    \vspace{0.15cm}
\small  O VI &\small 2.649202  &\small 0.000006 &\small 13.172 &\small 0.082  &\small 10.18  &\small 1.16 \\    \vspace{0.15cm}
\small  C IV &\small 2.649072  &\small 0.000014 &\small 12.414 &\small 0.036  &\small 17.29  &\small 1.87 \\    \vspace{0.15cm}
\small  C IV &\small 2.649486  &\small 0.000010 &\small 12.049 &\small 0.058  &\small  6.46  &\small 1.42 \\    \vspace{0.15cm}
\small  C IV &\small 2.649800  &\small 0.000006 &\small 12.223 &\small 0.033  &\small  6.15  &\small 0.83\\
\end{longtable}                                                                                            
\end{center} 

\begin{center}
\begin{longtable}{lccccrr}
\caption[Voigt Profile parameters of the absorption lines (J 2233-606)]{Voigt Profile parameters of the absorption lines (J 2233-606).} \\
\hline \\ [-3.ex]  \hline 
\multicolumn{7}{c}{$z = 2.198231$}  \\ 
\endfirsthead

\multicolumn{7}{c}%
{{\tablename\ \thetable{} -- continued from previous page}} \\
\hline 
\endhead

\hline \multicolumn{7}{c}{{Continued on next page}} \\ \hline
\endfoot

\hline \\ [-3.ex] \hline
\endlastfoot

\small $\rm Ion$ & \small $z$ &\small $\rm Error$ & log [N (\sqc)] &\small $\rm Error$ & \small $b$  [\kms] &\small $\rm Error$  \\  
\hline  \\  \vspace{0.15cm}
\small O VI&\small 2.198320  &\small 0.000022&\small 15.258&\small 0.088&\small 27.40 &\small  1.78  \\   \vspace{0.15cm}
\small O VI&\small 2.197800  &\small 0.000023&\small 14.314&\small 0.140&\small 11.79 &\small  4.08 \\    \vspace{0.15cm}
\small O VI&\small 2.197558  &\small 0.000055&\small 13.223&\small 0.701&\small  7.15 &\small  8.93 \\     \vspace{0.15cm}
\small O VI&\small 2.197311  &\small 0.000036&\small 13.919&\small 0.108&\small 18.66 &\small  4.54 \\     \vspace{0.15cm}
\small N V &\small 2.198264  &\small 0.000041&\small 14.068&\small 0.074&\small  29.94&\small  2.43 \\     \vspace{0.15cm}
\small N V &\small 2.197900  &\small 0.000029&\small 13.647&\small 0.173&\small  21.41&\small  2.65 \\
\end{longtable}
\end{center} 

\begin{center}
\begin{longtable}{lccccrr}
\caption[Voigt Profile parameters of the absorption lines (PKS 0237-233)]{Voigt Profile parameters of the absorption lines (PKS 0237-233)} \\
\hline  \\ [-3.ex] \hline 
\multicolumn{7}{c}{$z = 2.202783$}  \\ 
\endfirsthead

\multicolumn{7}{c}%
{{\tablename\ \thetable{} -- continued from previous page}} \\
\hline 
\endhead

\hline \multicolumn{7}{c}{{Continued on next page}} \\ \hline
\endfoot

\hline \\ [-3.ex] \hline
\endlastfoot

\small $\rm Ion$ & \small $z$ &\small $\rm Error$ & log [N (\sqc)] &\small $\rm Error$ & \small $b$  [\kms] &\small $\rm Error$  \\                                                                                  
\hline     \\  \vspace{0.15cm}                                                                           
\small O VI &\small 2.202783  &\small 0.000000&\small 13.262&\small 0.564&\small 12.31  &\small  6.77 \\    \vspace{0.15cm}
\small O VI &\small 2.202523  &\small 0.000000&\small 12.361&\small 1.557&\small  6.70  &\small 22.27 \\   \vspace{0.15cm}
\small N V  &\small 2.202772    &\small 0.000029&\small 11.908&\small 0.162&\small  7.88  &\small  4.59 \\   \vspace{0.15cm}
\small C II &\small 2.202816    &\small 0.000025&\small 12.665&\small 0.086&\small 14.24  &\small  3.83  \\   \vspace{0.15cm}
\small C II &\small 2.203126    &\small 0.000028&\small 12.056&\small 0.234&\small  4.44  &\small  5.33 \\    \vspace{0.15cm}
\small C III&\small 2.203121    &\small 0.000026&\small 13.847&\small 0.148&\small 11.08  &\small  1.67 \\
\end{longtable}
\end{center} 

\begin{center}
\begin{longtable}{lccccrr}
\caption[Voigt Profile parameters of the absorption lines (PKS 0329-255)]{Voigt Profile parameters of the absorption lines (PKS 0329-255)} \\
\hline \\ [-3.ex]  \hline 
\multicolumn{7}{c}{$z = 2.425241$}  \\ 
\endfirsthead

\multicolumn{7}{c}%
{{\tablename\ \thetable{} -- continued from previous page}} \\
\hline 
\endhead

\hline \multicolumn{7}{c}{{Continued on next page}} \\ \hline
\endfoot

\hline \\ [-3.ex]  \hline
\endlastfoot

\small $\rm Ion$ & \small $z$ &\small $\rm Error$ & log [N (\sqc)] &\small $\rm Error$ & \small $b$  [\kms] &\small $\rm Error$   \\                          
\hline   \\   \vspace{0.15cm}                                                                                                    
\small O VI  &\small 2.425241   &\small 0.000018 &\small 13.106&\small 0.178&\small 11.06  &\small   2.97 \\   \vspace{0.15cm}   
\small O VI  &\small 2.425625   &\small 0.000024 &\small 13.792&\small 0.049&\small 28.11  &\small   3.60 \\   \vspace{0.15cm}   
\small O VI  &\small 2.424821   &\small 0.000046 &\small 12.535&\small 0.205&\small 10.69  &\small   7.61 \\    \vspace{0.15cm}   
\small C IV  &\small 2.425259   &\small 0.000008 &\small 12.945&\small 0.027&\small 14.07  &\small   0.85  \\    \vspace{0.15cm}  
\small C IV  &\small 2.425548   &\small 0.000006 &\small 12.945&\small 0.027&\small 11.50  &\small   0.67  \\    \vspace{0.15cm}  
\small H I   &\small 2.424604   &\small 0.001015 &\small 13.411&\small 0.824&\small 72.54  &\small 103.29 \\
\end{longtable}                                                                                                      
\end{center}                                                                                                         
                                                                                                                     
\begin{center}                                                                                                       
\begin{longtable}{lccccrr}                                                                                          
\caption[Voigt Profile parameters of the absorption lines (PKS 2126-158)]{Voigt Profile parameters of the absorption lines (PKS 2126-158)} \\                                                                               
\hline \\ [-3.ex]  \hline                                                                                                               
\multicolumn{7}{c}{$z = 2.907378$}  \\                                                                       
\endfirsthead                                                                                                        
                                                                                                                     
\multicolumn{7}{c}%
{{\tablename\ \thetable{} -- continued from previous page}} \\                                                       
\hline 
\endhead

\hline \multicolumn{7}{c}{{Continued on next page}} \\ \hline
\endfoot

\hline \\ [-3.ex] \hline
\endlastfoot

\small $\rm Ion$ & \small $z$ &\small $\rm Error$ & log [N (\sqc)] &\small $\rm Error$ & \small $b$  [\kms] &\small $\rm Error$   \\ 
\hline  \\   \vspace{0.15cm}                        
\small O VI    &\small  2.907377    &\small 0.000007&\small  13.586&\small  0.033&\small  16.16 &\small  0.96 \\         \vspace{0.15cm}
\small C II    &\small  2.906836    &\small 0.000032 &\small 12.442 &\small 0.130 &\small 14.86  &\small 3.88 \\         \vspace{0.15cm}
\small C II    &\small  2.907119    &\small 0.000021 &\small 12.347 &\small 0.137 &\small 10.60  &\small 2.20 \\         \vspace{0.15cm}
\small C II    &\small  2.907362    &\small 0.000009 &\small 11.576 &\small 0.114 &\small  2.09  &\small 2.07 \\         \vspace{0.15cm}
\small C II    &\small  2.907619    &\small 0.000012 &\small 11.880 &\small 0.048 &\small  9.21  &\small 1.76  \\        \vspace{0.15cm}
\small C IV    &\small  2.906183    &\small 0.000015 &\small 12.120 &\small 0.053 &\small 12.86  &\small 1.90  \\
\end{longtable}                                                                                                           
\end{center}                                                                                                              
                                                                                                                          
\begin{center}                                                                                                            
\begin{longtable}{lccccrr}                                                                                               
\caption[Voigt Profile parameters of the absorption lines (Q 0002-422)]{Voigt Profile parameters of the absorption lines (Q 0002-422)} \\                                                                                    
\hline \\ [-3.ex] \hline                                                                                                                    
\multicolumn{7}{c}{$z = 2.464007$}  \\                                                                              
\endfirsthead                                                                                                             
                                                                                                                          
\multicolumn{7}{c}%
{{\tablename\ \thetable{} -- continued from previous page}} \\
\hline 
\endhead

\hline \multicolumn{7}{c}{{Continued on next page}} \\ \hline
\endfoot

\hline \\ [-3.ex] \hline
\endlastfoot

\small $\rm Ion$ & \small $z$ &\small $\rm Error$ & log [N (\sqc)] &\small $\rm Error$ & \small $b$  [\kms] &\small $\rm Error$   \\                       
\hline      \\   \vspace{0.15cm}                
\small O VI &\small 2.462091  &\small 0.000016 &\small13.831 &\small 0.067 &\small 16.98 &\small 1.88  \\               \vspace{0.15cm}
\small O VI &\small 2.462868  &\small 0.000016 &\small14.204 &\small 0.020 &\small 40.24 &\small 4.37  \\               \vspace{0.15cm}
\small O VI &\small 2.464191  &\small 0.000069 &\small14.272 &\small 0.078 &\small 49.62 &\small 4.58  \\               \vspace{0.15cm}
\small O VI &\small 2.464387  &\small 0.000017 &\small13.159 &\small 0.205 &\small 10.54 &\small 3.00  \\               \vspace{0.15cm}
\small O VI &\small 2.466035  &\small 0.000028 &\small13.572 &\small 0.158 &\small 20.05 &\small 3.16  \\               \vspace{0.15cm}
\small O VI &\small 2.461326  &\small 0.000022 &\small13.168 &\small 0.053 &\small 20.47 &\small 3.00 \\
\end{longtable}%
                                                                                                   
\end{center}                                                                                                      

\newpage
\section{Velocity spread}\label{velocity_spread}
\begin{longtable}{lrrrrrrrl}
\caption[Velocity spread for \Ovi, \Civ and \Hi\!\! absorbers.]{Velocity spread for \Ovi, \Civ and \Hi\!\! absorbers.} \\
\hline \\ [-3.ex]
\hline 
\multicolumn{8}{c}{Absorption systems}  \\ 
\endfirsthead

\multicolumn{8}{c}%
{{\tablename\ \thetable{} -- continued from previous page}} \\
\hline 
\endhead

\hline \multicolumn{3}{r}{{Continued on next page}} \\ \hline
\endfoot

\hline \\ [-3.ex] \hline
\endlastfoot

\scriptsize QSO name &\scriptsize $z_\text{abs}$(comp.)&\scriptsize $v_{\rm min,(O\,\text{\sc vi})}$ &\scriptsize $\delta v_{(O\,\text{\sc vi})}$ &\scriptsize $v_{\rm min,(C\,\text{\sc iv})}$ &\scriptsize $v_{\rm min,(N\,\text{\sc v})}$ &\scriptsize $v_{\rm min,(H\,\text{\sc i})}$ &\scriptsize$\delta v_{(H\,\text{\sc i})}$ \\
                     &                                 &\scriptsize $v_{\rm max,(O\,\text{\sc vi})}$  &                                       &\scriptsize $v_{\rm max,(C\,\text{\sc iv})}$  &\scriptsize $v_{\rm max,(N\,\text{\sc v})}$  &\scriptsize $v_{\rm max,(H\,\text{\sc i})}$  &                                     \\                         

~& ~& \tiny [\kms]&\tiny [\kms]&\tiny [\kms]&\tiny [\kms]&\tiny [\kms]&\tiny [\kms] \\
\hline \\
\scriptsize HE 1122-1648  &\scriptsize 2.4193(3) &\scriptsize $-$20  $+$70&\scriptsize 90 &\scriptsize   ---    &\scriptsize ---     &\scriptsize $-$70 $+$80  &\scriptsize 150\\ 
                          &                      &                     &               &                      &                     &\scriptsize$-$ 70 +180  &\scriptsize 250\\
\scriptsize HE 1347-2457  &\scriptsize 2.1162(2) &\scriptsize $-$20  $+$20&\scriptsize 40 &\scriptsize $-$60  $+$70  &\scriptsize $-$20  $+$20 &\scriptsize  $-$60 $+$70 &\scriptsize 130\\
                         &                      &                     &               &                      &                     &\scriptsize  $-$180 +220 &\scriptsize 400\\
\end{longtable}                                                                                                                                             


\chapter{Ionization modeling with Cloudy}\label{ionization_modeling_uves}

\section{Intervening \Ovi systems}
\subsection*{System at $z=$2.574499 towards HE 1347-2457} 

This is a single-component \Ovi system, which contains \Civ\!\!, \Ciii\!\!, \Cii\!\!, \Siiv\!\!, \Siiii\!\!, \Siii\!\!. The \Ovi line is weak and broad, with a column density similar to that of \Civ\!\!. \Ciii shows a strong and saturated two component feature. \Hi is also saturated. The column densities of \Ciii and \Hi are similar. In our first model we assumed that the \Ovi\!\!, \Civ and \Ciii components, which are aligned and arise in single gas phase. (The alignment between all of the components as good as $<1.0$\kms\!\!.) The observed and modeled column densities of the three ions did not match. Neglecting the value obtained for \Ovi\!\!, no match is achieved between \Ciii and \Civ either. 

We summed up the two \Ciii\!\!, the two \Civ and two of the \Hi components and obtained new model values, with the same result. Further, we assumed that the aligned \Ovi and \Civ components arise in a single gas phase, using the \Civ\!\!/\Ovi ratio for modeling. Now the modeled column densities matched well with the observed one. The results are given in Table~\ref{z2.574499he1347}. Here and in all tables, $n_{\rm H}$ is the hydrogen space density, $Z$ denotes the metallicity relative to solar, $L$ is the absorption pathlength, and $f_{\rm HI}$ the ionization fraction of \Hi\!\!.

\begin{table*}[th!]
 \caption{Ionization model for a \Civ\!\!/\Ovi absorber at $z = 2.574499$}
 \begin{small}
 \begin{tabular}{cccccccccc}
 \hline
 \hline
\scriptsize  $v$\,[\kms\!\!] &\scriptsize log [N(O\,{\sc vi)}\,(\sqc)] &\scriptsize log [N(C\,{\sc iv)}\,(\sqc)]&\scriptsize 
log [$n_{\rm H}\,$(\cc)] &\scriptsize log $Z$  &\scriptsize log [$T$(K)] &\scriptsize $L$ [kpc] &\scriptsize log $f_{\rm HI}$ \\
 \hline
\scriptsize 0.0 &\scriptsize 13.541 &\scriptsize13.122  &\scriptsize $-$3.759 &\scriptsize $-$1.925 &\scriptsize 4.628 &\scriptsize 106.6 &\scriptsize $-$4.857 \\
 \hline
 \end{tabular}
 \end{small}
 \label{z2.574499he1347}
 \end{table*}

\subsection*{System at $z=$2.234653 towards HE 1347-2457} 

The \Ovi feature shows three components, blended by Ly series lines (\Lyb up to Ly-$\zeta$) at higher redshift. There are five weak component \Civ lines at that redshift, 
which are blended by \Siiv at higher and Cr\,{\sc ii} at lower $z$. One of the \Civ components is well aligned with the strongest \Ovi component 
($\Delta v_{\rm comp.}$ = 0.9 \kms\!\!). Therefore we assumed in our model that both species live in one gas phase. The corresponding \Hi \Lya feature is strong (log $N$(\Hi\!\!)=14.8) 
and saturated, also blended by Fe\,{\sc ii} and Ca\,{\sc ii} at lower redshift. For our model we use as stopping criterion the column density of the closest 
\Hi component, shifted from \Ovi by 4.8 \kms\!\!. Our model matched well the observed \Civ and \Ovi column densities. The results are given in Table~\ref{z2.234653he1347}.

\begin{table*}[th!]
 \caption[Ionization model for a \Civ\!\!/\Ovi absorber at $z = 2.234653$]{Ionization model for a \Civ\!\!/\Ovi absorber at $z = 2.234653$ towards HE 1347-2457}
 \begin{small}
 \begin{tabular}{cccccccccc}
 \hline
 \hline
\scriptsize  $v$\,[\kms\!\!] &\scriptsize log [N(O\,{\sc vi)}\,(\sqc)] &\scriptsize log [N(C\,{\sc iv)}\,(\sqc)]&\scriptsize 
log [$n_{\rm H}\,$(\cc)] &\scriptsize log $Z$  &\scriptsize log [$T$(K)] &\scriptsize $L$ [kpc] &\scriptsize log $f_{\rm HI}$ \\
 \hline
\scriptsize 0.0 &\scriptsize 13.683 &\scriptsize 12.584  &\scriptsize $-$4.131 &\scriptsize $-$2.166 &\scriptsize 4.677 &\scriptsize 486.0 &\scriptsize $-$5.263 \\
 \hline
 \end{tabular}
 \end{small}
 \label{z2.234653he1347}
 \end{table*}





\subsection*{System at $z=$2.075462 towards HE 2217-2818}

This system is similar to the previous one-component strong \Ovi system. The oxygen line is blended by a Ly-$\delta$ feature at higher redshift and by \Lya at lower redshift. 
There is a broad \Civ line with 2 components. The closest \Civ component is displaced by 6.4 \kms respective to \Ovi\!\!. Therefore, we assumed that \Civ and \Ovi live in one gas phase ($N$(\Civ\!\!)/$N$(\Ovi\!\!) = $-1.516$). The corresponding five component \Lya feature is huge and saturated. One of the \Hi components well match with the oxygen (shifted by 1.0 \kms\!\!) and we used its column density (log $N$(\Hi\!\!)=14.1) as the stopping criterion for our model. The result is given in Table~\ref{z2.075462he2217}.

\begin{table*}[th!]
 \caption[Ionization model for a \Civ\!\!/\Ovi absorber at $z = 2.075462$]{Ionization model for a \Civ\!\!/\Ovi absorber at $z = 2.075462$ towards HE 2217-2818}
 \begin{small}
 \begin{tabular}{cccccccccc}
 \hline
 \hline
\scriptsize  $v$\,[\kms\!\!] &\scriptsize log [N(C\,{\sc iv)}\,(\sqc)]&\scriptsize log [N(O\,{\sc vi)}\,(\sqc)] &\scriptsize 
log [$n_{\rm H}\,$(\cc)] &\scriptsize log $Z$  &\scriptsize log [$T$(K)] &\scriptsize $L$ [kpc] 
&\scriptsize log $f_{\rm HI}$ \\
 \hline
\scriptsize 0.0 &\scriptsize 12.737 &\scriptsize 14.253 &\scriptsize $-$4.435 &\scriptsize $-$1.037 &\scriptsize 4.657 &\scriptsize 375.5 &\scriptsize $-$5.546 \\
 \hline
 \end{tabular}
 \end{small}
 \label{z2.075462he2217}
 \end{table*}

Our second assumption was that the carbon and the oxygen do not live in one gas phase. We modeled only \Ovi\!\!, fixing the temperature according to its $b$-value. 
($b$ = 12.15$\pm$0.55 \kms\!\!, $T$ = 141,936.2 K). The results are given in Table~\ref{z2.075462he2217_A}.

\begin{table*}[th!]
\begin{center}
\caption[Ionization models for an \Ovi absorber at $z = 2.075462$]{Ionization models for an \Ovi absorber at $z = 2.075462$ towards HE 2217-2818}
 \begin{small}
 \begin{tabular}{cccrcrcccc}
 \hline
 \hline
\scriptsize  $v$\,[\kms\!\!] &\scriptsize log [N(O\,{\sc vi)}\,(\sqc)] &\scriptsize 
log [$n_{\rm H}\,$(\cc)] &\scriptsize log $Z$  &\scriptsize log [$T$(K)] &\scriptsize $L$ [kpc] 
&\scriptsize log $f_{\rm HI}$ \\
 \hline
\scriptsize 0.0 &\scriptsize 14.252 &\scriptsize $-$3.105 &\scriptsize 0.000 &\scriptsize 5.152 &\scriptsize 9.6 &\scriptsize $-$5.283 \\
\scriptsize 0.0 &\scriptsize 14.252 &\scriptsize $-$3.890 &\scriptsize $-$1.000 &\scriptsize 5.152 &\scriptsize 123.2 &\scriptsize $-$5.607 \\
 \hline
 \end{tabular}
 \end{small}
 \label{z2.075462he2217_A}
\end{center}
 \end{table*}

\subsection*{System at $z=$2.711919 towards HE 2347-4342}

There are two weak \Ovi components blended by \Lya and \Lyb respectively at lower and at higher redshifts. There is also a very weak three component \Civ absorption, and a huge \Lya feature blended by \Siiii at higher redshift. The strongest \Ovi and \Civ components are well aligned to each other (shifted by $-$0.9 \kms\!\!), therefore we assumed that they live in one gas phase and modeled them together ($N$(\Civ\!\!)/$N$(\Ovi\!\!) = $-1.516$). The closest \Hi component is displaced by $-$4.1 \kms\!\!. We used it in the model for defining the stopping criterion in CLOUDY (log $N$(\Hi\!\!) = 14.194). The results are given in Table~\ref{z2.711919he2347}.

\begin{table*}[th!]
\caption[Ionization model for a \Civ\!\!/\Ovi absorber at $z = 2.711919$]{Ionization model for a \Civ\!\!/\Ovi absorber at $z = 2.711919$ towards HE 2347-4342}
 \begin{small}
 \begin{tabular}{cccccccccc}
 \hline
 \hline
\scriptsize  $v$\,[\kms\!\!] &\scriptsize log [N(C\,{\sc iv)}\,(\sqc)]&\scriptsize log [N(O\,{\sc vi)}\,(\sqc)] &\scriptsize 
log [$n_{\rm H}\,$(\cc)] &\scriptsize log $Z$  &\scriptsize log [$T$(K)] &\scriptsize $L$ [kpc] 
&\scriptsize log $f_{\rm HI}$ \\
 \hline
\scriptsize 0.0 &\scriptsize 12.040 &\scriptsize 13.502 &\scriptsize $-$4.390 &\scriptsize $-$1.889 &\scriptsize 4.667 &\scriptsize 400.3 &\scriptsize $-$5.508 \\
 \hline
 \end{tabular}
 \end{small}
 \label{z2.711919he2347}
 \end{table*}

\subsection*{System at $z=$2.209873 towards J 2233-606}

This is two four-component \Ovi system, blended by one \Lya at lower redshift and by one \Lyb at higher $z$. The column density of the oxygen is about one order of magnitude higher than that of the hydrogen in the system. The first of the two \Hi components is well aligned with the first \Ovi component (shifted by 
$-$2.1 \kms\!\!). The second (the strongest) oxygen component has not corresponding hydrogen feature. The second \Hi component is shifted to the third \Ovi 
by 11.2 \kms\!\!. We modeled the first and the third \Ovi components, using the corresponding $b$-values to calculate the temperature, $T_{\rm first}$ = 458 193 K, 
$T_{\rm second}$ = 234 887 K. The high temperature and the very low \Hi column density is a hint for a collisional ionization. We modeled both, collisional and 
photoionization. The results are given in Table~\ref{z2.209873j2233}.

\begin{table*}[th!]
\begin{center}
\caption[Ionization models for an \Ovi absorber at $z = 2.209873$]{Ionization models for an \Ovi absorber at $z = 2.209873$ towards J 2233-606}
 \begin{small}
 \begin{tabular}{rccccrcccc}
 \hline
 \hline
\scriptsize  $v$\,[\kms\!\!] &\scriptsize log [N(O\,{\sc vi)}\,(\sqc)] &\scriptsize 
log [$n_{\rm H}\,$(\cc)] &\scriptsize log $Z$  &\scriptsize log [$T$(K)] &\scriptsize $L$ [kpc] 
&\scriptsize log $f_{\rm HI}$ \\
 \hline
\multicolumn{7}{c}{\it photoionization} \\ \hline
\scriptsize $-$20.3 &\scriptsize 13.492 &\scriptsize $-$3.860 &\scriptsize $-$0.030 &\scriptsize 5.661 &\scriptsize 9.4 &\scriptsize $-$6.331 \\
\scriptsize $-$20.3 &\scriptsize 13.495 &\scriptsize $-$1.500 &\scriptsize $-$0.210 &\scriptsize 5.661 &\scriptsize 0.03 &\scriptsize $-$6.150 \\
\scriptsize 23.9 &\scriptsize 13.562 &\scriptsize $-$3.700 &\scriptsize $-$0.330 &\scriptsize 5.371 &\scriptsize 0.4 &\scriptsize $-$5.844 \\
 \hline
\multicolumn{7}{c}{\it collisional ionization} \\ \hline
\scriptsize $-$20.3 &\scriptsize 13.494 &\scriptsize $-$5.000 &\scriptsize $-$0.030 &\scriptsize 5.661 &\scriptsize 85.6 &\scriptsize $-$6.150 \\
\scriptsize 23.9 &\scriptsize 13.562 &\scriptsize $-$5.000 &\scriptsize $-$0.170 &\scriptsize 5.371 &\scriptsize 5.4 &\scriptsize $-$5.642 \\
 \hline
 \end{tabular}
 \end{small}
 \label{z2.209873j2233}
\end{center}
 \end{table*}

Next, we modeled the sum of the column densities of the three \Ovi components, using the sum of the two \Hi column densities as a stopping criterion. The temperature 
was calculated according to the weighted mean $b$-value of the oxygen components, log $N$(\Ovi\!\!)$_{\rm sum}$ = 14.06$\pm$0.524, log $N$(\Hi\!\!)$_{\rm sum}$ = 12.352$\pm$0.363,
$<b>$ = 18.477$\pm$3.09 \kms\!\!, $T$ = 328 249,1 K. We modeled both, collisional and photoionization. The results of our models are given in Table~\ref{z2.209873j2233_A}. The last two photoionization models have higher metallicities than solar (log $Z/Z_{\odot}>$ 0) and very low densities (log $n_{\rm H} \approx -$5), and therefore have no reasonable physical meaning. Although, we included them as possible solutions of CLOUDY.  

\begin{table*}[th!]
\begin{center}
\caption[Another set of ionization models for an \Ovi absorber at $z = 2.209873$]{Another set of ionization models for an \Ovi absorber at $z = 2.209873$ towards J 2233-606}
 \begin{small}
 \begin{tabular}{cccrcrcccc}
 \hline
 \hline
\scriptsize  $v$\,[\kms\!\!] &\scriptsize log [N(O\,{\sc vi)}\,(\sqc)] &\scriptsize 
log [$n_{\rm H}\,$(\cc)] &\scriptsize log $Z$  &\scriptsize log [$T$(K)] &\scriptsize $L$ [kpc] 
&\scriptsize log $f_{\rm HI}$ \\
 \hline
\multicolumn{7}{c}{\it photoionization} \\ \hline
\scriptsize 0.0 &\scriptsize 14.060 &\scriptsize $-$3.430 &\scriptsize 0.000 &\scriptsize 5.516 &\scriptsize 2.0 &\scriptsize $-$6.003 \\
\scriptsize 0.0 &\scriptsize 14.060 &\scriptsize $-$2.300 &\scriptsize $-$0.100 &\scriptsize 5.516 &\scriptsize 0.1 &\scriptsize $-$5.918 \\
\scriptsize 0.0 &\scriptsize 14.061 &\scriptsize $-$3.800 &\scriptsize $-$0.900 &\scriptsize 5.516 &\scriptsize 5.8 &\scriptsize $-$6.103 \\
\scriptsize 0.0 &\scriptsize 14.061 &\scriptsize $-$3.946 &\scriptsize $-$0.700 &\scriptsize 5.516 &\scriptsize 9.3 &\scriptsize $-$6.161 \\
\scriptsize 0.0 &\scriptsize 14.060 &\scriptsize $-$4.800 &\scriptsize 0.010 &\scriptsize 5.516 &\scriptsize 236.4 &\scriptsize $-$6.711 \\
\scriptsize 0.0 &\scriptsize 14.061 &\scriptsize $-$5.000 &\scriptsize 0.110 &\scriptsize 5.516 &\scriptsize 532.9 &\scriptsize $-$6.864 \\
 \hline
\multicolumn{7}{c}{\it collisional ionization} \\ \hline
\scriptsize 0.0 &\scriptsize 14.064 &\scriptsize $-$3.300 &\scriptsize $-$0.092 &\scriptsize 5.516 &\scriptsize 1.2 &\scriptsize $-$5.910 \\
\scriptsize 0.0 &\scriptsize 14.060 &\scriptsize $-$3.000 &\scriptsize $-$0.102 &\scriptsize 5.516 &\scriptsize 0.6 &\scriptsize $-$5.910 \\
\scriptsize 0.0 &\scriptsize 14.060 &\scriptsize $-$1.500 &\scriptsize $-$0.108 &\scriptsize 5.516 &\scriptsize 0.02 &\scriptsize $-$5.910 \\ 
\hline
 \end{tabular}
 \end{small}
 \label{z2.209873j2233_A}
\end{center}
 \end{table*}

\subsection*{System at $z=$2.202783 towards PKS 0237-233}

This is a two-component \Ovi system, blended by \Lya line at lower redshift and by  \Lyb line at higher $z$. It contains very weak \Nv and strong \Civ and \Siiv lines. It also contains a low ions as \Cii, \Ciii, \Siii and \Siiii. \Hi shows a saturated strong  absorption (e.g., log $N$(\Hi)=15.728). One of \Civ components is shifted relative to \Ovi by $-$1.5 \kms and we modeled them assuming that they belong to one system. We used the closest \Hi component (shifted from \Ovi by $-$4.0 \kms\!\!) as a stopping criterion in the model. The result 
is shown in Table~\ref{z2.202783pks0237}.

\begin{table*}[th!]
\caption[Ionization model for a \Civ\!\!/\Ovi absorber at $z = 2.202783$]{Ionization model for a \Civ\!\!/\Ovi absorber at $z = 2.202783$ towards PKS 0237-233}
 \begin{small}
 \begin{tabular}{cccccccccc}
 \hline
 \hline
\scriptsize  $v$\,[\kms\!\!] &\scriptsize log [N(C\,{\sc iv)}\,(\sqc)]&\scriptsize log [N(O\,{\sc vi)}\,(\sqc)] &\scriptsize 
log [$n_{\rm H}\,$(\cc)] &\scriptsize log $Z$  &\scriptsize log [$T$(K)] &\scriptsize $L$ [kpc] 
&\scriptsize log $f_{\rm HI}$ \\
 \hline
\scriptsize 0.0 &\scriptsize 13.964 &\scriptsize 13.262 &\scriptsize $-$3.157 &\scriptsize $-$1.670 &\scriptsize 4.501 &\scriptsize 36.5 &\scriptsize $-$4.168 \\
 \hline
 \end{tabular}
 \end{small}
 \label{z2.202783pks0237}
 \end{table*}

Second, we assumed that \Siiv and \Ovi are in one gas phase, because they are also well aligned (\Siiv is shifted y $-$1.1 \kms\!\!) and modeled them 
using the same stopping criterion. The result is given in Table ~\ref{z2.202783pks0237_A}.

\begin{table*}[th!]
 \caption[Ionization model for a \Siiv\!\!/\Ovi \\ absorber at $z = 2.202783$]{Ionization model for a \Siiv\!\!/\Ovi \\ absorber at $z = 2.202783$ towards PKS 0237-233}
 \begin{small}
 \begin{tabular}{cccccccccc}
 \hline
 \hline
\scriptsize  $v$\,[\kms\!\!] &\scriptsize log [N(Si\,{\sc iv)}\,(\sqc)]&\scriptsize log [N(O\,{\sc vi)}\,(\sqc)] &\scriptsize 
log [$n_{\rm H}\,$(\cc)] &\scriptsize log $Z$  &\scriptsize log [$T$(K)] &\scriptsize $L$ [kpc] 
&\scriptsize log $f_{\rm HI}$ \\
 \hline
\scriptsize 0.0 &\scriptsize 13.151 &\scriptsize 13.262 &\scriptsize $-$2.890 &\scriptsize $-$0.947 &\scriptsize 4.357 &\scriptsize 8.2 &\scriptsize $-$3.783 \\
 \hline
 \end{tabular}
 \end{small}
 \label{z2.202783pks0237_A}
 \end{table*}

\subsection*{System at $z=$2.425241 towards PKS 0329-255}

This is a three-component \Ovi system, blended by \Lya at lower redshift and by \Lyb and $Ly \gamma$ features at higher $z$. The corresponding 
\Hi is strong and saturated. The closest hydrogen component is shifted by 7.6 \kms to the middle
(second strongest) \Ovi component. The system contains two \Civ components. The closest to the middle 
oxygen one is shifted by 1.6 \kms\!\!. We assumed that both, oxygen and carbon, live in one gas phase. 
The results for the CLOUDY model is given in Table~\ref{z2.425241pks0329}. 

\begin{table*}[th!]
\caption[Ionization model for a \Civ\!\!/\Ovi absorber at $z = 2.425241$]{Ionization model for a \Civ\!\!/\Ovi absorber at $z = 2.425241$ towards PKS 0329-255}
 \begin{small}
 \begin{tabular}{cccccccccc}
 \hline
 \hline
\scriptsize  $v$\,[\kms\!\!] &\scriptsize log [N(C\,{\sc iv)}\,(\sqc)]&\scriptsize log [N(O\,{\sc vi)}\,(\sqc)] &\scriptsize 
log [$n_{\rm H}\,$(\cc)] &\scriptsize log $Z$  &\scriptsize log [$T$(K)] &\scriptsize $L$ [kpc] 
&\scriptsize log $f_{\rm HI}$ \\
 \hline
\scriptsize 0.0 &\scriptsize 12.945 &\scriptsize 13.106 &\scriptsize $-$3.641 &\scriptsize $-$2.713 &\scriptsize 4.633 &\scriptsize 244.5 &\scriptsize $-$4.753 \\
 \hline
 \end{tabular}
 \end{small}
 \label{z2.425241pks0329}
 \end{table*}

\subsection*{System at $z=$2.649423 towards PKS 0329-255}

This is weak two-component \Ovi system, containing one additional \Ciii and two \Civ components.  \Ciii is well aligned with the stronger \Ovi component 
($-$0.2 \kms\!\!). The two \Civ components are shifted to the corresponding \Ovi by $-$3.2 \kms (the stronger oxygen component) and 
by $-$0.2 \kms\!\!. Respectively, the strongest \Hi component is shifted to both of the \Ovi components by $-$9.3 \kms and by $+$2.7 \kms\!\!. 
We assumed two different models: first, that \Civ and \Ovi components are connected, and second that \Ciii and one of the \Civ components are connected. Our stopping criterion 
in the models comes from the closest \Hi component (log $N$(\Hi\!\!) = 14.509). The results are given in Table~\ref{z2.649423pks0329} and Table~\ref{z2.649423pks0329_A}. For the second model (\Ciii 
and \Civ in one gas-phase) the modeled \Ovi column density is log $N$(\Ovi\!\!) = 11.64, for comparison the observed value is log $N$(\Ovi\!\!) = 13.021.

\begin{table*}[th!]
\caption[Ionization models for a \Civ\!\!/\Ovi absorber at $z = 2.649423$]{Ionization models for a \Civ\!\!/\Ovi absorber at $z = 2.649423$ towards PKS 0329-255}
 \begin{small}
 \begin{tabular}{rcccccrccc}
 \hline
 \hline
\scriptsize  $v$\,[\kms\!\!] &\scriptsize log [N(C\,{\sc iv)}\,(\sqc)]&\scriptsize log [N(O\,{\sc vi)}\,(\sqc)] &\scriptsize 
log [$n_{\rm H}\,$(\cc)] &\scriptsize log $Z$  &\scriptsize log [$T$(K)] &\scriptsize $L$ [kpc] 
&\scriptsize log $f_{\rm HI}$ \\
 \hline
\scriptsize 0.0 &\scriptsize 12.263 &\scriptsize 13.021 &\scriptsize $-$3.931 &\scriptsize $-$2.333 &\scriptsize 4.667 &\scriptsize 101.2 &\scriptsize $-$5.054 \\
\scriptsize $-$12.0 &\scriptsize 11.953 &\scriptsize 12.697 &\scriptsize $-$3.923 &\scriptsize $-$2.646 &\scriptsize 4.667 &\scriptsize 98.6 &\scriptsize $-$5.051 \\
\hline
 \end{tabular}
 \end{small}
 \label{z2.649423pks0329}
 \end{table*}

\begin{table*}[th!]
\caption[Ionization models for a \Ciii\!\!/\Civ absorber at $z = 2.649423$]{Ionization models for a \Ciii\!\!/\Civ absorber at $z = 2.649423$ towards PKS 0329-255}
 \begin{small}
 \begin{tabular}{cccccccccc}
 \hline
 \hline
\scriptsize  $v$\,[\kms\!\!] &\scriptsize log [N(C\,{\sc iii)}\,(\sqc)]&\scriptsize log [N(C\,{\sc iv)}\,(\sqc)] &\scriptsize 
log [$n_{\rm H}\,$(\cc)] &\scriptsize log $Z$  &\scriptsize log [$T$(K)] &\scriptsize $L$ [kpc] 
&\scriptsize log $f_{\rm HI}$ \\
 \hline
\scriptsize 0.0 &\scriptsize 12.430 &\scriptsize 12.263 &\scriptsize $-$3.185 &\scriptsize $-$2.179 &\scriptsize 4.536 &\scriptsize 2.6 &\scriptsize $-$4.217 \\
 \hline
 \end{tabular}
 \end{small}
 \label{z2.649423pks0329_A}
 \end{table*}

\subsection*{System at $z=$2.660974 towards PKS 0329-255}

This is weak two-component \Ovi system without any other metals. The corresponding hydrogen is strong, log $N$(\Hi\!\!) = 14.303. The shifts between the two oxygen and 
the strongest hydrogen components are $-$5.7 \kms and $+$3.03 \kms\!\!, respectively. We summed up the oxygen components and used the weighted 
mean $b$-value as an estimate for the temperature in the model, log $N$(\Ovi\!\!)$_{\rm sum}$ = 13.419$\pm$0.209, $<b>$ = 30.65$\pm$15.1 \kms\!\!, $T$ = 903 461 K. We modeled 
photoionization and collisional ionization cases. The results are given in Table~\ref{z2.660974pks0329}. Additionally to the results presented in the table, we also 
tried to model the sum of the oxygen column density, using as an input parameter a space density up to log $n_{\rm H}$ = $-4.71$. This results was not reasonable, because 
the output absorption pathlength was $L$ = 34 Mpc. The $<b>$-value error is big and  costs higher uncertainty. Therefore, we also modeled the two oxygen components separately. 
The input temperatures are very different for this two models, because of the difference in the $b$-values. The central, but weaker component has $b$ = 6.82$\pm$1.16 \kms\!\!, 
respectively $T$ = 44 721 K, the second (but with higher column density) component has $b$ = 40.23$\pm$5.22 \kms\!\!, respectively $T$ = 1 556 111 K. The results are given in Table~\ref{z2.660974pks0329_A}.

\begin{table*}[th!]
\begin{center}
\caption[Ionization models for an \Ovi absorber at $z = 2.660974$]{Ionization models for an \Ovi absorber at $z = 2.660974$ towards PKS 0329-255}
 \begin{small}
 \begin{tabular}{cccccrcccc}
 \hline
 \hline
\scriptsize  $v$\,[\kms\!\!] &\scriptsize log [N(O\,{\sc vi)}\,(\sqc)] &\scriptsize 
log [$n_{\rm H}\,$(\cc)] &\scriptsize log $Z$  &\scriptsize log [$T$(K)] &\scriptsize $L$ [kpc] 
&\scriptsize log $f_{\rm HI}$ \\
 \hline
\multicolumn{7}{c}{\it photoionization} \\ \hline
\scriptsize 0.0 &\scriptsize 13.419 &\scriptsize $-$3.500 &\scriptsize $-$1.575 &\scriptsize 5.956 &\scriptsize 905.0
 &\scriptsize $-$6.643 \\
\scriptsize 0.0 &\scriptsize 13.419 &\scriptsize $-$3.000 &\scriptsize $-$1.664 &\scriptsize 5.956 &\scriptsize 256.8 &\scriptsize $-$6.596 \\
\scriptsize 0.0 &\scriptsize 13.149 &\scriptsize $-$2.800 &\scriptsize $-$1.681 &\scriptsize 5.956 &\scriptsize 159.0 &\scriptsize $-$6.588 \\
\scriptsize 0.0 &\scriptsize 13.419 &\scriptsize $-$2.300 &\scriptsize $-$1.702 &\scriptsize 5.956 &\scriptsize 49.1 &\scriptsize $-$6.577 \\
\hline
\multicolumn{7}{c}{\it collisional ionization} \\ \hline
\scriptsize 0.0 &\scriptsize 13.419 &\scriptsize $-$3.500 &\scriptsize $-$1.711 &\scriptsize 5.956 &\scriptsize 768.5 &\scriptsize $-$6.572 \\
\scriptsize 0.0 &\scriptsize 13.419 &\scriptsize $-$3.000 &\scriptsize $-$1.711 &\scriptsize 5.956 &\scriptsize 243.0 &\scriptsize $-$6.572 \\
\scriptsize 0.0 &\scriptsize 13.419 &\scriptsize $-$2.300 &\scriptsize $-$1.712 &\scriptsize 5.956 &\scriptsize 48.5 &\scriptsize $-$6.572 \\
\hline
 \end{tabular}
 \end{small}
 \label{z2.660974pks0329}
\end{center}
 \end{table*}

\begin{table*}[th!]
\begin{center}
\caption[Another set of models for an \Ovi absorber at $z = 2.660974$]{Another set of models for an \Ovi absorber at $z = 2.660974$ towards PKS 0329-255}
\begin{small}
\begin{tabular}{rccccrcccc}
 \hline
 \hline
\scriptsize  $v$\,[\kms\!\!] &\scriptsize log [N(O\,{\sc vi)}\,(\sqc)] &\scriptsize 
log [$n_{\rm H}\,$(\cc)] &\scriptsize log $Z$  &\scriptsize log [$T$(K)] &\scriptsize $L$ [kpc] 
&\scriptsize log $f_{\rm HI}$ \\
 \hline
\multicolumn{7}{c}{\it photoionization} \\ \hline
\scriptsize 0.0 &\scriptsize 12.876 &\scriptsize $-$3.500 &\scriptsize $-$1.473 &\scriptsize 4.651 &\scriptsize 8.8 &\scriptsize $-$4.631 \\
\scriptsize 0.0 &\scriptsize 12.876 &\scriptsize $-$3.000 &\scriptsize $-$0.274 &\scriptsize 4.651 &\scriptsize 1.01 &\scriptsize $-$4.19 \\
\scriptsize 0.0 &\scriptsize 12.876 &\scriptsize $-$2.800 &\scriptsize $-$0.643 &\scriptsize 4.651 &\scriptsize 0.5 &\scriptsize $-$4.036 \\
\scriptsize $-$5.7 &\scriptsize 13.272 &\scriptsize $-$2.800 &\scriptsize $-$1.776 &\scriptsize 6.192 &\scriptsize 310.9 &\scriptsize $-$6.879 \\ 
\hline
\multicolumn{7}{c}{\it collisional ionization} \\ \hline
\scriptsize $-$5.7 &\scriptsize 13.272 &\scriptsize $-$2.800 &\scriptsize$-$1.801  &\scriptsize 6.192 &\scriptsize 300.0 &\scriptsize $-$6.846 \\
\hline
 \end{tabular}
 \end{small}
 \label{z2.660974pks0329_A}
\end{center}
 \end{table*}

\subsection*{System at $z=$2.708873 towards PKS 0329-255}

This is a complex multicomponent \Ovi system, containig \Civ and \Ciii\!\!. We modeled the strongest \Ovi component, connecting it to the corresponding first \Civ feature, 
shifted by $-$1.5 \kms\!\!. We also modeled the stronger of the two \Ciii components and the second \Civ component, without \Ovi\!\!. For both models the 
calculated absorption pathlengths were 8 and 4 Mpc, respectively, which is unrealistic for an IGM cloud. We have another two, more successful models. First, we modeled the sum 
of the first two \Civ components, which are close to each other (in a distance of 1.9 \kms\!\!), with the corresponding \Ovi one, 
log $N$(\Civ\!\!)$_{\rm sum}$ = 13.585$\pm$0.514. The shift between both carbon and oxygen components is 2.2 \kms\!\!. For a stopping criterion we used \Hi\!\!, 
shifted by 6.63 \kms\!\!. Second, we modeled the third \Ovi component with the corresponding \Civ\!\!, shifted by 0.9 \kms\!\!, and taking as a stopping criterion 
the closest \Hi\!\!, shifted by 3.6 \kms\!\!. The results are given in Table~\ref{z2.708873pks0329}.

\begin{table*}[th!]
\caption[Ionization models for an\\ absorber at $z = 2.708873$]{Ionization models for an\\ absorber at $z = 2.708873$ towards PKS 0329-255}
 \begin{small}
 \begin{tabular}{rcccccrccc}
 \hline
 \hline
\scriptsize  $v$\,[\kms\!\!] &\scriptsize log [N(C\,{\sc iv)}\,(\sqc)]&\scriptsize log [N(O\,{\sc vi)}\,(\sqc)] &\scriptsize 
log [$n_{\rm H}\,$(\cc)] &\scriptsize log $Z$  &\scriptsize log [$T$(K)] &\scriptsize $L$ [kpc] 
&\scriptsize log $f_{\rm HI}$ \\
 \hline
\scriptsize 0.0 &\scriptsize 13.585 &\scriptsize 14.194 &\scriptsize $-$3.868 &\scriptsize $-$0.797 &\scriptsize 4.506 &\scriptsize 29.9 &\scriptsize $-$4.859 \\
\scriptsize 37.9 &\scriptsize 12.466 &\scriptsize 13.865 &\scriptsize $-$4.352 &\scriptsize $-$1.739 &\scriptsize 4.636 &\scriptsize 524.3 &\scriptsize $-$5.446 \\
 \hline
 \end{tabular}
 \end{small}
 \label{z2.708873pks0329}
 \end{table*}

 







\subsection*{System at $z=$2.539533 towards Q 0002-422}

This is a one-component \Ovi system, blended by Fe\,{\sc ii} and \Lya at lower redshift, and \Lyb, Ly-$\gamma$ and Ly-$\kappa$ features at higher redshift. There is one \Civ component, 
shifted by $-$6.4 \kms\!\!. The corresponding \Hi is shifted by 4.2 \kms\!\!, which is also blended by Fe\,{\sc ii} at
lower redshift. We modeled assuming that both oxygen and carbon, are in one gas phase. The results are given in Table~\ref{z2.539533q0002}.

\begin{table*}[th!]
\caption[Ionization model for an absorber at $z = 2.539533$]{Ionization model for an absorber at $z = 2.539533$ towards Q 0002-422}
 \begin{small}
 \begin{tabular}{cccccccccc}
 \hline
 \hline
\scriptsize  $v$\,[\kms\!\!] &\scriptsize log [N(C\,{\sc iv)}\,(\sqc)]&\scriptsize log [N(O\,{\sc vi)}\,(\sqc)] &\scriptsize 
log [$n_{\rm H}\,$(\cc)] &\scriptsize log $Z$  &\scriptsize log [$T$(K)] &\scriptsize $L$ [kpc] 
&\scriptsize log $f_{\rm HI}$ \\
 \hline
\scriptsize 0.0 &\scriptsize 12.545 &\scriptsize 13.741 &\scriptsize $-$4.202 &\scriptsize $-$1.318 &\scriptsize 4.638 &\scriptsize 88.8 &\scriptsize $-$5.298 \\
 \hline
 \end{tabular}
 \end{small}
 \label{z2.539533q0002}
 \end{table*}

\subsection*{System at $z=$2.701121 towards Q 0002-422}

This is a single and an weak \Ovi line which does not show other corresponding metals. The corresponding \Hi feature has the same redshift as the 
oxygen, $z=$2.701121, even though its redshift was set as a free parameter in the fitting procedure. We used it as a stopping criterion when modeling the oxygen column density. We fixed the temperature, according the oxygen $b-$value, $b$ = 11.73$\pm$1.74 \kms\!\!, $T$ = 132 293 K. Because of the high $T$ we have modeled photoionization and collisional ionization. This results in several models, but in Table~\ref{q0002z2.701121} we show only those with reasonable pathlengths.

\begin{table*}[th!]
\begin{center}
 \caption[Ionization models for an \Ovi \\ absorber at $z = 2.701121$]{Ionization models for an \Ovi absorber at $z = 2.701121$ towards Q 0002-422}
 \begin{small}
 \begin{tabular}{cccrcrcccc}
 \hline
 \hline
\scriptsize  $v$\,[\kms\!\!] &\scriptsize log [N(O\,{\sc vi)}\,(\sqc)] &\scriptsize 
log [$n_{\rm H}\,$(\cc)] &\scriptsize log $Z$  &\scriptsize log [$T$(K)] &\scriptsize $L$ [kpc] 
&\scriptsize log $f_{\rm HI}$ \\
 \hline
\multicolumn{7}{c}{\it photoionization} \\ \hline
\scriptsize 0.0 &\scriptsize 12.696 &\scriptsize $-$2.140 &\scriptsize 0.000 &\scriptsize 5.121 &\scriptsize 0.4 &\scriptsize $-$5.108 \\
\scriptsize 0.0 &\scriptsize 12.696 &\scriptsize $-$3.000 &\scriptsize $-$1.035 &\scriptsize 5.121 &\scriptsize 3.5 &\scriptsize $-$5.186 \\
\hline
\multicolumn{7}{c}{\it collisional ionization} \\ \hline
\scriptsize 0.0 &\scriptsize 12.697 &\scriptsize $-$3.737 &\scriptsize 0.000  &\scriptsize 5.121 &\scriptsize 15.0 &\scriptsize $-$5.087 \\
\scriptsize 0.0 &\scriptsize 12.696 &\scriptsize $-$5.500 &\scriptsize$-$1.660  &\scriptsize 5.121 &\scriptsize 870.0 &\scriptsize $-$5.087 \\
\hline
 \end{tabular}
 \end{small}
 \label{q0002z2.701121}
\end{center}
 \end{table*}

\subsection*{System at $z=$2.062566 towards Q 0122-380}

This is a single narrow \Ovi line, whose 1037.62 transition is blended by a \Lya line at lower redshift. At the same redshift there are  one-component lines of \Nv and \Civ\!\!. 
The nitrogen line is weak, \Nv 1242.80 is partly blended by a \Lya feature at higher redshift. The carbon line is narrow. Both \Ovi and \Civ column densities are higher than the 
corresponding \Hi one. The latter is a very weak two-component line. The redshift of the second component correspond to the other elements.  All elements are tied and there is no 
velocity shift. Tabl.~\ref{z2.062566q0122} shows the modeled column densities of oxygen and carbon. 

\begin{table*}[th!]
\caption[Ionization model for an absorber at $z = 2.062566$]{Ionization model for an absorber at $z = 2.062566$ towards Q 0122-380}
 \begin{small}
 \begin{tabular}{cccccccccc}
 \hline
 \hline
\scriptsize  $v$\,[\kms\!\!] &\scriptsize log [N(C\,{\sc iv)}\,(\sqc)]&\scriptsize log [N(O\,{\sc vi)}\,(\sqc)] &\scriptsize 
log [$n_{\rm H}\,$(\cc)] &\scriptsize log $Z$  &\scriptsize log [$T$(K)] &\scriptsize $L$ [kpc] 
&\scriptsize log $f_{\rm HI}$ \\
 \hline
\scriptsize 0.0 &\scriptsize 12.798 &\scriptsize 13.217 &\scriptsize $-$3.785 &\scriptsize 0.000 &\scriptsize 4.356 &\scriptsize 0.4 &\scriptsize $-$4.658 \\
 \hline
 \end{tabular}
 \end{small}
 \label{z2.062566q0122}
 \end{table*}

The modeled column densities are matching in reasonable way only assuming solar metallicities. It could not model without photoionization assumed.

\subsection*{System at $z=$2.352071 towards Q 0329-385}

This is a strong and narrow one-component \Ovi system, containing \Nv and \Civ\!\!. The \Ovi line is blended by \Lya at lower
redshift and by \Lyb feature at higher $z$. \Civ is also strong and narrow, and it is not blended. \Hi is very weak, with 
log $N$(\Hi\!\!) = 13.11, which is less than the oxygen and carbon column density respectively. The strongest \Hi line is shifted toward the \Ovi by
7.33 \kms\!\!. \Ovi and the strongest \Civ are well aligned (within 6.2 \kms\!\!). \Ovi\!\!, \Nv\!\!, \Civ and \Hi
are extended within $\pm$30 \kms each.  We assumed that they belong to one gas phase and modeled the sum of the \Civ and \Hi column densities.
The results are given in Table~\ref{z2.352071q0329}.

\begin{table*}[th!]
\caption[Ionization model for an absorber at $z = 2.352071$]{Ionization model for an absorber at $z = 2.352071$ towards Q 0329-385}
 \begin{small}
 \begin{tabular}{ccccccccc}
 \hline
 \hline
\scriptsize  $v$\,[\kms\!\!] &\scriptsize log [N(C\,{\sc iv)}\,(\sqc)]&\scriptsize log [N(O\,{\sc vi)}\,(\sqc)] &\scriptsize 
log [$n_{\rm H}\,$(\cc)] &\scriptsize log $Z$  &\scriptsize log [$T$(K)] &\scriptsize $L$ [kpc] 
&\scriptsize log $f_{\rm HI}$ \\
 \hline
\scriptsize 0.0 &\scriptsize 13.221 &\scriptsize 13.882 &\scriptsize $-$3.913 &\scriptsize $-$0.044 &\scriptsize 4.409 &\scriptsize 2.3 &\scriptsize $-$4.825 \\
 \hline
 \end{tabular}
 \end{small}
 \label{z2.352071q0329}
 \end{table*}

\subsection*{System at $z$ = 2.363858 towards Q 0329-385}

This is a three-component \Ovi system. It additionally contains \Civ\!\!. The 2 strongest, almost overlapping \Ovi 
components are shifted toward the strongest \Hi by 11.14 \kms and toward the strongest \Civ by 5.53 \kms\!\!. The left weaker \Ovi component is shifted
toward the left weaker \Hi by 1.16 \kms  and toward the left weaker \Civ line by 2.32 \kms\!\!. The results of the modeled left weaker \Ovi component are
given in Table~\ref{z2.363858q0329}.

\begin{table*}[th!]
\caption[Ionization model for an absorber at $z = 2.363858$]{Ionization model for an absorber at $z = 2.363858$ towards Q 0329-385}
 \begin{small}
 \begin{tabular}{cccccccccc}
 \hline
 \hline
\scriptsize  $v$\,[\kms\!\!] &\scriptsize log [N(C\,{\sc iv)}\,(\sqc)]&\scriptsize log [N(O\,{\sc vi)}\,(\sqc)] &\scriptsize 
log [$n_{\rm H}\,$(\cc)] &\scriptsize log $Z$  &\scriptsize log [$T$(K)] &\scriptsize $L$ [kpc] 
&\scriptsize log $f_{\rm HI}$ \\
 \hline
\scriptsize 0.0 &\scriptsize 12.116 &\scriptsize 13.373 &\scriptsize $-$4.231 &\scriptsize $-$2.173 &\scriptsize 4.684 &\scriptsize 329.3 &\scriptsize $-$5.367 \\
 \hline
 \end{tabular}
 \end{small}
 \label{z2.363858q0329}
 \end{table*}
 
\subsection*{System at $z$ = 2.397850 towards Q 0453-423}

This is a multicomponent \Ovi system, blended by  \Lya at lower redshift, and \Lyb and Ly-$\gamma$ feature at higher $z$. The system contains
\Siiv and a weak \Nv\!\!. 
The strongest \Ovi component has exactly the same redshift, $z = 2.397850$, as one of the hydrogen components. The closest \Civ component to both
($z = 2.397814$) is shifted by 3.18 \kms\!\!. We modeled this components, assuming that \Ovi and \Civ live in the same gas phase. The results are
given in Table~\ref{z2.397850q0453}.

\begin{table*}[th!]
\caption[Ionization model for an absorber at $z = 2.397850$]{Ionization model for an absorber at $z = 2.397850$ towards Q 0453-423}
 \begin{small}
 \begin{tabular}{cccccccccc}
 \hline
 \hline
\scriptsize  $v$\,[\kms\!\!] &\scriptsize log [N(C\,{\sc iv)}\,(\sqc)]&\scriptsize log [N(O\,{\sc vi)}\,(\sqc)] &\scriptsize 
log [$n_{\rm H}\,$(\cc)] &\scriptsize log $Z$  &\scriptsize log [$T$(K)] &\scriptsize $L$ [kpc] 
&\scriptsize log $f_{\rm HI}$ \\
 \hline
\scriptsize 0.0 &\scriptsize 13.639 &\scriptsize 14.465 &\scriptsize $-$3.996 &\scriptsize $-$0.568 &\scriptsize 4.504 &\scriptsize 32.95 &\scriptsize $-$4.984 \\
 \hline
 \end{tabular}
 \end{small}
 \label{z2.397850q0453}
 \end{table*}

\subsection*{System at $z$ = 2.636236 towards Q 0453-423}

This is a four-component \Ovi system, blended by \Lya line at lower redshift. It contains \Nv\!\!, \Civ\!\!, \Siiv and \Siiii\!\!.
\Nv\!\!, \Siiv and \Civ are well aligned to each other. \Siiv and \Siiii are shifted by 2.06 \kms\!\!. 
The strongest \Ovi and \Civ lines are shifted by 5.52 \kms\!\!.
 
The strongest  \Ovi line shows two components close to each other. We sum their column densities. The shift between this \Ovi line and a weaker 
\Civ line, redward of the strongest one, is 0.9 \kms\!\!. The closest \Hi component to this \Ovi line is shifted by 4.53 \kms\!\!.
We modeled this components. In our first model, using the ratio $N_{\rm sum}$(\Ovi\!\!)/$N$(\Civ\!\!), we use as a stopping criterion the column density of the hydrogen line 
($z = 2.636183$) shifted by 4.53 \kms\!\!. 

Because the other elements are well aligned to each other, we computed a second model, using the ratio $N$(\Siiii\!\!)/$N$(\Siiv\!\!). 
The stopping criterion is the same as in the first model.

Our third model is for the \Civ component at $z = 2.636171$. The temperature is fixed to the corresponding $b$-value ($T = 76 673$  K).
The stopping criterion is the same as in the first model.

The results for the three models are given in Tables ~\ref{z2.636236q0453}, \ref{z2.636236q0453_A} and \ref{z2.636236q0453_B}.

\begin{table*}[th!]
\caption[Ionization model for an absorber at $z = 2.636236$]{Ionization model for an absorber at $z = 2.636236$ towards Q 0453-423}
 \begin{small}
 \begin{tabular}{cccccccccc}
 \hline
 \hline
\scriptsize  $v$\,[\kms\!\!] &\scriptsize log [N(C\,{\sc iv)}\,(\sqc)]&\scriptsize log [N(O\,{\sc vi)}\,(\sqc)] &\scriptsize 
log [$n_{\rm H}\,$(\cc)] &\scriptsize log $Z$  &\scriptsize log [$T$(K)] &\scriptsize $L$ [kpc] 
&\scriptsize log $f_{\rm HI}$ \\
 \hline
\scriptsize 0.0 &\scriptsize 12.382 &\scriptsize 13.813 &\scriptsize $-$4.365 &\scriptsize $-$2.112 &\scriptsize 4.673 &\scriptsize 125.9 &\scriptsize $-$5.491 \\
 \hline
 \end{tabular}
 \end{small}
 \label{z2.636236q0453}
 \end{table*}

\begin{table*}[th!]
\caption[Another ionization model for an absorber at $z = 2.636236$]{Another ionization model for an absorber at $z = 2.636236$ towards Q 0453-423}
 \begin{small}
 \begin{tabular}{cccccccccc}
 \hline
 \hline
\scriptsize  $v$\,[\kms\!\!] &\scriptsize log [N(Si\,{\sc iii)}\,(\sqc)]&\scriptsize log [N(Si\,{\sc iv)}\,(\sqc)] &\scriptsize 
log [$n_{\rm H}\,$(\cc)] &\scriptsize log $Z$  &\scriptsize log [$T$(K)] &\scriptsize $L$ [kpc] 
&\scriptsize log $f_{\rm HI}$ \\
 \hline
\scriptsize 0.0 &\scriptsize 11.767 &\scriptsize 12.354 &\scriptsize $-$3.400 &\scriptsize $-$0.477 &\scriptsize 4.342 &\scriptsize 8.1 &\scriptsize $-$4.267 \\
 \hline
 \end{tabular}
 \end{small}
 \label{z2.636236q0453_A}
 \end{table*}

\begin{table*}[th!]
\begin{center}
  \caption[Third ionization model for an absorber at $z = 2.636236$]{Third ionization model for an absorber at $z = 2.636236$ towards Q 0453-423}
 \begin{small}
 \begin{tabular}{ccccccccc}
 \hline
 \hline
\scriptsize  $v$\,[\kms\!\!] &\scriptsize log [N(C\,{\sc iv)}\,(\sqc)]&\scriptsize 
log [$n_{\rm H}\,$(\cc)] &\scriptsize log $Z$  &\scriptsize log [$T$(K)] &\scriptsize $L$ [kpc] 
&\scriptsize log $f_{\rm HI}$ \\
 \hline
\scriptsize 0.0 &\scriptsize 13.589 &\scriptsize $-$3.400$^{\rm a}$ &\scriptsize $-$1.430 &\scriptsize 4.885$^{\rm b}$ &\scriptsize 32.95 &\scriptsize $-$4.885 \\
 \hline
 \end{tabular}
 \end{small}
 \label{z2.636236q0453_B}\\
\end{center}
\tiny$^{\rm a}$ Lower limit \\
\tiny$^{\rm b}$ Fixed to $b$-value lower limit \\
 \end{table*}

\subsection*{System at $z$ = 2.640247 towards Q 0453-423}

This is a two-component \Ovi system which also contains one component \Civ\!\!. \Civ is shifted toward the strongest \Ovi by $-$4.36 \kms\!\!.
\Lya is blended. There are two overlapping \Hi components aligned with \Ovi\!\!. We used the sum of their column densities
as a stopping criterion in our model. 
The result for the model is given in Table~\ref{z2.640247q0453}.

\begin{table*}[th!]
\caption[Ionization model for an absorber at $z = 2.640247$]{Ionization model for an absorber at $z = 2.640247$ towards Q 0453-423}
 \begin{small}
 \begin{tabular}{cccccccccc}
 \hline
 \hline
\scriptsize  $v$\,[\kms\!\!] &\scriptsize log [N(C\,{\sc iv)}\,(\sqc)]&\scriptsize log [N(O\,{\sc vi)}\,(\sqc)] &\scriptsize 
log [$n_{\rm H}\,$(\cc)] &\scriptsize log $Z$  &\scriptsize log [$T$(K)] &\scriptsize $L$ [kpc] 
&\scriptsize log $f_{\rm HI}$ \\
 \hline
\scriptsize 0.0 &\scriptsize 12.408 &\scriptsize 13.238 &\scriptsize $-$3.917 &\scriptsize $-$2.290 &\scriptsize 4.885 &\scriptsize 212.6 &\scriptsize $-$5.271 \\
 \hline
 \end{tabular}
 \end{small}
 \label{z2.640247q0453}
 \end{table*}

\section{Associated \Ovi systems}

\subsection*{System at $z$ = 2.419294 towards HE 1122-1648}

This is a three-component \Ovi system, which contains very weak \Civ and \Nv\!\!. The oxygen lines are strong and narrow. On the contrary, the hydrogen is weak with column densities a bit lower then those of the oxygen. The shift between the strongest \Hi and  \Ovi components is $\Delta v_{\rm comp.} =$ 4.0 \kms\!\!. \Civ component, aligned with the corresponding \Hi has $\Delta v_{\rm comp.} =$ 1.67 \kms\!\!, \Nv - \Hi pair have $\Delta v_{\rm comp.} =$ 1.40 \kms\!\!.

First, we modeled only the strongest oxygen component($z=$2.419294). We calculated the temperature from the component's $b$-value ($b$ = 9.49 \kms\!\!), using Eq.~\ref{eq_temp_Doppler}, which leads to $T = 86,774$ K. For a stopping criterion we used the strongest component of \Hi\!\!, log $N$(\Hi)=13.65. The results are given in Table~\ref{z2.419294he1122}.

\begin{table*}[th!]
\begin{center}
\caption[Ionization models for an \Ovi absorber at $z = 2.419294$]{Ionization models for an \Ovi absorber at $z = 2.419294$ towards HE 1122-1648}
\begin{small}
\begin{tabular}{cccrcrcccc}
\hline
\hline
\scriptsize $v$\,[\kms\!\!] &\scriptsize log [N(O\,{\sc vi)}\,(cm$^{-2}$)] 
&\scriptsize log [$n_{\rm H}\,$(\cc)] & \scriptsize log $Z$ & 
\scriptsize log [$T$(K)] & \scriptsize$L$ [kpc] & \scriptsize log $f_{\rm HI}$ \\
\hline
\scriptsize0.0 & \scriptsize 13.991  & \scriptsize$-$4.160  &\scriptsize $-$0.800 &\scriptsize 4.940 &\scriptsize 73.4 &\scriptsize $-$5.540  \\
\scriptsize0.0 & \scriptsize 13.992  & \scriptsize$-$3.556 & \scriptsize$-$0.004 &\scriptsize 4.940 & \scriptsize 6.0 &\scriptsize$-$5.062 \\
\hline
\end{tabular}
\end{small}
\label{z2.419294he1122}
\end{center}
\end{table*}

Second, we modeled the sum of the column densities of the two weaker \Ovi components, $z_{\rm mean} = 2.4197385$. $N_{s\rm um}$=13.525$\pm$0.193. The temperature was calculated from the weighted mean $b$-value of the two components: $<b>$ = 24.22$\pm$7.54 \kms\!\!, $T$ = 564 013 K. For the stopping criterion we used the second strongest \Hi component, log $N$(\Hi\!\!) = 13.013. There are four different Cloudy models given in Table~\ref{z2.4197385he1122_A}.

\begin{table*}[]
\begin{center}
\caption[Ionization models for an \Ovi absorber at $z = 2.419294$]{Ionization models for an \Ovi absorber at $z_{\rm mean} = 2.4197385$ towards HE 1122-1648}
\begin{small}
\begin{tabular}{cccrcrcccc}
\hline
\hline
\scriptsize $v$\,[\kms\!\!] &\scriptsize log [N(O\,{\sc vi)}\,(\sqc)] 
&\scriptsize log [$n_{\rm H}\,$(\cc)] &\scriptsize log $Z$ & 
\scriptsize log [$T$(K)] &\scriptsize $L$ [kpc] &\scriptsize log $f_{\rm HI}$ \\
\hline
\scriptsize 39.0 &\scriptsize 13.525 &\scriptsize $-$4.579&\scriptsize 0.000     &\scriptsize 5.750 &\scriptsize 844.6 &\scriptsize $-$6.824 \\
\scriptsize 39.0 &\scriptsize 13.525 &\scriptsize $-$3.100 &\scriptsize $-$0.645&\scriptsize 5.750 &\scriptsize 8.8 &\scriptsize $-$6.321 \\
\scriptsize 39.0 &\scriptsize 13.525 &\scriptsize $-$1.500  &\scriptsize $-$0.681&\scriptsize 5.750 &\scriptsize 0.2 &\scriptsize $-$6.288 \\
\scriptsize 39.0 &\scriptsize 13.525 &\scriptsize $-$0.300  &\scriptsize $-$0.682&\scriptsize 5.750 &\scriptsize 0.01 &\scriptsize $-$6.287 \\
\hline
\end{tabular}
\end{small}
\label{z2.4197385he1122_A}
\end{center}
\end{table*}

After that we modeled only the third \Ovi component (log $N$(\Ovi\!\!) = 12.837) with $T$ calculated from the its $b$-value. ($b$ = 7.58 \kms\!\!, $T$ = 55 243 K). The stopping criterion, log $N$(\Hi\!\!) = 13.013. The result is given in Table~\ref{z2.419846he1122}. 

\begin{table*}[]
\begin{center}
\caption[Ionization model for an \Ovi absorber at $z = 2.419846$]{Ionization model for an \Ovi absorber at $z = 2.419846$ towards HE 1122-1648}
\begin{small}
\begin{tabular}{cccrcrcccc}
\hline
\hline
\scriptsize$v$\,[\kms] &\scriptsize log [N(O\,{\sc vi)}\,(\sqc)] 
&\scriptsize log [$n_{\rm H}\,$(\cc)] &\scriptsize log $Z$ & 
\scriptsize log [$T$(K)] &\scriptsize $L$ [kpc] &\scriptsize log $f_{\rm HI}$ \\
\hline
\scriptsize48.4 &\scriptsize 12.837  &\scriptsize $-$3.395 &\scriptsize 0.000 &\scriptsize 4.742 &  \scriptsize0.4 &\scriptsize $-$4.646 \\
\hline
\end{tabular}
\end{small}
\label{z2.419846he1122}
\end{center}
\end{table*}

\subsection*{System at $z$ = 2.862508 towards HE 2347-4342}

This is a three-component weak \Ovi system blended by \Lya absorption from lower redshift. There is also a weak three-component \Civ absorption feature, blended by \Siii at lower $z$. 
\Lya is saturated and has also three components. The \Civ component is shifted to the strongest \Ovi by $-$3.4 \kms\!\!. The closest \Hi component to \Ovi 
is shifted by $-$12.7 \kms\!\!. On the other hand, one of the weaker \Hi and \Ovi components have the same redshift and show no displacement. Because of the 
blending effect we do not have clear information about how real is the shift between some of the components. Therefore, we decided that it is better to sum the 
column densities, assuming that all components belong to one and the same gas phase, log $N$(\Ovi\!\!)$_{\rm sum}$ = 13.616$\pm$0.433, log $N$(\Civ\!\!)$_{\rm sum}$ = 12.935$\pm$0.551 
and log $N$(\Hi\!\!)$_{\rm sum}$ = 14.315$\pm$0.439. The result of the modeling is given in Table~\ref{z2.862508he2347}.

\begin{table*}[th!]
\caption[Ionization model for a \Civ\!\!/\Ovi absorber at $z = 2.862508$]{Ionization model for a \Civ\!\!/\Ovi absorber at $z = 2.862508$ towards HE 2347-4342}
 \begin{small}
 \begin{tabular}{cccccccccc}
 \hline
 \hline
\scriptsize  $v$\,[\kms\!\!] &\scriptsize log [N(C\,{\sc iv)}\,(\sqc)]&\scriptsize log [N(O\,{\sc vi)}\,(\sqc)] &\scriptsize 
log [$n_{\rm H}\,$(\cc)] &\scriptsize log $Z$  &\scriptsize log [$T$(K)] &\scriptsize $L$ [kpc] 
&\scriptsize log $f_{\rm HI}$ \\
 \hline
\scriptsize 0.0 &\scriptsize 12.935 &\scriptsize 13.616 &\scriptsize $-$3.895 &\scriptsize $-$1.490 &\scriptsize 4.617 &\scriptsize 49.9 &\scriptsize $-$4.978 \\
 \hline
 \end{tabular}
 \end{small}
 \label{z2.862508he2347}
 \end{table*}

\subsection*{System at $z$ = 2.213517 towards PKS 0237-233}

This is a three-component weak \Ovi system, containing a very weak one component \Civ feature, shifted by 5.8 \kms to the strongest oxygen component. The corresponding 
\Hi is strong and saturated, shifted by 9.2 \kms to the first (strongest) \Ovi component. We assumed that both, carbon and oxygen live in one gas phase and we 
modeled them together. The result is given in Table~\ref{z2.213517pks0237}.

\begin{table*}[th!]
\caption[Ionization model for a \Civ\!\!/\Ovi absorber at $z = 2.213517$]{Ionization model for a \Civ\!\!/\Ovi absorber at $z = 2.213517$ towards PKS 0237-233}
 \begin{small}
 \begin{tabular}{cccccccccc}
 \hline
 \hline
\scriptsize  $v$\,[\kms\!\!] &\scriptsize log [N(C\,{\sc iv)}\,(\sqc)]&\scriptsize log [N(O\,{\sc vi)}\,(\sqc)] &\scriptsize 
log [$n_{\rm H}\,$(\cc)] &\scriptsize log $Z$  &\scriptsize log [$T$(K)] &\scriptsize $L$ [kpc] 
&\scriptsize log $f_{\rm HI}$ \\
 \hline
\scriptsize 0.0 &\scriptsize 11.835 &\scriptsize 13.350 &\scriptsize $-$4.425 &\scriptsize $-$2.104 &\scriptsize 4.688 &\scriptsize 536.8 &\scriptsize $-$5.56 \\
 \hline
 \end{tabular}
 \end{small}
 \label{z2.213517pks0237}
 \end{table*}




\subsection*{System at $z$ = 2.237771 towards PKS 0237-233}

This is a three-component narrow \Ovi system, which shows no any other metals. The corresponding 
\Hi is weaker than the oxygen and shifted by $-$5.5 \kms to the first (strongest) \Ovi 
component. The \Ovi lines are blended by \Lya line at lower redshift. The first 2 components are close 
ot each other (distance of 2.7 \kms\!\!) and we modeled their sum, taking as a stopping criterion the \Hi 
column density, and fixing the temperature according the weighted mean $b$-value of the oxygen component, 
log $N$(\Ovi\!\!)$_{\rm sum}$ = 13.547, $<b>$=  9.557$\pm$3.276 \kms\!\!, $T$ = 87 824 K. The results 
for the photoionized and collisional ionized models are given in Table~\ref{z2.237771pks0237}. The only collisional and the last photoionization models have higher than solar metallicities (log $Z/Z_{\odot} >$ 0) and, in the first case, very low space density (log $n_{\rm H} \approx -$5.4). Therefore, they do not have a reasonable physical meaning, but we included them as possible solutions.

\begin{table*}[th!]
\begin{center}
\caption[Ionization models for an \Ovi absorber at $z = 2.237771$]{Ionization models for an \Ovi absorber at $z = 2.237771$ towards PKS 0237-233}
 \begin{small}
 \begin{tabular}{cccrcrcccc}
 \hline
 \hline
\scriptsize  $v$\,[\kms\!\!] &\scriptsize log [N(O\,{\sc vi)}\,(\sqc)] &\scriptsize 
log [$n_{\rm H}\,$(\cc)] &\scriptsize log $Z$  &\scriptsize log [$T$(K)] &\scriptsize $L$ [kpc] 
&\scriptsize log $f_{\rm HI}$ \\
 \hline
\multicolumn{7}{c}{\it photoionization} \\ \hline
\scriptsize 0.0 &\scriptsize 13.547 &\scriptsize $-$5.304 &\scriptsize $-$0.090 &\scriptsize 4.944 &\scriptsize 576.3 &\scriptsize $-$6.631 \\
\scriptsize 0.0 &\scriptsize 13.547 &\scriptsize $-$3.800 &\scriptsize $-$0.350 &\scriptsize 4.944 &\scriptsize 0.7 &\scriptsize $-$5.245 \\
\scriptsize 0.0 &\scriptsize 13.547 &\scriptsize $-$3.500 &\scriptsize 0.004 &\scriptsize 4.944 &\scriptsize 0.23 &\scriptsize $-$5.028 \\
 \hline
\multicolumn{7}{c}{\it collisional ionization} \\ \hline
\scriptsize 0.0 &\scriptsize 13.547 &\scriptsize $-$5.395 &\scriptsize 0.040 &\scriptsize 4.944 &\scriptsize 6.5 &\scriptsize $-$4.593 \\
\hline
 \end{tabular}
 \end{small}
 \label{z2.237771pks0237}
\end{center}
 \end{table*}

We also modeled the two strongest components separately. The temperatures in the two models, according the $b$-values, are significantly different, 
$b_{1}$ = 4.23$\pm$2.82 \kms\!\!, $T$ = 17 204 K, and $b_{2}$ = 10.2$\pm$2.69 \kms\!\!, $T$ = 100 032 K. We used the same stopping criterion as for the  sum of the components. 
The results are given in Table~\ref{z2.237771pks0237_A}.

\begin{table*}[th!]
\begin{center}
\caption[Another set of ionization models for an \Ovi absorber at $z = 2.237771$]{Another set of ionization models for an \Ovi absorber at $z = 2.237771$ towards PKS 0237-233}
 \begin{small}
 \begin{tabular}{rccrcrcccc}
 \hline
 \hline
\scriptsize  $v$\,[\kms\!\!] &\scriptsize log [N(O\,{\sc vi)}\,(\sqc)] &\scriptsize 
log [$n_{\rm H}\,$(\cc)] &\scriptsize log $Z$  &\scriptsize log [$T$(K)] &\scriptsize $L$ [kpc] 
&\scriptsize log $f_{\rm HI}$ \\
 \hline
\multicolumn{7}{c}{\it photoionization} \\ \hline
\scriptsize $-$2.7 &\scriptsize 13.114 &\scriptsize $-$3.947 &\scriptsize 0.000 &\scriptsize 4.236 &\scriptsize 0.32 &\scriptsize $-$ 4.728\\
\scriptsize $-$2.7 &\scriptsize 13.114 &\scriptsize $-$4.357 &\scriptsize $-$0.400 &\scriptsize 4.236 &\scriptsize 2.1 &\scriptsize $-$5.139 \\
\scriptsize 0.0 &\scriptsize 13.348 &\scriptsize $-$3.950 &\scriptsize 0.000 &\scriptsize 5.000 &\scriptsize 1.6 &\scriptsize $-$5.439 \\
\scriptsize 0.0 &\scriptsize 13.349 &\scriptsize $-$4.400 &\scriptsize $-$0.150 &\scriptsize 5.000 &\scriptsize 11.3 &\scriptsize $-$5.826 \\
 \hline
\multicolumn{7}{c}{\it collisional ionization} \\ \hline
\scriptsize 0.0 &\scriptsize 13.347 &\scriptsize $-$5.810 &\scriptsize $-$0.002 &\scriptsize 5.000 &\scriptsize 25.2 &\scriptsize $-$4.765 \\
\hline
 \end{tabular}
 \end{small}
 \label{z2.237771pks0237_A}
\end{center}
 \end{table*}

\subsection*{System at $z$ = 2.401175 towards Q 0109-3518}

This is a single very weak \Ovi system which does not contain other metals. The corresponding \Hi feature is weak and has three components. 
The strongest hydrogen component is shifted by 3.1 \kms\!\!. We used it as a stopping criterion in our model. We fixed the temperature, according the oxygen 
$b-$value, $T$ = 68 166 K. The results are given in Table~\ref{z2.401175q0109}.

\begin{table*}[th!]
\begin{center}
\caption[Ionization models for an \Ovi absorber at $z = 2.401175$]{Ionization models for an \Ovi absorber at $z = 2.401175$ towards Q 0109-3518}
 \begin{small}
 \begin{tabular}{cccrcrcccc}
 \hline
 \hline
\scriptsize  $v$\,[\kms\!\!] &\scriptsize log [N(O\,{\sc vi)}\,(\sqc)] &\scriptsize 
log [$n_{\rm H}\,$(\cc)] &\scriptsize log $Z$  &\scriptsize log [$T$(K)] &\scriptsize $L$ [kpc] 
&\scriptsize log $f_{\rm HI}$ \\
 \hline
\scriptsize 0.0 &\scriptsize 12.892 &\scriptsize $-$3.062 &\scriptsize 0.000 &\scriptsize 4.834 &\scriptsize 0.8 &\scriptsize $-$4.568 \\
\scriptsize 0.0 &\scriptsize 12.891 &\scriptsize $-$4.000 &\scriptsize $-$1.847 &\scriptsize 4.834 &\scriptsize 36.9 &\scriptsize $-$6.055 \\
 \hline
 \end{tabular}
 \end{small}
 \label{z2.401175q0109}
\end{center}
 \end{table*}







\newpage
\chapter{Some examples of the observed and simulated absorption spectra}\label{examples_all}
\section{UVES absorption spectra - examples}\label{examples_uves}

\begin{figure}[h!]
\begin{center}
\resizebox{0.6 \hsize}{!}{\includegraphics{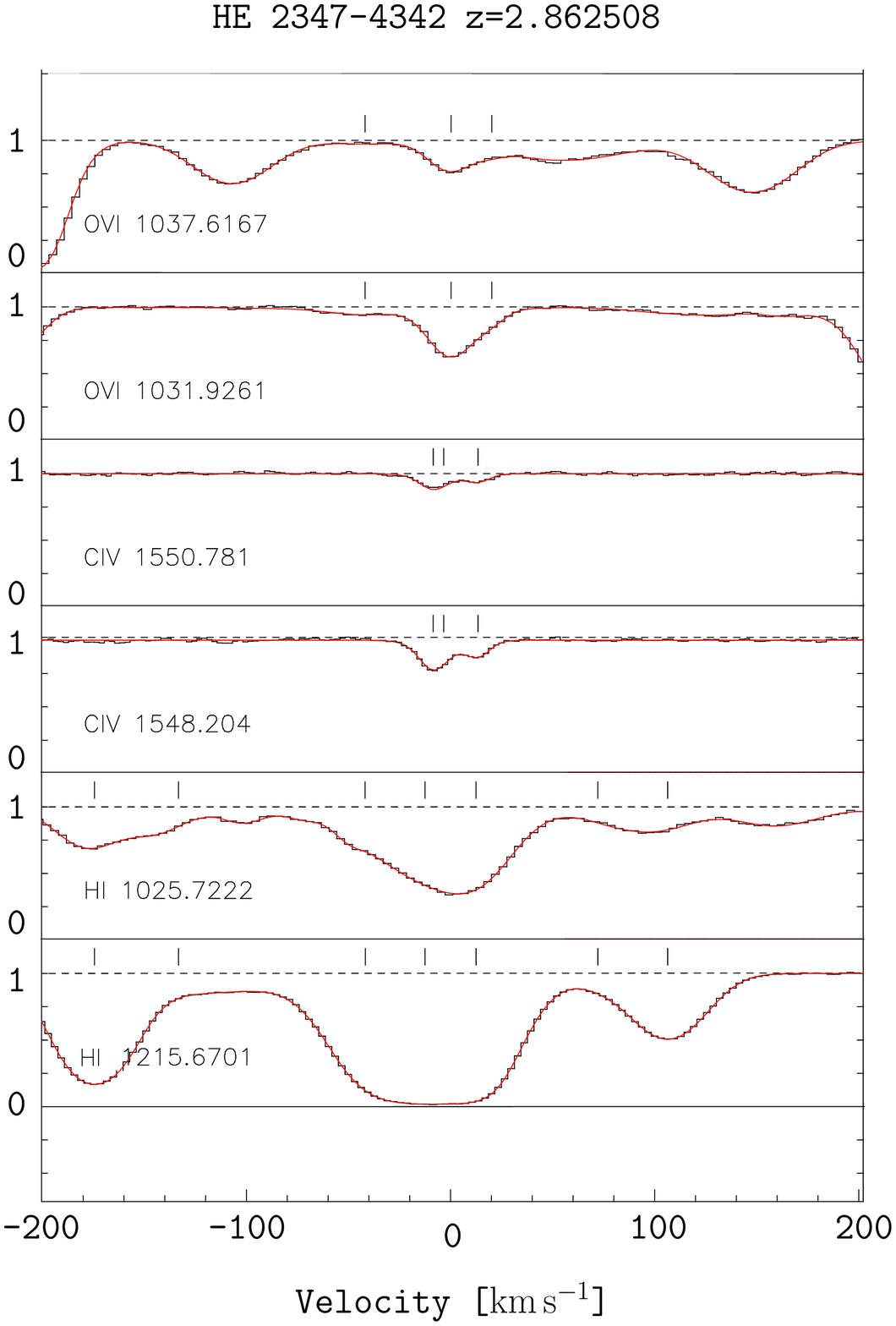}}
\caption[HE 2347-4342.]{HE 2347-4342 $z=2.862508$. This system is classified as category $2$.}
\label{fig_he2347}
\end{center}
\end{figure}

\begin{figure}[h!]
\begin{center}
\resizebox{0.6 \hsize}{!}{\includegraphics{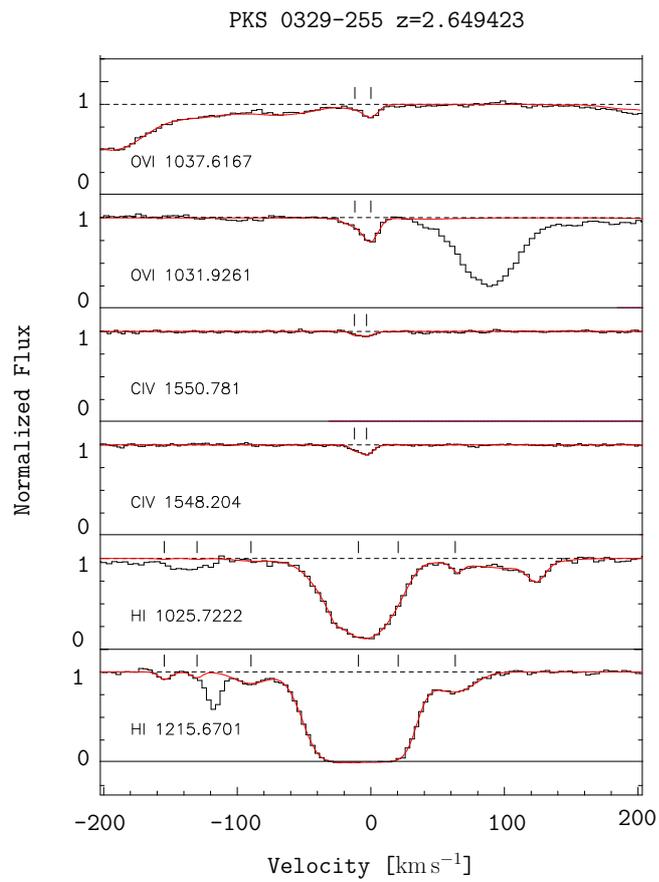}}
\caption[PKS 0329-255.]{PKS 0329-255 $z=2.649423$. This system is classified as category $2$.}
\label{fig_pks0329}
\end{center}
\end{figure}

\begin{figure}[h!]
\begin{center}
\resizebox{0.7 \hsize}{!}{\includegraphics{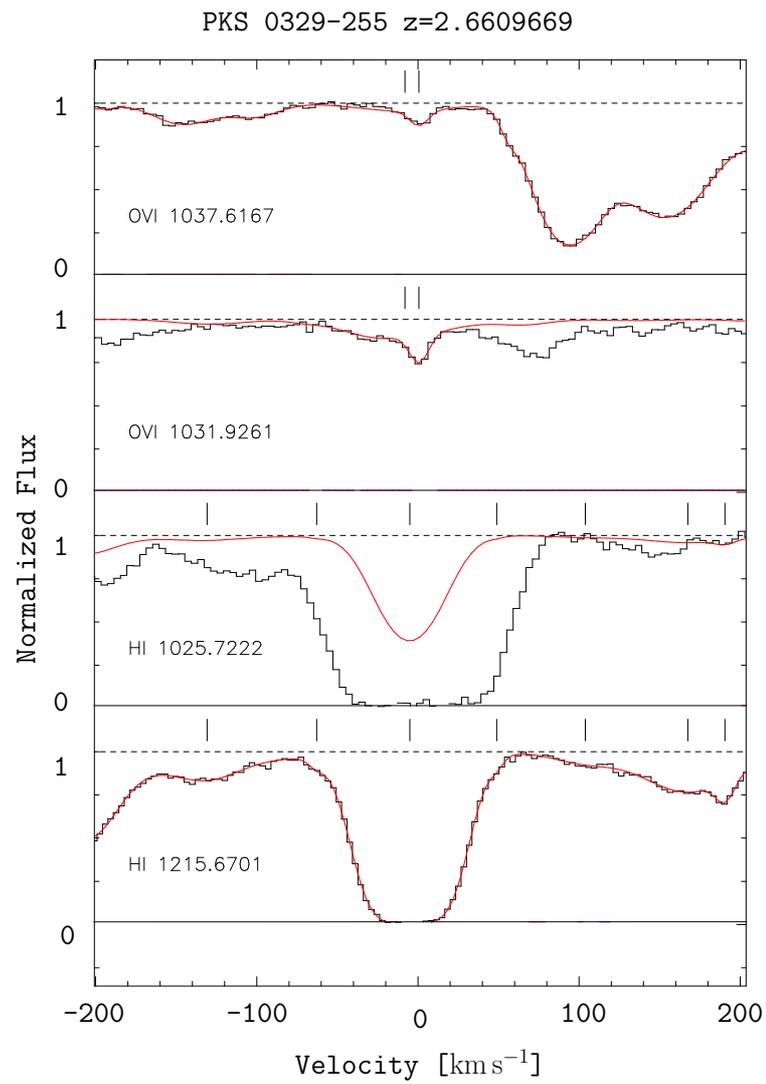}}
\caption[PKS 0329-255.]{PKS 0329-255 $z=2.6609669$. This system is classified as category $2$.}
\label{fig_pks0329a}
\end{center}
\end{figure}

\begin{figure}[h!]
\begin{center}
\resizebox{0.7 \hsize}{!}{\includegraphics{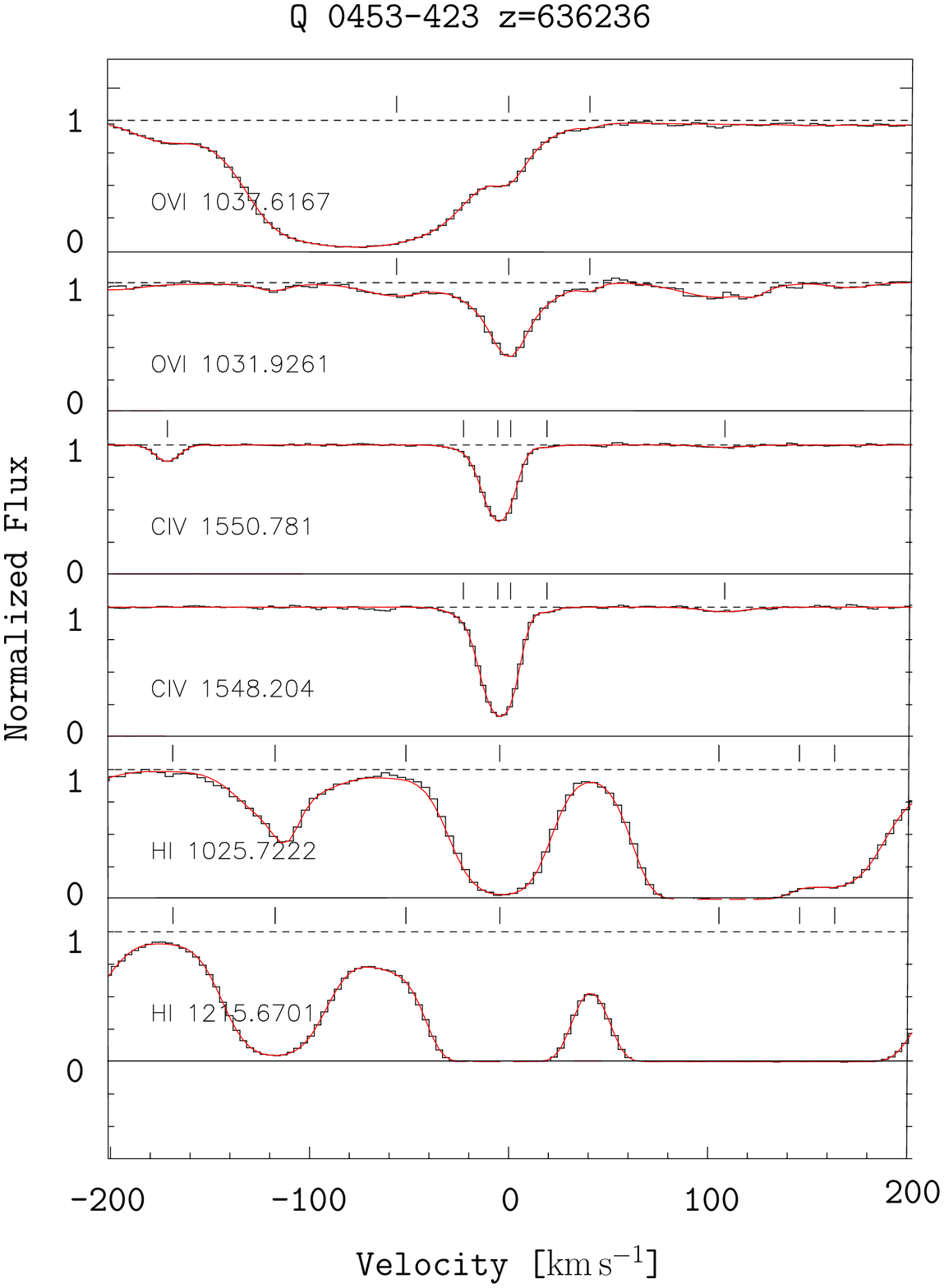}}
\caption[Q 0453-423.]{Q 0453-423 $z=2.636236$. This system is classified as category $1$.}
\label{fig_q0453}
\end{center}
\end{figure}

\begin{figure}[h!]
\begin{center}
\resizebox{0.7 \hsize}{!}{\includegraphics{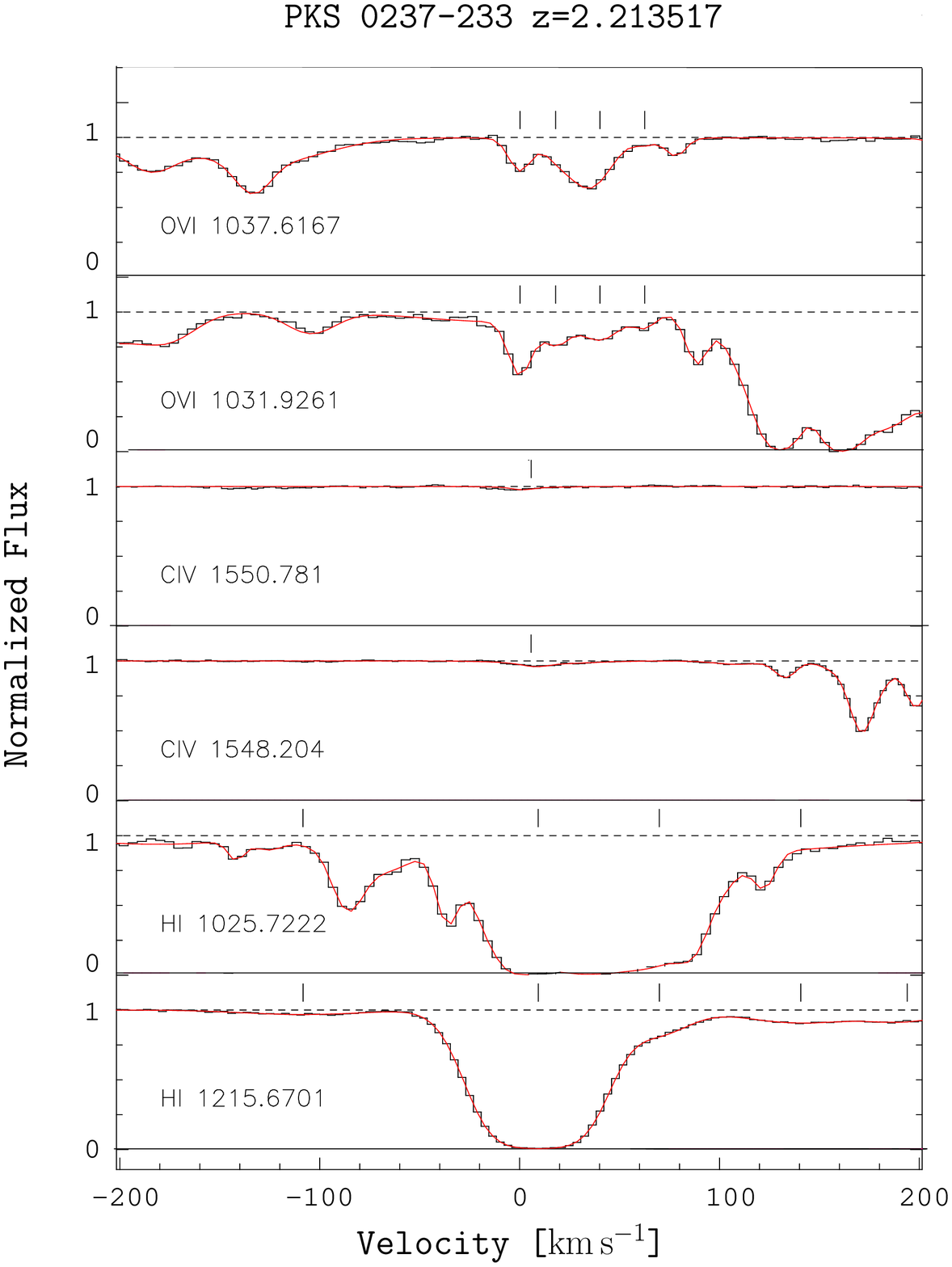}}
\caption[PKS 0237-233.]{PKS 0237-233 $z=2.213517$. This system is classified as category $1$.}
\label{fig_pks0237}
\end{center}
\end{figure}

\begin{figure}[h!]
\begin{center}
\resizebox{0.6 \hsize}{!}{\includegraphics{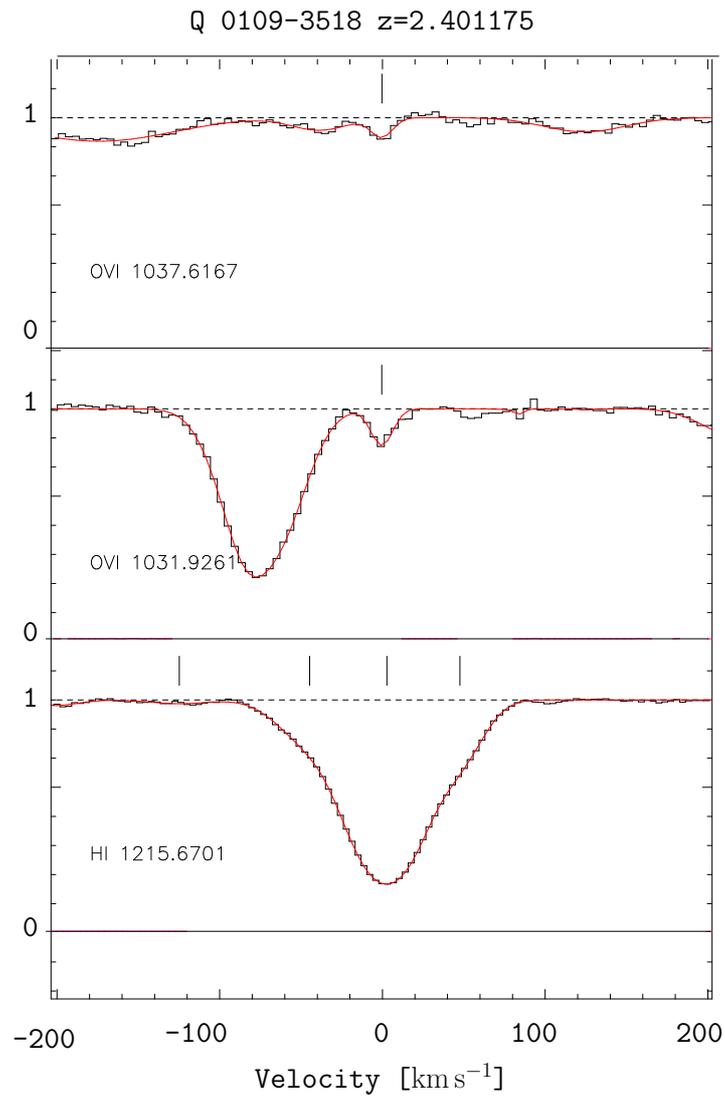}}
\caption[Q 0109-3518.]{Q 0109-3518 $z=2.401175$. This system is classified as category $1$.}
\label{fig_q0109}
\end{center}
\end{figure}

\begin{figure}[h!]
\begin{center}
\resizebox{0.7 \hsize}{!}{\includegraphics{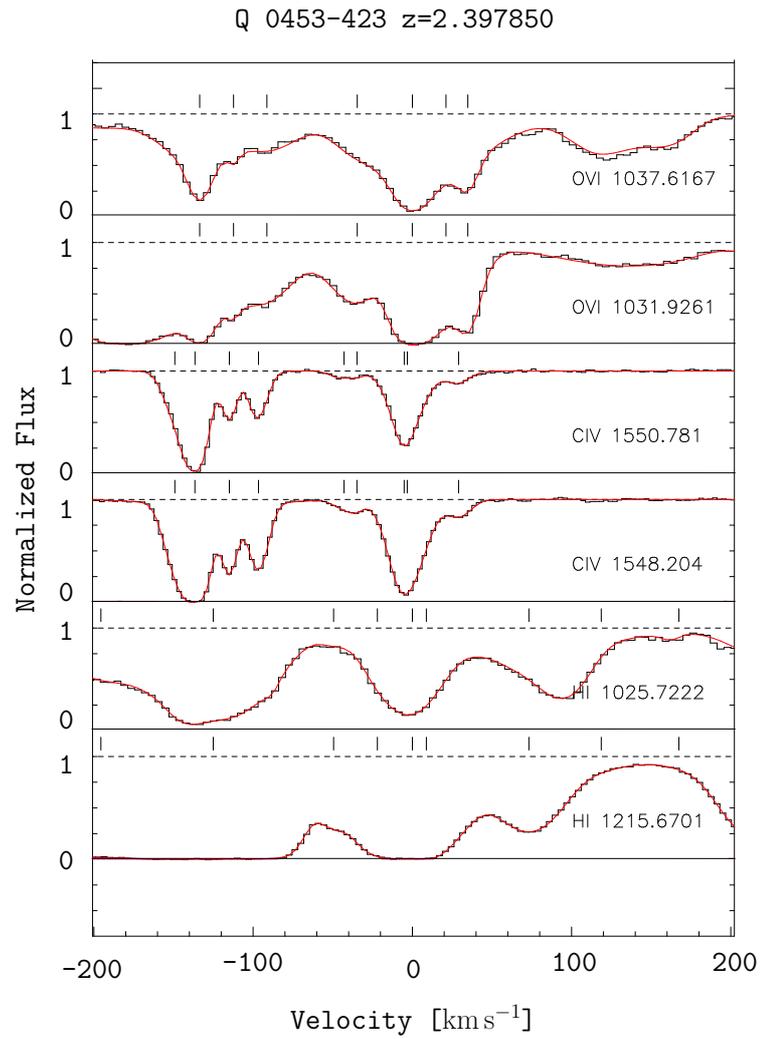}}
\caption[Q 0453-423.]{Q 0453-423 $z=2.397850$. This system is classified as category $0$.}
\label{fig_q0453a}
\end{center}
\end{figure}

\begin{figure}[h!]
\begin{center}
\resizebox{0.7 \hsize}{!}{\includegraphics{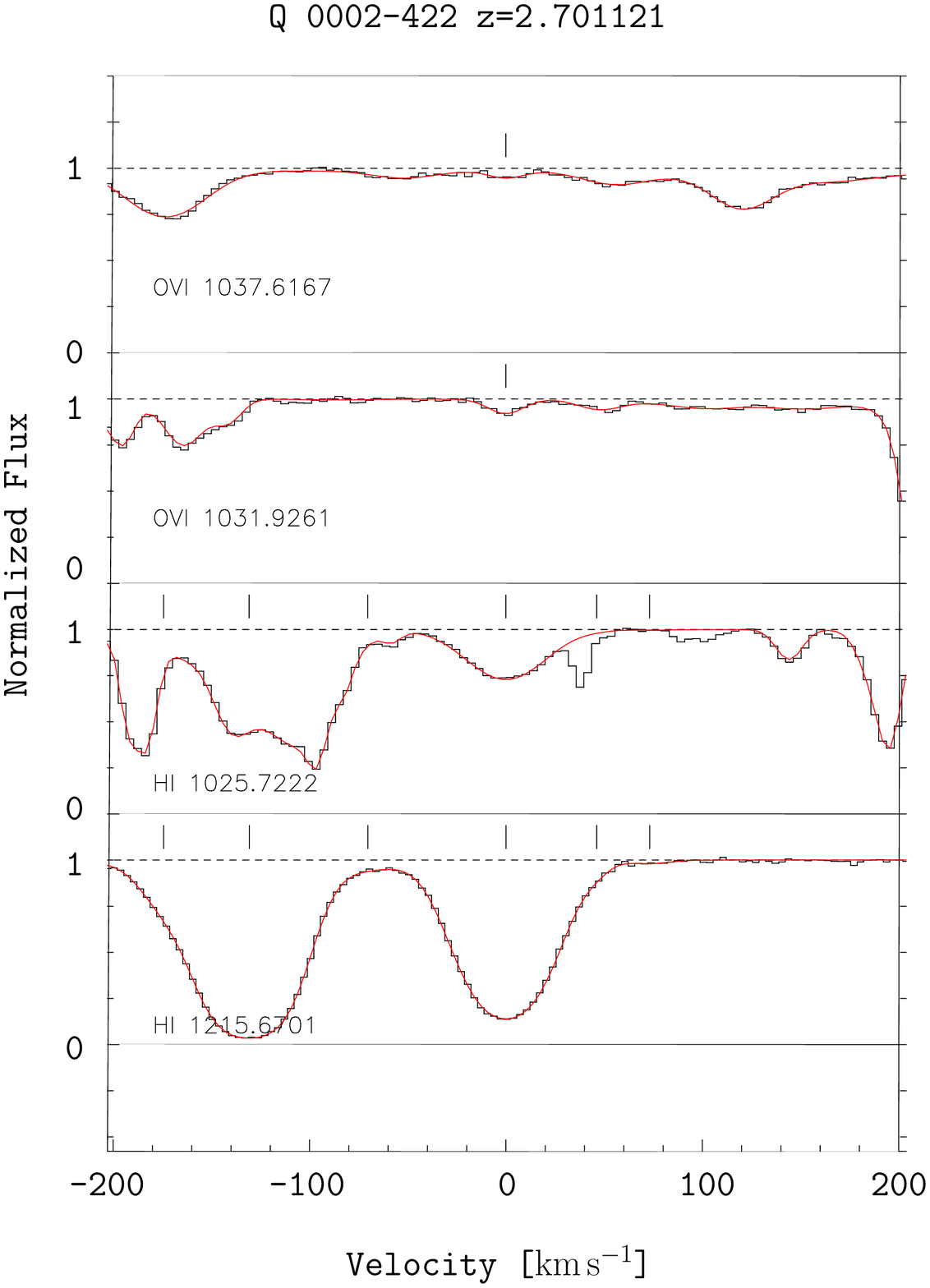}}
\caption[Q 0002-422.]{Q 0002-422 $z=2.701121$. This system is classified as category $0$.}
\label{fig_q0002}
\end{center}
\end{figure}

\begin{figure}[h!]
\begin{center}
\resizebox{0.7 \hsize}{!}{\includegraphics{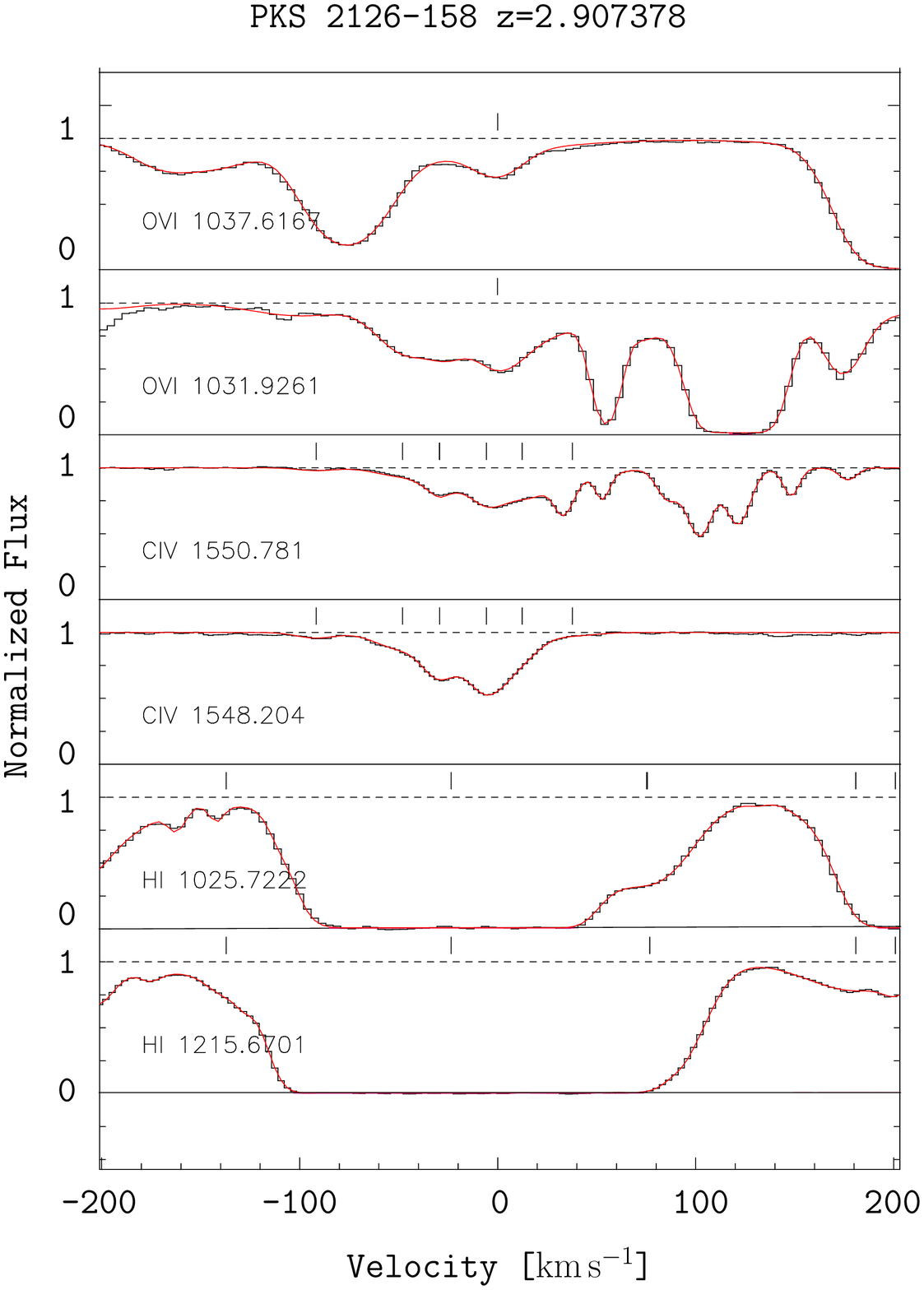}}
\caption[PKS 2126-158.]{PKS 2126-158 $z=2.907378$. This system is classified as category $0$.}
\label{fig_q0002}
\end{center}
\end{figure}

\section{Simulated (OWLS) absorption spectra - examples}\label{examples_owls}

\begin{figure}[h!]
\begin{center}
\resizebox{0.7 \hsize}{!}{\includegraphics{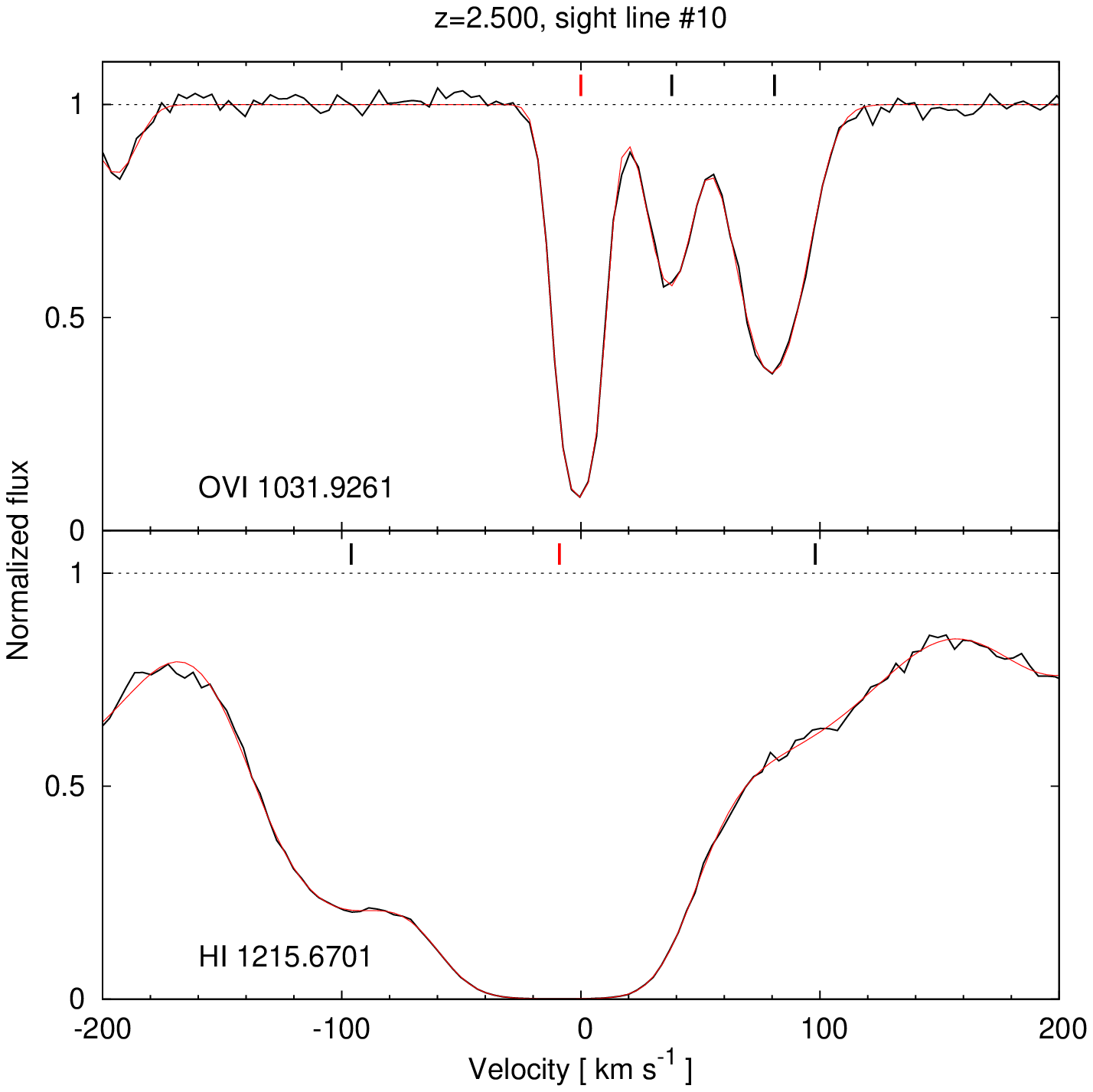}}
\caption[Example No.1 of synthetic spectra.]{Synthetic spectra of sightline No.10 at $z=2.50$. The (red and black) tick marks show the main components in the spectra. With red are shown the components with velocity alignment between \Ovi and \Hi ($\le 10$ km~$\rm s^{-1}$).}
\label{fig_N1}
\end{center}
\end{figure}

\begin{figure}[h!]
\begin{center}
\resizebox{0.7 \hsize}{!}{\includegraphics{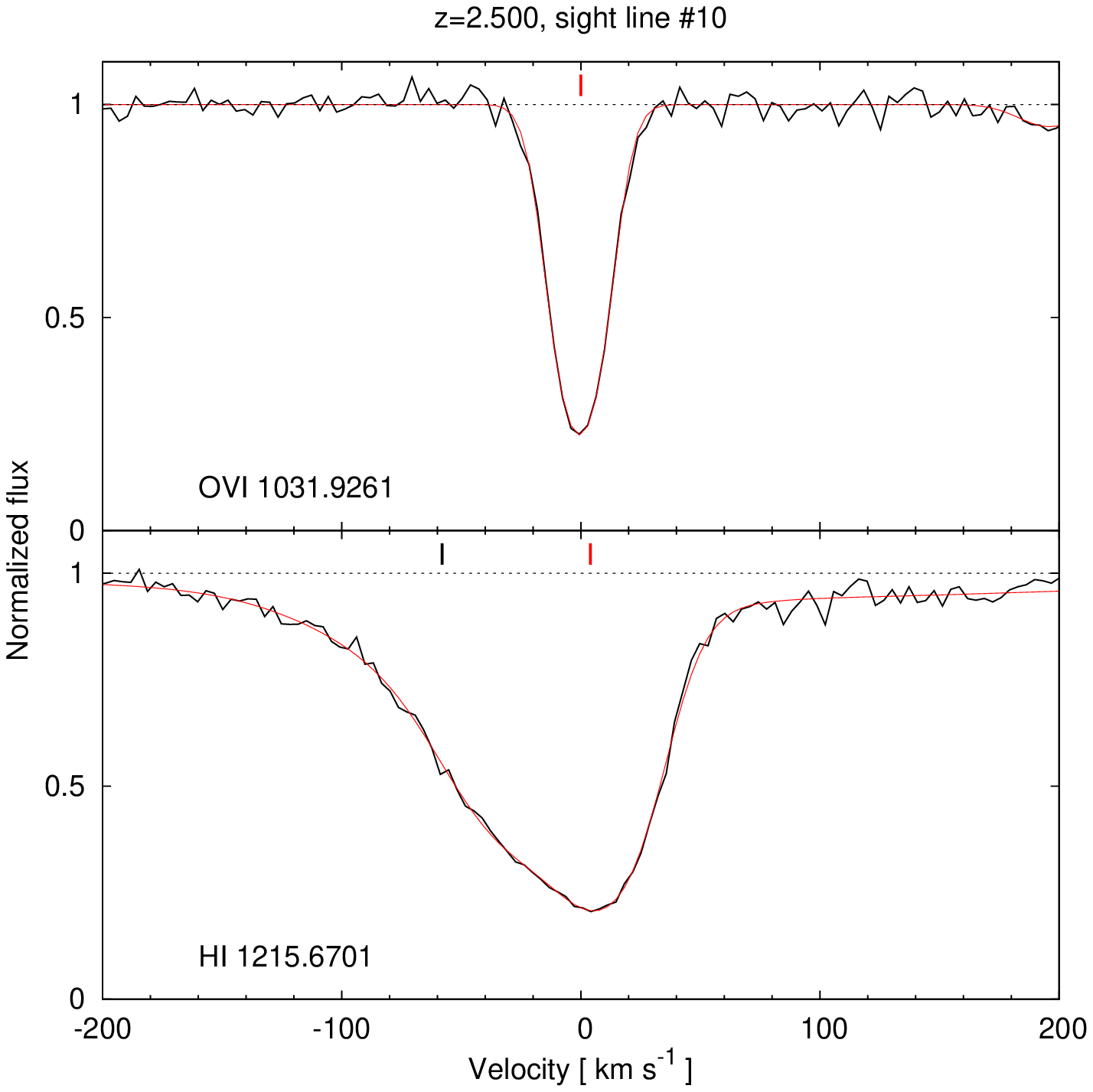}}
\caption[Example No.2 of synthetic spectra.]{Synthetic spectra of sightline No.10 at $z=2.50$. The tick marks are the same as in Fig.~\ref{fig_N1}.}
\label{fig_N2}
\end{center}
\end{figure}

\begin{figure}[h!]
\begin{center}
\resizebox{0.7 \hsize}{!}{\includegraphics{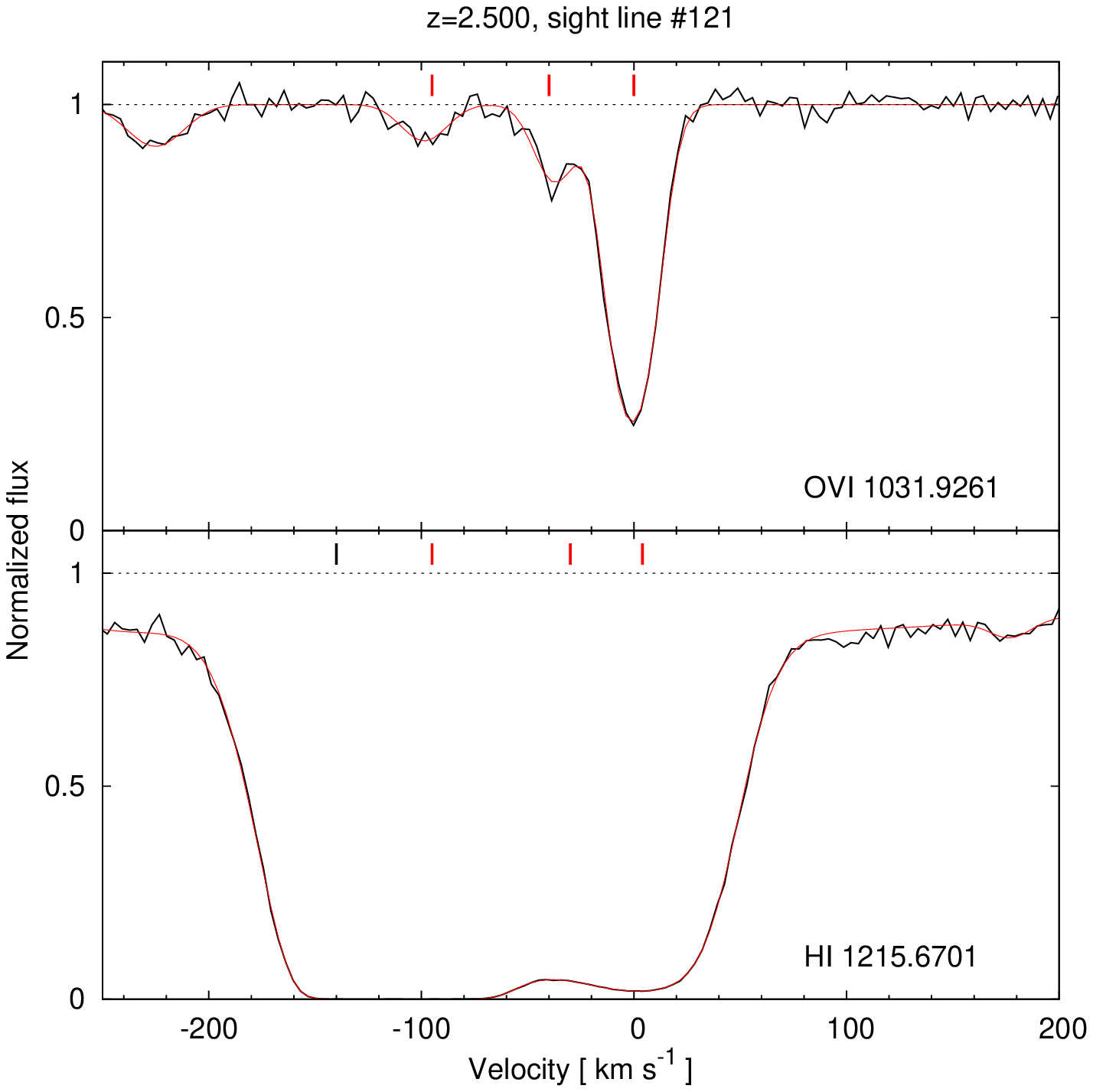}}
\caption[Example No.3 of synthetic spectra.]{Synthetic spectra of sightline No.121 at $z=2.50$. The tick marks are the same as in Fig.~\ref{fig_N1}.}
\label{fig_N3}
\end{center}
\end{figure}

\begin{figure}[h!]
\begin{center}
\resizebox{0.7 \hsize}{!}{\includegraphics{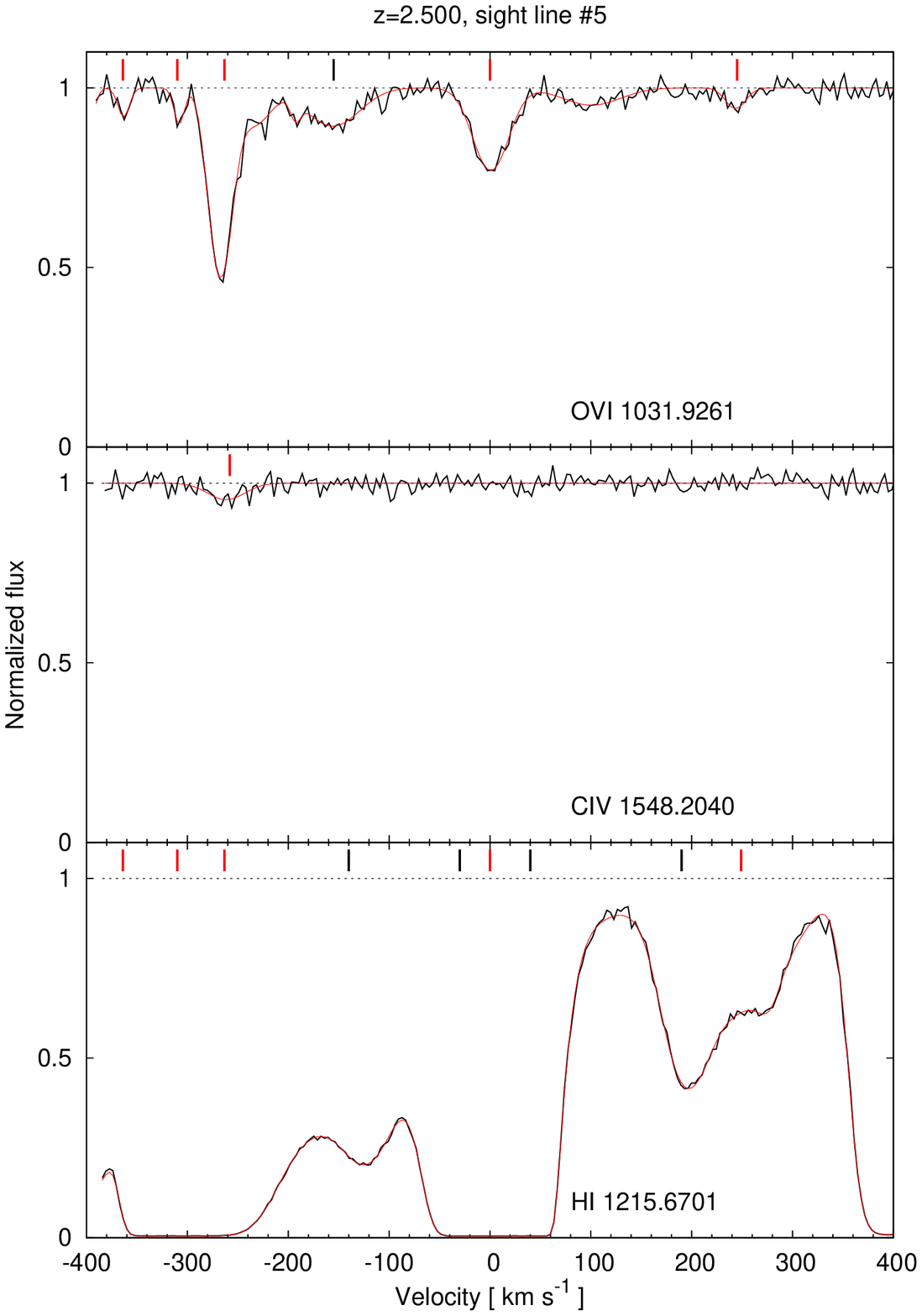}}
\caption[Example No.4 of synthetic spectra.]{Synthetic spectra of sightline No.5 at $z=2.50$. The black tick marks are the same as in Fig.~\ref{fig_N1}. The red show the components with velocity alignment between \Ovi\!\!, (in some cases) \Civ and \Hi ($\le 10$ km~$\rm s^{-1}$).}
\label{fig_N4}
\end{center}
\end{figure}

\begin{figure}[h!]
\begin{center}
\resizebox{0.7 \hsize}{!}{\includegraphics{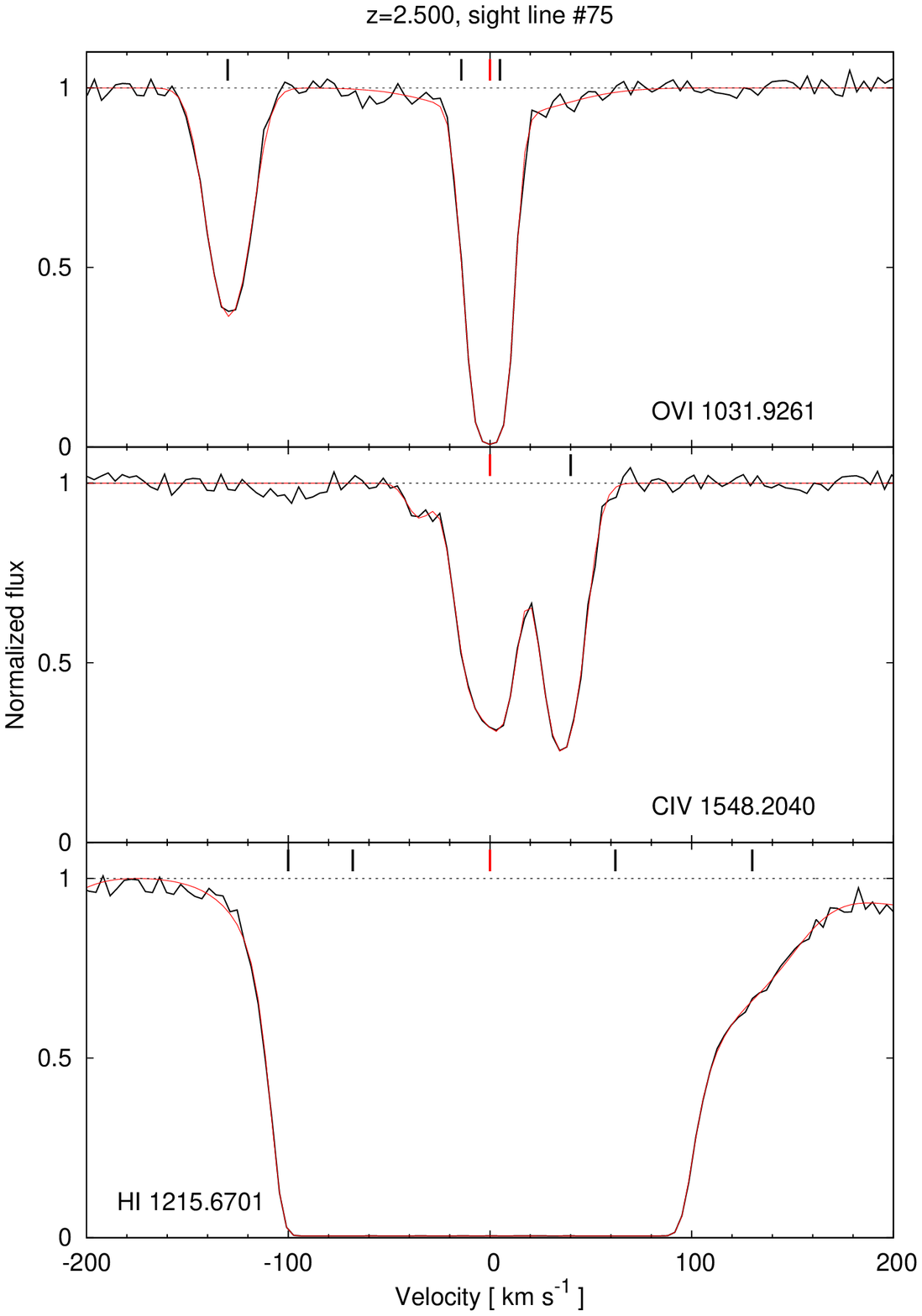}}
\caption[Example No.5 of synthetic spectra.]{Synthetic spectra of sightline No.75 at $z=2.50$. The tick marks are the same as in Fig.~\ref{fig_N4}.}
\label{fig_N5}
\end{center}
\end{figure}

\newpage

\end{appendix}

\bibliographystyle{aa} 
\bibliography{papers}
     

\newpage
\textbf{\Large{Acknowledgments}}\vspace{2cm}\\

First, I would like to thank my supervisor Prof. Philipp Richter for giving me the freedom and the support to accomplish this Thesis. I was glad to be a member of your group.

I am grateful to Thorsten Tepper-Garc{\'{\i}}a for his guidance and care for my work. I appreciate it very much.

I would like to thank as well Cora Fechner, Martin Wendt and Andrea Brockhaus who were always there when I needed scientific or organizational advice.

Special thanks to Dominik Hildebrandt for the discussions about physics and life and for giving me great personal support.

I would like to thank also Prof. Matthias Steinmetz for his response to an e-mail from me when I was looking for an opportunity to continue my education in astrophysics in Germany. 

Special thanks to Todor Veltchev for his full personal support throughout the years and for editing the language in this Thesis.

I am very thankful to my son for his understanding in moments when I had to work and had to leave him alone.

\end{document}